%% file: main.tex
\documentclass[12pt,twoside]{book}
\usepackage[a4paper,width=150mm,top=30mm,bottom=28mm,bindingoffset=6mm]{geometry}
\usepackage{amsmath,amssymb,amsthm}
\usepackage{mathrsfs}
\usepackage{bm}
\usepackage{floatrow}
\usepackage{amsmath}
\usepackage{amssymb}
\usepackage{amsthm}
\usepackage{setspace}
\usepackage{graphicx}
\usepackage{braket}
\usepackage{mathrsfs}
\usepackage{lscape,graphicx}
\usepackage{amsfonts}
\usepackage{longtable}
\usepackage{epsfig}
\usepackage{float}
\usepackage{rotating}
\usepackage[T1]{fontenc}
\usepackage{fancyhdr}
\usepackage[english]{babel}
\usepackage{fancyhdr}
\usepackage{parskip}
\usepackage{cite}
\usepackage{setspace}
\usepackage{caption}
\usepackage[hidelinks]{hyperref}
\usepackage{braket}
\usepackage{romannum}
\usepackage{bigints}
\usepackage[toc]{appendix}
\usepackage{longtable}
\usepackage[noabbrev]{cleveref}
\usepackage[acronym]{glossaries}
\usepackage{enumitem}
\usepackage{booktabs}
\usepackage{lastpage}

\usepackage{newtxtext}

\usepackage[raggedright]{titlesec}

\titlespacing\section{0pt}{12pt plus 4pt minus 2pt}{0pt plus 2pt minus 2pt}
\titlespacing\subsection{0pt}{12pt plus 4pt minus 2pt}{0pt plus 2pt minus 2pt}
\titlespacing\subsubsection{0pt}{12pt plus 4pt minus 2pt}{0pt plus 2pt minus 2pt}

\usepackage[font=small]{caption}
\usepackage{graphics,epsfig,amsmath,float,wrapfig}
\addto{\captionsenglish}{}

\newenvironment{changemargin}[2]{%
\begin{list}{}{%
\setlength{\leftmargin}{#1}%
\setlength{\rightmargin}{#2}%
}%

\item[]}{\end{list}}
\hypersetup{
    pdfauthor = {Addepalli Lavakumar}, 
    pdftitle = {}, 
    colorlinks,
    linkcolor={blue},
    citecolor={blue},
    linkcolor={blue},
    citecolor={blue}
}
\PassOptionsToPackage{unicode}{hyperref}
\begin{document}
\include{title}

\pagenumbering{roman}
\cfoot{ \thepage \hspace{1pt}} 
\baselineskip=18pt
\include{declaration_scholar}
\include{declaration_advisor}

\include{Dedication}

\include{acknowledgement}

\include{abstract}

\clearpage

\baselineskip=18pt
\doublespacing
\tableofcontents
\clearpage
\doublespacing
\listoffigures
\clearpage
\clearpage
\pagenumbering{arabic}


\pagestyle{fancy}
\fancyhf{}
\fancyhead[LE]{\em\leftmark}
\fancyhead[RO]{\em\rightmark}

\renewcommand{\headrulewidth}{0.1pt}
\cfoot{ \thepage \hspace{1pt}} 
\baselineskip=18pt
\doublespacing
\include{chap1}
\include{chap2}
\include{chap3}
\include{chap4}
\include{chap5}
\include{Conclusion}
\include{Publications}
\appendix
\addcontentsline{toc}{chapter}{Appendices} 
\chapter*{Appendices}
\include{app_chap2}
\include{app_chap3}
\include{app_chap4}

\include{app_chap5}
\bibliography{ref}
\end{document}

%% file: title.tex
\begin{changemargin}{0cm}{0cm}
\thispagestyle{empty}
\baselineskip25pt
\begin{center}
{\Large {\bf Multi-Photon Lasing Phenomena in Quantum Dot–Cavity QED}}\\
\end{center}

\vfill
\baselineskip15pt
\begin{center}
{ \large \em \textbf{A Thesis}}
\vfill
\begin{center}
    \textit{\large Submitted for the award of the degree}
\end{center}
\large\textit{of}

\vskip .90
\baselineskip
{\large{\bf\em Doctor of Philosophy}}
\end{center}
\baselineskip15pt
\vfill
\begin{center} {\bf {\em by}} \\
{\large{\bf Addepalli Lavakumar}\\ (Reg. No. D20034)} \\
\end{center}

\vfill
\begin{center}
\begin{figure}[h!]
\centering
\includegraphics[scale=0.6]{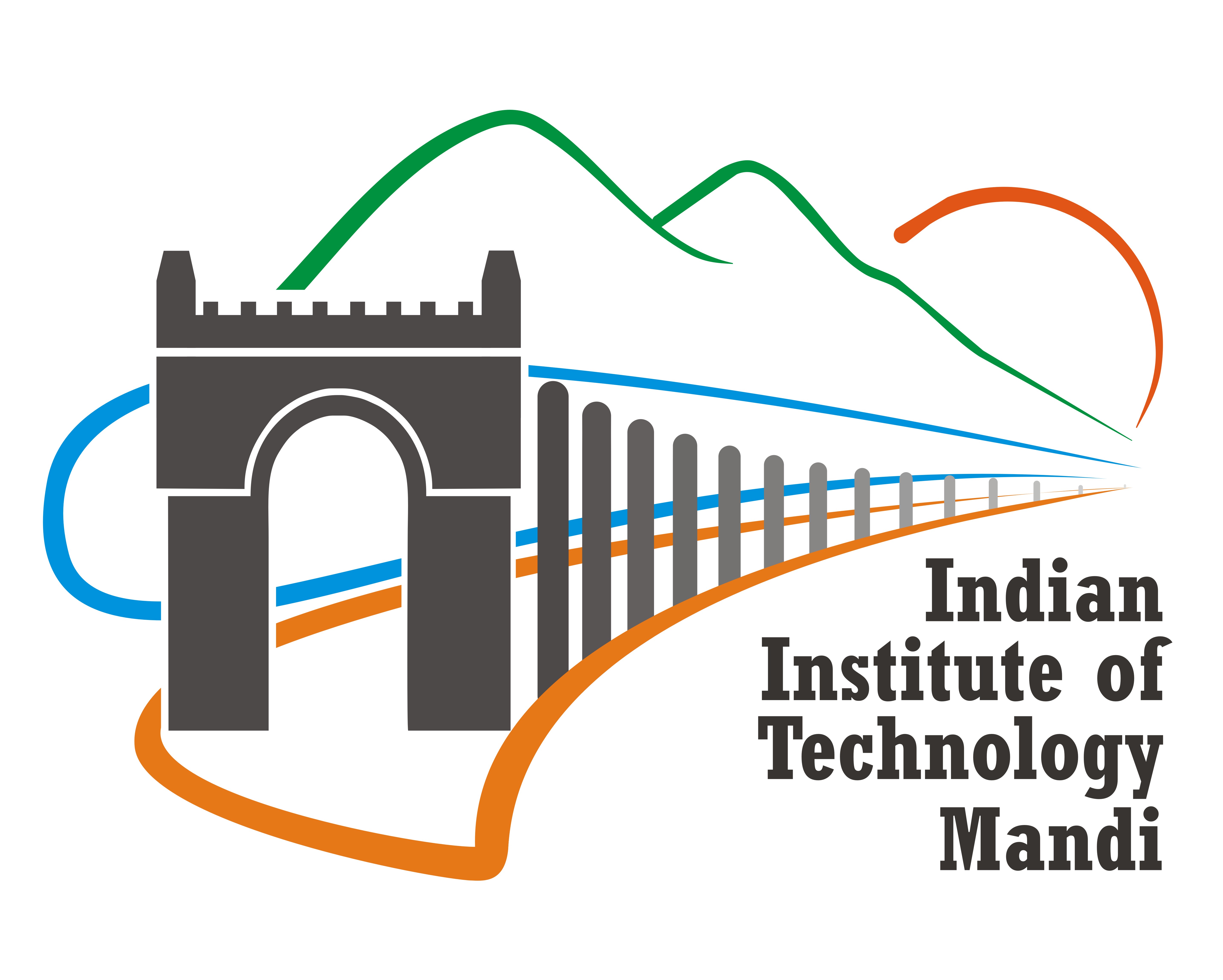}
\end{figure}
 {\bf {\large\em to the }} \\
 \vspace{0.25cm}
{\bf {\large School of Physical Sciences}} \\
{\bf {\large Indian Institute of Technology Mandi}} \\
{\bf \large Kamand, Himachal Pradesh (175075), India} \\
{\bf 11 December, 2025} 
\end{center}
\end{changemargin}

%% file: declaration_scholar.tex
\vspace*{-2.5cm}
\begin{minipage}{0.2\textwidth}		
\includegraphics[width=1\textwidth]{IIT_Mandi_logo.jpg}
\end{minipage}
\begin{minipage}{0.75\textwidth}
\centering
		{\bf{INDIAN INSTITUTE OF TECHNOLOGY, MANDI}}	\\ [0.1em]
		{\bf{Kamand, Himachal Pradesh - 175075, INDIA}}						\\ [0.1em]
		www.iitmandi.ac.in								
	\end{minipage}\\\
	\rule{\textwidth}{1pt}\\\
	
\vspace*{1cm}
\centerline{\large \bf \underline{\smash{Declaration by the Research Scholar}}}
\vspace*{1cm}
\noindent 
I hereby declare that the entire work embodied in this Thesis is the result of investigations carried out by \textbf{Addepalli Lavakumar} in the \textbf{School of Physical Sciences}, Indian Institute of Technology Mandi, under the supervision of  \textbf{Dr. Pradyumna Kumar Pathak},  and  that  it  has  not  been  submitted  elsewhere  for  any  degree  or diploma. In keeping with the general practice, due acknowledgements have been made wherever the work described is based on finding of other investigators.\\[1cm]

Place: IIT Mandi  \hspace{6cm} Signature: \\\\
\vspace{1cm}
\hspace{-0.15cm}Date: \hspace{8cm} Name: \textbf{Addepalli Lavakumar} 

%% file: declaration_advisor.tex
\vspace*{-2.5cm}
\begin{minipage}{0.2\textwidth}
		
		\includegraphics[width=0.9\textwidth]{IIT_Mandi_logo.jpg}
	\end{minipage}
	\begin{minipage}{0.75\textwidth}
		\centering
		{\bf{INDIAN INSTITUTE OF TECHNOLOGY, MANDI}}			\\ [0.1em]
		{\bf{Kamand, Himachal Pradesh - 175075, INDIA}}						\\ [0.1em]
		www.iitmandi.ac.in								
	\end{minipage}\\\
	\rule{\textwidth}{1pt}\\\
	
\vspace*{3cm}
\centerline{\large \bf \underline{\smash{Declaration by the Thesis Advisor}}}
\vspace*{1cm}
\noindent 
I hereby certify that the entire work in this thesis has been carried out by \textbf{Addepalli Lavakumar}, under my supervision in the \textbf{School of Physical Sciences, Indian Institute of Technology Mandi,}  and that no part of it has been submitted elsewhere for any Degree or Diploma.\\[3em]

Signature:\\[1em]
Name of the Guide: \textbf{Dr. Pradyumna Kumar Pathak}\\[1em]
\hspace*{0.05cm}Date:

%% file: Dedication.tex

\begin{center}
\vspace*{\fill}
\large \textit{Dedicated to my parents \\for their love}
\vspace*{\fill}
\end{center}

%% file: acknowledgement.tex
\begin{center}
{\bf ACKNOWLEDGEMENTS} 
\end{center}

I am deeply grateful to my supervisor, Dr. Pradyumna Kumar Pathak, for his invaluable ideas and insightful discussions throughout my Ph.D. journey. I have always aspired to work in an environment that encourages freedom of thought rather than compulsion, and I sincerely thank him for providing such a healthy atmosphere and for constantly motivating me to pursue research independently. 

I would also like to thank my doctoral committee members, Dr. Suman Kalyan Pal, Prof. Hari Varma, Dr. Srikant Sugavanam, and Dr. Gargee Sharma, for their valuable inputs and constructive suggestions during the DC meetings, which greatly helped me refine the presentation of my research. I would like to extend my sincere thanks to my collaborators, Dr. Samit Hazra and Dr. Tarak Nath Dey, for their valuable insights, engaging discussions and patience to work with me.

I would like to thank the administration of IIT Mandi for providing the research scholarship, contingency support, and travel grants that have greatly helped my research journey and enabled me to showcase my work. I wish to thank Anusandhan National Research Foundation (ANRF) for the financial support, to present my work at the FQMT’24 conference in Prague. I would further like to acknowledge the housekeeping staff and security personnel of IIT Mandi for their dedicated efforts in maintaining a safe and supportive campus environment.

I am grateful to my young lab-mates, Arvind, Chayan, Akshay, Dinesh, and Mech, for the engaging and fruitful discussions. I would like to thank my friends, Shubhash bhai, Ganesh, Sathosh, Sandeep, Devansu, Bhuvan bhai, Shivam, Shamshad, Shahin, Bheem, Shraddha, Ipsitha, Prashant, Ashutosh, and Devesh, for the memories at IIT Mandi.

I am grateful to my high school teachers, Elumalai Sir and Chengalraju Sir, whose inspiring physics classes sparked my interest and made me fall in love with physics. 

I am forever indebted to my father for his unwavering support, encouragement, and motivation to pursue whatever I aspired to in life, and to my mother, who has always made me feel that even my smallest achievement is her proudest moment. My brother, Shyam, has been a constant source of inspiration, humility, and motivation. His influence has guided me both personally and professionally. I would also like to thank my sister, Ramya, Harsha Vadina, Pradeep Bavagaru, and other family members for their constant support throughout my life. I am grateful to have my bestiee, Swati, whose positive outlook towards every situation has always inspired me. Her optimism and kindness have influenced me to be more compassionate, grateful, and content, and it all made the difference.

Finally, I would like to thank all those not mentioned here by name, whose goodwill, support, and encouragement have directly or indirectly contributed to my journey.

\vspace*{\fill} 

%% file: abstract.tex
\begin{center}
{\large {\bf  ABSTRACT }}
\end{center}

Multi-photon lasing has been realized in systems with strong nonlinear interactions between emitters and cavity modes, where single-photon processes are suppressed. Coherence between the internal states of a quantum emitter, or among multiple emitters, plays a key role. Such continuous nonclassical sources of light can find applications in quantum computation, quantum sensing, quantum metrology, and quantum communication. 

This thesis explores the multi-photon lasing phenomena in various quantum dot-photonic crystal cavity quantum electrodynamic (QED) setups. Exciton-phonon interactions are inevitable in such systems and are incorporated using the polaron polaron-transformed master equation. The Born-Markov approximation is employed to obtain the reduced density matrix rate equation. Using quantum laser theory, we derived the Scully-Lamb laser rate equations and evaluated the single- and multi-photon excess emission rates defined as the difference between emission and absorption rates into the cavity mode without mean-field approximations. First, we investigate the cooperative two-photon lasing with two quantum dots strongly coupled to the single-mode photonic crystal cavity. Cooperativity between N quantum emitters coupled to common radiation modes can lead to enhanced emission in comparison to independent emitter behavior with scaling as $\propto N^2$ in Dicke superradiance or $>N^2$ for Hyperradiance in the strongly coupled emitter-cavity QED system. We show that such a system shows superradiant or hyperradiant behavior for the incoherently pumped quantum dots and subradiant behavior for the coherent pumping scenario, quantified by the Radiance witness parameter. Extending this analysis, we consider two identical quantum dots incoherently pumped and coupled off-resonantly to the bimodal cavity. The phonon-induced effects play a dominant role in populating the cavity modes, leading to the two-mode Hyperradiant lasing phenomenon. In superradiant lasers, the emission intensity scales $\propto N^2$ and the linewidth narrows as $1/N^2$. We demonstrate a similar linewidth suppression in the emission spectrum using the quantum regression theorem.

Further, we study a system with a single quantum dot coupled to a bimodal cavity, where the coherence between quantum dot excitonic states $|x\rangle$ and $|y\rangle$ is generated using an external coherent drive. The transfer of this coherence from the emitter states to the coupled cavity modes can lead to quenching or squeezing the relative or average phase fluctuations of the cavity modes, enabling the system to behave as a ``Correlated Emission Laser" (CEL). We analyze this behavior by evaluating the variances of Hermitian operators corresponding to average and relative phase, along with the associated drift and diffusion coefficients using the Fokker-Planck equation in Glauber-Sudarshan `$P$' representation. Our results show that the fluctuations in the system approach the Vacuum Noise Limit (VNL). In continuation, we investigate the system with a QD-biexciton ($|u\rangle$) driven using  $x$-polarized incoherent or coherent pump ($|g\rangle\leftrightarrow|x\rangle\leftrightarrow|u\rangle$) and $|u\rangle\leftrightarrow|y\rangle\leftrightarrow|g\rangle$ transitions are coupled to a non-degenerate bimodal cavity. We showed that such a system can act as a non-degenerate two-photon laser at low temperatures and pumping rates. By evaluating excess emission rates from the laser rate equations, we demonstrate that the cavity modes are dominantly populated via two-mode two-photon processes. Additionally, we explored the generation of continuous variable entanglement between the cavity modes when tuned to the QD transitions to satisfy the two-photon resonance condition. Finally, our results suggest that the semiconductor QD-cavity QED systems are promising candidates for realizing multi-photon lasing phenomena such as cooperative two-photon lasing, hyperradiant lasing, correlated emission laser, and non-degenerate two-mode two-photon lasing.

%% file: chap1.tex
\chapter{Introduction}\label{intro}

Invention of pulsed LASER (Light Amplification by Stimulated Emission of Radiation) in 1960 by Maiman\cite{Maiman1960} and the first continuous wave laser by Javan \cite{Javan1961} paved the way for advancements in science and technology. Conventional lasers, owing to their coherence and directional properties, found a wide range of applications in optical communication (Internet, GPS), high resolution microscopy, medicine, metrology, etc \cite{Thyagarajan2010lasers}.
These systems typically work on the principle of population inversion in the macroscopic gain medium (billions or trillions of atoms/$cm^3$) coupled to a cavity resonator having dimensions of the order centimeters to meters, with a threshold power of the order of $\sim 10-100$ mW. However, these lasers are constrained by phase fluctuations and the fundamental Schawlow-Townes linewidth \cite{Schawlow1958} limit to find a place in this era of quantum technology.

In contrast, quantum lasers operate with a single or a few emitters coupled to a micro- or nano-cavity. The light-matter interactions in these systems are described by a full quantum-mechanical treatment as developed by Dirac \cite{Dirac1927} in 1927, where both emitters and electromagnetic field are quantized. Thereafter, in the early 1960s, H. Haken \cite{Haken1973}, WE. Lamb \cite{Lamb1964} and M. Lax \cite{Lax1969} developed the quantum theory of lasers incorporating cavity quantum electrodynamic (QED) effects \cite{Kimble1998CavityQED}. Early demonstrations of quantum lasers were primarily realized in atomic systems \cite{Haroche2006}, showing phenomena such as vacuum Rabi oscillations \cite{Jaynes1963,Agarwal1985vacuum}, inhibiting spontaneous emission\cite{Kleppner1981}, enhanced spontaneous emission \cite{Goy1983}, lasing without inversion \cite{Arimondo1996}, correlated emission lasers \cite{Bergou1988QBL,Bergou1988hanle,Zaheer1988QBL,Scully1988cascade,Lu1990JOSAB} suppressing the spontaneously emission noise, thresholdless lasing\cite{Protsenko1999thresholdless}, deterministic single \cite{Kuhn2010singlePhoton,Cui2006singlePhoton} and multi-photon emission \cite{Zhu1987,Zakrzewski1991}, photon blockade \cite{Hamsen2017photonBlockade}, entangled photon pair emission \cite{Raimond2001}, squeezed states of light \cite{Loudon1987squeezed,Raizen1987squeezed,Lvovsky2015squeezed}, and quasi-continuous superradiant lasers \cite{Yu2010,Meiser2009,Norcia2016} with mHz linewidth etc are realized. In the last two decades, a huge interest in solid-state lasers such as superconducting circuit QED and semiconductor cavity QED has increased significantly. In contrast to atomic systems having complex experimental setups limiting their scalability, solid-state quantum lasers, specifically semiconductor quantum dot (QD) lasers, show promise due to their scalability and integrability with quantum photonic circuits, leading to on-chip quantum technology. Realizing the above mentioned quantum phenomena in these semiconductor QD-photonic crystal (PhC) cavity QED systems\cite{Yablonovitch1987,Li2009, Ota2017thresholdless,Prieto2015thresholdless,Stace2003twoPhoton,del2013photonpairs,Leymann2015,scheibner2007,Lukin2023,Peng2015,Pascale2017, Dousse2010ultrabright} can find their applications in quantum communication \cite{Gisin2007quantumcomm}, quantum computing \cite{Imamoglu1999,Kiraz2004}, quantum metrology \cite{Giovannetti2006}, sensing \cite{Zhou2016}.

Moreover, in these semiconductor QD-PhC cavity systems, the QD exciton interaction with the surrounding phonon bath is inevitable, forming an open quantum system \cite{roy2011,Mahan1990,Wilson2002}. At cryogenic temperatures, the QD exciton mainly interacts with longitudinal acoustic phonons \cite{Ramsay2010excitondamping,Ramsay2010,Krummheuer2002} lead to energy shifts, incoherent excitations that result in cavity mode feeding and excitation-induced de-phasing phenomena \cite{roy2011}, linewidth broadening\cite{Besombes2001linebroadening}, modifying fluorescence spectrum, damping Rabi oscillations\cite{Ramsay2010excitondamping,Forstner2003excitondamping}, population inversion in two level systems \cite{Hughes2013}, excitation transfer \cite{roy2011} etc. These exciton-phonon interactions are incorporated mainly via the polaron transformation technique, followed by the Born-Markov approximation in deriving the master equation. In this thesis, we investigate cooperative two-photon lasing, hyperradiant lasing, correlated emission lasing, and non-degenerate two-mode two-photon lasing in systems with one or two QDs coupled to a single or bimodal photonic crystal cavity.

\section{Quantization of electromagnetic field}

In 1865, James Clerk Maxwell's seminal work on ``\textit{A Dynamical Theory of the Electromagnetic Field}" provides a unified theory of electricity and magnetism explaining all of classical electromagnetic phenomena. Maxwell's equations in the differential form for the electromagnetic field in free space (without charges or currents) are given as,

\begin{subequations}
	\begin{align}
	\nabla.\textbf{E}=0\label{subeq:delE}\\ \nabla\times \textbf{E}=-\frac{\partial\textbf{B}}{\partial t}\label{subeq:curlE}\\
	\nabla.\textbf{B}=0\label{subeq:delB}\\ \nabla\times\textbf{B}=\frac{1}{\mu_0 c^2}\frac{\partial\textbf{E}}{\partial t}\label{subeq:curlB}.
	\end{align}
	\label{eq:MaxEq}
\end{subequations}

Here, $\mu_0$, $\epsilon_0$ are the magnetic permeability, electric permittivity of the free space, respectively, and $\mu_0\epsilon_0=c^{-2}$ where \textit{c} is the speed of light in vacuum. Using Eq. \ref{subeq:curlE} and Eq. \ref{subeq:delB}, the equation for the electric field, $\textbf{E}(\textbf{r},t)$ is given by,

\begin{equation}
	\nabla^2\textbf{E}-\frac{1}{c^2}\frac{\partial^2 \textbf{E}}{\partial t^2}=0
\end{equation}

Although Maxwell's equations give a complete description of classical electromagnetic phenomena, they cannot explain several quantum phenomena. For example, using semiclassical theory (SCT), where matter is treated quantum mechanically and the electromagnetic field is treated classically, one can understand phenomena such as Rabi oscillations, resonance fluorescence, and absorption-emission processes. SCT, along with vacuum fluctuations, can account for phenomena such as the Lamb shift, spontaneous emission, Casimir effect. However, to explain the occurrence of sub-Poissonian photon statistics (anti-bunching), ``Quantum Beat" phenomenon, squeezed light, etc, the quantization of the electromagnetic field is necessary \cite{ScullyZubairyQOT}.

The derivation of the quantized EM field presented here closely follows the approach given in ``\textit{Quantum Optics}" by M. O. Scully and M. S. Zubairy \cite{ScullyZubairyQOT}. Let us consider that, EM field is localized in a box of length, \textit{L}, and volume, $V=L^3$. The mode expansion of the linearly polarized electric field in \textit{x}-direction is given by, 
\begin{equation*}
    E_x=\Sigma_n A_n q_n(t) \sin(k_n z)
    \label{eqn:Ex}
\end{equation*}
 where the coefficient,
 \begin{equation*}
     A_n=(\frac{2m_n\omega_n^2}{V\epsilon_0})^{1/2}
 \end{equation*}

 and $q_n(t)$ is the generalized coordinate associated with the mode. $m_n$ is the constant having dimensions of mass, and $\omega_n$ denotes frequency, and $k_n$ is the wavevector.
 
 The corresponding magnetic field obtained from Eq. \ref{subeq:curlB} is along \textit{y}-direction, given by,

 \begin{equation*}
     B_y=\Sigma_n A_n (\frac{\dot{q_n} \epsilon_0}{c^2k_n})\cos(k_nz)
     \label{eqn:By}
 \end{equation*}
 
  The total energy of the field is described by the classical Hamiltonian, 
  \begin{equation}
      H=\frac{1}{2}\int_V d\tau (\epsilon_0 E_x^2+\frac{1}{\mu_0} B_y^2)
      \label{eqn: HamCEM}
  \end{equation}
   
After the substitution of $E_x$ and $B_y$ into Eq. \ref{eqn: HamCEM} and performing the integration over the volume of the box, we obtain,

\begin{equation}
	H=\frac{1}{2}\sum_n (m_n\omega_n^2q_n^2+m_n \dot{q_n}^2)=\frac{1}{2}\sum_n (m_n\omega_n^2q_n^2+\frac{p_n^2}{m_n})
	\label{eq:HamQEM}
\end{equation}

where $p_n=m_n\dot{q_n}$ the canonical momentum corresponding to the $n$th mode. The Eq. \ref{eq:HamQEM} clearly shows that the electromagnetic field is nothing but a collection of harmonic oscillators.

Further, we quantize the field by introducing operators for $q_n$, $p_n$ variables as $Q_n$, $P_n$, following the standard procedure of quantization of the harmonic oscillator. The commutation relation between the operators is,

\begin{equation}
    [Q_i, P_j]=i\hbar\delta_{ij}
\end{equation}

It is convenient to introduce the ladder operators, $a_n$, $a_n^\dagger$ annihilation and creation operators in terms of $Q_n$ and $P_n$

\begin{subequations}
	\begin{align}
	a_n=\frac{1}{\sqrt{2m_n\hbar\omega_n}}(m_n\omega_nQ_n+iP_n) \label{annhOp}\\
	a^\dagger_n=\frac{1}{\sqrt{2m_n\hbar\omega_n}}(m_n\omega_nQ_n-iP_n) \label{creatOp}
	\end{align}
\end{subequations}

The operators $a_n$ and $a_n^\dagger$ satisfy the bosonic commutation relation,

\begin{equation}
    [a_m,a_n^\dagger]=\delta_{mn}
\end{equation}

The Hamiltonian for the single-mode EM field in terms of $a$, $a^\dagger$ is given by,

\begin{equation}
    H=\hbar \omega(a^\dagger a+\frac{1}{2})
\end{equation}

 and action on the energy eigenstates, $|n\rangle$, known as photon number states or Fock basis states, is given by, 

 \begin{equation}
     H|n\rangle=E_n|n\rangle,\, \text{where}\, E_n=\hbar\omega(n+\frac{1}{2})
 \end{equation}
 
The action of the ladder operators on the eigenstates, $|n\rangle$, 

\begin{equation}
    a|n\rangle=\sqrt{n}|n-1\rangle; \qquad a^\dagger|n\rangle=\sqrt{n+1}|n+1\rangle
\end{equation}

The `\textit{n}' photon Fock state is obtained from the vacuum state ($|0\rangle$) by repeated application of $a^\dagger$ `$n$' times, $|n\rangle=\frac{(a^\dagger)^n}{n!}|0\rangle$ equivalent to creation of \textit{n} photons.  

Using Eqs. \ref{eqn:Ex} $\&$ \ref{eqn:By}, the electric and magnetic fields can be written in terms of the ladder operators $a$, $a^\dagger$,

\begin{align}
    &E_x(z,t)=\mathcal{E} (ae^{i\omega t}+a^\dagger e^{-i\omega t})\sin(kz) \label{eqn:QEx}\\
    &B_y(z,t)=-\frac{i}{c}\mathcal{E}(ae^{i\omega t}-a^\dagger e^{-i\omega t})\cos(kz) \label{eqn:QBy}
\end{align}

where, $\mathcal{E}=(\frac{\hbar \omega}{\epsilon_0 V})^{1/2}$ having dimensions of electric field.

\section{Quantum confinement}\label{sec:QuantumConf}

Quantum confinement occurs when the characteristic dimensions of the crystal, $\Delta x$, $\Delta y$, $\Delta z$, are of the order of the de'Broglie wavelength, $\lambda$, of the charge carrier.
\begin{equation}
    \lambda=\hbar/p
\end{equation}

where $p=\sqrt{3m_{eff}k_BT}$ is the momentum at temperature, T. Therefore, the de'Broglie's wavelength becomes,

\begin{equation}
    \lambda=\frac{\hbar}{\sqrt{3m_{eff}k_BT}}
\end{equation}

The spatial localization of charge carriers increases the electron's kinetic energy, defined as confinement energy, $E_{conf}$. We know from uncertainty principle, $\Delta p_x \approx \hbar/\Delta x$, 

\begin{equation}
    E_{conf} = \frac{(\Delta p_x)^2}{2m} = \frac{\hbar^2}{2m(\Delta x)^2}
\end{equation}

The size effects are significant for $E_{conf}\geq \frac{1}{2} k_B T$ i.e., $\Delta x \leq\frac{\hbar}{\sqrt{mk_B T}}$. For example, in a semiconductor crystal with an effective mass of the charge carrier (electron), $m=m_{eff}=0.1m_e$, where $m_e$ is the free electron mass, $m_e=6.1\times 10^{-31}kg$. For the sample dimensions are $\Delta x \approx 5nm$ at room temperature, $T=373K$, the quantum confinement effects become significant, modifying the electronic density of states. The density of states (DOS) is defined as, $\rho=\frac{dN(E)}{dE}$, the number of available states per energy, $E$ per unit volume of the real space.

The DOS of bulk crystal (3D), quantum wells (2D), quantum wires (1D), and quantum dots (0D) structures are obtained by approximating electrons as a free electron gas with periodic boundary conditions, having no electron-electron and electron-crystal potential interaction. The corresponding electron wavefunction is given by, $\psi(\textbf{r})=e^{i\textbf{k}.\textbf{r}}$ satisfying the boundary conditions,

\begin{equation}
	\psi(x,y,z)=\psi(x+L_x,y,z)=\psi(x,y+L_y,z)=\psi(x,y,L_z+z)
\end{equation}

This implies, the components of $\textbf{k}$ vector should satisfy, $k_x=\frac{2\pi}{L}n_x$, $k_y=\frac{2\pi}{L}n_y$ and $k_z=\frac{2\pi}{L}n_z$, where $n_x,\, n_y,\, n_z$ are integers. 

\begin{figure}
    \centering
    \includegraphics[width=\linewidth]{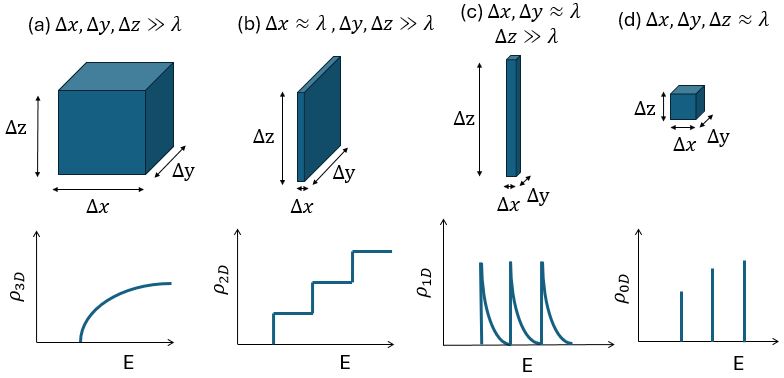}
    \caption{Density of states for (a) Bulk crystal (3D) (b) Quantum well (2D) (c) Quantum wire (1D) (d) Quantum dot (0D) }
    \label{fig:QDDOS}
\end{figure}

\textbf{3D bulk crystal}: For a bulk crystal, whose dimensions are much larger than the electron (hole) de'Broglie wavelength, Fig. \ref{fig:QDDOS} (a). The total number of states available in $k$-space per unit volume of real space is given by,

\begin{equation}
	N_{3D}=2\frac{4\pi k^3}{3}\frac{1}{(2\pi/L)^3}\frac{1}{L^3}=2\frac{4\pi k^3}{3(2\pi)^3}
	\label{eq:N3D}
\end{equation}
where the factor of `2' is due to spin degeneracy. The DOS is given by,
\begin{equation}
	\rho_{3D}=\frac{dN(E)}{dE}=\frac{dN(E)}{dk}\frac{dk}{dE}
\end{equation}
From Eq. \ref{eq:N3D}, we obtain, 

\begin{equation}
	\frac{dN}{dk}=\frac{k^2}{\pi^2}
\end{equation}

In accordance with the assumption of free electron gas, the dispersion relation is parabolic, and the electron energy is given by $E=\frac{\hbar^2 k^2}{2m_{eff}}$. The Fermi wavevector is, $k=\frac{\sqrt{2m_{eff}E}}{\hbar}$. Therefore,

\begin{equation}
	\frac{dk}{dE}=\frac{\sqrt{2m_{eff}}E^{-1/2}}{2\hbar}
	\label{eq:dkdE}
\end{equation}

Finally, the DOS of a 3D bulk crystal takes the form,
\begin{equation}
	\rho_{3D}(E)=\frac{1}{2\pi^2}\Big(\frac{2m_{eff}}{\hbar^2}\Big)^{3/2}E^{1/2}
\end{equation}

\textbf{Quantum well}:
For 2D layered structures, also known as `\textit{quantum wells}', DOS is modified due to confinement along one of the directions of the crystal as shown in Fig. \ref{fig:QDDOS} (b). Let us consider the confinement along the `\textit{x}' axis leads to quantized states with energy, $E_{n_x}$, and the electron motion is quasi-continuous states in `\textit{y}', `\textit{z}' directions.
\begin{equation}
	E=\Big(\frac{n_x}{L_x}\Big)^2+k_y^2+k_z^2
\end{equation}
For 2D structures, the number of states per unit volume is given by, 
\begin{equation}
	N_{2D}=2\frac{\pi k^2}{(2\pi/L)^2}\frac{1}{L^2}=\frac{2\pi k^2}{(2\pi)^2}
\end{equation}
The dispersion relation is parabolic in the YZ-plane, and using \ref{eq:dkdE}, the DOS is given by,
\begin{equation}
	\rho_{2D}(E)=\frac{dN_{2D}}{dE}=\frac{m_{eff}}{\pi \hbar^2}
\end{equation}
If there are `\textit{n}' confined subbands in the \textit{quantum well}, then DOS for a particular energy, E, is given by the sum over all the subbands below $E$.
\begin{equation}
	\rho_{2D}=\sum_{i=1}^n \frac{m_{eff}}{\pi \hbar^2} \Theta(E-E_i)
\end{equation}
where, `$\Theta(E)$' is the Heaviside step fucntion.

\textbf{Quantum wire}:
For 1D systems, known as `\textit{quantum wires}' where the charge carriers, electrons (holes), are restricted to move along `\textit{x}' and `\textit{y}' directions as shown in Fig. \ref{fig:QDDOS} (c). As a result, the energy spectrum is discrete along `\textit{x}' and `\textit{y}'  and takes quasi-continuous values along `\textit{z}' direction. The total energy is given by,
\begin{equation}
	E=\Big(\frac{n_x}{L}\Big)^2+\Big(\frac{n_y}{L}\Big)^2 + k_z^2
\end{equation}

The total number of states available per unit volume of the real space is given by,
\begin{equation}
N_{1D}=2\frac{2k}{(2\pi/L)}\frac{1}{L}
\end{equation}
and
\begin{equation}
\frac{dk_z}{dE}=\frac{\sqrt{2m_{eff}}}{(2\hbar)}E^{-1/2}
\end{equation}

Density of states for ``\textit{quantum wires}" (1D) is given by,
\begin{equation}
	\rho_{1D}=\frac{2}{\pi}\frac{\sqrt{2m_{eff}}E^{-1/2}}{2\hbar}=\Big(\frac{2m_{eff}}{\hbar^2}\Big)^2\frac{1}{\pi E^{1/2}}
\end{equation}
Similar to 2D systems, if there are $n$ occupied subbands, the total DOS at energy $E$ is given by summing over all the possible contributions from subbands below the energy $E$.
\begin{equation}
\rho_{1D}=\sum_{i=1}^n\Big(\frac{2m_{eff}}{\hbar^2}\Big)^2\frac{\Theta(E-E_i)}{\pi (E-E_i)^{1/2}}
\end{equation}

\textbf{Quantum dot}:
Eventually, when the movement of the charge carriers is restricted in all three directions, the system becomes 0-dimensional and is known as `\textit{quantum dot}' (QD), Fig. \ref{fig:QDDOS} (d). In such systems, the density of states is discrete, resembling the atomic energy spectrum, and hence the quantum dots are also called `\textit{artificial atoms}'.

\begin{equation}
	\rho_{0D}=2\sum_{i=1}^n \delta(E-E_i)
\end{equation}

Upon optical or electrical excitation of QD, either coherently or incoherently, leads to the generation of quasi-particles, namely, an exciton consisting of a single electron ($e$) in the conduction band (CB) and a hole ($h$) in the valence band (VB). Beyond neutral excitons, QDs also host positively (1$e$-2$h$) or negatively (2$e$-1$h$) charged excitons, and also biexcitons, which are the bound states of two $e-h$ pairs with binding energy, $\Delta$. There is an important distinction between the excitons in bulk, quantum wells or wires and the quantum dots, where in the latter case the confinement potential dominates the Coulomb interaction potential. 

\begin{figure}
    \centering
    \includegraphics[width=0.75\linewidth]{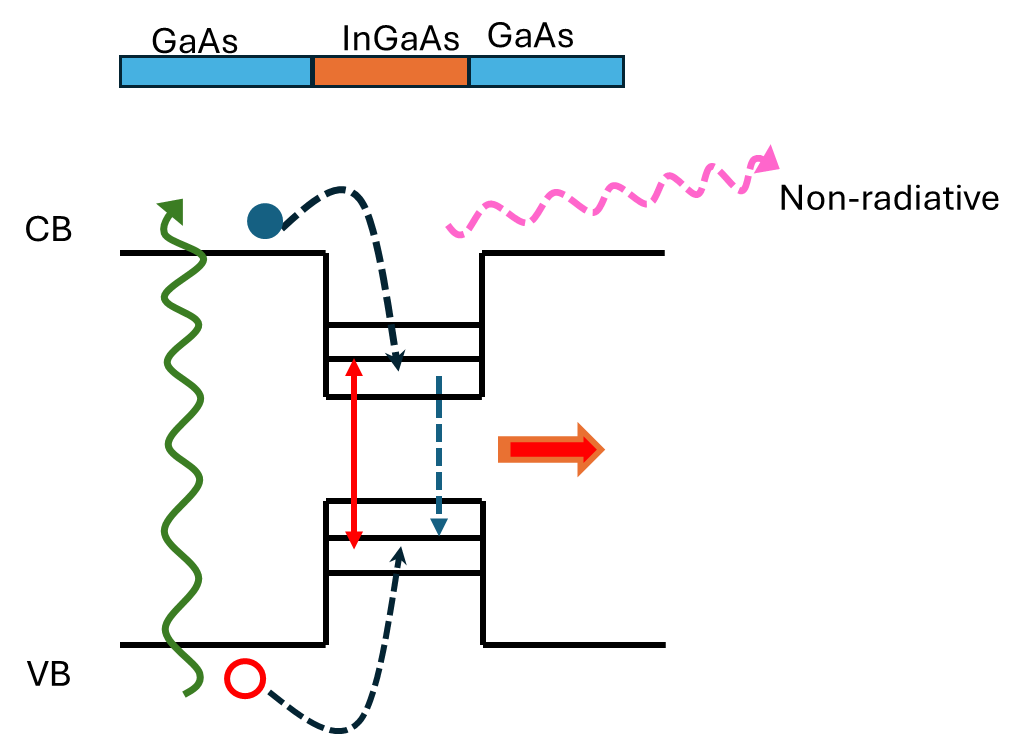}
    \caption{A schematic figure showing a quantum dot excited either incoherently (green) or coherently (red) forms an electron–hole pair (exciton) after non-radiative relaxation due to interaction with the surrounding phonon bath. Electron-hole pair whose recombination leads to photon emission.}
    \label{fig:QDExct}
\end{figure}

Further, this electron-hole radiative recombination leads to the emission of photons as shown in Fig. \ref{fig:QDExct}. The quantum state of an electron in the conduction band, a hole in valence band can be expressed as,

\begin{equation}
    |\Phi_{c/v}\rangle = |\psi_{c/v}\rangle|u_{c/v}\rangle|s_{c/v}\rangle
\end{equation}

The total wavefunction contains, envelope function $|\psi_{c/v}\rangle$, the Bloch function, $|u_{c/v}\rangle$, due to crystal periodicity, and the spin part of the wave function is $|s_{c/v}\rangle$.

We obtain the envelope function from the effective mass Schr\"{o}dinger equation.

\begin{equation}
    \Big[-\frac{\hbar^2}{2m^*}\nabla^2+V_{c/v}(\textbf{r}) \Big] |\psi_{c/v}\rangle = (E-E_{c/v}) |\psi_{c/v}\rangle
\end{equation}

where, $m^*$ is the effective mass of the charge carries, $V_{c/v}(\textbf{r})$ is the confinement potential, $E$ is the energy of the carrier and $E_{c/v}$ is the band edge energy. For the case of spherically symmetric parabolic potentials, the ground state envelope function takes the Gaussian profile, 
\begin{equation}
    \psi_{c/v}(r)=e^{-r^2/2d_{c/v}^2}
\end{equation}
where \textbf{k} is the carrier wavevector and $d_{c/v}$ is the confinement length.

The transitions between these quasi particles are either dipole allowed or dipole forbidden, classified as ``\textit{bright}" or "\textit{dark}" excitons. Denoting the Bloch functions and the spins of electron and hole by pseudospin states, $|\uparrow\rangle=|u_c\rangle|\uparrow_e\rangle$, $|\downarrow\rangle = |u_c\rangle|\downarrow_e\rangle$, and for the hole, $|\Uparrow\rangle=|u_h\rangle|\uparrow_h\rangle$, $|\Downarrow\rangle=|u_h\rangle|\downarrow_h\rangle$. For the QD with quantization along the z-axis, electron spin state $|\uparrow_e\rangle$ corresponds to $S_z=1/2$ and for heavy holes states in the conduction band $|\uparrow_h\rangle\,(|\downarrow_h\rangle)$  to $S_z=+3/2\,(-3/2)$. In total, four electron-hole states are possible, and among them $|\uparrow\Downarrow\rangle$, $|\downarrow\Uparrow\rangle$ participate in dipole allowed optical transitions and $|\uparrow\Uparrow\rangle$, $|\downarrow\Downarrow\rangle$ are dipole forbidden. Further, their linear combination of these states form a set of bright and dark excitons with polarization along $x$ and $y$ directions given by,

\begin{align*}
    &|X_b\rangle=\frac{|\uparrow\Downarrow\rangle-|\downarrow\Uparrow\rangle}{\sqrt{2}};
    \qquad |Y_b\rangle=\frac{|\uparrow\Downarrow\rangle+|\downarrow\Uparrow\rangle}{\sqrt{2}}
    \\& |X_d\rangle=\frac{|\uparrow\Uparrow\rangle-|\downarrow\Downarrow\rangle}{\sqrt{2}};
    \qquad |Y_d\rangle=\frac{|\uparrow\Uparrow\rangle+|\downarrow\Downarrow\rangle}{\sqrt{2}}
\end{align*}

Since biexciton consists of two bound electron-hole pairs, denoted by $|\uparrow\Uparrow\downarrow\Downarrow\rangle$ or $|XX\rangle$. The biexciton can have two decay channels, $|XX\rangle\rightarrow|X_b\rangle$, $X_b\rightarrow|g\rangle$ and $|XX\rangle\rightarrow|Y_b\rangle$, $|Y_b\rightarrow|g\rangle$ emitting horizontal or vertically polarized photons . These cascaded emissions from the biexcitonic state in QDs can act as a source of polarization entangled photons by controlling the fine-structure splitting $\delta$ between $X_b$ and $Y_b$, which arises due to QD in-plane asymmetry \cite{Akopian2006, Johne2008, Juska2013, Huber2018, Dousse2010ultrabright}.

\begin{figure}
    \centering
    \includegraphics[width=\linewidth]{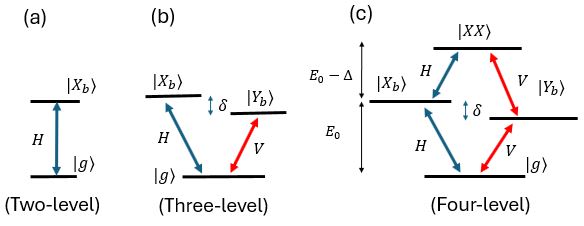}
    \caption{(a) Exciton (b) $|X_b\rangle$, $|Y_b\rangle$ fine structure splitting, $\delta$, and along with $|g\rangle$ forming a 3-level system (c) $|XX\rangle$, $|X_b\rangle$, $|Y_b\rangle$, and $|g\rangle$ forming a  4-level system showing transitions between biexcitonic and excitonic states.}
    \label{fig:QDTrans}
\end{figure}

Since QDs are nano-crystals, the electronic excitation from the valence band to the conduction band redistributes the electron density, leading to the displacement of ion equilibrium positions. This leads to the coupling between the electron and the phonons. Also, embedding these ``\textit{Quantum Dots}" inside a semiconductor photonic crystal (PhC) cavity, one can realize various cavity QED phenomena. The exciton-phonon interactions and the semiconductor cavity QED systems are briefly discussed in the next sections.

\section{Phonons} \label{sec:Phonons}

The vibrational modes of the crystal are nothing but the collective motion of the ions about their equilibrium position. The quantized vibrational modes are called ``\textit{Phonons}".
A simple introduction to these normal modes is given by the one-dimensional harmonic chain model. In this model, we mainly make two assumptions, 

(i) The mean equilibrium position of each ion is a Bravais lattice site, R, and oscillates about this mean position.

(ii) The ion oscillates back and forth, and the amplitude is very small compared to the crystal lattice constant. The oscillations can be approximated by a sinusoidal function \cite{Ashcroft}.

\begin{figure}
	\centering
	\includegraphics{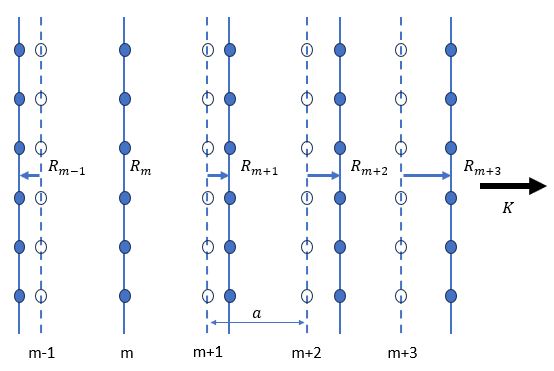}
	\caption{Dashed lines represent planes of atoms in equilibrium, while solid lines show their displacement when a longitudinal wave propagates through the medium.}
	\label{fig:longPhonon}
\end{figure}

\begin{figure}
	\centering
	\includegraphics{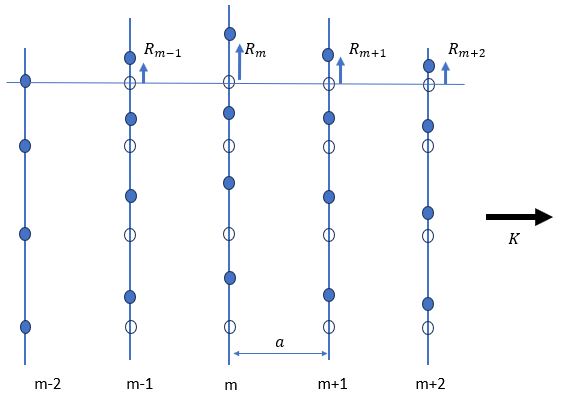}
	\caption{Displacement of planes of atoms (solid circles) when a transverse wave passes through the medium.}
	\label{fig:TransvPhonon}
\end{figure}

A schematic diagram of the atomic displacement in longitudinal and transverse phonon cases is given in Fig. \ref{fig:longPhonon} and Fig. \ref{fig:TransvPhonon}, respectively.

Let us consider the position of the m-th ion, 
\begin{equation}
    R_m=R_m^0+Q_m
\end{equation}
where $R_m^0$ is the equilibrium position and $Q_m$ is the displacement about this equilibrium position. The potential energy depends on the distance between the neighboring atoms, $R_m - R_{m+1}$, and the total potential energy is the sum of the contributions from distinct pairs of atoms.  In a simple case of a monoatomic chain, with nearest-neighbor interactions, the total energy including kinetic and potential energy terms is written as,
\begin{equation}
    H = \sum_m \frac{P_m^2}{2M} + \frac{K}{2}\sum_m (R_m-R_{m+1})^2
    \label{eqn:HOHam}
\end{equation}
where $M$ is the atomic (ionic) mass and $K$ is the effective spring constant of the restoring force between neighboring atoms.
The $m$-th ion's equation of motion (EOM) is given by,
\begin{equation}
    -M\Ddot{R_m} = M\omega^2R_m=K(R_m-R_{m+1}-R_{m-1}) 
\end{equation}
We assume a trial solution for $R_m$ to be of the form, 
\begin{equation}
    R_m=R_0\, \cos(kam)
\end{equation}
where $a$ is the lattice spacing and $k$ is the wavevector. Substituting in the EOM gives,

\begin{align}
    R_m-R_{m+1}-R_{m-1} &= R_0[2\cos(kam)-\cos(kam+ka)-\cos(kam-ka)]
                        \\&=2R_0\cos(kam)[1-\cos(ka)]
\end{align}
The solutions for the normal modes are given by,
\begin{equation}
    \omega_k^2 = \frac{2K}{M}[1-\cos(ka)]=\frac{4K}{M}\sin^2(\frac{ka}{2})
\end{equation}

\begin{figure}
    \centering
    \includegraphics[width=0.75\linewidth]{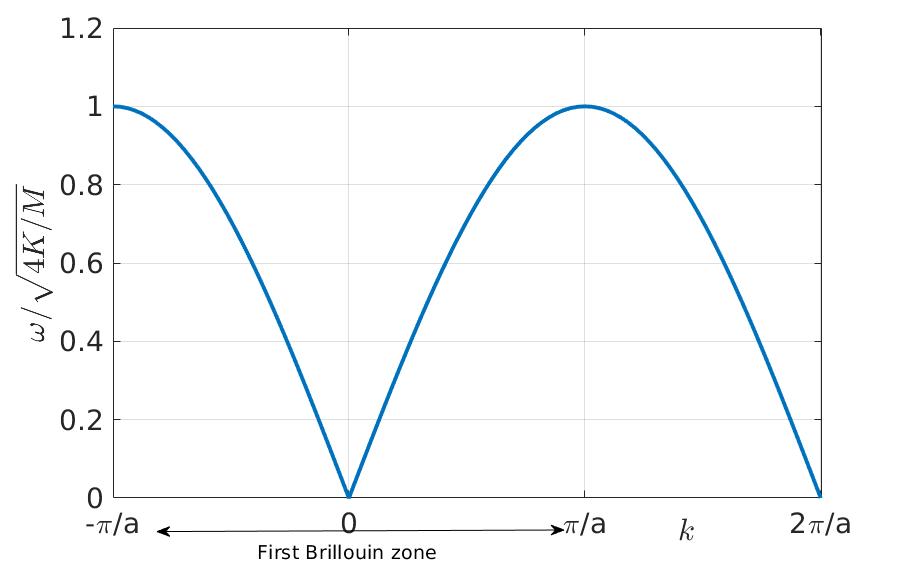}
    \caption{Plot of $\omega$ vs $k$. In the long-wavelength limit, $k \ll 1/a$ (or equivalently $\lambda \gg a$), the system can be described using the continuum approximation.}
    \label{fig:dispersionCurve}
\end{figure}

The range of values of $k$ is $-\pi/a\leq k \leq \pi/a$; the values within the first Brillouin zone (BZ) only are significant. If the value of $k$ lies outside this range, it can be related to the value within the first BZ, $k'$ by $k'=k-2\pi n/a$ where $n$ is an integer. Then the displacement $R_m=R_0\, \cos(kam)=R_0\, \cos(k'am+2\pi mn)=R_0\, \cos(k'am)$, ( $m,\, n \in \mathcal{Z}$) can be expressed in terms of wavevector in the first BZ.

In the long wavelength limit $ka\ll 1$, we can expand $\cos(ka)\simeq 1-(ka)^2/2$, so
\begin{equation*}
    \omega^2\simeq \frac{2K}{M}[1-(1-\frac{(ka)^2}{2}]\simeq \frac{K}{M}(ka)^2 \implies \omega\simeq c|k|
\end{equation*}
Here, $c=a\sqrt{\frac{K}{M}}$ is the speed of sound. Thus, in the long wavelength limit, the dispersion relation is linear, Fig.\ref{fig:dispersionCurve}. For a monoatomic chain, there is only this single acoustic branch. Optical branches appear only for multi-atomic bases \cite{Kittel,Ashcroft,Mahan1990}, resulting in longitudinal acoustic (LA), transverse acoustic (TA), longitudinal optical (LO), and transverse optical (TO) branches. For a system with $N$ atoms in its primitive cell, there are $3N$ branches of the dispersion relation, namely, $3$ ``acoustic branches" and $3N-3$ ``optical branches".

\indent In continuation, we define the normal coordinates of the real space in terms of the wave-vector space coordinates by Fourier series expansion, satisfying the periodic boundary conditions,

\begin{align}
    &R_l= \frac{1}{\sqrt{N}}\sum_k e^{ikal}R_k
    \\&P_l= \frac{1}{\sqrt{N}}\sum_k e^{-ikal}P_k
\end{align}

The kinetic and potential terms in Eq. \ref{eqn:HOHam} can be rewritten as,
\begin{align}
    &\sum_m P_m^2 = \sum_k P_k P_{-k}
    \\&\frac{K}{2}\sum_m (R_m-R_{m+1})^2=\frac{K}{2}\sum_k R_k R_{-k}(2-e^{ika}-e^{-ika})= \frac{M}{2}\sum_k \omega_k^2 R_k R_{-k}
\end{align}
The total Hamiltonian in wave-vector space is,
\begin{equation}
    H=\frac{1}{2M}\sum_k [P_kP_{-k}+M^2\omega_k^2 R_{k}R_{-k}]
\end{equation}
Finally, similar to the quantization of the electromagnetic field, by introducing the annihilation and creation operators, we obtain the quantized version of lattice vibrations called phonons.
\begin{align}
    &a_k = (\frac{M\omega_k}{2\hbar})^{1/2}(R_k+\frac{i}{M\omega_k}P_{-k})
    \\&a_k^\dagger = (\frac{M\omega_k}{2\hbar})^{1/2}(R_k-\frac{i}{M\omega_k}P_{-k})
\end{align}
follow the commutation relations, $[a_k,a_{k'}^\dagger]=\delta_{k,k'},\, [a_k,a_{k'}]=0, \, [a_k^\dagger, a_{k'}^\dagger]=0$. The Hamiltonian in terms of $a_k,\, a_k^\dagger$ is given by,

\begin{equation}
    H=\sum_k \hbar \omega_k [a_k^\dagger a_k+\frac{1}{2}]
\end{equation}

 The expectation of the number operator, $\hat{n}=a_k^\dagger a_k$ provides the number of phonons in the wave-vector state \textit{k}.

For 3D solids, it is straight forward extension. The potential energy between the ions has the form, $\sum_{ij} V(\textbf{R}_i-\textbf{R}_j)$. As earlier, we can write the ion position about the mean position, $\textbf{R}_m=\textbf{R}_m^0+\textbf{Q}_m$.. The final expression for the Hamiltonian is,

\begin{align}
    H&=\frac{1}{2M} \sum_{\textbf{k}, \zeta} (\textbf{P}_{\textbf{k}, \zeta}. \textbf{P}_{-\textbf{k}, \zeta}+M\omega_{\textbf{k}, \zeta}^2 \textbf{Q}_{\textbf{k}, \zeta}. \textbf{Q}_{-\textbf{k}, \zeta}
    \\& = \sum_{\textbf{k}, \zeta} \hbar \omega_{\textbf{k}\zeta}(a_{\textbf{k},\zeta}^\dagger a_{\textbf{k},\zeta}+\frac{1}{2})
\end{align}

Here, $\zeta$ corresponds to the polarization of the \textbf{k}-th phonon bath mode.

\subsection{Exciton-phonon interactions}

As mentioned earlier, the excitation of the electron from the valence band to the conduction band leads to the redistribution of the lattice ion equilibrium position. This results in the coupling between the electrons and acoustic or optical phonons. The electron-phonon interaction Hamiltonian is given by,
\begin{equation}
H_{int}=\int_V d\textbf{r} \rho(\textbf{r}) \sum_m V_{int}(\textbf{r}-\textbf{R}_m)
\label{eqn:EPIntHam}
\end{equation}
$\rho(\textbf{r})$ is the charge density operator in the solid, \textbf{r} is the electron position, $\textbf{R}_m$ denotes the $m$-th ion position. Here, let us make it explicit the ion displacement by writing, $\textbf{R}_m=\textbf{R}_m^0 + \textbf{Q}_m$, where $\textbf{R}_m^0$ is the equilibrium position and $\textbf{Q}_m$ is the displacement about this equilibrium position.

\begin{equation}
V_{int}(\textbf{r}-\textbf{R}_m)\approx V(\textbf{r}-\textbf{R}_m^0)-\textbf{Q}_m . \nabla V(\textbf{r}-\textbf{R}_m^0)
\end{equation}
In the above expansion of the interaction potential, $V_{int}$, we considered terms up to first order in $\textbf{Q}_m$ and ignored the terms of the order $\textbf{Q}_m^2$ and higher. The zeroth-order term in the above equation is the periodic potential experienced by the electrons, which gives rise to the Bloch eigenstates. The first-order term corresponds to the linear coupling between the electron and phonon. We expand the lattice potential in reciprocal space,
\begin{equation}
    V(\textbf{r})=\frac{1}{N}\sum_\textbf{q} v(\textbf{q}) e^{i\textbf{r}.\textbf{q}}
\end{equation}
The interaction potential is given by,
\begin{align}
V_{ep} &= -\sum_m \textbf{Q}_m. \nabla V(\textbf{r}-\textbf{R}_m^0)\\
&= -\frac{i}{N}\sum_\textbf{q} \sum_m \textbf{Q}_m e^{-i\textbf{q}.\textbf{R}_m^0}.\textbf{q} v(\textbf{q}) e^{i\textbf{r}.\textbf{q}}\\
&=-\frac{i}{\sqrt{N}}\sum_\textbf{q} \textbf{Q}_{\textbf{q}}.\textbf{q} v(\textbf{q}) e^{i\textbf{r}.\textbf{q}}
\end{align}

Here, $\textbf{Q}_{\textbf{q}}=\frac{1}{\sqrt{N}}\sum_m \textbf{Q}_m e^{-i\textbf{q}.\textbf{R}_m^0}$. As mentioned earlier, the phonon wavevectors, \textbf{k}, are defined in the first Brillouin zone, but the \textbf{q} vectors are defined everywhere. In the low-energy excitation regime, coupling to the first Brillouin states can be considered. Therefore, we can replace \textbf{q} by \textbf{k} and the displacement operator, $\textbf{Q}_\textbf{k}$ can be written in terms of the phonon creation and annihilation operators, $\textbf{Q}_{\textbf{k}}=i\sqrt{\frac{\hbar}{2M\omega_\textbf{k}}}\zeta_{\textbf{k}}(b_{\textbf{k}}+b_{-\textbf{k}}^\dagger)$ where $\zeta_\textbf{k}$ is the polarization vector. The subscript $\zeta$ has been omitted for brevity. Therefore, the interaction potential takes the form,

\begin{equation}
    V_{ep}=\sum_{\textbf{k}} \sqrt{\frac{\hbar}{2MN\omega_{\textbf{k}}}}(b_{\textbf{k},\zeta}+b_{-\textbf{k}}^\dagger)\nu(\textbf{k})\textbf{k}.\zeta_\textbf{k}e^{i\textbf{k}.\textbf{r}}
\end{equation}

and the interaction Hamiltonian is given by,

\begin{equation}
    H_{int}=\sum_k M(\textbf{k})\hat{\rho}(\textbf{k}).\zeta_\textbf{k} (b_\textbf{k}+b_{-\textbf{k}}^\dagger)
\end{equation}

\textbf{Types of electron-phonon coupling}:

The electrons (holes) are coupled to different types of phonon modes via different coupling mechanisms, resulting in exciton dephasing. Deformation potential coupling is dominant in the case of LA phonons, while piezoelectric coupling is relevant for TA phonons. In addition, dephasing of the exciton also occurs due to the interaction with LO phonons, dominantly via polar optical coupling.  

1) \textit{Polar optical coupling} of LO phonons with coupling strength given by,

\begin{equation}
M_{LO,ph}^{e/h}(\textbf{k})=i\Big[\frac{2\pi e^2 \hbar \omega_{LO}(\textbf{k})}{4\pi \epsilon_0 V}\Big(\frac{1}{\epsilon_s}-\frac{1}{\epsilon_\infty}\Big)\Big]^{1/2} \frac{1}{|\textbf{k}|}
\end{equation}

where, $\omega_{LO}(\textbf{k})$ is the frequency of phonon mode, $V$ is the total volume, $\epsilon_s$ and $\epsilon_\infty$ are the static and the high frequency dielectric constants.

2) \textit{Piezoelectric coupling} of TA and LA phonons, \textit{Deformation potential coupling} to LA phonons. The coupling strength to acoustic phonons is given by,
\begin{equation}
M_{AC,ph}^{e/h}(\textbf{k}) = \sqrt{\frac{\hbar}{2\rho\omega_\zeta(\textbf{k}) V}}[|\textbf{k}|D_{\zeta}^{e/h}+iM_\zeta(\textbf{k})]
\end{equation}

Here, $\zeta$ corresponds to the polarization of the acoustic phonon, whether it is longitudinal or transverse, $\rho$ is the mass density ($\rho V=MN$). The exciton coupling to acoustic phonons is both via deformation potential coupling (real) and piezoelectric coupling (imaginary).

Recent experiments at cryogenic temperatures ($T<75K$), have shown that InGaAs/GaAs QD systems electron interaction with LA phonons via deformation potential coupling dominates over the other exciton-phonon coupling mechanisms \cite{Ramsay2010, Krummheuer2002}.

In this work, we consider an isotropic system such that the only contribution is from the coupling of the electrons to the longitudinal phonons, i.e., the contribution to the interaction potential, $V_{ep}$, is only from longitudinally polarized phonons (i.e., polarization parallel to \textbf{k}). Specifically, we consider that electrons are coupled to the LA phonons via deformation potential coupling.

Therefore, electron-phonon interaction Hamiltonian (Eq. \ref{eqn:EPIntHam}) can be written as,

\begin{equation}
    H_{int}=\sum_\textbf{k} M(\textbf{k}) \hat{\rho}(\textbf{k})(b_\textbf{k}+b_{-\textbf{k}}^\dagger)
    \label{eqn:genIntHam}
\end{equation}

where, $\hat{\rho}(\textbf{k})=\int d\textbf{r}\hat{\rho}(\textbf{r})e^{i\textbf{k}.\textbf{r}}$ and $M(\textbf{k})=|\textbf{k}|\nu(\textbf{k})\sqrt{\hbar/2MN\omega_\textbf{k}}$. For the deformation potential coupling to LA phonons scenario, the electron-ion potential, $\nu(\textbf{k})$, can be approximated by deformation constants, $D_c$ and $D_v$ corresponding to the conduction and valence bands. These deformation constants are obtained by measuring how the conduction and valence bands shift with an increase in pressure, i.e., the rate of change of band energy with pressure \cite{Mahan1990}.

Let us consider the simplest case of a two-level QD with $|0\rangle$ and $|X\rangle$ as the ground and excited states. Assuming spherically symmetric parabolic potentials for the valence and the conduction bands, the envelope functions for QD states take the form $\psi_j(\textbf{r})=(d_j\sqrt{\pi})^{-3/2} e^{-r^2/2d_j^2}$ for $j=0,X$ where $d_j$ is the size of the wavefunction. $d_j$ depends on the confinement length of electrons/holes in the QD. The charge density operator is given by,

\begin{equation}
    \hat{\rho}(\textbf{k})=\rho_{00}(\textbf{k})|0\rangle\langle 0|+\rho_{XX}|X\rangle\langle X|
\end{equation}

Here the form factors are given by, $\rho_{jj}(\textbf{k})=\int d\textbf{r} |\psi_j(\textbf{r})|^2e^{i\textbf{k}.\textbf{r}}$. Therefore, the interaction Hamiltonian, Eq. \ref{eqn:genIntHam} is simplified to 

\begin{align}
    H_{int}&=\sum_\textbf{k} (M_\textbf{k}^{(0)}\rho_{00}(\textbf{k})|0\rangle\langle 0|+M_\textbf{k}^{(X)}\rho_{XX}(\textbf{k})|X\rangle\langle X|)(b_\textbf{k}+b_{-\textbf{k}}^\dagger)
    \\& = \sum_\textbf{k} (M_\textbf{k}^{(0)}\rho_{00}(\textbf{k})(|0\rangle\langle 0|+|X\rangle\langle X|)+(M_\textbf{k}^{(X)}\rho_{XX}(\textbf{k})-M_\textbf{k}^{(0)}\rho_{00}(\textbf{k}))|X\rangle\langle X|(b_\textbf{k}+b_{-\textbf{k}}^\dagger)
\end{align}
Eliminating the term proportional to identity, $I=|0\rangle\langle 0|+|X\rangle\langle X|$, which can be absorbed in the global displacement of the phonon bath, the interaction Hamiltonian takes the form,
\begin{equation}
    H_{int}= |X\rangle\langle X|\sum_\textbf{k}(M_\textbf{k}^{(X)}\rho_{XX}(\textbf{k})-M_\textbf{k}^{(0)}\rho_{00}(\textbf{k}))(b_\textbf{k}+b_{-\textbf{k}}^\dagger)
\end{equation}

Using the envelope functions defined earlier, we obtain the form factors, $\rho_{jj}=e^{-d^2|\textbf{k}|^2/4}$. Here, we considered $d_0=d_X=d$ for simplification and writing the interaction Hamiltonian, $H_{int}$ as,

\begin{equation}
    H_{int}=|X\rangle\langle X|\sum_\textbf{k} \hbar \lambda_\textbf{k} (b_\textbf{k} + b_{-\textbf{k}}^\dagger)
\end{equation}

where, $\lambda_\textbf{k}=\frac{|\textbf{k}|}{\sqrt{2MN\omega_\textbf{k}}}[D_c-D_v]e^{-d^2|\textbf{k}|^2/4}$ is the coupling strength. For the deformation potential coupling, in the long wavelength limit (acoustic phonon coupling), we can consider the phonon bath having linear dispersion, $\omega_\textbf{k}=c|\textbf{k}|$ where $c$ is the speed of sound. 

The phonon spectral density is given by,

\begin{equation}
    J_{ph}(\omega)=\sum_\textbf{k} \lambda_\textbf{k}^2 \delta(\omega-\omega_\textbf{k})=\alpha \omega^3e^{-\omega^2/\omega_c^2}
\end{equation}

The final expression in the above equation is obtained in the continuum limit, after converting the summation into an integral over the phonon bath modes, \textbf{k} as $\sum_\textbf{k}\rightarrow V/(2\pi)^3 \int d\textbf{k}$. The coupling strength is proportional to $\alpha=V(D_c-D_v)^2/(4\pi^2 MNc^5)$ and the phonon-bath cut-off frequency is given by $\omega_c=\sqrt{2}c/d$. In general, depending on the form of the phonon spectral density, $J(\omega)\propto \omega^s$, the phonon bath is classified as sub-Ohmic ($s<1$), Ohmic ($s=1$) or super-Ohmic ($s>1$). In this work, we consider a super-Ohmic phonon bath arising from the deformation potential coupling to longitudinal acoustic (LA) phonons in semiconductor QDs, which in general is valid \cite{Wilson2002}.

\section{Photonic crystal cavity}

\begin{figure}
    \centering
    \includegraphics[width=\linewidth]{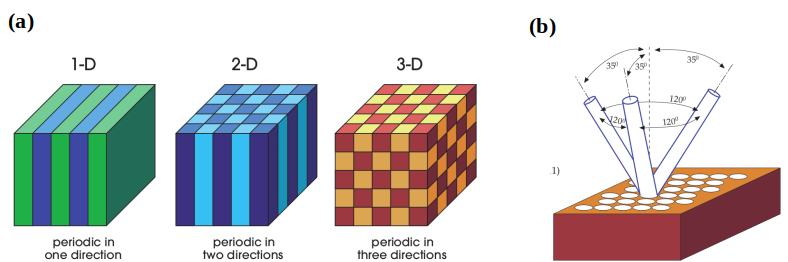}
    \caption{Source \cite{Yablonovitch1991,JoannopoulosTB} : (a) Schematic figure of 1D, 2D, 3D photonic crystals (b) First 3D photonic band gap material, Yablonovite.}
    \label{fig:Phcs}
\end{figure}

Photonic crystals (PhCs) are Distributed Bragg Reflectors (DBRs) consisting of a periodic arrangement of two different refractive index materials. Such a periodic arrangement leads to strong reflection of particular wavelengths of light, allowing PhCs to act as nearly perfect band-pass filters. Extending this periodic arrangement along two directions leads to 2D PhC and in three directions forms a complete photonic band gap structure, 3D PhC, Fig.\ref{fig:Phcs}

First successful realization of 3D photonic crystal, known as``Yablonovite", was formed by drilling air holes in a ceramic material, Fig. \ref{fig:Phcs} (b). This 3D PhC blocks the radio waves between 13-16 GHz.

\begin{figure}
    \centering
    \includegraphics[width=0.5\linewidth]{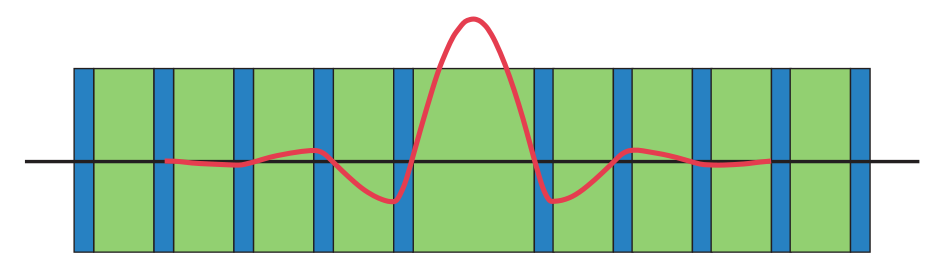}
    \caption{Source \cite{JoannopoulosTB} : By introducing a defect in a one-dimensional distributed Bragg reflector (DBR), the localization of an electromagnetic field mode within the structure can be achieved.}
    \label{fig:1DPhcC}
\end{figure}

\begin{figure}
    \centering
    \includegraphics[width=0.5\linewidth]{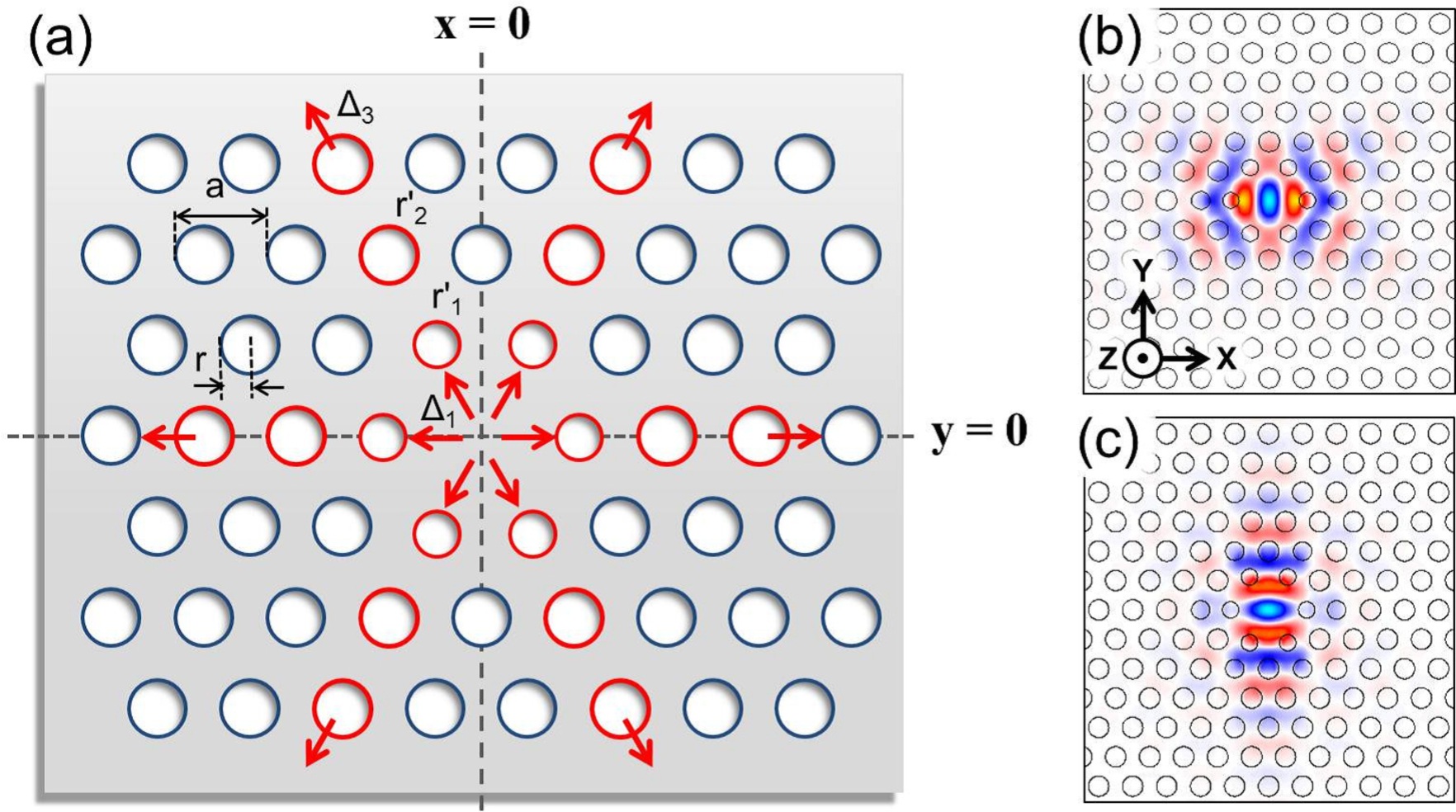}
    \caption{Source \cite{Takagi2012}: The removal of the central air hole in a photonic crystal lattice creates an H1 cavity, which can support bimodal confined optical modes.}
    \label{fig:H1PhcC}
\end{figure}

Introducing a defect into the DBR structure enables the localization of the electromagnetic (EM) field within extremely small mode volumes, often on a scale smaller than a cubic optical wavelength. Defects in photonic band-gap structures can be compared to doping in electronic semiconductors. By modifying the periodic arrangement, for example, by increasing the thickness of one of the dielectric slabs in 1D DBR as shown in Fig. \ref{fig:1DPhcC} or removing air holes in the thin film 2D photonic crystal, introducing a defect leads to the formation of a cavity. First 2D PhC cavity, known as H1 cavity \cite{Painter1999}, O. Painter et al., showed that by removing a single air hole from the center of the 2D PhC as shown in Fig. \ref{fig:H1PhcC}. 

Unlike other microcavities, such as microspheres or microdisks, which confine light via Total Internal Reflection (TIR), PhCs exploit the Distributed Bragg Reflection (DBR) mechanism for confinement. Further, to realize quantum optical phenomena in the semiconductor cavity QED systems, it is required to strongly couple the emitter to the cavity mode, which depends on the quality factor and the mode volume of the cavity.

\textbf{Quality factor of the cavity}:

The quality factor, Q of the cavity characterizes how efficiently the light is confined inside the cavity. It is defined as,

\begin{equation}
    Q=\omega\frac{\langle U\rangle}{\langle P\rangle}
\end{equation}

where $\omega$, $U$, and $P$ are the frequency, the average energy of the confined EM field mode, and the far-field radiation intensity, respectively \cite{Englund2005,Fabre2017}. The energy stored inside the cavity decays in time ($t$) as, $I(t)=I_0 e^{-\omega t/Q}=I_0 e^{-2\kappa t}$. Here, $\kappa=\frac{\omega}{2Q}$ is the cavity decay rate. The lower the value of $\kappa$, the higher the quality of the cavity.

\textbf{Mode volume}:

Mode volume of the cavity quantifies the cavity space in which the EM field is concentrated and is given by,

\begin{equation}
    V=\frac{\int d\textbf{r} \varepsilon(\textbf{r})|E(\textbf{r})|^2}{max(\varepsilon(\textbf{r})|E(\textbf{r})|^2)}
\end{equation}

\begin{figure}[h]
    \centering
    \includegraphics[width=\linewidth]{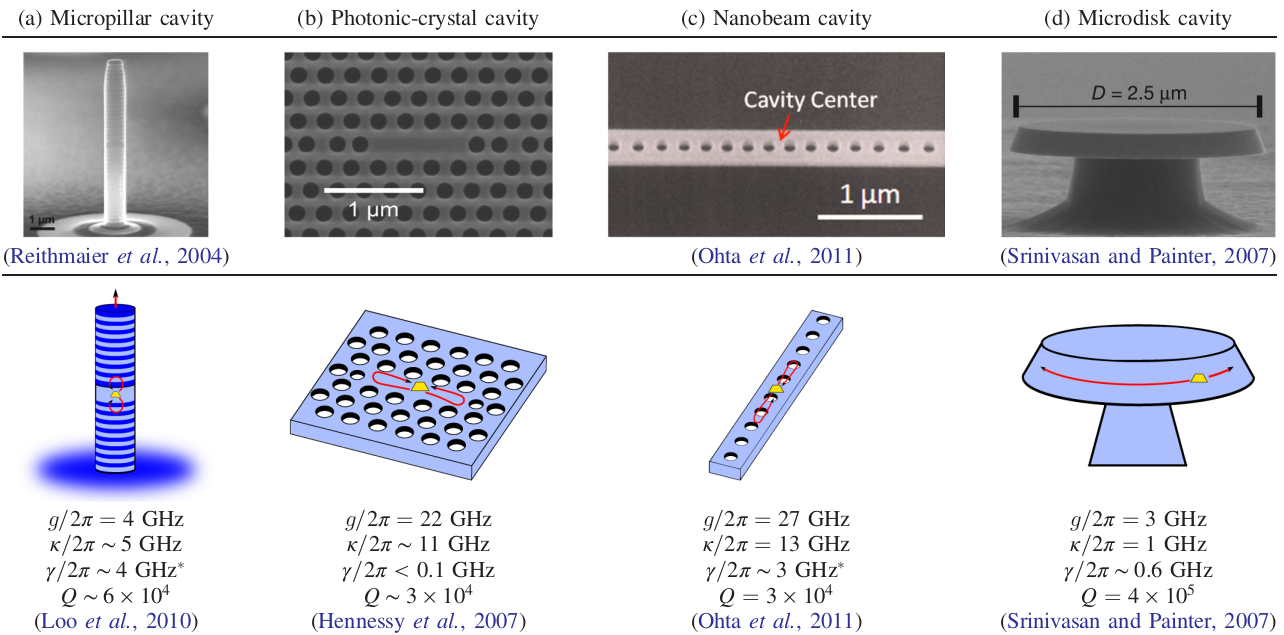}
    \caption{Source \cite{Lodahl2015}: Overview of nanophotonic cavities. Each panel displays a scanning electron micrograph of a real device along with a sketch illustrating the operational principle for a quantum emitter coupling to the structure. (a) Micropillar cavity. The Bragg stack above and below the center of the pillar confines light to the central region, as shown in the inset \cite{Reithmaier2004, Loo2010}. (b) Modified photonic-crystal L3 cavity implemented in a membrane. The photonic band gap localizes light in the defect region, and the schematic shows how a quantum dot preferentially emits into the cavity mode \cite{Hennessy2007}. (c) A nanobeam cavity. The cavity mode is confined by 1D Bragg diffraction in the high-refractive-index material of the nanorod \cite{Ohta2011}. (d) Microdisk cavity. The emitter couples to optical modes that travel circularly around the microdisk \cite{Srinivasan2007}.}
    \label{fig:QDPhc}
\end{figure}

Thus, it is desirable to have a high $Q$ and a small mode volume $V$, for confining a photon for long times inside the cavity with increased localization up to a single photon. Such cavities provide an ideal platform to realize strong light-matter interaction by embedding semiconductor QDs. In particular, photonic crystal cavities of a very high quality factor of the order of $\sim10^6-10^7$ have been demonstrated \cite{Asano2017,Muhammad2024}, providing an excellent platform for realizing cavity QED phenomena. Fig. \ref{fig:QDPhc} shows various ways to couple the QD to micropillar cavities, photonic crystal cavities, nanobeam cavities, and microdisk resonators.

Following various proposals for the generation of semiconductor QD lasers, there has been significant progress in cavity fabrication and also the tuning methods including temperature variation of the lattice \cite{Reithmaier2004, Press2007}, the application of magnetic fields \cite{Reitzenstein2009}, electrical tuning \cite{Laucht2009}, strain tuning \cite{SchmidtPRL2011, Plumhof2011, Sun2013}, and thin-film condensation techniques \cite{Mosor2005, Hennessy2006, Hennessy2007}. Therefore, with current technologies, it is possible to implement the systems discussed in this thesis. We considered single and two QDs coupled to single and bimodal cavities to study the generation of various quantum states of light. In the following section, we discuss light–matter interactions in cavity quantum electrodynamic (QED) systems.

\newpage
\section{Light matter interaction in QD-cavity system}

In this section, we consider the simplest case of a two-level atom interacting with a single-mode quantized electromagnetic field. We can extend the analysis to multi-mode and multi-level atoms.

Considering an electron of charge \textit{q} and mass \textit{m} interacting with an external electromagnetic field, is described by a minimal-coupling Hamiltonian,

\begin{equation}
	\mathcal{H}=\frac{1}{2m}[\textbf{p}-e\textbf{A(r,t)}]^2+eU(r,t)+V(r)
\end{equation}

where \textbf{A}(\textbf{r},t) and $U(\textbf{r},t)$ are the vector and scalar potentials of the external field, and \textbf{r} is the electron position vector from the center of the atom. The electrostatic binding potential between electron and nucleus located at \textbf{$r_0$} is given by \textit{V(r)}. In the scenarios where the size of the atom is small compared to the wavelength of the EM field, the dipole approximation can be applied,

\begin{equation*}
    \textbf{k}.\textbf{r}<<1
\end{equation*}

where \textbf{k} is the EM field mode wavevector. Using this approximation, the vector potential takes the form, 

\begin{equation}
	\begin{split}	
	\textbf{A}(\textbf{r}_0+\textbf{r},t) &=\textbf{A}(t)\exp[i\textbf{k}.(\textbf{r}_0+\textbf{r})]\\& \approxeq \textbf{A}(t)\exp[i\textbf{k}.\textbf{r}_0]
	\end{split}
\end{equation}

The Schr\"{o}dinger equation in the dipole approximation and in the radiation gauge, $U(\textbf{r},t)=0,\, \nabla.\textbf{A}=0 $ is given by,

\begin{equation}
    \Big[-\frac{\hbar^2}{2m}\Big(\nabla^2-\frac{ie}{\hbar}\textbf{A}(\textbf{r}_0,t)\Big)^2+V(\textbf{r)}\Big ] \psi(\textbf{r},t)=i\hbar \frac{\partial \psi(\textbf{r},t)}{\partial t}
\end{equation}

After applying the gauge transformation, $\chi(\textbf{r},t)=-e\textbf{A}(\textbf{r}_0,t).\textbf{r}/\hbar$,

\begin{equation}
    \textbf{A}(\textbf{r}_0,t)\rightarrow\textbf{A}(\textbf{r}_0,t)+\frac{\hbar}{e}\nabla\chi(\textbf{r},t) \,; \quad V(\textbf{r}) \rightarrow V(\textbf{r}); \quad \psi(\textbf{r},t) \rightarrow e^{-\frac{ie}{\hbar}\textbf{A}(\textbf{r}_0,t).\textbf{r}}\psi(\textbf{r},t)
\end{equation}

we can write the total Hamiltonian, $\mathcal{H}$ as,

\begin{equation}
	\mathcal{H}=H_0+H_I
\end{equation}

where, $H_0=\frac{p^2}{2m}+V(r)$ is the non-interacting part of the Hamiltonian and $H_I=-e\textbf{r}.E(\textbf{r}_0,t)$ is the interaction part.

\textbf{Jaynes-Cummings Model}:

The interaction between a two-level atom and a single-mode quantized electromagnetic field is described by a fully quantum mechanical treatment. From Eq. \ref{eqn:QEx}, the electric field operator is given by, $E = \hat{\epsilon}\mathcal{E} (a+a^\dagger)$ where $\mathcal{E}=(\frac{\hbar \omega_c}{2\epsilon_o V})^{1/2}$. The total Hamiltonian is given by,

\begin{equation}
	H = \hbar \omega_c a^\dagger a + \hbar \omega_0 \sigma^+\sigma^- + \hbar g(\sigma^+ + \sigma^-).(a + a^\dagger)
\end{equation}

where $\omega_c$ and $\omega_0$ are the cavity mode and atomic transition frequencies, respectively. $a$ ($a^\dagger$) is the photon annihilation (creation) operator, $\sigma^+=|e\rangle\langle g|$ is the atomic operator. 

The coupling strength between the atom and the cavity mode is given by, 

\begin{equation}
    g=\frac{(\textbf{p}.\hat{\epsilon})\mathcal{E}}{\hbar}
\end{equation}

where $\textbf{p}=\langle e|\hat{\textbf{p}}|g\rangle$ is the atomic dipole moment matrix element. In the rotating wave approximation, where the fast oscillating terms or energy non-conserving terms are neglected, the Jaynes-Cummings Hamiltonian is given by,

\begin{equation}
	H = \hbar \omega_c a^\dagger a + \hbar \omega_0 \sigma^+\sigma^- + \hbar g(\sigma^+ a+ \sigma^- a^\dagger)
\end{equation}

Further, we can consider the quantum emitter pumped coherently or incoherently using external drives.

\textbf{Coherent pumping}:

A coherent pumping scheme involves a classical laser field that resonantly or near-resonantly drives the emitter transition, thereby inducing coherence between the emitter states \cite{ScullyZubairyQOT}. The Hamiltonian term corresponding to the interaction between a two-level emitter and a monochromatic laser is given by,

\begin{equation}
    H_{drive}=\frac{\eta}{2}(\sigma^+ e^{i\phi}+\sigma^- e^{-i\phi})
\end{equation}

where, $\eta$ is the coherent pumping strength, $\phi$ is the laser phase.

\textbf{Incoherent pumping}:

Incoherent pumping refers to driving population into excited states without inducing a fixed phase relation between emitter states. The incoherent pumping mechanism can be realized via broadband (incoherent) optical pumping \cite{ScullyZubairyQOT, mu1992, Santori2002}, electrical injection \cite{Böckler2008}, Non-resonant optical pumping \cite{Ulrich2011} (e.g., pumping into higher bands or wetting layer and thereafter relaxation to the excitonic state), and interaction with thermal bath \cite{carmichael1999}.

The incoherent processes are usually described using the open quantum system formalism discussed in the next section. The term corresponding to the incoherent pumping of the emitter in the master equation is given by a Lindblad superoperator \cite{carmichael1999},

\begin{equation}
    L[\sigma^+]=\eta(\sigma^-\sigma^+\rho - 2\sigma^+\rho\sigma^-+\rho\sigma^-\sigma^+)
\end{equation}

In a semiconductor QD-cavity QED system, along with exciton-photon interaction, the interaction between exciton-phonons, as described in the earlier section, is also required to be considered. The interaction with the surrounding phonon bath leads to dissipation and decoherence, altering the system dynamics. Such an open quantum system is analyzed using master equation techniques presented in the next section.
 
\section{Open quantum systems}

The quantum state can either be a pure or a mixed state and can be described by a density matrix,

\textbf{Pure state}:
\begin{equation}
    \rho = |\Psi\rangle\langle\Psi|;\quad \text{where} \quad |\Psi\rangle=\sum_i \alpha_i|\psi_i\rangle
\end{equation}

\textbf{Mixed state}:

\begin{equation}
    \rho=\sum_i P_i |\psi_i\rangle\langle\psi_i| 
\end{equation}

where $P_i$ denotes classical probability with which the mixed quantum state is prepared in the bare state, $\psi_i$.

For example, a two-level system with basis states $|g\rangle$, $|e\rangle$. The density matrix is given by,

\begin{equation}
    \rho=\begin{pmatrix}
        \rho_{11} & \rho_{12}\\ \rho_{21} & \rho_{22}
    \end{pmatrix}
\end{equation}

where, $\rho_{11}=|g\rangle\langle g|$, $\rho_{12}=|g\rangle\langle e|$, $\rho_{21}=|e\rangle\langle g|$, $\rho_{22}=|e\rangle\langle e|$. The diagonal elements give information about the populations, and the off-diagonal elements represent the coherence. The density matrix is positive semi-definite and has properties, $Tr(\rho)=1$, preserving the normalization condition, and $Tr(\rho^2)=1$ or $T(\rho^2)<1$ for pure or mixed states.

The evolution of the quantum state is given by the Schr\"{o}dinger equation. In terms of the state vector, $|\psi\rangle$

\begin{equation}
    \frac{d|\psi\rangle}{dt}=-\frac{i}{\hbar}H|\psi\rangle 
    \label{eqn: SchrodEq}
\end{equation}

 and in the density matrix, $\rho$ terms,

\begin{equation}
    \frac{d\rho}{dt} = -\frac{i}{\hbar}[H,\rho] 
    \label{eqn:LiouvilleEq}
\end{equation}

Eq. \ref{eqn:LiouvilleEq} is known as the Liouville or von-Neumann equation. The evolution of populations and coherences of the quantum system can be obtained using the above equation. However, Eq. \ref{eqn:LiouvilleEq} applies for a ``\textit{closed system}" where the system doesn't interact with the environment.

When a quantum system $\bm{S}$ interacts with a bath $\bm{B}$ having infinitely many degrees of freedom, they form an ``\textit{open quantum system}.” The evolution of the system state is then governed not only by coherent dynamics but also by decoherence and dissipation induced by the interaction with the bath. In general, such dissipative dynamics are described by the Kraus operator-sum representation (OSR), where the system evolves according to completely positive trace-preserving (CPTP) maps \cite{Andersson2007}. From the Kraus OSR, one can obtain the Schr\"{o}dinger equation representing coherent evolution and the Lindblad equation for incoherent processes as special cases. In this work, we derive the reduced density matrix rate equation for the system interacting with the environment, following closely ``\textit{Statistical Methods in Quantum Optics 1}" by H. J. Carmichael \cite{carmichael1999}.

The total Hamiltonian for the system and bath is written as,

\begin{equation}
    H=H_s+H_b+H_{sb}
\end{equation}

consisting of system Hamiltonian, $H_s$, Bath Hamiltonian, $H_b$ and the system-bath interaction Hamiltonian, $H_{sb}$. The Schrodinger equation for the total density matrix is given by, 

\begin{equation}
\dot{\rho}= -\frac{i}{\hbar}[H,\rho(t)]
\end{equation}

In the interaction picture, the density matrix can be written as, 

\begin{equation}
\tilde{\rho}=e^{i/\hbar(H_s+H_b)t}\rho(t)e^{-i/\hbar(H_s+H_b)t}
\end{equation}
whose time evolution is given by,

\begin{align}
\dot{\tilde{\rho}}&=i/\hbar(H_s+H_b)\tilde{\rho}-i/\hbar\tilde{\rho}(H_s+H_b)+e^{i/\hbar(H_s+H_b)t}\dot{\rho}e^{-i/\hbar(H_s+H_b)t}\\
\dot{\tilde{\rho}}&=-i/\hbar[\tilde{H}_{sb},\tilde{\rho}]
\end{align}

Upon formal integration,

\begin{equation}
\tilde{\rho}(t)=\tilde{\rho}(0)-i/\hbar\int_0^t dt' [H_{sb}(t'),\tilde{\rho}(t')]
\end{equation}

We apply the first important approximation to obtain the system dynamics, the Born approximation for the system-bath interaction part of the Hamiltonian. Accordingly, we consider terms up to $\mathcal{O}(H_{sb}^2)$ in the density matrix rate equation. We assume, the bath is not affect due to its large degrees of freedom and its short correlation time, therefore, writing $\tilde{\rho}(t)=\tilde{\rho}_s (t)(\rho_b)_0+\mathcal{O}(H_{SR})$, neglecting the second and other higher order terms, the density matrix is written as, 

\begin{equation}
\tilde{\rho}(t)=\tilde{\rho}(0)-i/\hbar\int_0^t dt' [H_{sb}(t'),\tilde{\rho}_s(t')(\rho_b)_0]
\end{equation}

The second important approximation in the analysis of open quantum systems is the Markov approximation, which is valid for a large number of quantum optics problems. In the Markovian approximation, we consider that there is no effect of past state on the present evolution of the system density matrix. This approximation is usually valid when the system and bath are weakly coupled. Also, the system decay rates are smaller than the inverse correlation time of the fluctuating forces acting on the system due to interaction with the bath. Therefore, we replace $\rho_s(t')$ by $\rho_s(t)$.

\begin{equation}
\tilde{\rho}(t)=\rho(0)-i/\hbar\int_0^t dt' [H_{sb}(t'),\tilde{\rho}_s(t)(\rho_b)_0]
\end{equation}

To obtain the master equation for the reduced density matrix of the system, we trace over the bath states,

\begin{equation}
\tilde{\rho}_s(t)=\tilde{\rho}_s(0)-i/\hbar\int_0^t dt' Tr_b [H_{sb}(t'),\tilde{\rho}_s(t)(\rho_b)_0]
\end{equation}

\begin{equation}
\dot{\tilde{\rho}}_s(t)=-i/\hbar Tr_b[\tilde{H}_{sb}(t),\tilde{\rho}(0)]-i/\hbar\int_0^t dt' Tr_b [\tilde{H}_{sb}(t),\tilde{H}_{sb}(t'),\tilde{\rho}_s(t)(\rho_b)_0]
\end{equation}

Depending on the form of the interaction Hamiltonian, $H_{sb}$, we obtain bath-induced two-time correlation functions.

For example, the master equation for the two-level system coupled to a single-mode cavity with coupling strength `$g$'. Also, consider that both atom and cavity modes are interacting with surrounding electromagnetic field modes, leading to the decay of atomic excitation and the leakage of cavity mode photons. The system, bath, and the system-bath interaction Hamiltonian are given by,

\begin{align}
    &H_s = \hbar \omega_a \sigma^+\sigma^- + \hbar\omega_c a^\dagger a + \hbar g(\sigma^+a+\sigma^-a^\dagger)
    \\& H_b = \sum_{\textbf{k},\zeta_\textbf{k}} \hbar \omega_\textbf{k} b_{\textbf{k},\zeta_\textbf{k}}^\dagger b_{\textbf{k},\zeta_\textbf{k}}
    \\& H_{sb} = \sum_{\textbf{k},\zeta_\textbf{k}}\lambda_{\textbf{k},\zeta_\textbf{k}}^a( \sigma^+b_{\textbf{k},\zeta_\textbf{k}}+\sigma^-b^\dagger_{\textbf{k},\zeta_\textbf{k}}) + \lambda_{\textbf{k},\zeta_\textbf{k}}^c( a^\dagger b_{\textbf{k},\zeta_\textbf{k}}+ab^\dagger_{\textbf{k},\zeta_\textbf{k}})
\end{align}

Here the atomic operators, $\sigma^+=|e\rangle\langle g|$, $\sigma^-=|g\rangle\langle e|$, the cavity mode annihilation (creation) operator is denoted by $a\,(a^\dagger)$ and the bath (electromagnetic field modes) annihilation (creation) operator for mode \textbf{k} and polarization, $\zeta_\textbf{k}$ is $b_{\textbf{k},\zeta_\textbf{k}}\, b_{\textbf{k},\zeta_\textbf{k}}^\dagger$. The coupling strength of atom to the cavity mode, $g$, the coupling strength of atom to the bath modes, $\lambda_{\textbf{k},\zeta_\textbf{k}}^a$ and the cavity mode with the bath modes is denoted by $\lambda_{\textbf{k},\zeta_\textbf{k}}^c$

We can obtain the master equation (Eq. \ref{eqn:dampedAHO}) for this system following the procedure presented above \cite{carmichael1999}.

\begin{equation}
    \begin{split}
    \dot{\rho_s} = &-\frac{i}{\hbar}[H,\rho_s]-\frac{\kappa}{2}L[a]\rho_s-\frac{\gamma}{2}L[\sigma_i^-]\rho_s
    \end{split}
    \label{eqn:dampedAHO}
\end{equation}

where $L[O]$ is the Lindblad superoperator having the form,

\begin{equation}
    L[O]=O^\dagger O\rho-2O\rho O^\dagger+\rho O^\dagger O 
\end{equation}

The spontaneous emission rate:

\begin{equation}
    \gamma=2\pi \int d^3k\, D(\textbf{k})\,|\lambda^a(\textbf{k},\zeta)|^2 \delta(kc-\omega_a)
\end{equation}

The cavity decay rate:

\begin{equation}
    \kappa=2\pi \int d^3k\, D(\textbf{k})\,|\lambda^c(\textbf{k},\zeta)|^2 \delta(kc-\omega_c)
\end{equation}

Here, `$\gamma$' is spontaneous emission rate and `$\kappa$' is the photon leakage rate which depend on the coupling strengths $\lambda_{\textbf{k},\zeta_\textbf{k}}^a, \, \lambda_{\textbf{k},\zeta_\textbf{k}}^c$ and the spectral density of the surrounding electromagnetic modes, $D(\textbf{k})$. In this thesis, we phenomenologically include incoherent processes such as spontaneous emission, incoherent pumping, cavity decay, and pure dephasing using Lindblad superoperators.

\newpage
\section{Superradiance, subradiance and Hyperradiance}\label{sec:Superradiance}

\begin{figure}[h]
    \centering
    \includegraphics[width=0.75\linewidth]{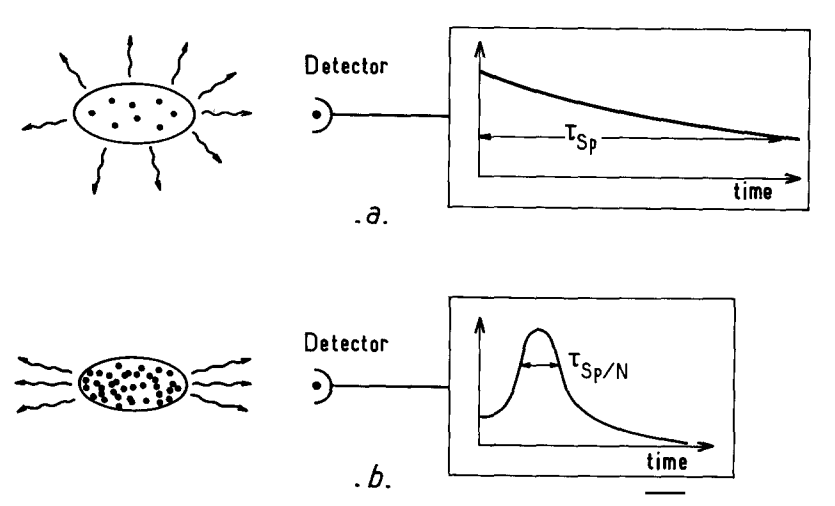}
    \caption{Source: \cite{gross1982} Comparison between ordinary fluorescence and superradiance (a) Ordinary spontaneous emission
is essentially isotropic with an exponentially decaying intensity (time constant $\tau_{sp}$) (b) Superradiance is anisotropic with an emission occurring in a short burst of duration $\sim \tau_{sp}/N$.}
    \label{fig:superRadiance}
\end{figure}

R. H. Dicke introduced the phenomenon of ``\textit{Superradiance}" \cite{Dicke1954}, which describes a coherent spontaneous emission from an ensemble of emitters. This is a result of the correlation between the emitters due to dipole-dipole coupling or coupling to the same radiation mode. The emitted radiation is stronger and faster than the emission from the individual emitters, and its intensity scales $\propto N^2$ compared to the $\propto N$ for the independent case \cite{gross1982}. The comparison between the ordinary fluorescence and superradiant emission from an ensemble of atoms is qualitatively shown in Fig. \ref{fig:superRadiance}

Let us consider the qualitative description of the superradiance in a system of $N$ two-level atoms. The ground and excited states of the two-level atom are denoted by $|g\rangle$ and $|e\rangle$, and the spontaneous emission rate, $\gamma$. These atoms are symmetrically coupled to the same radiation mode, i.e., there is no difference upon the exchange of atoms. The raising and lowering operators are $\sigma_i^+=|e\rangle\langle g|$ and $\sigma_i^-=|g\rangle\langle e|$ respectively obeying the commutation relation, $[\sigma_i^+,\sigma_i^-]=2\sigma_i^z$, where $\sigma_i^z=\frac{1}{2}(|e\rangle\langle e|-|g\rangle\langle g|)$.

The collective raising and lowering operators for an $N$ two-level atomic system are given by,

\begin{align}
    &S^+ = \sum_i \sigma_i^+ ; \qquad S^- = \sum_i \sigma_i^-
    \\& S^z = \sum_i \sigma_i^z; \qquad S^2 = \frac{1}{2}(S^+S^-+S^-S^+)+(S^z)^2
\end{align}

There are in total $N+1$ symmetric states for this collection of $N$ two-level (spin-$1/2$) systems, which are invariant under the permutation. They are given by $|\psi\rangle=|e, e,.....e\rangle$, and the others are generated upon the action of the de-excitation operator,

\begin{equation}
    |jm\rangle = \sqrt{\frac{(j+m)!}{N!(j-m)!}}\sum_i (\sigma_i^-)^{j-m}|e,e,......e\rangle
\end{equation}

Here, $j=N/2$ and $-j\leq M\leq j$. The state $|jm\rangle$ is the fully symmetric state with a total of $j+m$ atoms excited and $j-m$ in the ground state. $|jm\rangle$ is also the eigenstate of the operators $S^2$ and $S^z$.

\begin{equation}
    S^2|jm\rangle = j(j+1)|jm\rangle; \qquad S^z|jm\rangle = M|jm\rangle
\end{equation}

and,

\begin{equation}
    \langle jm|S^+S^-|jm\rangle=j+m; \qquad \langle jm|S^-S^+|jm\rangle = j-m
\end{equation}

We know that the emission intensity from a single atom is $I_0 \propto \gamma \langle \sigma_i^+\sigma_i^-\rangle$. Similarly, for the $N$ atomic system, 

\begin{equation}
    I_N \propto \gamma\langle S^+ S^-\rangle
\end{equation}

For the system in the symmetric state, $|JM\rangle$, the total intensity, $I_N\propto \gamma(j+m)(j-m+1)$. For the case of $m=0$, 
\begin{equation}
    I_N\propto \gamma \frac{N}{2}(\frac{N}{2}+1)
\end{equation}
i.e., $I_N \propto N^2$ in a superradiant emission.

Apart from superradiance, collective effects in the multi-atomic systems can also show subradiance, $<NI_0$ and ``\textit{Hyperradiance}", $>N^2 I_0$.

\subsection{Superradiant laser}

In contrast to the superradiant pulses produced by the $N$ emitters prepared in Dicke state, which is a transient behavior, a steady-state output can be realized with multiple emitters coupled to a common cavity mode, resulting in superradiant lasing. In this regime, the mean cavity photon number scaling $\propto N^2$ and the linewidth narrows as $1/N^2$ \cite{Haake1993, Meiser2009}. Here, the emitters are coupled to the cavity mode with equal strength, establishing cooperativity between them, leading to strong suppression of intensity fluctuations compared to an ordinary laser without cooperativity between the emitters \cite{Oh2024,Norcia2016}.

For a two-emitter coupled to a cavity mode system, a parameter ``\textit{Radiance Witness}"(RW) is defined to know whether the system shows subradiance, superradiance, or hyperradiant behavior.

\begin{equation}
    RW=\frac{\langle a^\dagger a\rangle_2 -2\langle a^\dagger a\rangle_1}{2\langle a^\dagger a\rangle_1}
\end{equation}

Here, $\langle a^\dagger a\rangle_{1(2)}$ corresponds to the mean cavity photon number when 
one (two) quantum dot(s) are coupled to the cavity mode. Radiance witness is a good quantifier stating whether the cooperative effects between two quantum emitters can lead to suppression or enhancement in the coupled cavity mode field. Therefore, $RW<0$ corresponds to subradiance, $0<RW<1$ implies enhanced emission, $RW=1$, superradiance, and for ``\textit{Hyperradiance}", $RW>1$ \cite{Pleinert2017}.

In the case of coherently pumped two-emitter coupled to a single-mode cavity \cite{Pleinert2017}, the authors investigated the possibility of generation of Hyperradiance in the cavity field. Furthermore, in the off-resonant pumping scheme \cite{Xu2017}, the occurrence of Hyperradiance accompanied by anti-bunching behavior is studied. In this thesis, we extend these ideas and present the results for systems showing single-mode and two-mode Hyperradiant lasing in the incoherently pumped two QD-cavity QED systems.

\section{Entanglement}\label{sec:Entang}

In the course on quantum mechanics, we learn that a state of a single quantum system can be denoted by a state vector, $|\psi\rangle$, which in general can be a superposition of basis states. Lets say, a two-level system with basis states $|g\rangle$ and $|e\rangle$, the general state-vector is written as,

\begin{equation}
	|\psi\rangle = \alpha_1|g\rangle + \alpha_2 |e\rangle
\end{equation}

where $\alpha_1,\, \alpha_2 \in \mathbb{C}$ and $|\alpha_1|^2+|\alpha_2|^2=1$. Extending this idea, suppose we consider two such two-level systems, and then the basis states for this composite system are given as \cite{Plenio2014},

\begin{equation}
	|g_1\rangle|g_2\rangle, |e_1\rangle|g_2\rangle, |g_1\rangle|e_2\rangle, |g_1\rangle|g_2\rangle
\end{equation}

which are the direct product of the individual subsystem states. We can always define a general state of the composite system as a superposition of these basis states. For example,

\begin{equation}
	|\psi\rangle = \frac{|g_1\rangle |g_2\rangle +|e_1\rangle |e_2\rangle}{\sqrt{2}}
	\label{eqn:entState}
\end{equation}

It is also a typical form of an entangled state corresponding to two quantum systems, as this pure state cannot be written in the product form of individual subsystems. An intuitive analogy with the classical probability case of a collection of coins with $50\%$ ``\textit{hh}" and $50\%$ ``\textit{tt}". Such a collection cannot be written as a product of individual distributions, showing the presence of classical correlations. The quantum state, $|\psi\rangle$ (Eq. \ref{eqn:entState}), shows quantum correlations, and these quantum correlations are different from classical correlations. \textit{Entanglement} is indeed a form of quantum correlation and has a counterintuitive aspect recognized by Schr\"{o}dinger, i.e., the information provided by the entangled state about the whole system is much greater than that of the subsystems. This quantum information is related to the von Neumann entropy \cite{Schumacher1995} 

\begin{equation}
	S(\rho) = -\mathrm{Tr}\,\rho\, \log \,\rho
\end{equation}

and is later shown by Horodecki \cite{Horodecki1994}, that the entropy of the total system in an entangled state is less than the sum of the entropy of the subsystems, which is truly non-classical. Entanglement is a resource having a wide range of applications in various quantum technologies such as quantum computation, quantum metrology \cite{Giovannetti2006}, quantum cryptography \cite{Gisin2007quantumcomm}, etc. Entanglement can be realized between the discrete variables \cite{Horodecki2009}, continuous variables \cite{Eisert2003,Adesso2007}, or between a discrete and a continuous variable of the system \cite{Jeong2014, Takeda2015}.

\textbf{Discrete variable entanglement}:

For the simplest case of pure bipartite state, $|\psi_{AB}\rangle\, \in\, \mathcal{H}_{AB}=\mathcal{H}_A\otimes \mathcal{H}_B$ is said to be separable (inseparable) if the $|\psi_{AB}\rangle$ can (cannot) be written as product of subsystem states: 

\begin{equation}
	|\psi_{AB}\rangle=|\phi_A\rangle |\phi_B\rangle
\end{equation}

For bipartite systems satisfying the positive partial transpose (PPT) criterion as a necessary condition for separability is given by Peres \cite{Peres1996} and is shown to be also a sufficient condition for the $2\otimes 2,\, 2\otimes 3$ systems by Horodecki et.al., \cite{Horodecki1996}. Let $\rho_{AB}$ be the total density matrix of the bipartite system. The PPT criterion states that $\rho_{AB}$ is separable if the partially transposed matrix, $\rho_{AB}^{T_B}$, 

\begin{equation}
	\langle m|\langle \mu|\rho_{AB}^{T_B}|n\rangle |\nu \rangle=\langle m|\langle \nu|\rho_{AB}|n\rangle |\mu \rangle
\end{equation}

has non-negative eigenvalues and can also be stated in terms of partial transposition with respect to subsystem A, $\rho_{AB}^{T_A}$. Violation of the above condition suggests the presence of discrete-variable (DV) entanglement, which arises in systems with spin qubits such as atom-atom entanglement \cite{Antman2024}, trapped ions \cite{Turchette1998}, spontaneous parametric down conversion \cite{Bouwmeester1999},  polarization entangled photons from quantum dots \cite{Chen2018}, orbital angular momentum \cite{Mair2001}, and flux qubits in superconducting circuits \cite{Dicarlo2010}. 

Apart from the PPT criterion, more general entanglement witnesses are defined in terms of linear positive maps \cite{Horodecki1996} and quantifiers including distillable entanglement, Entanglement Cost, Concurrence, Entanglement of Formation, Relative entropy of entanglement, Logarithmic negativity, etc. For comprehensive discussions on quantum entanglement, see articles by Horodecki et.al.,\cite{Horodecki2009}, Plenio et.al., \cite{Plenio2014}.

\textbf{Continuous variable entanglement}:

Further, it is shown that the PPT criterion is also a necessary and sufficient condition for separability of $1\otimes 1$ Gaussian states \cite{DGCZ1,DGCZ2,Simon2000} and extended to $1\otimes n$ Gaussian states by \cite{Werner2001}. A widely used separability criterion for the entanglement between any bipartite continuous variable states is given by Duan et.al., \cite{DGCZ1} in terms of uncertainty relations. Consider a pair of continuous variables, position and momenta $q_A,\, p_A$ and $q_B, \, p_B$ following the commutation relations $[q_j,p_{j'}]=i\delta_{jj'}$ and define Einstein-Poldosky-Rosen (EPR) like variables, 

\begin{align}
	u &= |a|q_A+\frac{1}{a}q_B\\
	v &= |a|p_A-\frac{1}{a}p_B
\end{align}

where $a \in \mathbb{R}$ and $a\neq 0$. Then for any separable state, $\rho_{AB}$, the sum of the variances of the EPR-like variables satisfies the condition,

\begin{equation}
	\langle (\Delta u)^2\rangle + \langle (\Delta v)^2\rangle \geq a^2+\frac{1}{a^2}
\end{equation}

Violation of the above DGCZ criterion suggests the presence of CV entanglement in the system. CV entanglement arises in systems with infinite-dimensional Hilbert spaces where the relevant observables are position-momentum \cite{Kumar2023continuousvariable}, quadratures of the cavity fields \cite{Vendromin2021continuousvaraible}. Unlike the DV entanglement, there is no universal measure for CV entanglement; however, there are some proposals for generalized measures for pure and mixed CV entanglement \cite{Swain2022}. In this thesis, we explore the generation of CV entanglement in three-level or four-level single quantum dots coupled to bimodal photonic crystal cavities.

\section{Outline of the thesis}

Multi-photon lasing process is highly non-linear, and realizing such phenomena in a semiconductor quantum dot (QD) cavity quantum electrodynamic (QED) system is beneficial in realizing on-chip quantum technology. In earlier works, the laser rate equations obtained using Scully-Lamb quantum theory of lasers employed mean field approximations, neglecting emitter-cavity mode correlations in deriving the single and multi-photon emission and absorption rates. In the strong coupling regime, the correlations between emitter and cavity mode need to be considered to obtain the correct results for single and multi-photon emission, absorption rates. Moreover, in the semiconductor cavity QED systems, the interaction between the QD exciton and lattice phonons is inevitable, which can lead to exciton dephasing and other coherent and incoherent processes. In this thesis, we investigate multi-photon lasing in systems with one or two QDs strongly coupled to single or bimodal cavities. The laser rate equations are obtained without mean field approximations, and the exciton-phonon interactions are treated using the polaron transformation approach. The chapter-wise summary of the thesis is given below.

\textbf{Chapter 2}: In this chapter, we explore the cooperative two-photon lasing phenomenon in a system of incoherently or coherently driven two quantum dots strongly coupled to a single-mode photonic crystal cavity. The coupling of QDs to the same cavity mode can show ``\textit{Superradiance}" or ``\textit{Hyperradiance}" due to the quantum correlation establishment through photon exchange between emitters. We investigate the steady-state dynamics of the system using the polaron-transformed master equation. Thereafter, we derive a simplified master equation and obtain laser rate equations without mean field approximations, and evaluate the single and multi-photon emission and absorption rates.

\textbf{Chapter 3}: In this chapter, we highlight the effect of exciton-phonon interactions leading to two-mode two-photon lasing in the system with incoherently driven quantum dots (QDs) are coupled off-resonantly to the cavity modes. We study the steady-state dynamics using the polaron-transformed master equation and obtain the cavity photon statistics. We define a parameter, ``Radiance Witness (RW)", quantifying the enhancement in the cavity field. We also show that the phonon-induced two-mode two-photon process leads to ``two-mode Hyperradiant Lasing" and support this result by obtaining the emission and absorption rates from laser rate equations without mean field approximations for the two-mode two-photon process. We analyze the effect of various system parameters on RW and compare the results for with and without the coupling of the QDs to the second cavity mode. Finally, we present the results of RW for the case of resonant coupling of QDs to the cavity modes.

\textbf{Chapter 4}: In this chapter, we investigate the phenomenon of correlated emission lasing in a coherently driven single quantum dot coupled to a bimodal photonic crystal cavity. In a correlated emission laser (CEL), the transfer of coherence from the emitter states to coupled electromagnetic modes leads to suppression of quantum noise in the laser, driving it towards the vacuum noise limit (VNL). We derive a polaron-transformed master equation and analyze the steady-state dynamics. We show the noise quenching in the relative or average phase of the cavity modes by analyzing the variances of the corresponding Hermitian operators. We obtain the relative and average phase drift and diffusion coefficients from a Fokker-Planck equation in Glauber-Sudarshan `P' representation and show quenching in the diffusion coefficients. Apart from CEL, our findings also reveal the possibility of the generation of continuous variable entanglement between the cavity modes.

\textbf{Chapter 5}: This chapter explores the generation of non-degenerate two-photon lasing and continuous variable entanglement. Considering a single quantum dot (QD) biexciton is coupled to non-degenerate cavity modes, we examine the possibility of two-mode two-photon emission when the QD biexciton is incoherently or coherently pumped. We show that the cavity modes are dominantly populated via two-mode two-photon emission at low incoherent pumping and low temperatures. Further, in the coherent pumping scenario, at two-photon resonance conditions, we show that the single photon emission can be completely suppressed and the system acts like a non-degenerate two-photon laser, leading to suppression of cavity mode fluctuations. Finally, we considered two-photon coherent pumping of the QD biexciton, and the cascaded decay from the biexcitonic state to the ground state leads to the generation of continuous variable entanglement between the cavity modes.

\textbf{Chapter 6}: This chapter provides a summary of the thesis work and future outlook. 

%% file: chap2.tex
\chapter{Cooperative two-photon lasing}\label{chap2}
{\small This chapter is based on our works \\ ``\textsc{Cooperative two-photon lasing in two quantum dots embedded inside a photonic microcavity}"; \textbf{Lavakumar Addepalli} and P. K. Pathak; \textit{Phys. Rev. B} \textbf{110}, 085408 (2024) \\
``\textsc{Hyperradiant lasing in the incoherently driven two quantum dots coupled to a single mode cavity}";\textbf{Lavakumar Addepalli} and P. K. Pathak; \textit{The European Physical Journal Special Topics 1-4 (2025).}}

\section{Introduction}

In this chapter, we propose cooperative two-photon lasing in the system with incoherently or coherently driven two quantum dots coupled to a single-mode photonic crystal cavity. The coupling of QDs to the same cavity mode can show ``\textit{Superradiance}" or ``\textit{Hyperradiance}" due to the quantum correlation establishment through photon exchange between emitters. Superradiance has been observed in various quantum systems such as trapped atoms\cite{goban2015,devoe1996}, superconducting qubits\cite{mlynek2014,Lambert2016}, including quantum dots\cite{scheibner2007,kim2018}. In a superradiant laser, steady-state superradiance has been achieved using an incoherent pump such that the correlation between emitters is maintained. 

On the other hand, in the coherently pumped QDs case, the coherent pump dresses the excitonic states, leading to some fascinating effects changing the absorption and emission properties of the system. For finite detuning between pump and emitter states, the population inversion is achieved between these dressed states. A dressed state laser has been realized by placing an ensemble of pump dressed emitters in a high-quality cavity that is resonant with the transition between the dressed states\cite{Zhu1987,Boone1989,Boone1990,Davidovich1987}. Also, two-photon gain in dressed-state lasers has also been realised \cite{Lewenstein1990,Law1991,Zakrzewski1991,Gauthier1992}. Recently, Neuzner \textit{et al.,} observed cooperative two-photon emission from cavity-dressed states of two identical atoms coupled with optical cavity\cite{Rempe2016}. Further, phonon-mediated cooperative two-photon emission from two-QDs embedded in a photonic microcavity \cite{verma2018} and photon-mediated cooperative two-photon emission from two-QDs into free space\cite{Koong2022}, using various pumping schemes, has been investigated. Highly efficient single QD laser in weak \cite{Reitzenstein2008,Strauf2006,Nomura2009} and strong coupling \cite{Nomura2010laser, Elena2009} schemes has been demonstrated. 

Unlike in the atomic systems, exciton-phonon interactions (EPI) are inevitable in semiconductor QDs that lead to decoherence, and the phonon-mediated cavity mode feeding \cite{Hennessy2007,Kaniber2008,Senellart2009,Dalacu2010,Calic2011}, and excitation-induced dephasing phenomena\cite{hughes2011mollow} predominantly. Various other phenomena, such as population inversion in two-level QDs, have also been realised due to exciton-phonon interactions\cite{Quilter2015,Ardelt2014,Hughes2013}. Thereby, the exciton-phonon coupling at low temperatures is essential to be included in the dynamics of QD-CQED systems\cite{nazir2016,roy2011}. We also note that QDs are tuned to the resonance of the coupled cavity mode employing tuning methods using electric field or magnetic field\cite{Laucht2010, Kim2011, Calic2017}.

In earlier studies on cooperative two-photon emission, photon bunching has been observed in second-order photon-photon correlation. However, cooperative two-photon lasing has not been explored in previous works, even in atomic systems. Here, we propose cooperative two-photon lasing in two QDs strongly coupled to a single-mode photonic crystal cavity, with the QDs driven either incoherently or coherently using an external pump. Additionally, we show that the system exhibits ``\textit{Hyperradiant lasing}" under incoherent pumping and subradiant behavior under coherent pumping, due to the cooperative effects present in the system. We obtain the single-photon, multi-photon emission and absorption rates in steady-state exactly without mean field approximations using the reduced system density matrix master equation.

\section{\label{sec:TPLIncoh}Incoherently pumped two-QDs-cavity system}

\begin{figure*}
    \centering
    \includegraphics[width=\textwidth]{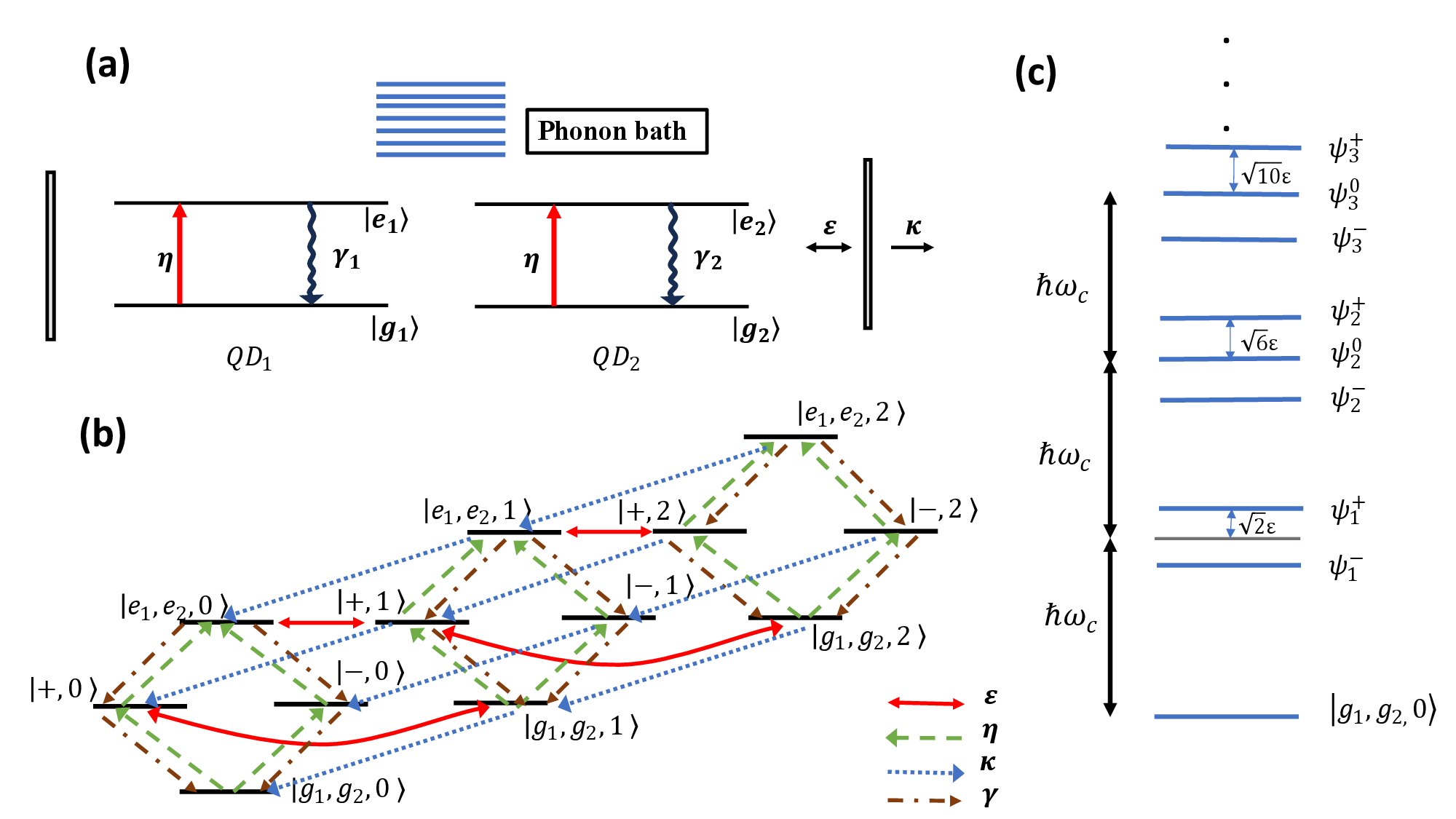}
    \caption{a) Schematic for energy levels of two-QD embedded in a single-mode microcavity with external pumping. b) Transitions between collective QDs-cavity states induced by cavity coupling ($\varepsilon)$, incoherent pumping($\eta$), cavity decay($\kappa$) and spontaneous emission($\gamma$) are shown (Phonon induced processed are not labeled). c) The cavity dressed two-QD states for $\Delta=0$. The dressed state are given by, $\psi_1^\pm=\frac{\ket{+,0}\pm\ket{g_1,g_2,1}}{\sqrt{2}}$, $\psi_2^0=\frac{-\sqrt{2}\ket{e_1,e_2,0}+\ket{g_1,g_2,2}}{\sqrt{3}}$, $\psi_2^\pm=\frac{\ket{e_1,e_2,0}\pm\sqrt{3}\ket{+,1}+\sqrt{2}\ket{g_1,g_2,2}}{\sqrt{6}}$, $\psi_3^0=\frac{-\sqrt{3}\ket{e_1,e_2,1}+\sqrt{2}\ket{g_1,g_2,3}}{\sqrt{5}}$, $\psi_3^\pm=\frac{\sqrt{2}\ket{e_1,e_2,1}\pm\sqrt{5}\ket{+,2}+\sqrt{3}\ket{g_1,g_2,3}}{\sqrt{10}}$ \cite{Addepalli2024}.}
    \label{fig:chap2/Fig1}
\end{figure*}

We consider two separate QDs coupled to a single-mode photonic-crystal cavity. In Fig. \ref{fig:chap2/Fig1}(a), the schematic diagram of exciton states in two QDs coupled to a single-mode cavity and a common phonon bath is shown. The QDs are pumped incoherently using external fields.

\subsection{Model system}

The Hamiltonian for the system in the rotating frame is given by \cite{Laucht2010, roy2011},                        
\begin{equation}
    \begin{split}
     H =& \hbar\delta_1 \sigma_1^+ \sigma_1^- + \hbar\delta_2\sigma_2^+\sigma_2^- + \hbar(\varepsilon_1\sigma_1^+ a+\varepsilon_2\sigma_2^+ a+H.C)\\& + H_{ph},
    \end{split}
 \label{eqn:Ham}
\end{equation}
 where, the detuning $\delta_i = \omega_i - \omega_c$, with $\omega_i$ and $\omega_c$ are the transition frequency between ground state $\ket{g_i}$ to excitonic state $\ket{e_i}$ for $i-{th}$ QD and cavity mode frequency, respectively. The lowering and raising operators for QDs are given by $\sigma_i^+ = \ket{e_i}\bra{g_i}$, $\sigma_i^- = \ket{g_i}\bra{e_i}$ and $\varepsilon_i$ is the exciton-cavity mode coupling constant, $a$ is the cavity field operator. The last term in the Hamiltonian (\ref{eqn:Ham}) represents the exciton and longitudinal acoustic phonon interaction , $H_{ph} = \hbar\Sigma_k \omega_k b_k^\dagger b_k + \hbar\Sigma_{k,i} \lambda_k^i \ket{e_i}\bra{e_i}(b_k + b_k^\dagger)$. Here, $b_k$ is the field operator of phonon mode of frequency $\omega_k$ and $\lambda_k^i$ is the coupling strength of exciton $\ket{e_i}$ to the phonon mode. We derive the polaron transformed master equation \cite{Mahan1990} to include exciton-phonon interaction in QD-cavity dynamics. The charge carries (electrons) dressed up by the surrounding phonons, resulting in a quasiparticle known as``\textit{Polaron}". This transformation allows us to include the multi-phonon effects into the system by changing the basis to polaron frame. Polaron transformation for the Hamiltonian(\ref{eqn:Ham}) is given by $H'=e^S H e^{-S}$, with $S=\Sigma_i \sigma_i^+\sigma_i^-\Sigma_k \frac{\lambda_k^{i}}{\omega_k}(b_k^\dagger-b_k)$.  The transformed Hamiltonian can be written as $H' = H_s+H_b+H_{sb}$, where $H_s$ is  QD-cavity system Hamiltonian, $H_b$ is the phonon-bath Hamiltonian and $H_{sb}$ is the system-bath interaction Hamiltonian. 
 
\begin{equation}
H_s = \hbar\Delta_1\sigma_1^+\sigma_1^- + \hbar\Delta_2\sigma_2^+\sigma_2^- + \langle B \rangle X_g 
\label{eqn:Ham2}
\end{equation}
\begin{equation}
H_b = \hbar\Sigma_k\omega_k b_k^\dagger b_k
\end{equation}
\begin{equation}
H_{sb} = \zeta_g X_g + \zeta_u X_u 
\end{equation}

The electron-phonon bath interaction results in a polaron shift in the QDs states energy, $\Sigma_k \frac{(\lambda_k^i)^2}{\omega_k}$ which are absorbed in $\Delta_1, \Delta_2$. The phonon displacement operators are given by, $B_{\pm} = \exp[\pm \Sigma_k \frac{\lambda_k^{i}}{\omega_k}(b_k - b_k^\dagger]$, with $\langle B_{\pm} \rangle = \langle B \rangle$. For simplification, we have considered the QD-phonon bath coupled strength is identical for both the QDs, $\lambda_k^{1} = \lambda_k^{2}$. The system operators are, $X_g = \hbar(\varepsilon_1\sigma_1^+a + \varepsilon_2\sigma_2^+a)+H.C.$, $X_u = i\hbar(\varepsilon_1\sigma_1^+a+\varepsilon_2\sigma_2^+a)+H.C.$ and bath fluctuation operators around the bath mean equilibrium displacement, $\langle B\rangle$ are given by $\zeta_g = \frac{1}{2}(B_++B_- -2\langle B \rangle)$, $\zeta_u = \frac{1}{2i}(B_+ - B_-)$. Using the polaron transformed Hamiltonian, $H'$, and Born-Markov approximation, we derive the master equation for the QDs-cavity system \cite{roy2011} after tracing over the phonon bath states.

\subsection{Master equation}

The master equation for the density matrix of QDs-cavity system is given by,
\begin{equation}
    \begin{split}
   \dot{\rho_s} = &-\frac{i}{\hbar}[H_s,\rho_s]-L_{ph}\rho_s-\frac{\kappa}{2}L[a]\rho_s-\Sigma_{i=1,2}(\frac{\gamma_i}{2}L[\sigma_i^-]\\&+\frac{\gamma_i'}{2}L[\sigma_i^+\sigma_i^-]+\frac{\eta_i}{2}L[\sigma_i^+])\rho_s,
\end{split}
\label{eqn:incohME}
\end{equation}
where $L[\hat{O}]\rho = \hat{O^\dagger}\hat{O}\rho - 2\hat{O}\rho\hat{O^\dagger}+\rho\hat{O^\dagger}\hat{O}$ is the Lindblad super-operator. The second term in the master equation $L_{ph}\rho_s$ describes the phonon-induced processes changing the system dynamics is written as 
\begin{equation}
\begin{split}
    L_{ph}\rho_s = &\frac{1}{\hbar^2}\int_{0}^{\infty}d\tau \Sigma_{j=g,u}G_j(\tau)\\&\times[X_j(t),X_j(t,\tau)\rho_s(t)]+H.C.
\end{split}
\end{equation}
where $X_j(t,\tau)=e^{-iH_s\tau/\hbar}X_j(t)e^{iH_s\tau/\hbar}$, and bath-bath two time correlation functions, $G_j(\tau)=\langle\zeta_j(t)\zeta_j(t,\tau)\rangle_{bath}$, $G_g(\tau)=\langle B \rangle^2{\cosh(\phi(\tau)-1)}$, $G_u(\tau)=\langle B \rangle^2\sinh(\phi(\tau))$ also called polaron Green's functions. The phonon correlation function is given by,
\begin{equation}
    \phi(\tau)=\int_{0}^{\infty}d\omega\frac{J(\omega)}{\omega^2}[\coth(\frac{\hbar\omega}{2k_BT})\cos(\omega\tau)-i\sin(\omega\tau)],
\end{equation}
\par
where $k_B$ and $T$ are the Boltzmann constant and temperature of the phonon bath, respectively. The spectral density function of phonon bath is given by,
\begin{equation}
    J(\omega)=\Sigma_k(\lambda_k^{i})^2\delta(\omega-\omega_k)=\alpha_p\omega^3\exp[-\frac{\omega^2}{2\omega_b^2}]
\end{equation}
takes the latter form in the continuum limit. The electron-phonon coupling strength $\alpha_p$, depends on the deformation potential, and the cut-off frequency $\omega_b$ depends on the speed of sound and phonon wave-function profile. We considered spherically symmetric parabolic potentials for both the conduction and valence band Chapter 1, Sec. \ref{sec:QuantumConf}. In our calculations, we consider,

\begin{equation}
    \alpha_p=2.36\, ps^2; \qquad \omega_b=1\,meV
\end{equation}

which provide experimentally compatible values of $\langle B \rangle$=0.92, 0.9, 0.84 and 0.73 for $T$= 0K, 5K, 10K and 20K, respectively\cite{hughes2011mollow}. We also include Lindblad terms corresponding to cavity damping with decay rate $\kappa$, spontaneous exciton decay with rate $\gamma_i$, pure dephasing with rate $\gamma_i'$, and incoherent pumping with rate $\eta_i$. The master equation (\ref{eqn:incohME}) is then numerically integrated using the quantum optics toolbox \cite{SMTan1999} to obtain the steady-state populations (SSP) and cavity photon statistics.

\subsection{Steadystate and transient dynamics}

\begin{figure}
    \centering
    \includegraphics[width=\columnwidth]{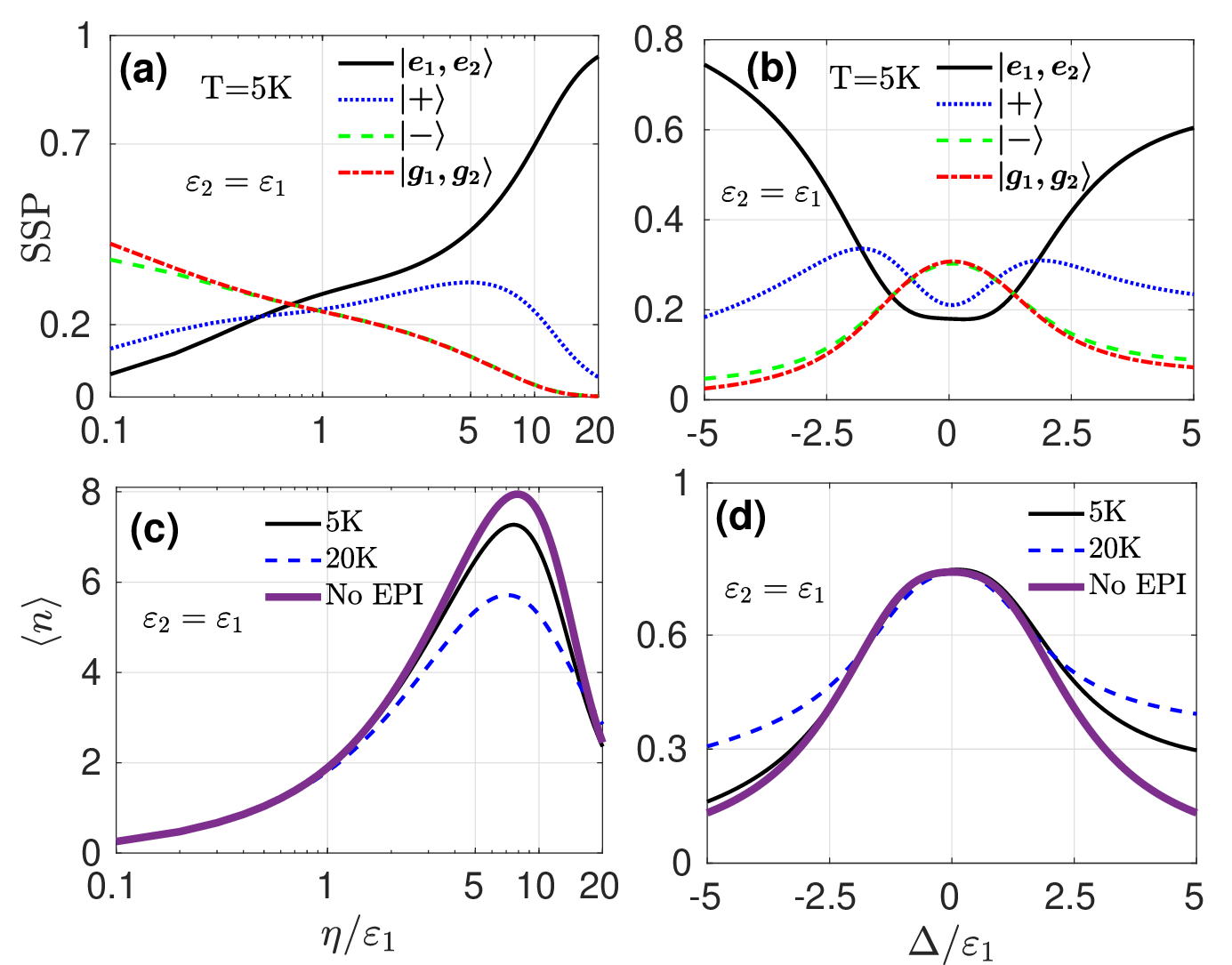}
    \caption{The steady-state populations (SSP) of the symmetrically coupled system ($\varepsilon_2=\varepsilon_1$) in collective QD states, $\ket{e_1,e_2}$ (solid black), $\ket{+}=(\ket{e_1,g_2}+\ket{g_1,e_2})/\sqrt{2}$ (dotted blue), $\ket{-}=(\ket{e_1,g_2}-\ket{g_1,e_2})/\sqrt{2}$ (dashed green) and $\ket{g_1,g_2}$ (dash-dotted red) in (a) $\&$ (b). In subplots (c) $\&$ (d), the cavity mean photon number, $\langle n\rangle$ for T=5K (solid black), T=20K (dashed blue), No EPI(thick solid purple). In (b) $\&$ (d) $\Delta_1 = \Delta_2=\Delta$, incoherent pumping rate, $\eta_1=\eta_2=0.35\varepsilon_1$.   In (a) $\&$ (c) incoherent pumping rate $\eta_1=\eta_2=\eta$ and detuning, $\Delta_1=\Delta_2=\Delta=0.0\varepsilon_1$. Other parameters, cavity decay rate $\kappa=0.5\varepsilon_1$, spontaneous emission rate $\gamma_1=\gamma_2=0.01\varepsilon_1$, pure dephasing rate $\gamma_1'=\gamma_2'=0.01\varepsilon_1$.}
    \label{fig:chap2/Fig2}
\end{figure}

The SSP in two-QD states and average photons in cavity mode are shown in Fig.\ref{fig:chap2/Fig2}.  In Fig.\ref{fig:chap2/Fig2}(a) and (c), the steady-state populations and mean cavity photon number with respect to incoherent pumping rate, $\eta_1=\eta_2=\eta$  are presented. The QDs are resonantly coupled to the cavity mode with coupling strengths, $\varepsilon_1=\varepsilon_2=100\mu eV$. In Fig.\ref{fig:chap2/Fig2} (b) and (d), the results with respect to detuning, $\Delta_1=\Delta_2=\Delta$ for a fixed value of incoherent pumping rate $\eta_1=\eta_2=\eta$ are shown. 

 Fig. \ref{fig:chap2/Fig2}(a) $\&$ (c) show the results corresponding to variation in the incoherent pumping rate, considering both QDs are driven equally, $\eta_1=\eta_2=\eta$ and are resonantly coupled to cavity mode. From Fig. \ref{fig:chap2/Fig2}(a), we find that the steady-state population in state $\ket{e_1,e_2}$ start dominating for very small pump rate, $\eta<\varepsilon_1$ and increases on increasing pump rate. For the symmetric coupling case,  $\varepsilon_1=\varepsilon_2$, the state $\ket{+}=(\ket{e_1,g_2}+\ket{g_1,e_2})/\sqrt{2}$ is coupled with cavity mode and the state $\ket{-}=(\ket{e_1,g_2}-\ket{g_1,g_2})/\sqrt{2}$ remains uncoupled with cavity mode. The population in $\ket{+}$ increases monotonically on increasing pump rate. The state $\ket{-}$ is populated due to incoherent pumping and spontaneous decay of $\ket{e_1,e_2}$.
 
 Further, the steady-state populations in collective QD states, $\ket{g_1,g_2}$ and $\ket{-}$ remain equal and decrease monotonically on increasing pump strength. The collective QD states $\ket{e_1,e_2}$, $\ket{+}$, and $\ket{g_1,g_2}$ get dressed with the cavity field as shown in Fig.\ref{fig:chap2/Fig1}(c) for $\Delta=0$. It should also be noted that at a higher pump rate, the dressed states with the higher number of cavity photons get more populated. Various possible transitions between these dressed states can lead to the single and multi-photon emission into cavity mode as shown in Fig.\ref{fig:chap2/Fig1}(b). 
 
 The average photon number in cavity mode, $\langle n\rangle$, increases and attains a peak value on increasing pump rate, as shown in Fig. \ref{fig:chap2/Fig2}(c). On further increasing the incoherent pump rate, the mean photon number decreases and the intensity fluctuations increase due to the destruction of coherence in the system and leading to the \textit{self-quenching effect} \cite{mu1992}. The self-quenching also leads to rapid increase and decrease in the populations of $\ket{e_1,e_2}$ and $\ket{+}$ states, respectively. For higher temperature, T=20K, steady-state populations and cavity photon statistics follow a similar fashion. In Fig. \ref{fig:chap2/Fig2}(c), the peak in $\langle n\rangle$ (dashed blue), is smaller for T=20K than for T=5K (solid black) due to renormalization of cavity QD coupling by a factor $\langle B\rangle$. 
 
In Fig. \ref{fig:chap2/Fig2}(b) $\&$ (d), we plot steady-state populations and average cavity photon by varying the QDs detunings $\Delta_1=\Delta_2=\Delta$, for the typical value of pump rate, $\eta_1=\eta_2=\eta=0.35\varepsilon_1$.  We choose the incoherent pumping rate value corresponding to the single-photon excessive emission rate becoming zero (c.f. Fig. \ref{fig:chap2/Fig5}(a)) for $\Delta=0$.  In Fig. \ref{fig:chap2/Fig2}(b), we find that when the QDs are tuned to resonance with the cavity mode, $\Delta_1=\Delta_2=\Delta=0$, there is population transfer from the excited states $\ket{e_1,e_2}, \ket{+}$ to the population in ground state $\ket{g_1,g_2}$ resulting in the mean cavity photon number ($\langle n\rangle$) peak, Fig. 2(d). The asymmetry in the mean photon number curve is due to phonon-induced cavity mode feeding, which is more pronounced for positive detunings. Clearly, it suggests that QDs undergo transitions, $\ket{e_1,e_2} \rightarrow \ket{+} \rightarrow\ket{g_1,g_2}$ accompanied with two-photon emission into cavity mode. Since, there are no cavity assisted transitions, the anti-symmetric state, $\ket{-}$ is more populated than $\ket{+}$ when QDs are at resonance, $\Delta=0$ for low pumping rate. 

Comparing the results for the T=20K case with T=5K, the peak value of average photon number $\langle n\rangle$ is slightly reduced when the QDs are resonant, but the values are higher at higher temperature for off-resonant QDs due to phonon-assisted transitions. Without exciton phonon interaction, the $\langle n\rangle$ curve is symmetric with maxima at $\Delta=0$. It is to be noted that the system shows similar behavior for anti-symmetric coupling, $\varepsilon_2=-\varepsilon_1$ case with the fact that $|-\rangle$ state plays the role of $|+\rangle$.

\begin{figure}
    \centering
    \includegraphics[width=\columnwidth]{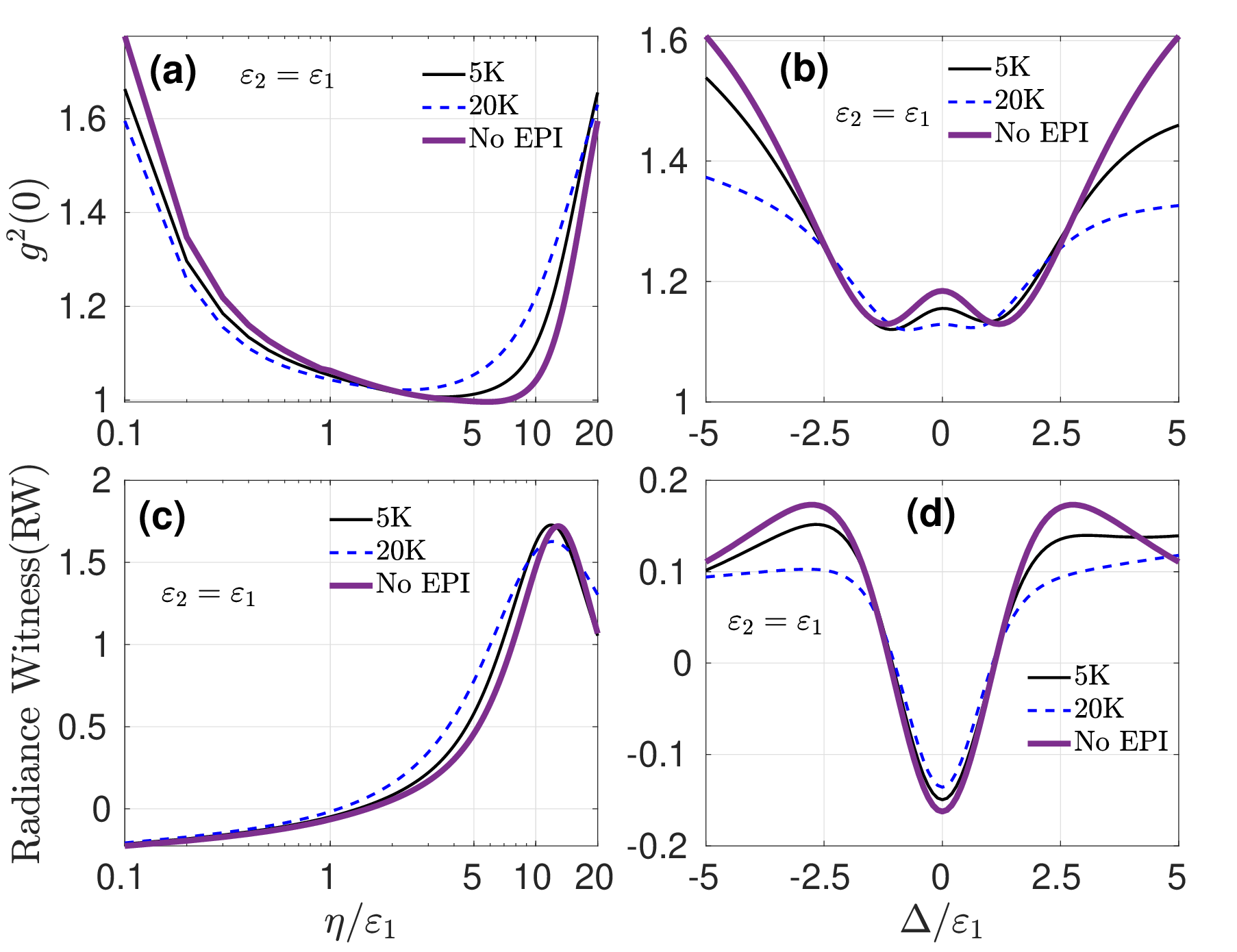}
    \caption{The zero time delay second-order photon correlation function, $g^2(0)$, in (a, b) and the radiance-witness (RW) in (c, d). The parameters are the same as in Fig.\ref{fig:chap2/Fig2}. The color scheme is the same as in Fig. \ref{fig:chap2/Fig2} (c,d).}
    \label{fig:chap2/Fig3}
\end{figure}

In Fig. \ref{fig:chap2/Fig3}, we present the results for the cavity photon statistics. In Fig. \ref{fig:chap2/Fig3} (a) $\&$ (b), we show the results for the zero-time delay second order cavity photon correlation function, $g^2(0)=\frac{\langle n^2\rangle - \langle n\rangle}{\langle n\rangle^2}$ varying the incoherent pumping rate, $\eta$ and QD-cavity detuning, $\Delta$ respectively. In Fig. \ref{fig:chap2/Fig3}(a), it is observed that $g^2(0)$ remains greater than one and less than two for all pumping rates and acquires a minimum value close to one when an average number of photons in the cavity mode $\langle n\rangle$ is maximum. This suggests an incoherently pumped system shows dominant multi-photon emission when both QDs are coupled resonantly with cavity mode. For a much stronger pump, QDs get decoupled with cavity mode leading to self-quenching and cavity field changes to thermal state, thus $g^2(0)$ approaches two. Earlier, Reid and Walls have shown that for a two-photon laser, $g^2(0)-1$ remains inversely proportional to average cavity photon $\langle n\rangle$\cite{Reid1983}. Further, when both QDs are resonantly coupled with the cavity mode, exciton-phonon interactions are not very significant; therefore, $g^2(0)$ remains relatively unchanged with temperature variations. In Fig. \ref{fig:chap2/Fig3} (b), variation of $g^2(0)$ with QD-cavity detuning is shown for a low incoherent pumping rate, $\eta=0.35\varepsilon_1$. We observe that when QDs are far-detuned from cavity mode $g^2(0)$ approaches two, indicating emission in the cavity mode in a thermal state. On decreasing detuning $\Delta_1=\Delta_2=\Delta$, $g^2(0)$ decreases indicating a reduction in cavity photon fluctuations. When both QDs become resonant with cavity mode, cavity mode-induced correlations lead to a small local peak in $g^2(0)$, which diminishes on increasing temperature due to phonon-induced dephasing. Earlier, similar behavior of $g^2(0)$ has been reported in single QD two-photon laser\cite{Elena2010}. Here we emphasize that correlation $g^2(0)$ in bad cavity limit, which is expressed in terms of emitter operators, for cooperative emission from two-atom\cite{temnov2009} and from two-QDs\cite{Koong2022} show quite different behavior and can only be comparable for pump strength much smaller than cavity coupling\cite{Rempe2016}. Additionally, for a two-photon coherent state, it has been shown that both $g^2(0)\geq 1$ and $g^2(0)<1$ are possible\cite{Yuen1976}. 

To examine the cooperative superradiant or subradiant behavior we plot ``Radiance Witness" in Fig. \ref{fig:chap2/Fig3} (c) $\&$ (d). The Radiance-Witness as defined in chapter 1, Sec. \ref{sec:Superradiance}, RW$=\frac{\langle a^\dagger a\rangle_2-2\times\langle a^\dagger a\rangle_1}{2\times\langle a^\dagger a\rangle_1}$ where $\langle a^\dagger a\rangle_1$ is the mean cavity photon number when only a single QD is coupled to cavity-mode, $\langle a^\dagger a\rangle_2$ is mean cavity photons when both QDs are coupled with cavity mode \cite{Pleinert2017}. RW$=0$ indicates uncorrelated (independent) emission, RW$>0$ indicates correlated superradiant emission, RW$<0$ indicates correlated subradiant emission from QDs in the cavity mode. In Fig.  \ref{fig:chap2/Fig3}(c), when the QDs are tuned to resonance with cavity mode ($\Delta=0$), we find that the cooperative emission into cavity mode makes a transition from subradiant(RW$<0$) to superradiant regime (RW$>0$) as the incoherent pumping rate increases. In fact, emission shows higher order dependence for large pumping than $N^2$, termed as hyperradiance\cite{Pleinert2017}. The results provided for different temperatures T= 5K, 20K, and without exciton-phonon interactions show similar behaviour. In Fig. \ref{fig:chap2/Fig3}(d), for a small pump rate, on decreasing cavity exciton detunings $\Delta$, the cooperative emission can change from superradiant to subradiant. In fact, we show in the latter section that the the cooperative two-photon emission into the cavity mode contribution leads to ``\textit{Hyperradiant lasing}" in this system.

\begin{figure}
    \centering
    \includegraphics[width=\columnwidth]{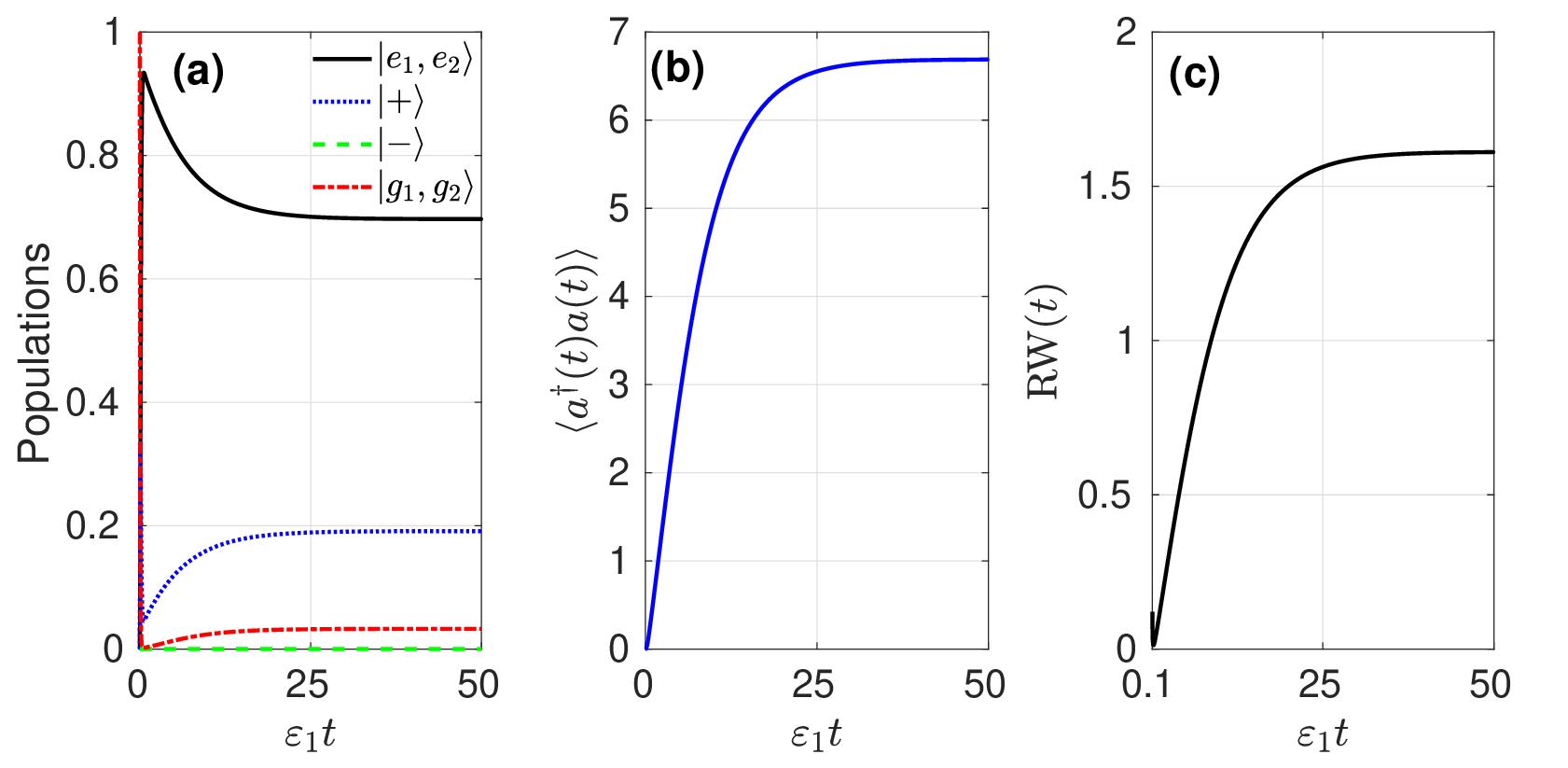}
    \caption{The transient behavior of (a) collective QD populations, (b) mean cavity photon number and (c) the radiance-witness (RW) for incoherent pumping rate, $\eta=10.0\varepsilon_1$, temperature, $T=5K$. The other parameters are the same as in Fig.\ref{fig:chap2/Fig2}.}
    \label{fig:chap2/transient_incoh}
\end{figure}

In Fig. \ref{fig:chap2/transient_incoh}, the transient dynamics of the system are shown for the incoherent pumping rate, $\eta=0.75\varepsilon_1$ and temperature T=5K. In Fig. \ref{fig:chap2/transient_incoh} (a), the population transfer from the state $|e_1,e_2\rangle$ to the states $|+\rangle$ and $|g_1,g_2\rangle$ leads to the cavity mode population, Fig. \ref{fig:chap2/transient_incoh} (b). Further, the result for radiance witness, RW$(t)$ is plotted in Fig. \ref{fig:chap2/transient_incoh} (c) and it gradually increases with time. The radiance witness of the cavity field attains a value of RW$\approx 1.6$ (\textit{Hyperradiant}) in the steadystate.

\subsection{Phonon-induced scattering rates}

\begin{figure}
    \centering
    \includegraphics[width=0.75\columnwidth]{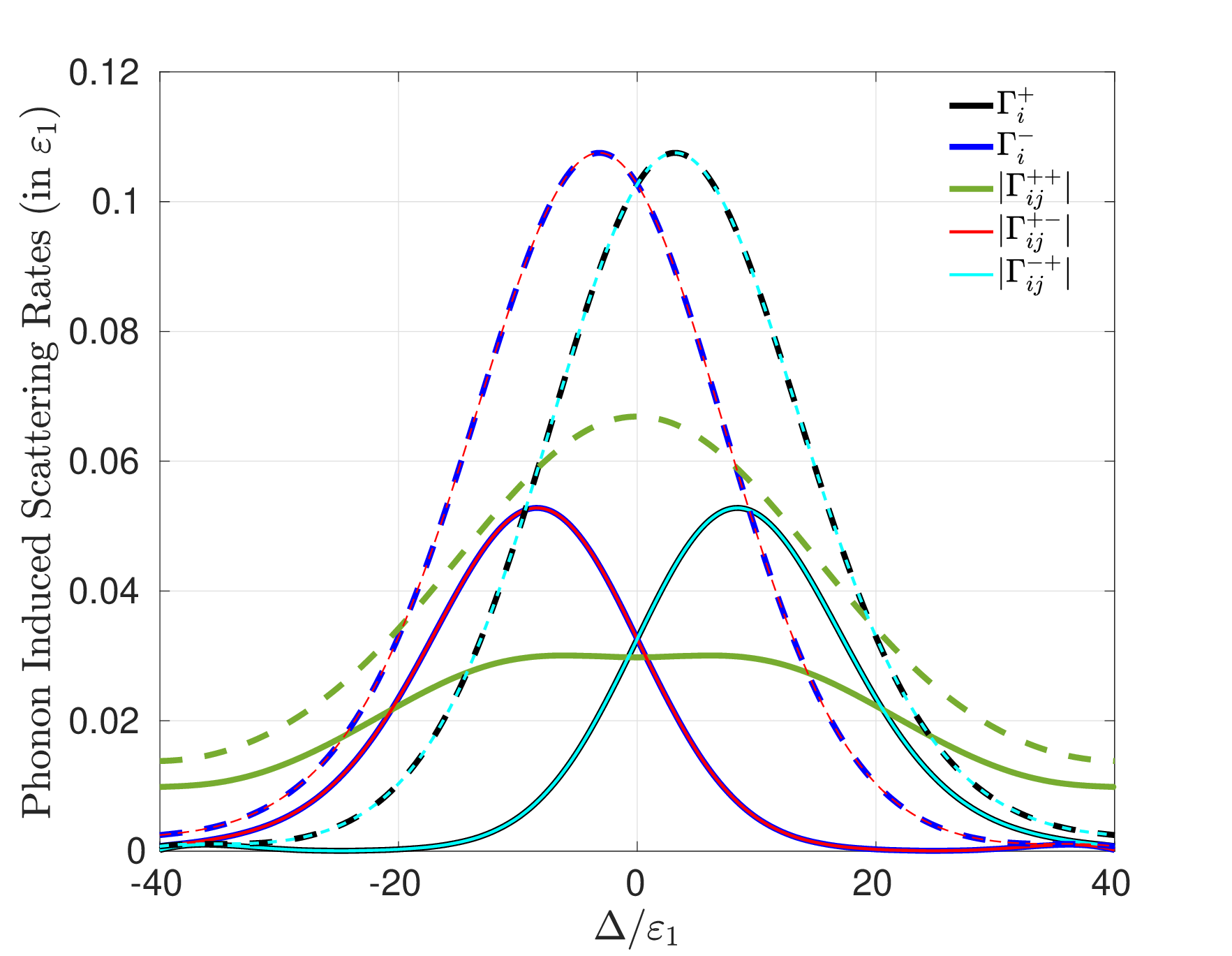}
    \caption{(Color Online)Phonon induced scattering rates for T=5K(solid) and T=20K(dashed). Coloring scheme, $\Gamma_i^+$(black), $\Gamma_i^-$(blue), $|\Gamma_{ij}^{++}|$(green), $|\Gamma_{ij}^{+-}|$(red), $|\Gamma_{ij}^{-+}|$(cyan). The parameters are same as in Fig. \ref{fig:chap2/Fig2}(b).}
    \label{fig:chap2/Fig4}
\end{figure}

In Fig. \ref{fig:chap2/Fig4} we plot the phonon-induced scattering rates for T=5K (solid) and T=20K (dashed). Considering both the QDs are identical and equally detuned from the cavity mode, it is clear from the definitions that $\Gamma_i^+$ and $|\Gamma_{ij}^{-+}|$ ($\Gamma_i^-$ and $|\Gamma_{ij}^{+-}|$) assume equal values and at $\Delta=0$ all four of them are equal. Further, the asymmetry in the scattering rates is anticipated and is more visible at low temperatures. The rates $\Gamma_i^+$, $|\Gamma_{ij}^{-+}|$  ( $\Gamma_i^-$, $|\Gamma_{ij}^{+-}|$ ) are larger when the QD-cavity detuning is negative(positive). The absolute value of the phonon-assisted two-photon process, $|\Gamma_{ij}^{++}|$ is symmetric w.r.t. $\Delta$. At higher temperatures, the scattering rates increase and are broadened.

\subsection{Simplified master equation}

To quantify photon emission into the cavity mode from single and multi-photon processes, we derive a laser rate equation using a simplified master equation(SME). The SME is an approximated Lindblad form of polaron transformed master equation (\ref{eqn:incohME}), which provides a clear picture of the various exciton-phonon scattering processes involved in the system. Phonon effects dominate when the quantum dot states are significantly detuned from the cavity mode. Therefore, we derive the SME under the condition, $\Delta_1,\ \Delta_2>> \varepsilon_1,\ \varepsilon_2$, and expand $L_{ph}\rho_s$, after making approximation $H_s\approx\hbar\Delta_1 \sigma_1^+\sigma_1^-+\hbar\Delta_2\sigma_2^+\sigma_2^-$ in $X_j(t,\tau)=e^{-iH_s\tau/\hbar}X_j(t)e^{iH_s\tau/\hbar}$. We arrange terms in the Lindblad form, which are proportional to $\varepsilon_1^2, \varepsilon_2^2$ and $\varepsilon_1\varepsilon_2$ in $L_{ph}\rho_s$\cite{verma2018}.
The simplified master equation is given by,
\begin{equation}
 \begin{split}
  \dot{\rho_s}=& -\frac{i}{\hbar}[H_{eff},\rho_s]-\frac{\kappa}{2}L[a]\rho_s
     -\Sigma_{i=1,2}(\Big[\frac{\eta_i}{2}L[\sigma_i^+])
    \\&+\frac{\gamma_i}{2}L[\sigma_i^-]
    +\frac{\gamma_i'}{2}L[\sigma_i^+\sigma_i^-]
    +(\frac{\Gamma_i^-}{2}L[\sigma_i^+a]+\frac{\Gamma_i^+}{2}L[\sigma_i^-a^\dagger])\Big]\rho_s
    \\&-\Big[\frac{\Gamma_{12}^{--}}{2}L[\sigma_2^+a,\sigma_1^+a]\rho_s
    +\frac{\Gamma_{12}^{++}}{2}L[a^\dagger\sigma_2^-,a^\dagger\sigma_1^-]\rho_s
    \\&+\frac{\Gamma_{12}^{-+}}{2}L[a^\dagger\sigma_2^-,\sigma_1^+a]\rho_s
    +\frac{\Gamma_{12}^{+-}}{2}L[\sigma_2^+a,a^\dagger\sigma_1^-]\rho_s
    \\&-\Omega_{11}^{--}\sigma_1^+ a\rho_s\sigma_1^+a -\Omega_{11}^{++}\sigma_1^-a^\dagger\rho_s\sigma_1^-a^\dagger
    +1 \leftrightarrow 2 \Big]
    \label{eqn:incohSME}
    \end{split}
 \end{equation}
Here, $L[\hat{O_1},\hat{O_2}]=\hat{O_2}\hat{O_1}\rho_s-2\hat{O_1}\rho_s\hat{O_2}+\rho_s\hat{O_1}\hat{O_2}$. The coherent evolution of the system density matrix, $\rho_s$ is given by the effective Hamiltonian, $H_{eff}$
\begin{equation}
\begin{split}
    H_{eff} = &H_s+\hbar\Sigma_{i=1,2}(\delta_i^-a^\dagger\sigma_i^-\sigma_i^+a+\delta_i^+\sigma_i^+a a^\dagger\sigma_i^-)
    \\&-(i\hbar\Omega_{2ph}\sigma_1^+\sigma_2^+a^2+H.C.)-(i\hbar\Omega_+\sigma_1^+a a^\dagger\sigma_2^- \\&+i\hbar\Omega_-a^\dagger\sigma_1^-\sigma_2^+a+H.C.)
\end{split}
\end{equation}
where $\delta_i^\pm$, $\Omega_{2ph}$, $\Omega_\pm$ correspond to Stark shifts, phonon mediated two QDs excitation(de-excitation) via two photon absorption(emission) into cavity mode and exciton exchange between QDs respectively. Further, $\Gamma_i^+$, $\Gamma_i^-$, $\Gamma_{ij}^{++}$, $\Gamma_{ij}^{--}$, $\Gamma_{ij}^{+-}$, $\Gamma_{ij}^{-+}$ are phonon-induced scattering rates of the incoherent processes such as phonon mediated QD excitation by absorption of cavity photon, QD de-excitation by emission of photon into cavity mode, phonon induced two-photon processes such as two QD excitation(de-excitation) by absorption(emission) from(into) cavity mode, exciton exchange, respectively. We also include terms including $\Omega_{ii}^{\pm\pm}$ that do not have Lindblad form. The terms discussed above are given below. 
\begin{equation}
    \delta_i^\pm=\varepsilon_i^2Im\left[\int_0^\infty d\tau G_+e^{\pm i\Delta_i\tau}\right]
\end{equation}
\begin{equation}
    \Omega_{2ph} = \frac{\varepsilon_1\varepsilon_2}{2}\int_0^\infty d\tau(G_- - G_-^*)(e^{-i\Delta_1\tau}+e^{-i\Delta_2\tau})
\end{equation}
\begin{equation}
    \Omega_\pm = \frac{\varepsilon_1\varepsilon_2}{2}\int_0^\infty d\tau(G_+e^{\pm i\Delta_2\tau}-G_+^*e^{\mp i\Delta_1\tau})
\end{equation}
\begin{equation}
    \Gamma_i^\pm = \varepsilon_i^2\int_0^\infty d\tau(G_+e^{\pm i\Delta_i\tau}+G_+^*e^{\mp i\Delta_i\tau})
\end{equation}
\begin{equation}
    \Gamma_{ij}^{\pm\pm} = \varepsilon_i\varepsilon_j\int_0^\infty d\tau(G_-e^{\pm i\Delta_j\tau}+G_-^*e^{\pm i\Delta_i\tau})
\end{equation}
\begin{equation}
    \Gamma_{ij}^{\pm \mp} = \varepsilon_i\varepsilon_j\int_0^\infty d\tau(G_+e^{\mp i\Delta_j\tau}+G_+^*e^{\pm i\Delta_i\tau})
\end{equation}
\begin{equation}
    \Omega_{ii}^{\pm\pm} = \varepsilon_i^2 \int_0^\infty d\tau(G_- + G_-^*)e^{\pm i\Delta_i\tau},
\end{equation}
where, $G_\pm=G_g\pm G_u$.

We compare steady-state populations and $\langle n\rangle$ calculated using the master equation, (\ref{eqn:incohME}) and SME (\ref{eqn:incohSME}), we find that SME is valid entirely in the range of detunings and pumping rate considered here (c.f. Fig.\ref{fig:chap2/Fig12}(a), (b)). We compare the results between ME and SME in Appendix \ref{sec:chap2_Appendix1}.

\begin{figure}
    \centering
    \includegraphics[width=\columnwidth]{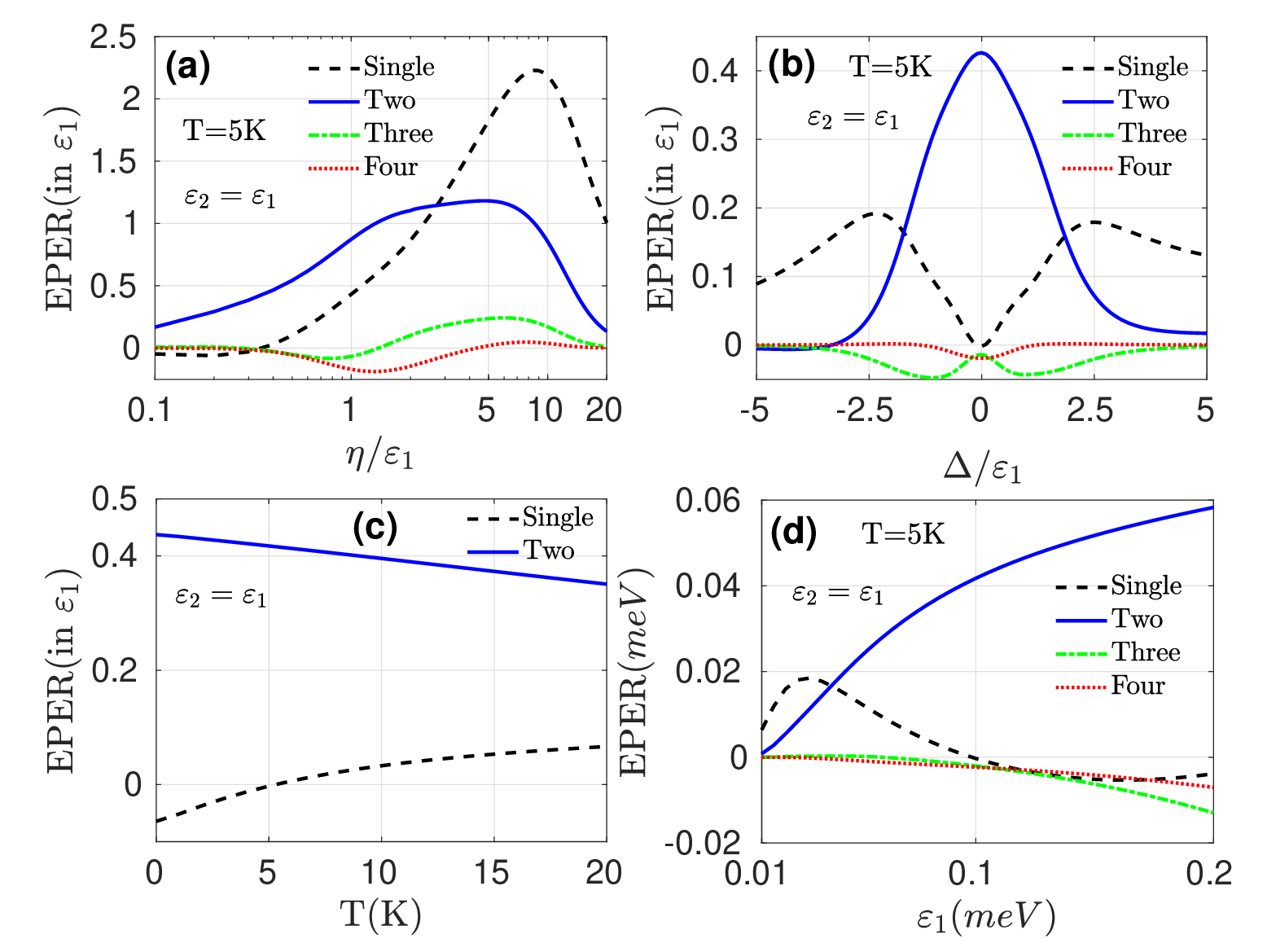}
    \caption{Excess photon emission rate (EPER) by varying (a) incoherent pumping rate, $\eta$, (b) detuning, $\Delta$, (c) temperature, T and (d) QD-cavity coupling strength, $\varepsilon_1$. The parameters are the same as in Fig.\ref{fig:chap2/Fig2}(a) and Fig.\ref{fig:chap2/Fig2}(b) for subplots (a, c, d) and (b), respectively. Color scheme: Single photon excess emission rate(dashed black), Two-photon excess emission rate (solid blue), Three photon excess emission rate (dotted red), Four photon excess emission rate (dash-dotted green).}
    \label{fig:chap2/Fig5}
\end{figure}

\subsection{Laser rate equations}

To obtain the quantum laser rate equation for the cavity field, we write the rate equations for both diagonal and off-diagonal QDs-cavity density matrix elements using SME (\ref{eqn:incohSME}). Following the quantum theory of lasers developed by Scully and Lamb \cite{Scully1967,sargent1974}, we express off-diagonal elements in terms of diagonal elements under steady-state conditions. After tracing over collective QD states, we obtain a rate equation for the probability of having `n' photons in the cavity $P_n=P_n^{ee}+P_n^{eg}+P_n^{ge}+P_n^{gg}$, where $P_n^{ab}=\langle a_1,b_2,n|\rho_s|a_1,b_2,n\rangle$; with $a,b=e,g$. Lasing action in the system is observed when the emission rate exceeds absorption and other losses. Therefore, we present ``Excessive Photon Emission Rate (EPER)" in the cavity mode, i.e., the difference between emission and absorption in cavity mode. We calculate EPER separately for single-photon and multi-photon processes. The laser rate equation is given by

\begin{equation}
    \begin{split}
    \dot{P_n}=&-[\alpha_n^{ee}P_n^{ee}+\alpha_n^{eg}P_n^{eg}+\alpha_n^{ge}P_n^{ge}+\alpha_n^{gg}P_n^{gg}]\\&+\sum_{k=1}^m (\Gamma_{n+k}^{ee(k)}P_{n+k}^{ee}+\Gamma_{n+k}^{eg(k)}P_{n+k}^{eg}+\Gamma_{n+k}^{ge(k)}P_{n+k}^{ge}+\Gamma_{n+k}^{gg(k)}P_{n+k}^{gg})\\&+\sum_{k=1}^m (G_{n-k}^{ee(k)}P_{n-k}^{ee}+G_{n-k}^{eg(k)}P_{n-k}^{eg}+G_{n-k}^{ge(k)}P_{n-k}^{ge}+G_{n-k}^{gg(k)}P_{n-k}^{gg})\\&-\kappa nP_n+\kappa (n+1)P_{n+1}
    \end{split}
    \label{eqn:cavityRE}
\end{equation}

\begin{figure}
    \centering
    \includegraphics[width=0.4\linewidth]{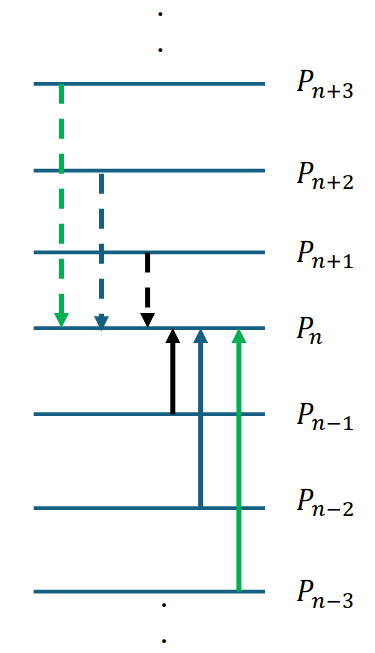}
    \caption{Transition between $n$ and $n\pm k$ photon states resulting emission or absorption.}
    \label{fig:emissAbs}
\end{figure}

The second term in Eq.\eqref{eqn:cavityRE} corresponds to k-photon absorption, and the third term corresponds to k-photon emission in the cavity mode. The coefficients $\alpha^{ab}_n$, $\Gamma^{ab(k)}_n$, and $G^{ab(k)}_n$ are calculated numerically. Emission and absorption transitions occurring between $P_n$ and $P_{n\pm k}$ states are shown in Fig. \ref{fig:emissAbs} . We consider up to four-photon processes in the laser rate equation; therefore, truncate the summation in Eq.\eqref{eqn:cavityRE} for $m=4$. The terms for $m>4$ remain negligible. Here, we \textit{do not} use mean-field approximation to separate QD-cavity correlations. 

Further, we find in steadystate 

\begin{equation}
    \sum_{a}\sum_{b}\alpha_n^{ab}P_n^{ab}=\sum_{k=1}^m\sum_{a}\sum_{b}[\Gamma_n^{ab(k)}P_n^{ab}+G_n^{ab(k)}P_n^{ab}]
    \label{eqn:emissAbsEquiv}
\end{equation}
The photon emission into cavity mode occurs via both stimulated emission and spontaneous emission similar to high $\beta$ laser\cite{Li2009}. Using the laser rate equation \eqref{eqn:cavityRE}, the steady-state mean photon number in cavity mode is given by

\begin{equation}
    \begin{split}
    \sum_nn\dot{P_n}=&\sum_n-n[\alpha_n^{ee}P_n^{ee}+\alpha_n^{eg}P_n^{eg}+\alpha_n^{ge}P_n^{ge}+\alpha_n^{gg}P_n^{gg}]\\&+\sum_n\sum_{k=1}^m n(\Gamma_{n+k}^{ee(k)}P_{n+k}^{ee}+\Gamma_{n+k}^{eg(k)}P_{n+k}^{eg}+\Gamma_{n+k}^{ge(k)}P_{n+k}^{ge}+\Gamma_{n+k}^{gg(k)}P_{n+k}^{gg})\\&+\sum_n\sum_{k=1}^m n(G_{n-k}^{ee(k)}P_{n-k}^{ee}+G_{n-k}^{eg(k)}P_{n-k}^{eg}+G_{n-k}^{ge(k)}P_{n-k}^{ge}+G_{n-k}^{gg(k)}P_{n-k}^{gg})\\&-\sum_n\kappa n^2P_n+\sum_n\kappa n(n+1)P_{n+1}
    \end{split}
\end{equation}

We know, in steadystate LHS goes to zero. Next we simplify the RHS by changing the summation over $n$,

\begin{equation}
    \begin{split}
    0=&\sum_n-n[\alpha_n^{ee}P_n^{ee}+\alpha_n^{eg}P_n^{eg}+\alpha_n^{ge}P_n^{ge}+\alpha_n^{gg}P_n^{gg}]\\&+\sum_n\sum_{k=1}^m (n-k)(\Gamma_{n}^{ee(k)}P_{n}^{ee}+\Gamma_{n}^{eg(k)}P_{n}^{eg}+\Gamma_{n}^{ge(k)}P_{n}^{ge}+\Gamma_{n}^{gg(k)}P_{n}^{gg})\\&+\sum_n\sum_{k=1}^m (n+k)(G_{n}^{ee(k)}P_{n}^{ee}+G_{n}^{eg(k)}P_{n}^{eg}+G_{n}^{ge(k)}P_{n}^{ge}+G_{n}^{gg(k)}P_{n}^{gg})\\&-\sum_n\kappa n^2P_n+\sum_n\kappa (n-1)(n)P_{n}
    \end{split}
\end{equation}

and using Eq. \ref{eqn:emissAbsEquiv}, we obtain the expression for the cavity mean photon number in terms of single and multi-photon emission and absorption rates,

\begin{align}
    \langle n\rangle = \frac{1}{\kappa}\sum_{n}\sum_{k} k&\Big[(G_n^{ee(k)}-\Gamma_n^{ee(k)})P_n^{ee}+(G_n^{eg(k)}-\Gamma_n^{eg(k)})P_n^{eg}+(G_n^{ge(k)}-\Gamma_n^{ge(k)})P_n^{ge} \nonumber\\&
    +(G_n^{gg(k)}-\Gamma_n^{gg(k)})P_n^{gg})\Big].
    \label{eqn:meanphotonNumber}
\end{align}

Here $\kappa\langle n\rangle$ gives an average rate of photons coming out from the cavity mode, thus $\sum_n k[G_n^{ee(k)}P_n^{ee}+G_n^{eg(k)}P_n^{eg}+G_n^{ge(k)}P_n^{ge}+G_n^{gg(k)}P_n^{gg})]$ provides k-photon emission rate into cavity mode and $\sum_n k[\Gamma_n^{ee(k)}P_n^{ee}+\Gamma_n^{eg(k)}P_n^{eg}+\Gamma_n^{ge(k)}P_n^{ge}+\Gamma_n^{gg(k)}P_n^{gg}]$ provides k-photon absorption rate from cavity mode.
In Eq.\eqref{eqn:meanphotonNumber}, the terms corresponding to $k=1,2,3,4$ in the average cavity photon number equation represent the single photon excessive emission rate, two-photon excessive emission rate, three-photon excessive emission rate, four-photon excessive emission rates into the cavity mode, respectively. The positive value of $k$-th term implies there is net emission into cavity mode through the $k$-photon process and the negative value implies absorption from cavity mode through the $k$-photon process. We have also compared the average photon numbers in cavity mode, $\langle n\rangle$ obtained from \eqref{eqn:meanphotonNumber} considering up to four-photon processes and the values obtained from the SME \eqref{eqn:incohSME} in the steady-state; the values match very well (c.f. Fig.\ref{fig:chap2/Fig14}(a)).

We plot the excess photon emission rate (EPER) into the cavity mode via single and multi-photon processes in Fig.\ref{fig:chap2/Fig5}. Considering QDs are resonantly coupled with cavity mode, in Fig.\ref{fig:chap2/Fig5} (a), the results of excess photon emission into cavity mode up to four-photon processes for T=5K and 20K, respectively, with increasing incoherent pumping rate $\eta$, are presented. The transitions between dressed states of QDs with more than one photon (Fig.\ref{fig:chap2/Fig1}(c)) lead to multi-photon absorption and emission in the cavity mode.

For low pumping rate ($\eta\leq 2\varepsilon_1$), the two-photon excessive emission rate into the cavity mode dominates the single-photon excessive emission rate. A small three-photon and four-photon excessive absorption and emission are also present. We find that when pump strength is varied, the single and multi-photon excess emission rates change from negative to positive values at different pump values. Especially for $\eta=0.35\varepsilon_1$, excess single photon emission is zero, the cavity mode is populated only due to two-photon emission, and the system behaves as a "\textit{two-photon laser}". This behavior at a low pumping rate with dominant excess two-photon emission leads to bunching in cavity mode (c.f. Fig.\ref{fig:chap2/Fig3}(a)). 

As the pump rate increases, the single-photon excessive emission rate grows rapidly compared to the multi-photon excess emission rate in the cavity mode. This domination of single photon emission over others shows that emission from individual QDs dominates over cooperative emission at higher pump strength. In the superradiant regime, $\eta>\varepsilon_1$, both single and two-photon excess emission rates majorly contribute to the cavity mode population. In this region, for the considered parameters system shows hyperradiant lasing. Upon further increasing the pumping rate, the excessive emission rates into cavity mode decrease as a result of self-quenching, which is also evident in the mean photon number in cavity mode, in Fig. \ref{fig:chap2/Fig2}(c). At higher temperatures, phonon-induced dephasing rises, leading to a decrease in the emission into cavity mode and the single-photon excess emission rate dominates over the two-photon excess emission rate at smaller incoherent pumping rate. Therefore, by controlling the pump rate, emission in the cavity mode predominantly occurs from cooperative two-photon processes.

We present the results by varying the QDs detuning, $\Delta$ with respect to the cavity mode in Fig. \ref{fig:chap2/Fig5}(b), for incoherent pumping rate, $\eta=0.35\varepsilon_1$ when single-photon excess emission rate in cavity mode becomes zero. It is observed that when both QDs are coupled resonantly with the cavity mode, the two-photon excess emission rate dominates, indicating dominant two-photon lasing in the system. However, when QDs are off-resonant, the single-photon excessive emission rate dominates. Further, a small three-photon and four-photon absorption also appear at low temperature when QDs are resonant to the cavity mode. In Fig.\ref{fig:chap2/Fig5}(c) with an increase in the temperature, at $\Delta=0$, the two-photon excessive emission rate decreases, but the single-photon excessive emission rate increases as discussed earlier. We also observed with at higher temperatures three-photon and four-photon emissions (not shown) also diminishes. We also noticed the broadening of excess photon emission curves similar to the $\langle n\rangle$ curves in Fig. \ref{fig:chap2/Fig2}(c).

In Fig. \ref{fig:chap2/Fig5}(d), we show the effect of cavity coupling strength on the excess single and multi-photon emission into the cavity mode. Here, the cavity decay rate, $\kappa=50\mu eV$, incoherent pumping rate, $\eta=35\mu eV$ and the detuning, $\Delta = 0$ are considered. For QD-cavity coupling strengths, $\varepsilon_1=\varepsilon_2=\varepsilon < \kappa$, the system being in a bad-cavity regime shows a higher excess single-photon emission rate. With the increase in coupling strength, the excess two-photon emission rate dominates and increases rapidly. We have considered coupling strength, $\varepsilon$ up to $0.2 meV$. Since the simplified master equation (\ref{eqn:incohSME}) used in deriving laser rate equation converges with ME (\ref{eqn:incohME}) for $0<\varepsilon\leq0.2 meV$ (c.f., Fig.\ref{fig:chap2/Fig13}(a) $\&$ (b)).

\section{Coherently pumped two-QDs-cavity system}

\subsection{Master equation}

In this section, we consider coherently pumped two-QD coupled with a single-mode photonic microcavity. The Hamiltonian of the system in a rotating frame with pump frequency is written as,
\begin{equation}
    \begin{split}
    \hat{H}=&\hbar\Delta_{cp}a^\dagger a+\hbar\Delta_{1p}\sigma_1^+\sigma_1^-+\hbar\Delta_{2p}\sigma_2^+\sigma_2^-\\&+\hbar(\varepsilon_1\sigma_1^+a+\varepsilon_2\sigma_2^+a+H.C)\\&+\hbar(\eta_1\sigma_1^++\eta_2\sigma_2^++H.C)+\hat{H}_{ph}
    \end{split}
\end{equation}
where $\eta_i$ is the coupling strength of coherent field with $i-{th}$ QD. The detuning of the cavity mode and the detunings of $i-{th}$ QD with respect to the pump frequency are $\Delta_{cp}=\omega_c-\omega_p$, and $\Delta_{ip}=\omega_i-\omega_p$, respectively.
 As discussed in the previous section, for including exciton-phonon interaction non-perturbatively, we construct the polaron transformed master equation (5), where the system Hamiltonian and the system operators are given by,
\begin{equation}
    H_s=\hbar\Delta_{cp}a^\dagger a+\hbar\Delta_{1p}\sigma_1^+\sigma_1^-+\hbar\Delta_{2p}\sigma_2^+\sigma_2^-+\langle B \rangle X_g,
\label{hs_coh}
\end{equation}
\begin{equation}
X_g=\hbar(\varepsilon_1\sigma_1^+a+\varepsilon_2\sigma_2^+a+\eta_1\sigma_1^+ +\eta_2\sigma_2^+)+H.C,
\label{xg_coh}
\end{equation}
\begin{equation}
X_u=i\hbar(\varepsilon_1\sigma_1^+a+\varepsilon_2\sigma_2^+a+\eta_1\sigma_1^+ +\eta_2\sigma_2^+)+H.C.
\label{xu_coh}
\end{equation}

\subsection{Steadystate and transient dynamics}

We use master equation (5), with system operators (\ref{hs_coh}), (\ref{xg_coh}), and (\ref{xu_coh}) to calculate the steady-state population in QDs and the average photons in cavity mode.
\begin{figure}
    \centering
    \includegraphics[width=\columnwidth]{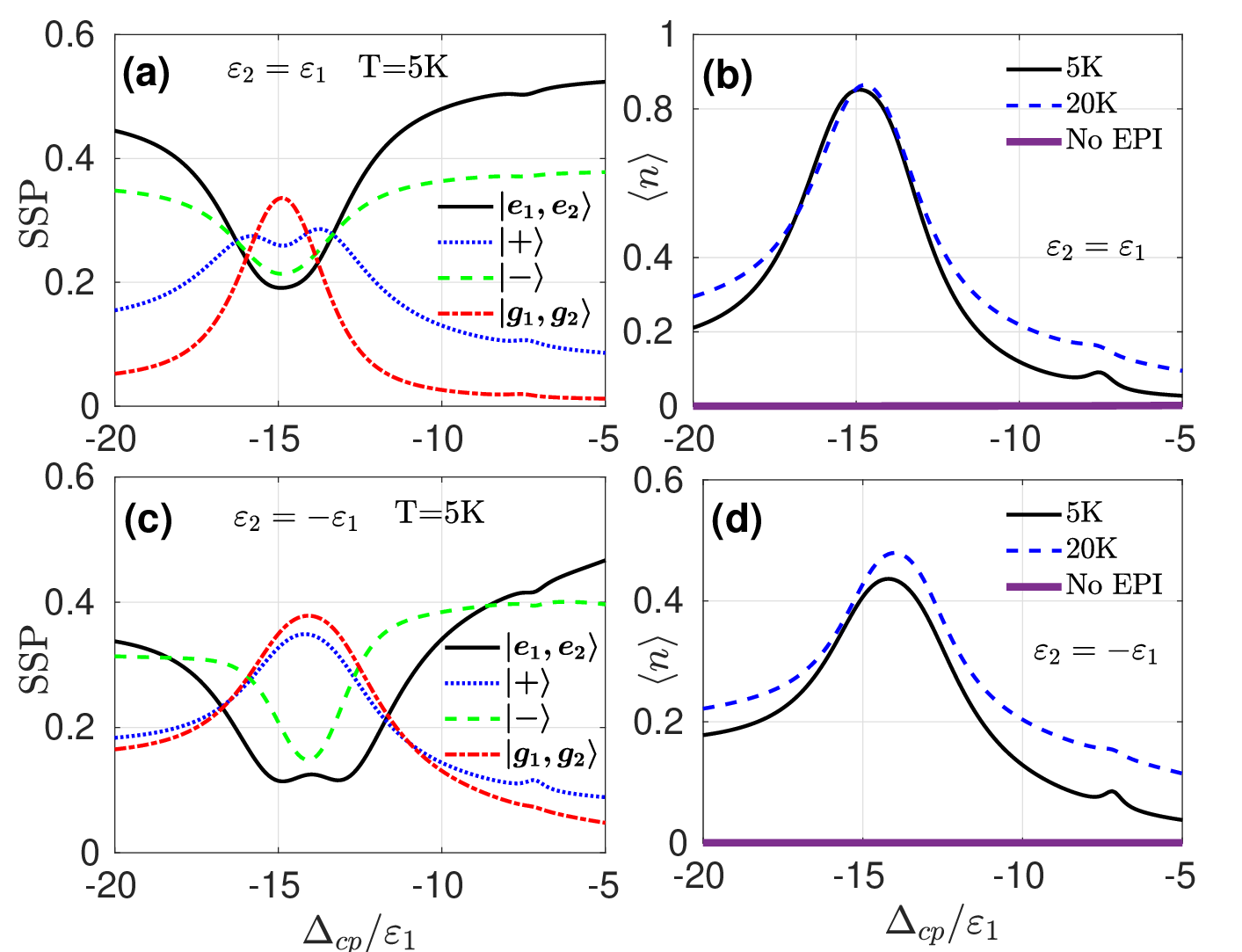}
    \caption{The steady-state populations in QD states and cavity photon number for coherently driven system, (a), (b) for the case of $\varepsilon_2=\varepsilon_1$ and (c), (d) for $\varepsilon_2=-\varepsilon_1$. The QD-cavity coupling strength considered is $\varepsilon_1=100\mu eV$. (a) $\&$ (c) show steady-state populations for T=5K, (b) $\&$ (d) show mean cavity photon number, $\langle n\rangle$ for the cases T=5K, T=20K, No EPI by varying the cavity detuning $\Delta_{cp}$, considering the other parameters $\kappa=0.5\varepsilon_1, \Delta_{1p}=\Delta_{2p}=-13.5 \varepsilon_1$, $\eta_1=\eta_2=\eta=3.0 \varepsilon_1$ in (a),(b), $\eta_1=\eta_2=\eta=1.9\varepsilon_1$ in (c),(d) $\gamma_1=\gamma_2=0.01\varepsilon_1, \gamma'_1=\gamma'_2=0.01\varepsilon_1 $. Color scheme is same as in Fig.\ref{fig:chap2/Fig2} (a) $\&$ (b).}
    \label{fig:chap2/Fig6}
\end{figure}

The results are shown in Fig.\ref{fig:chap2/Fig6}(a), (b) when both QDs are symmetrically ($\varepsilon_2= \varepsilon_1$) coupled and in Fig.\ref{fig:chap2/Fig6}(c),(d) when both QDs are asymmetrically ($\varepsilon_2=-\varepsilon_1$) coupled to the cavity mode. Earlier, the population inversion in QDs has been demonstrated for strong blue detuned coherent pump due to phonon-assisted transitions\cite{Quilter2015}. Therefore, we consider off-resonant coherent pump $\Delta_{1p}=\Delta_{2p}=\Delta=-13.5\varepsilon_1$. We find that resonant transitions in cavity mode occur when $\Delta_{cp}=-\Omega'$ where  $\Omega'=\sqrt{\Delta_{ip}^2+4\eta^2}$, generalized Rabi frequency and is the difference between coherent pump dressed QD states frequency. We have chosen optimal pumping rates $\eta=3.0\varepsilon_1$, and $\eta=1.9\varepsilon_1$ for symmetric, and anti-symmetric cases respectively for the system to act as two-photon laser as shown later in this subsection. Therefore, $\Omega'=14.8\varepsilon_1$ in (a) $\&$ (b) for $\eta=3.0\varepsilon_1$ and $\Omega'=14.0\varepsilon_1$ in (c) $\&$ (d) for $\eta=1.9\varepsilon_1$.

In Fig.\ref{fig:chap2/Fig6}(a) $\&$ (c) we see a dip in the steady-state population of $\ket{e_1,e_2}$ and a peak in the population of $\ket{g_1,g_2}$ at $\Delta_{cp}=-\Omega'$ indicating resonant transitions in cavity mode. Also, there is a dip in the populations of state $\ket{-}$, and the population of state $\ket{+}$ is greater than the population of state $\ket{-}$. The changes in the populations of states $\ket{+}$ and $\ket{-}$ occur due to single-photon coherent and incoherent transitions. Further, for $\varepsilon_2=\varepsilon_1$ the state $\ket{+}$ is coupled with cavity mode, and the state $\ket{-}$ remains uncoupled. Therefore, change in the population of $\ket{-}$ occurs due to incoherent single-photon transitions only, leading to a broader dip around lasing transitions (Fig.\ref{fig:chap2/Fig6}(a)). Similarly, for $\varepsilon_2=-\varepsilon_1$ the state $\ket{+}$, is coherently coupled with the pump but remains uncoupled with the cavity mode leading to a broader peak around the resonance (Fig.\ref{fig:chap2/Fig6}(c)). We also observe a tiny dip in the steady-state population of $\ket{e_1,e_2}$ and a tiny peak in the population of $\ket{+}$ for $\Delta_{cp}=-\Omega'/2$, where the cavity is resonant with the two-photon transition between pump dressed states of individual QDs. Such transitions have been utilized to generate two-photon dressed state laser\cite{Gauthier1992}. 

Corresponding to the cavity detuning $\Delta_{cp}=-\Omega'$, in Fig.\ref{fig:chap2/Fig6}(b) and (d), there is a maximum in the average photon number in the cavity mode. There is also a small peak in the average number of cavity photons corresponding to $\Delta_{cp}=-\Omega'/2$. Further, there is no major difference in cavity photons whether QDs are symmetrically ($\varepsilon_1=\varepsilon_2$) or anti-symmetrically ($\varepsilon_1=-\varepsilon_2$) coupled with the cavity mode. With the increase in temperature, at T=20K, there is more emission in cavity mode due to off-resonant phonon-assisted cavity mode feeding \cite{Hohenester2010}, and the peaks are broadened. To compare the results with and without exciton-phonon interaction, we also include the plot for no exciton-phonon interaction (No EPI curve). We can see, the emission in cavity mode is negligible for ``No EPI" case because, without phonon interactions, the off-resonant pump could not generate a significant population in the exciton state.

\begin{figure}
    \centering
    \includegraphics[width=\columnwidth]{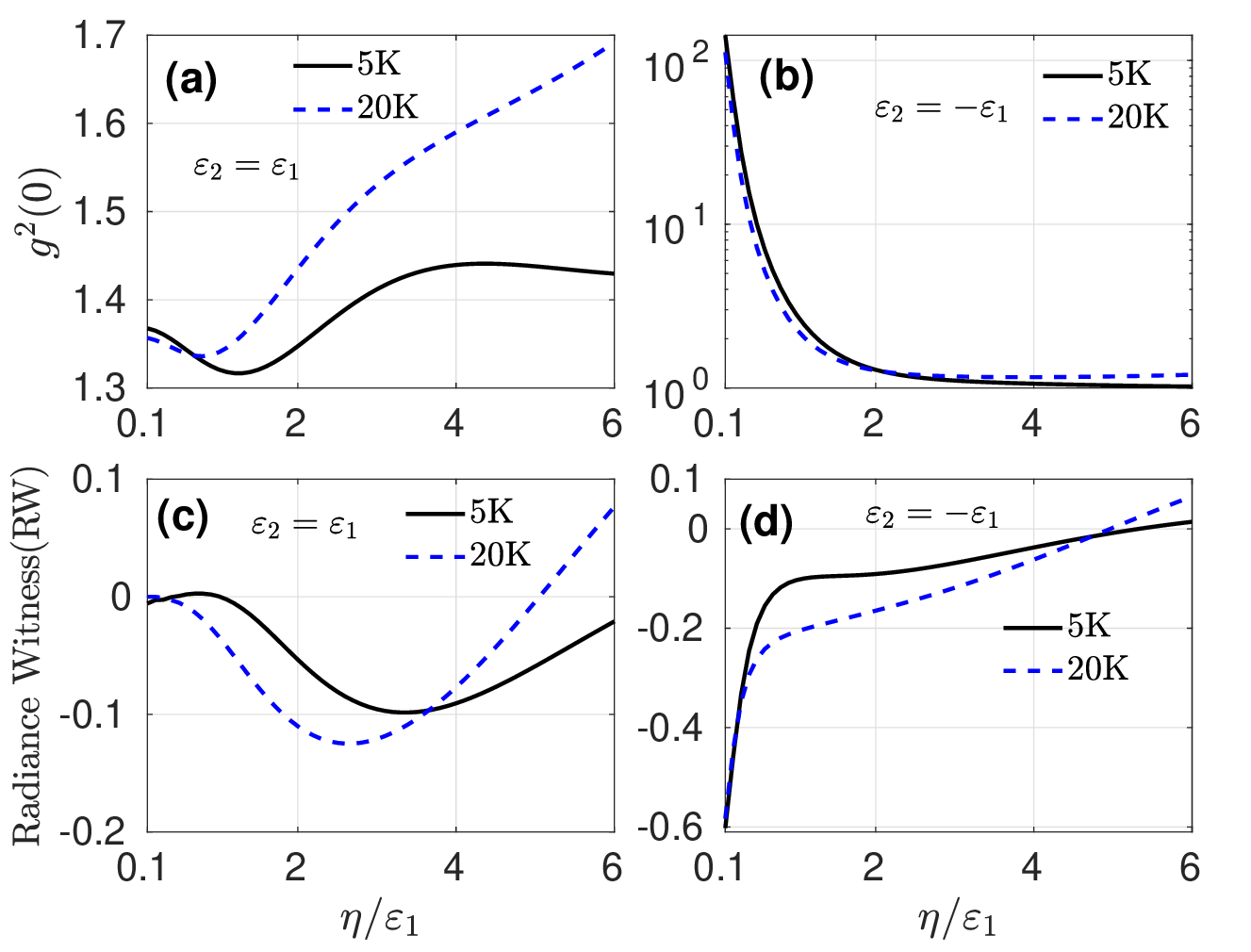}
    \caption{The zero time delay second-order photon correlation function, $g^2(0)$, in (a, b) and the radiance-witness (RW) in (c, d) for $\Delta_{1p}=\Delta_{2p}=-13.5 \varepsilon_1, \Delta_{cp}=-\Omega'$ with varying coherent pumping rate, $\eta_1=\eta_2=\eta$ and other parameters are same as in Fig.\ref{fig:chap2/Fig6}. Color scheme is same as in Fig. \ref{fig:chap2/Fig3}.}
    \label{fig:chap2/Fig7}
\end{figure}

In Fig. \ref{fig:chap2/Fig7}, we show the results for $g^2(0)$ and radiance witness (RW) when the cavity is tuned such that $\Delta_{cp}=-\Omega'$ for symmetric ($\varepsilon_2=\varepsilon_1$) and anti-symmetric ($\varepsilon_2=-\varepsilon_1$) cases. In Fig.\ref{fig:chap2/Fig7} (a), the correlation function  $g^2(0)$ has values greater than one and changes slightly on changing pump strengths when the cavity is coupled symmetrically. This can be understood as follows. The pump excites symmetric state $|+\rangle$ through one photon excitation and $|e_1,e_2\rangle$ through two-photon excitation. When $\varepsilon_2=\varepsilon_1$, the cavity couples both symmetric states $|+\rangle$ and $|e_1,e_2\rangle$ through one and two-photon transitions, respectively. Therefore, both one-photon and two-photon emissions into cavity mode take place. However, when $\varepsilon_2=-\varepsilon_1$, the symmetric state $|+\rangle$ remains decoupled with cavity mode and state $|e_1,e_2\rangle$ decays to $|g_1,g_2\rangle$ by emitting two-photons via anti-symmetric state $|-\rangle$. Therefore, a very large value of $g^2(0)$ results for low pump strength as shown in Fig.\ref{fig:chap2/Fig7} (b). On increasing pump strength, the average cavity photons $\langle n\rangle$ increases, and $g^2(0)$ decreases, similar to the incoherent pumped case. In Fig. \ref{fig:chap2/Fig7} (c) $\&$ (d), the radiance witness (RW) plots show that the cooperative effects in the system lead to subradiant emission (RW$<0$) for $\Delta_{cp}=-\Omega'$ in both cases. It has been demonstrated that in the strong coupling regime, the cavity field and pump field interfere destructively at the emitters' location leading to subradiant emission\cite{Reimann2015}. Therefore, in the coherent pump case, subradiant behavior is more pronounced, and superradiant or hyperradiant behavior, as seen in the incoherent pump case, is absent. As the temperature increases, the interference effect decreases due to phonon-induced dephasing, which leads to a decrease in subradiance.

\begin{figure}
    \centering
    \includegraphics[width=\columnwidth]{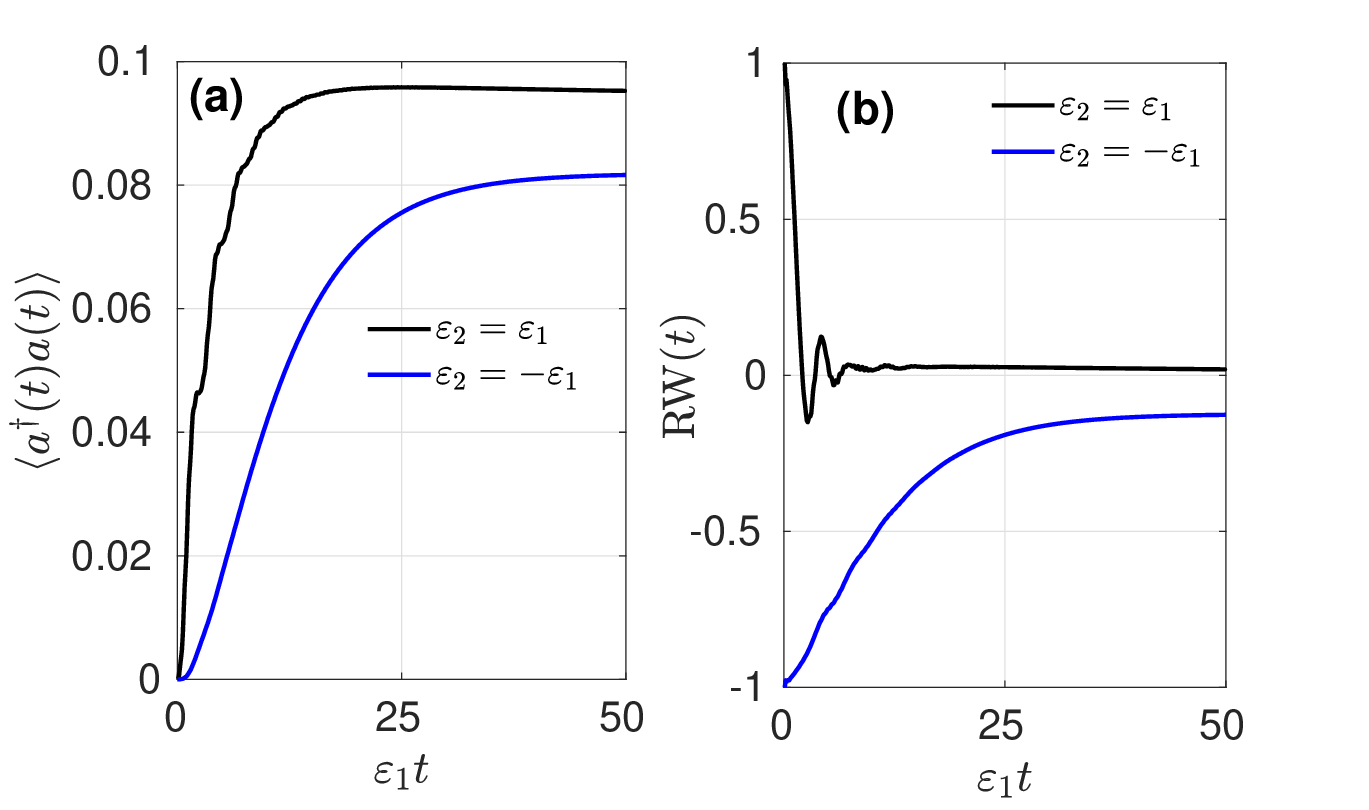}
    \caption{The transient behavior of (a) mean cavity photon number and (b) the radiance-witness (RW) for symmetric, $\varepsilon_2=\varepsilon_1$ (black) and anti-symmetric, $\varepsilon_2=-\varepsilon_1$ (blue) cases. The coherent pumping rate, $\eta=0.75\varepsilon_1$, temperature, $T=5K$ and other parameters are same as in Fig.\ref{fig:chap2/Fig6}.}
    \label{fig:chap2/transient_coh}
\end{figure}

In Fig. \ref{fig:chap2/transient_coh}, the results of the transient dynamics of the cavity field are shown for both the symmetric, $\varepsilon_2=\varepsilon_1$, and anti-symmetric, $\varepsilon_2=-\varepsilon_1$ cases. Since the coherent pump drives collective QD transitions via $|+\rangle$ state, Fig. \ref{fig:chap2/transient_coh} (a) shows that for the symmetric case, initially there are small oscillations in the cavity mean-photon number and is due to the interference between the coherent pump and the cavity mode induced transitions. These oscillations subside and the cavity field attains a steady-state value. These oscillations are absent for the anti-symmetric case, as the cavity field couples via the $|-\rangle$ state. This is manifested in the results for radiance witness as shown in Fig. \ref{fig:chap2/transient_coh} (b).  

\begin{figure}[h]
    \centering
    \includegraphics[width=0.75\columnwidth]{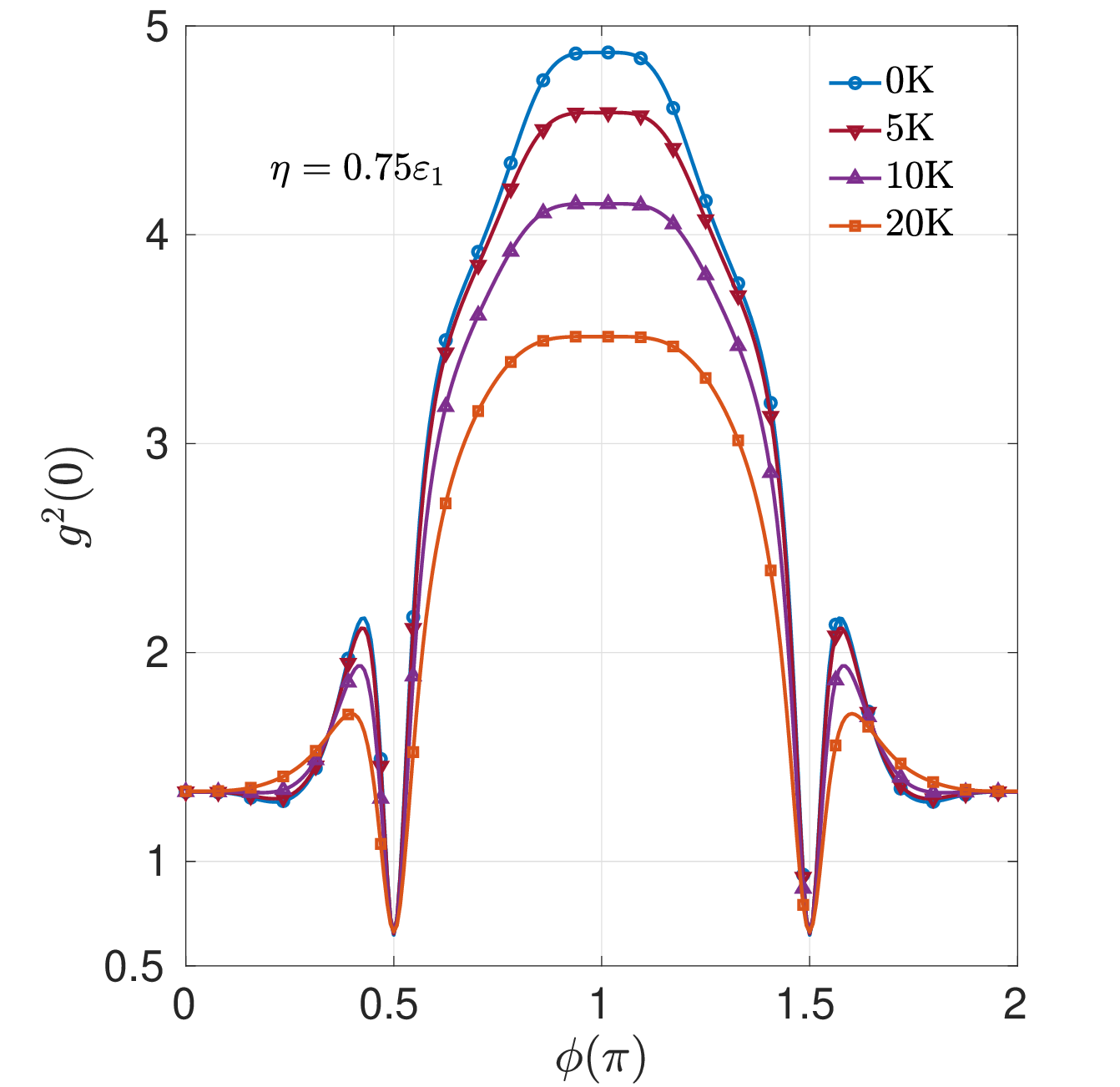}
    \caption{Coherent pumping: Variation in the zero-time delay second order photon correlation function, $g^2(0)$ with change in QD coupling, $\varepsilon_2/\varepsilon_1 = cos\phi$ for different temperatures, T for $\eta=0.75\varepsilon_1$. Here, $\Delta_{cp}=-\Omega'$. The other parameters same as in Fig. \ref{fig:chap2/Fig6}.}
    \label{fig:chap2/Fig8}
\end{figure}

In Fig. \ref{fig:chap2/Fig8}, we show the results for $g^2(0)$ function with change in the relative phase, $\phi$, between cavity coupling strength with QDs, $\varepsilon_2/\varepsilon_1 = cos\phi$. It is observed that at a low pumping rate, the system shows super bunching when QDs are coupled anti-symmetrically($\phi=\pi$). This suggests that cavity mode is populated majorly via two-photon emission\cite{Rempe2016}. It is clear from the plots the cavity field shows anti-bunching for $\phi=\pi/2,3\pi/2$, which corresponds to the scenario of single QD coupled with cavity mode. For other values of $\phi$, we can see bunching in the cavity mode. Further, with an increase in the temperature, $g^2(0)$ decreases due to increased phonon-induced incoherent processes leading to decreased cooperative multi-photon emission.

\subsection{Simplified master equation}

The steady-state mean photon number in cavity mode can be analyzed by considering the single and multi-photon excessive emission rate in cavity mode using the laser rate equation. Following the similar approach as discussed in the previous section, we derive laser rate equation \eqref{eqn:cavityRE}. We use a simplified master equation to derive the laser rate equation. In order to construct the simplified master equation, we approximate system Hamiltonian, $H_s\approx\hbar\Delta_{cp}a^\dagger a+\hbar\Delta_{1p}\sigma_1^+\sigma_1^-+\hbar\Delta_{2p}\sigma_2^+\sigma_2^-$ in $L_{ph}\rho_s$ and arranging terms proportional to $\varepsilon_1^2$, $\varepsilon_2^2$, $\varepsilon_1\varepsilon_2$, $\eta_1^2$,$\eta_2^2$, and $\eta_1\eta_2$ is given by

\begin{equation}
    \begin{split}
    \dot{\rho_s}=&-\frac{i}{\hbar}[H_{eff},\rho_s]-\frac{\kappa}{2}L[a]\rho_s-\Sigma_{i=1,2}(\frac{\gamma_i}{2}L[\sigma_i^-]+\frac{\gamma_i'}{2}L[\sigma_i^+\sigma_i^-])\rho_s\\&+\Sigma_{i=1,2}(\frac{\Gamma_i^-}{2}L[\sigma_i^+a]+\frac{\Gamma_i^+}{2}L[\sigma_i^-a]+\frac{\Gamma_p^{\sigma_i^+}}{2}L(\sigma_i^+)+\frac{\Gamma_p^{\sigma_i^-}}{2}L(\sigma_i^-))\rho_s
    \\&-\Big[(\frac{\Gamma_{12}^{--}}{2}L[\sigma_2^+a,\sigma_1^+a]+\frac{\Gamma_{12}^{++}}{2}L[\sigma_2^-a^\dagger,\sigma_1^-a^\dagger]+\frac{\Gamma_{12}^{-+}}{2}L[a^\dagger\sigma_2^-,\sigma_1^+a]
    \\&+\frac{\Gamma_{12}^{+-}}{2}L[\sigma_2^+a,a^\dagger\sigma_1^-])\rho_s
    +\Omega_{11}^{++}\sigma_1^+ a\rho_s\sigma_1^+ a+\Omega_{11}^{--}\sigma_1^-a^\dagger\rho_s\sigma_1^-a^\dagger
    +1 \leftrightarrow 2 \Big]
    \\&-\Big[\sum\limits_{k,l=\pm}\frac{\Gamma_p^{\sigma_2^k\sigma_1^l}}{2}L[\sigma_2^k,\sigma_1^l]
    +\Omega_p^{\sigma_1^+\sigma_1^+}\sigma_1^+\rho_s\sigma_1^+ + H.C.
    +1 \leftrightarrow 2 \Big]
    \end{split}
    \label{eqn:cohSME}
\end{equation}

The coherent evolution of the system is given by $H_{eff}$, which includes phonon-mediated stark shifts, $\delta_i^\pm$, $\delta_p^{\sigma_i^\pm}$, two-photon processes corresponding to $\Omega_{2ph}$, $\Omega_p^{++}$, phonon-assisted exciton exchange processes corresponding to $\Omega_\pm$, $\Omega_p^\pm$.
The effective Hamiltonian of the system is given by
\begin{equation}
\begin{split}
    H_{eff} = &H_s+\hbar\Sigma_{i=1,2}(\delta_i^-a^\dagger\sigma_i^-\sigma_i^+a+\delta_i^+\sigma_i^+a a^\dagger\sigma_i^-)
    \\&-(i\hbar\Omega_{2ph}\sigma_1^+\sigma_2^+a^2+H.C.)-(i\hbar\Omega_+\sigma_1^+a a^\dagger \\&+i\hbar\Omega_-a^\dagger\sigma_1^-\sigma_2^+a+H.C.)+\hbar\Sigma_{i=1,2}\delta_p^{\sigma_i^+}\sigma_i^+\sigma_i^-\\&+i\hbar\Sigma_{i=1,2}\delta_p^{\sigma_i^-}\sigma_i^-\sigma_i^+ +(i\hbar\Omega_p^{++}\sigma_1^+\sigma_2^++H.C.)\\&-(i\hbar\Omega_p^+\sigma_1^+\sigma_2^-+i\hbar\Omega_p^-\sigma_1^-\sigma_2^++H.C.),
\end{split}
\end{equation}
with
\begin{equation}
    \delta_p^{\sigma_i^\pm}=\eta_i^2Im[\int_0^\infty d\tau G_+e^{\pm i\Delta_{ip}\tau}],
\end{equation}
\begin{equation}
    \Omega_p^{++} = \frac{\eta_1\eta_2}{2}\int_0^\infty d\tau(G_- - G_-^*)(e^{-i\Delta_{1p}\tau}+e^{-i\Delta_{2p}\tau}),
\end{equation}
\begin{equation}
    \Omega_p^\pm = \frac{\eta_1\eta_2}{2}\int_0^\infty d\tau(G_+e^{\pm i\Delta_{2p}\tau}-G_+^*e^{\mp i\Delta_{1p}\tau}).
\end{equation}
The additional terms in the simplified master equation (\ref{eqn:cohSME}) given above, apart from the ones present in the incoherent case (\ref{eqn:incohSME}) as a consequence of coupling between coherent pump drive and phonon bath, includes phonon-assisted incoherent excitation corresponding to $\Gamma_p^{\sigma_i^+}$, enhanced exciton decay process corresponding to $\Gamma_p^{\sigma_i^-}$, double exciton creation and annihilation terms proportional to $\Gamma_p^{\sigma_2^+\sigma_1^+}$, $\Gamma_p^{\sigma_2^-\sigma_1^-}$ and exciton transfer proportional to $\Gamma_p^{\sigma_2^-\sigma_1^+}$, $\Gamma_p^{\sigma_2^+\sigma_1^-}$. The expressions of rates corresponding to these processes are 
\begin{equation}
    \Gamma_p^{\sigma_i^\pm} = \eta_i^2 \int_0^\infty d\tau(G_+e^{\pm i\Delta_{ip}\tau}+G_+^*e^{\mp i\Delta_{ip}\tau}),
\end{equation}
\begin{equation}
    \Gamma_p^{\sigma_i^\pm \sigma_j^\pm} = \eta_i\eta_j\int_0^\infty d\tau(G_-e^{\pm i\Delta_{ip}\tau} + G_-^*e^{\pm i\Delta_{jp}\tau}),
\end{equation}
\begin{equation}
    \Gamma_p^{\sigma_i^\mp \sigma_j^\pm} = \eta_i\eta_j\int_0^\infty d\tau(G_+e^{\pm i\Delta_{ip}\tau}+ G_+^* e^{\mp i\Delta_{jp}\tau}),
\end{equation}
\begin{equation}
    \Omega_p^{\sigma_i^\pm \sigma_i^\pm} = \eta_i^2 \int_0^\infty d\tau(G_- + G_-^*)e^{\mp i\Delta_{ip}\tau}.
\end{equation}
We have also retained some of the terms proportional to $\Omega_p^{\sigma_i^\pm \sigma_i^\pm}$, which do not have Lindblad form. However, these terms are necessary for better approximation. The comparison between polaron transformed master equation \eqref{eqn:incohME} and simplified master equation \eqref{eqn:cohSME} is relegated to Appendix \ref{sec:chap2_Appendix1}.

\subsection{Emission and absorption rates}

\begin{figure}[h]
    \centering
    \includegraphics[width=\columnwidth]{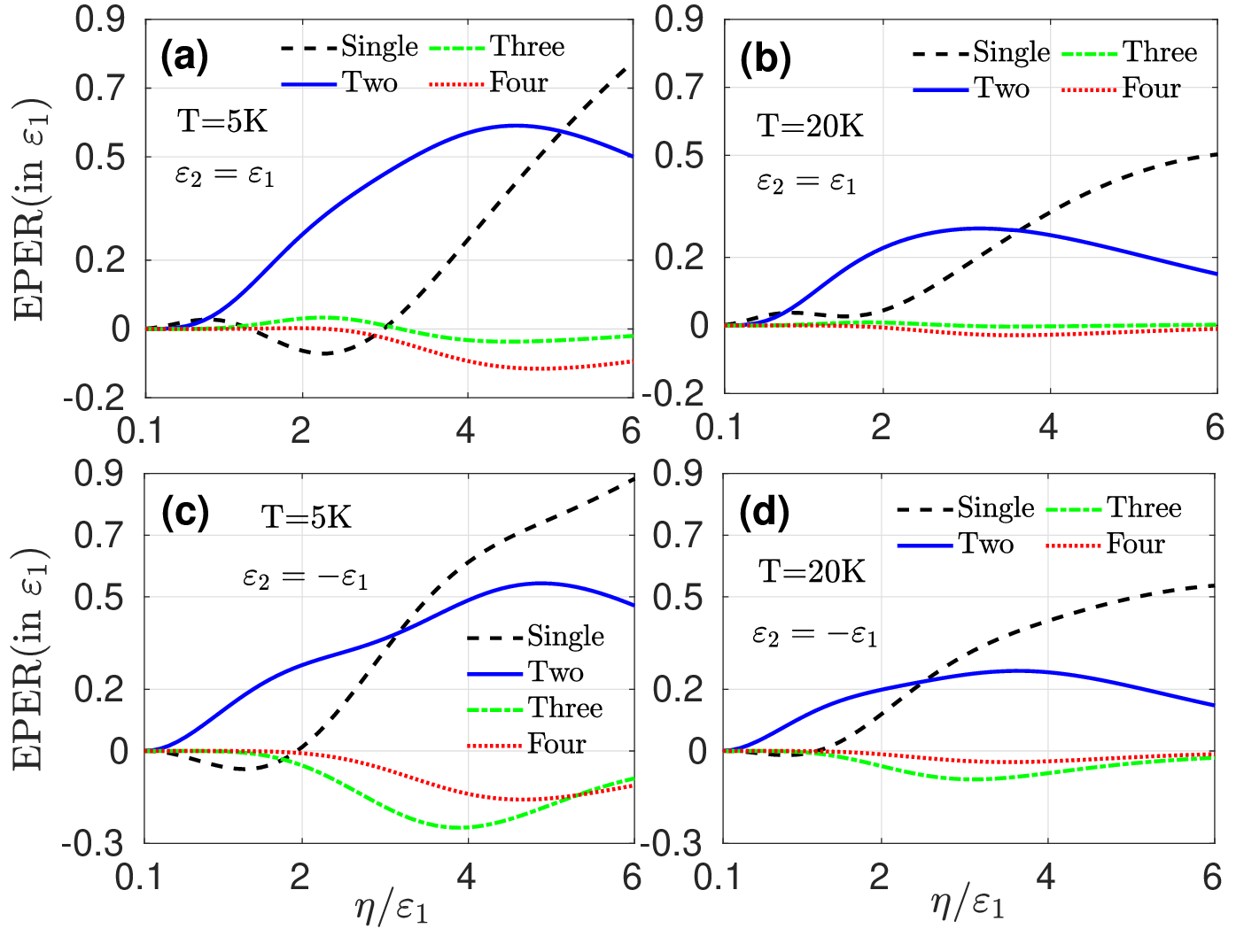}
    \caption{Excess photon emission rate (EPER) for $\Delta_{1p}=\Delta_{2p}=-13.5 \varepsilon_1, \Delta_{cp}=-\Omega'$ with varying coherent pumping rate, $\eta_1=\eta_2=\eta$ and other parameters are same as in Fig.\ref{fig:chap2/Fig6} for the cases, (a) $\varepsilon_2=\varepsilon_1$, T=5K (b) $\varepsilon_2=\varepsilon_1$, T=20K (c) $\varepsilon_2=-\varepsilon_1$, T=5K (d) for $\varepsilon_2=-\varepsilon_1$, T=20K. Color scheme is same as in Fig. \ref{fig:chap2/Fig5}.}
    \label{fig:chap2/Fig9}
\end{figure}

\begin{figure}[h]
    \centering
    \includegraphics[width=\columnwidth]{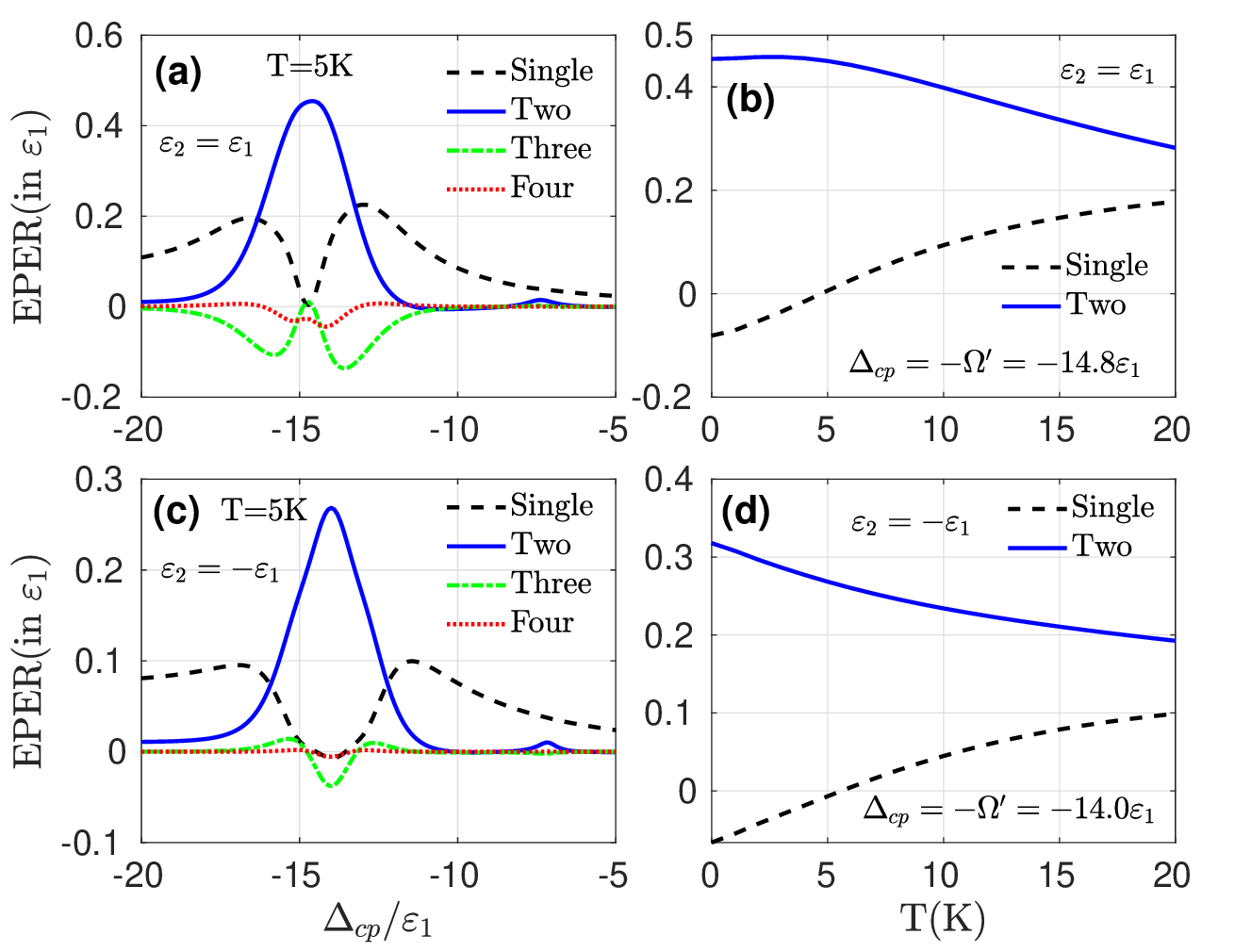}
    \caption{Excess photon emission rate (EPER) for symmetric case, $\varepsilon_2=\varepsilon_1$ in (a) $\&$ (b) and anti-symmetric case, $\varepsilon_2=-\varepsilon_1$ in (c) $\&$ (d). In (a) $\&$ (c) T=5K and $\Delta_{cp}$ is varied. In (b), (d) $\Delta_{cp}=-\Omega'$ for $\eta=3.0\varepsilon_1,1.9\varepsilon_1$ respectively and temperature, T is varied. Rest all parameters are same as in Fig. \ref{fig:chap2/Fig6}(a). Color scheme is same as in Fig.\ref{fig:chap2/Fig5}.}
    \label{fig:chap2/Fig10}
\end{figure}

 Following the laser rate equation \eqref{eqn:cavityRE}, we present single and multi-photon excessive emission rates into the cavity mode in Fig.\ref{fig:chap2/Fig9} and Fig.\ref{fig:chap2/Fig10}.
 In Fig. \ref{fig:chap2/Fig9}, we show the results by varying the coherent pumping rate, $\eta_1=\eta_2=\eta$. We consider far off-resonant blue detuned coherent pump and cavity detuning, $\Delta_{cp}=-\Omega'$ so that the cavity becomes resonant with the transition between pump-dressed QD states. In Fig. \ref{fig:chap2/Fig9}(a) $\&$ (b), we consider the QDs are coupled symmetrically ($\varepsilon_1=\varepsilon_2$) and in Fig. \ref{fig:chap2/Fig9}(c) $\&$ (d) QDs are coupled anti-symmetrically ($\varepsilon_2=-\varepsilon_1$) to the cavity mode. The symmetric and anti-symmetric coupling of QDs with cavity mode leads to different behaviour as both coherent pump and cavity mode couple to $\ket{+}$ state in the first case, whereas coherent pump couples to $\ket{+}$ state but cavity couples to $\ket{-}$ state resulting in distinct interference effects between transitions \cite{Pleinert2017}. 
 
 We find that the two-photon excess emission rate dominates the single-photon excess emission rate into cavity mode for low pumping rates which implies that photon emission in cavity mode occurs mostly due to cooperative two-photon emission by QDs. Such cooperative emission occurs when a correlation between QDs is established by photon exchange between QDs and cavity mode.
 Initially, on increasing pump strength single-photon, three-photon and four-photon excess emission rate into cavity mode changes from positive to negative and vice versa due to interference between different transitions corresponding to these processes.
 At higher pump strength the single-photon excess emission rate dominates the two-photon excess emission rate. For the symmetric case, the single-photon excess emission rate starts dominating at higher pump strength than in the anti-symmetric case, and the emission from individual QDs dominates. Three-photon and four-photon excess emission rates attain a slightly positive value for moderate pumping strength in the symmetric coupling case; otherwise, they show absorption. 
 Eventually, with a further increase in pumping strength, the excess photon emission rate into cavity mode is suppressed due to self-quenching effect. 
 
 In Fig. \ref{fig:chap2/Fig9}(b) $\&$ (d) the results corresponding to T=20K are plotted. On increasing temperature, single-photon emission into the cavity mode increases while the two-photon emission decreases, and other multi-photon emissions are largely suppressed. The single-photon excess emission rate increases with the rise in temperature only for the low pumping rates and is suppressed at higher rates. The decline in excess emission in cavity mode is in agreement with the notion that an increase in phonon scattering processes results in dephasing and hinders cavity exciton coupling. 

 In Fig. \ref{fig:chap2/Fig10} we present the results by varying the cavity detuning, $\Delta_{cp}$ and increase in temperature. We consider the typical value of coherent pump strength such that single-photon excess emission rate in cavity mode is zero, i.e. $\eta=3.0\varepsilon_1$ for symmetric coupling in Fig. \ref{fig:chap2/Fig10}(a) and $\eta=1.9\varepsilon_1$ for anti-symmetric coupling in Fig. \ref{fig:chap2/Fig10}(c). In Fig. \ref{fig:chap2/Fig10} (a,c), the two-photon excessive emission rate curve shows a peak at $\Delta_{cp}=-\Omega'$, where the single-photon excessive emission rate has a dip. The cooperative effects between the QDs lead to predominant two-photon emission into cavity mode, as mentioned earlier. Other higher-order photon excessive emission rates are not significant in this off-resonantly coupled system and they show very small negative values implying multi-photon absorption from the cavity mode. In Fig. \ref{fig:chap2/Fig10} (b) $\&$ (d), considering $\Delta_{cp}=-\Omega'$ for both symmetric and anti-symmetric QD-cavity coupling cases, with an increase in the system temperature, the single photon emission into the cavity mode increased and the two-photon emission decreased due to enhanced exciton-phonon scattering rates. Additionally, at $\Delta_{cp}=-\Omega'/2$ corresponding to the two-photon transition between pump-dressed states of individual QDs, we can see small peaks in two-photon excess emission rate and four-photon excess emission rate in Fig. \ref{fig:chap2/Fig10}(a) $\&$ (c). This aspect is also seen in mean photon number plots, Fig. \ref{fig:chap2/Fig6} (b) $\&$ (d). The appearance of this small peak is due to transitions induced by cavity photons between laser-dressed states, where QDs are emitting two photons independently. 

\begin{figure}[h]
    \centering
    \includegraphics[width=\columnwidth]{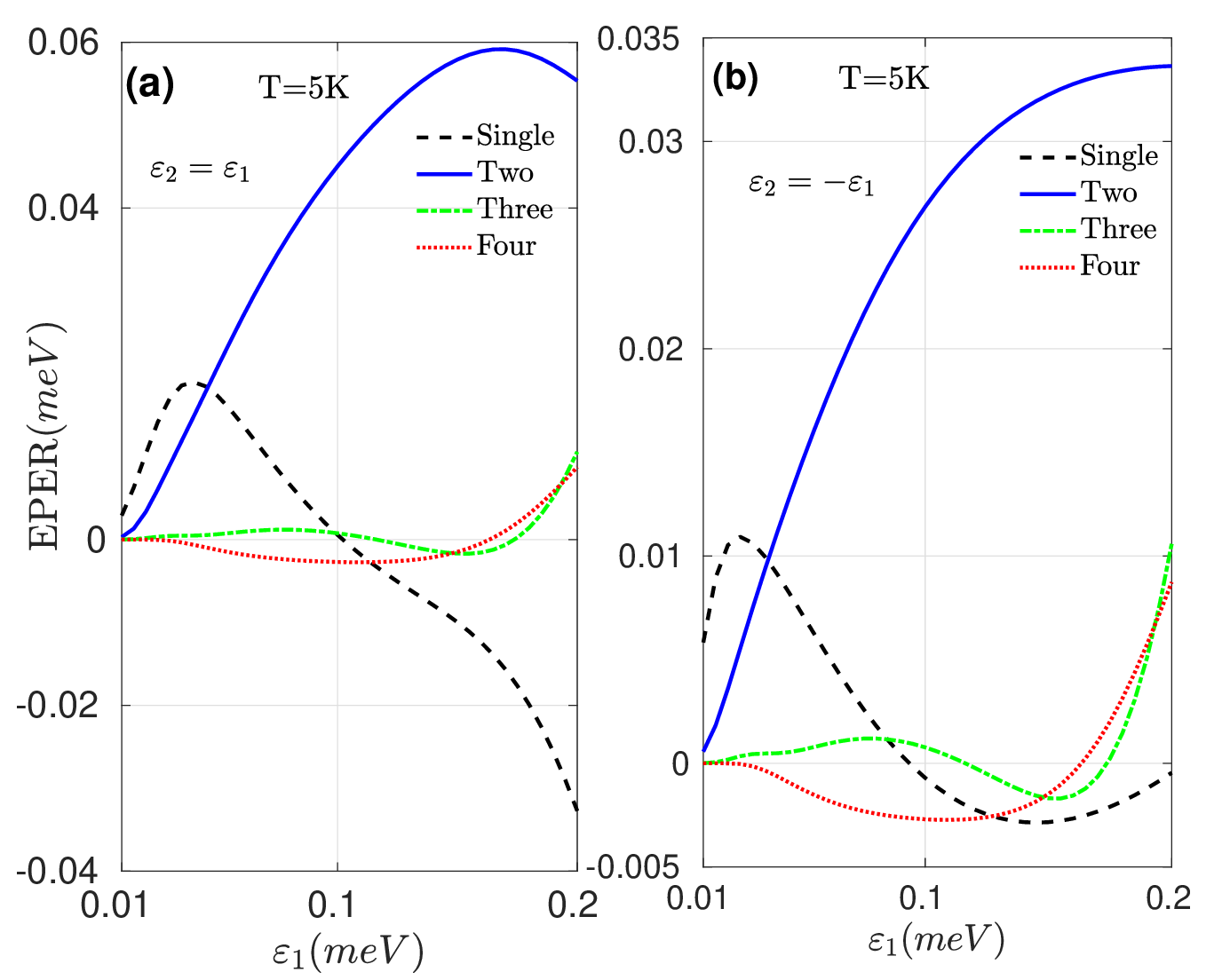}
    \caption{Excess photon emission rate (EPER) (a) symmetric case, $\varepsilon_2=\varepsilon_1$ and (b) anti-symmetric case, $\varepsilon_2=-\varepsilon_1$ by varying QDs-cavity coupling strength, $\varepsilon_1$, rest all parameters are same as in Fig. \ref{fig:chap2/Fig6}(a) and (c) respectively. Color scheme is same as in Fig.\ref{fig:chap2/Fig5}.}
    \label{fig:chap2/Fig11}
\end{figure}

In Fig. \ref{fig:chap2/Fig11}, we show the effect of QD-cavity coupling strength, $\varepsilon$ in the coherently pumped system coupled symmetrically (a) and anti-symmetrically (b). For low $\varepsilon$ values, the system being in the bad-cavity limit, the single-photon excess emission rate is dominant as observed for the incoherent pumping case, Fig. \ref{fig:chap2/Fig5} (c). Further, with an increase in the coupling strength, $\varepsilon$, the system transits to a strong coupling regime, and multi-photon processes are dominant. This results in a rapid increase in the excess two-photon emission rate and a decrease in the excess single-photon emission rate. The other multi-photon excess emission rates are comparatively negligible in both symmetric and anti-symmetric coupling scenarios. Especially for $\varepsilon=0.1meV$, the excess-single photon emission rate is zero for the parameters considered. For small pump strength and larger values of QDs cavity couplings, the single photon process shows a negative excess emission rate (absorption), and the multi-photon processes show a higher excess emission rate. The maximum $\varepsilon$ considered is $0.2meV$ owing to the validity of approximations made in deriving SME, Eq. \ref{eqn:cohSME}(c.f., Appendix \ref{sec:chap2_Appendix1} , Fig. \ref{fig:chap2/Fig13}(c) $\&$ (d)). Therefore, it is appropriate to consider cavity detuning $\Delta_{cp}=-\Omega'$ and the pump rates where the single-photon excess emission rate is negligible to realise coherently pumped cooperative two-QDs two-photon laser.
 
\section{Conclusions}
To conclude, we have considered two quantum dots(QDs) coupled to a single-mode photonic microcavity. The QDs are driven incoherently and coherently in a strong coupling regime. We incorporated exciton-phonon coupling using polaron transformed master equation. We have derived laser rate equation and investigated single and multi-photon lasing in both incoherently and coherently pumped systems. We have explicitly calculated the contribution from single and multi-photon excess emission rates into the cavity mode by exactly solving the rate equation with high photon number truncation for convergence in the numerical results. In the case of incoherent pumping, resonantly coupled QDs with $\eta<2.0\varepsilon_1$ show a two-photon excess emission rate greater than single and other multi-photon emissions into cavity mode. In the coherent pumping case, we have shown that cooperative effects lead to a significant two-photon excess emission rate for cavity detuning $\Delta_{cp}=-\Omega'$. We have shown the behaviour of the system for symmetric($\varepsilon_2=\varepsilon_1$) and anti-symmetric($\varepsilon_2=-\varepsilon_1$) coupling to the cavity mode. In both incoherent and coherent pumping cases, by selecting pump strength properly such that single-photon emission becomes negligible, the photons in the cavity mode are due to the two-photon excess emission rate and the system behaves as a two-photon laser.

%% file: chap3.tex
\chapter{Two-mode hyperradiant Lasing}\label{chap3}
{\small This chapter is based on our work ``\textsc{Two-mode hyperradiant lasing in a system of two quantum dots embedded in a bimodal photonic crystal cavity,}";  \textbf{Lavakumar Addepalli}, P. K. Pathak, arXiv:2506.21202 [quant-ph].}

In Chapter 2, we showed the possibility of cooperative two-photon lasing accompanied by ``\textit{Hyperradiance}" in a two QDs coupled to single-mode photonic crystal cavity. Moreover, systems with emitters coupled to bimodal cavities exhibit inter-mode correlations that facilitate enhanced emitted radiation. In atomic systems, such interactions with bimodal cavities have been shown to lead to two-photon scattering processes \cite{Richter2025} and pronounced two-atom, two-photon Rabi oscillations \cite{Pathak2004}. Further, sub-Poissonian light generation in QD systems coupled to bimodal planar photonic crystal cavities \cite{majumdar2012} demonstrated, and also photon blockade and enhancement of anti-bunching effects in microdisk resonators where emitters are coupled to whispering gallery modes (WGMs) \cite{Barak2008, Liu2016, Xie2016}. Also, unconventional photon blockade is realized in the system with a QD coupled to two orthogonally polarized micropillar cavity modes \cite{snijders2018}. Lukin et al. \cite{Lukin2023} reported superradiance and explored multi-mode interference effects in a system of two silicon carbide color centers weakly coupled to the WGMs of a microdisk resonator. In an earlier study, Verma et al. \cite{Verma2020} investigated phonon-induced cooperative two-mode two-photon emission in a system of two quantum dots coupled to a bimodal photonic crystal cavity. In this chapter, we show that cooperative two-mode two-photon lasing leads to ``Hyperradiant lasing" in the system where two quantum dots are incoherently pumped and strongly coupled to a bimodal cavity.

We investigate the enhancement in cavity field by analyzing the cavity photon statistics, including the mean photon number, radiance witness, and both inter- and intra-mode zero-time-delay second-order photon correlations. Following the approach detailed in earlier chapters, we derive the laser rate equations to evaluate the contributions of single-mode and two-mode two-photon processes to the cavity mode population. We discuss the effect of radiance witness on the cavity emission linewidth. We also studied the effect of the second mode coupling to QDs, considering both the resonant and off-resonant cases.

\section{Model system}

\begin{figure}
    \centering
    \includegraphics[width=0.5\columnwidth]{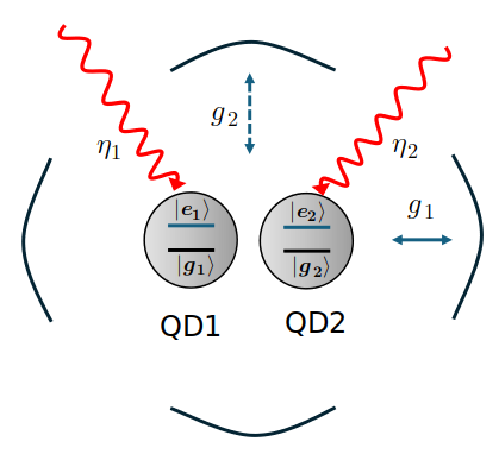}
    \caption{ Schematic figure showing two quantum dots (QDs) coupled to a bimodal cavity with coupling strength, $g_1$ to mode 1, $g_2$ to mode 2 and are incoherently pumped with strengths, $\eta_1$, $\eta_2$ \cite{Addepalli2025twoModeHyper}.}
    \label{fig:chap3/Fig1}
\end{figure}

Here, we consider two identical QDs, incoherently driven and coupled to a bimodal photonic crystal cavity, where the cavity modes interact via QDs, with no direct coupling between the modes. The Hamiltonian for the system in the rotating frame of the QD transition frequency is given as,

\begin{equation}
    \begin{split}
         H = & -\hbar \delta_1 a_1^\dagger a_1-\hbar \delta_2 a_2^\dagger a_2
         \\&+\hbar \sum_{i=1,2} g_i(\sigma_1^+a_i+ a_i^\dagger \sigma_1^-)+g_i(\sigma_2^+a_i+ a_i^\dagger\sigma_2^-)+ H_{ph},
    \end{split}
\end{equation}

Here, $\delta_i=\omega_D-\omega_{c_i}$ where $\omega_D$ is the QD exciton transition frequency and $\omega_{c_i}$ is the \textit{i}th cavity mode frequency. $a_i$ is the creation operator for \textit{i}th cavity mode and $\sigma_i^-=|g_i\rangle\langle e_i|$, the \textit{i}th QD operator. At cryogenic temperatures, exciton-phonon interactions in these semiconductor cavity QED systems are predominantly via deformation potential coupling. The exciton-phonon interaction Hamiltonian, $H_{ph}=\hbar \Sigma_k (\omega_k a_k^\dagger a_k+\Sigma_i\lambda_k^i\sigma_i^+\sigma_i^-(b_k +b_k^\dagger))$. Here, $\lambda_k^i$ is the coupling strength between the \textit{i}th QD exciton and the \textit{k}th phonon bath mode. $b_k$ is the annihilation operator of the \textit{k}th phonon bath mode. Further, we make a polaron transformation, taking the system to the polaron frame, $\Tilde{H}=e^S H e^{-S}$ where $S=\Sigma_k \Sigma_{i=1,2} \sigma_i^+\sigma_i^- \frac{\lambda_k^i}{\omega_k} (b_k-b_k^\dagger)$. The transformed Hamiltonian, $\Tilde{H}=H_s+H_b+H_{sb}$.

\begin{equation}
    \begin{split}
    \text{System Hamiltonian}: 
        H_s =&-\hbar \Delta_{1} a_1^\dagger a_1 - \hbar \Delta_{2} a_2^\dagger a_2 +\langle B\rangle X_g 
    \end{split}
\end{equation}

\begin{equation}
\text{Bath Hamiltonian}:\\
    H_b = \hbar \Sigma_k \omega_k b_k^\dagger b_k
\end{equation}

\begin{equation}
\text{System-Bath Hamiltonian}: 
    H_{sb} = \zeta_g X_g + \zeta_u X_u
\end{equation}

The phonon bath induced polaron shifts, $\Sigma_k \frac{(\lambda_k^i)^2}{\omega_k}$ are absorbed in the detunings, $\Delta_1$ and $\Delta_2$. Throughout our calculation, we assume $\lambda_k^1=\lambda_k^2$, i.e., both the QDs are equally coupled to the \textit{k}th phonon bath mode. The system operators are given by, $X_g= \hbar \Sigma_{i=1,2} (g_i\sigma_1^+a_i + g_i\sigma_2^+a_i)+H.C.$ and $X_u=i\hbar\Sigma_{i=1,2}(g_i\sigma_1^+a_i+g_i\sigma_2^+a_i)+H.C.$ and the bath fluctuation operators are, $\zeta_g=\frac{1}{2}(B_+ + B_- - 2\langle B\rangle)$ and $\zeta_u=\frac{1}{2i}(B_+ - B_-)$. Here, $B_\pm = e^{\pm\Sigma_k \frac{\lambda_k}{\omega_k}(b_k-b_k^\dagger)}$ are the phonon displacement operators. The mean phonon displacement is, $\langle B\rangle = \langle B_+\rangle = \langle B_-\rangle = \exp{[-\frac{1}{2} \int_0^\infty \frac{J(\omega)}{\omega}coth(\beta \hbar\omega/2)]}$ assuming bath is in thermal equilibrium at temperature, T having Bose-Einstein distribution. Here we consider super-ohmic phonon spectral function ,$J(\omega)=\Sigma_k(\lambda_k^{i})^2\delta(\omega-\omega_k)=\alpha_p\omega^3\exp[-\frac{\omega^2}{2\omega_b^2}]$, takes the latter form in continuum limit\cite{Wilson2002,roy2011}. $\langle B\rangle$ is equal to 0.9, 0.84 and 0.72 for T=5K, T=10K and T=20K respectively.

\section{Master equation}

Further, we derive the master equation for the system by including the residue term after making the polaron transformation, $H_{sb}$ is treated using the Born-Markov approximation \cite{nazir2016, Mahan1990}. We also phenomenologically incorporated incoherent processes such as cavity decay, QD exciton spontaneous emission, incoherent pumping, and pure dephasing via Lindblad superoperators \cite{carmichael1999}. The master equation for the QDs-cavity system after tracing over the phonon bath modes is given by,

\begin{equation}
    \begin{split}
    \dot{\rho_s} = &-\frac{i}{\hbar}[H_s,\rho_s]-L_{ph}\rho_s-\Sigma_{j=1,2}\frac{\kappa_j}{2}L[a_j]\rho_s\\&-\Sigma_{i=1,2}(\frac{\gamma_i}{2}L[\sigma_i^-]+\frac{\eta_i}{2}L[\sigma_i^+]+\frac{\gamma_i'}{2}L[\sigma_i^+\sigma_i^-])\rho_s
    \end{split}
    \label{eqn:MEtwoModeHyper}
\end{equation}

Here, the second term on the right-hand side of Eq. \ref{eqn:MEtwoModeHyper}, corresponds to exciton phonon-induced processes' effect on system dynamics given by,

\begin{equation}
    \begin{split}
        L_{ph}\rho_s = &\frac{1}{\hbar^2}\int_{0}^{\infty}d\tau \Sigma_{j=g,u}G_j(\tau)\\&\times[X_j(t),X_j(t,\tau)\rho_s(t)]+H.C.
    \end{split}
\end{equation}

where, $X_j(t,\tau)=e^{iH_s\tau}X_j(t)e^{-iH_s\tau}$. $G_j(\tau)=\langle \zeta_j(0)\zeta_j(\tau)\rangle_{bath}$ are the bath correlation functions $G_g(\tau)=\langle B \rangle^2{\cosh(\phi(\tau)-1)}$, $G_u(\tau)=\langle B \rangle^2\sinh(\phi(\tau))$ 

\section{Off-resonantly coupling}

We consider both the QDs to be equally coupled to the cavity modes within the strong coupling regime, $\gamma_i,\kappa_i<g_i$ with $\gamma_i=0.01g_1$ and $\kappa_1=\kappa_2=0.5g_1$. We study the two cases where the QDs are off-resonantly and resonantly coupled to the cavity modes. In the off-resonant case, the phonon-induced effects play a significant role on the dynamics of the QD state populations and cavity photon statistics. 

\subsection{Steadystate population and cavity photon statistics}

\begin{figure}
    \centering
    \includegraphics[width=\columnwidth,height=\columnwidth]{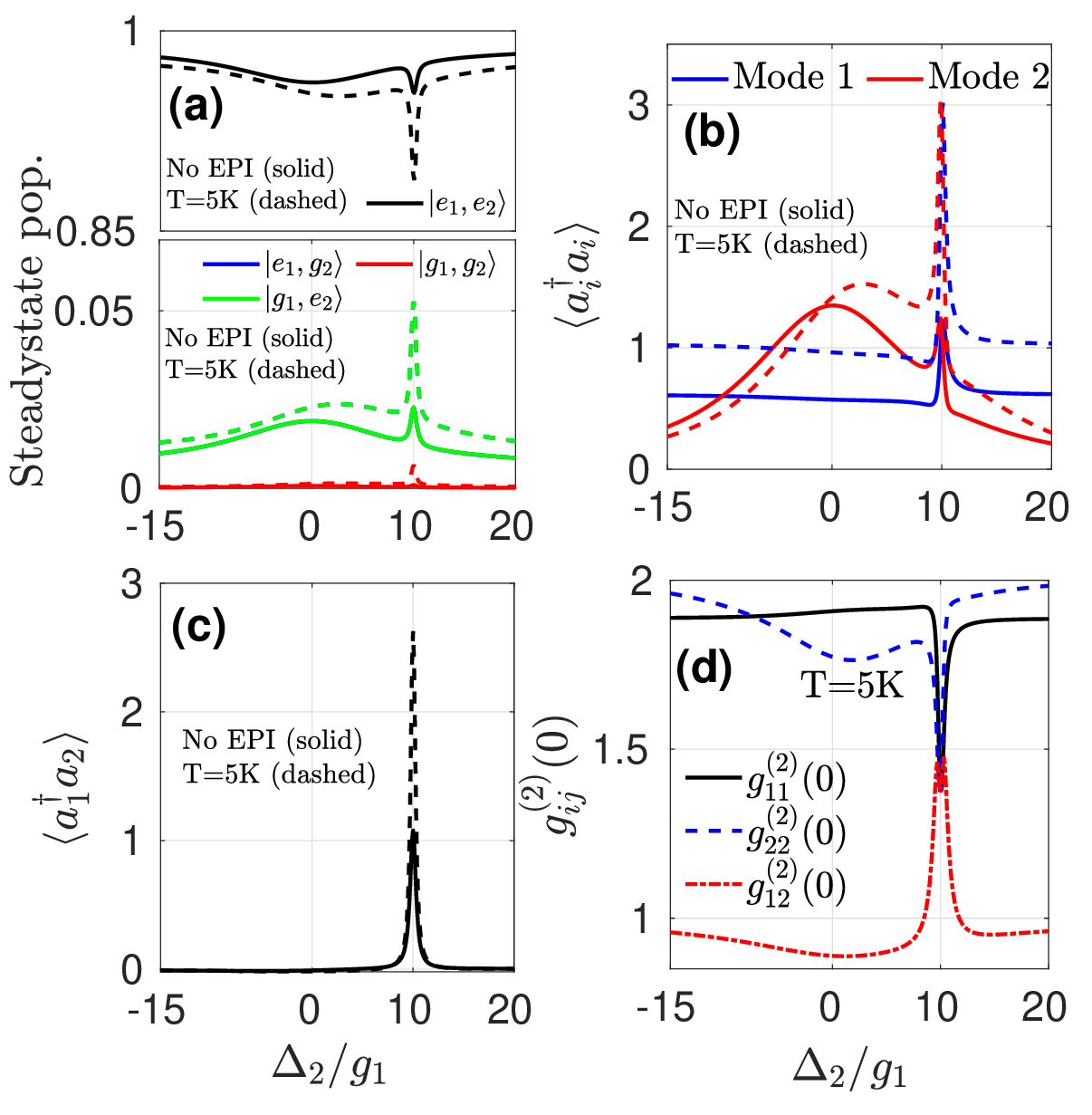}
    \caption{(Color online) Varying second cavity mode detuning w.r.t. QDs, $\Delta_2$ for $\Delta_1=10.0g_1$ and pumping rate, $\eta_1=\eta_2=\eta=25.0g_1$.(a) Steady-state population of collective QD states, $|e_1,e_2\rangle$ (black), $|e_1,g_2\rangle$ (blue), $|g_1,e_2\rangle$ (green) and $|g_1,g_2\rangle$ (red). Under the consideration of identical QDs, the population of $|e_1,g_2\rangle$ and $|g_1,e_2\rangle$ are overlapped. (b) mean photon number of mode 1, $\langle n\rangle_1$ (red) and mode 2, $\langle n\rangle_2$ (blue) (c) correlation between the cavity modes, $\langle a_1^\dagger a_2\rangle$ for ``No EPI" (dashed) and T=5K (solid) (d) intra and inter-mode zero-time delay second order correlation function for T=5K case. Considered system parameters, mode 2 coupling strength, $g_2=1.0g_1$, cavity decay rates, $\kappa_1=\kappa_2=\kappa=0.5g_1$, spontaneous decay rate of the QDs are $\gamma_1=0.01g_1$, $\gamma_2=0.01g_1$ respectively and the pure dephasing rates are $\gamma_1'=\gamma_2'=0.01g_1$.}
    \label{fig:chap3/Fig2}
\end{figure}

In Fig. \ref{fig:chap3/Fig2} we consider cavity mode 1 detuning $\Delta_1=10.0g_1$ and compare the results with and without exciton-phonon interaction (No EPI). We show the results for steady-state populations of the collective QD states along with cavity photon statistics for the incoherent pumping rate, $\eta=25.0g_1$, while the detuning of the cavity mode 2, $\Delta_2$ is varied. From Fig. \ref{fig:chap3/Fig2}(a), it is evident that when both the modes are equally detuned i.e., $\Delta_2=\Delta_1=10.0g_1$, there is a noticeable transfer of population from the state $|e_1,e_2\rangle$ showing dip to the states $|e_1,g_2\rangle$, $|g_1,e_2\rangle$ and $|g_1,g_2\rangle$ showing sharp peak. This leads to the population of the cavity modes ($\langle n\rangle_i$, $i=1,2$) as shown in the  Fig. \ref{fig:chap3/Fig2}(b). The population of the states $|e_1,g_2\rangle$ and $|g_1,e_2\rangle$ states overlap since we consider QDs are identical and their coupling strength to the cavity modes are equal, $g_2=g_1$. 

The mean photon number in mode 1, $\langle n\rangle_1$ remains nearly constant until the detuning $\Delta_2$ approaches $10.0g_1$, where there is a sharp peak. In contrast, the mean photon number in mode 2, $\langle n\rangle_2$ increases gradually as the cavity mode is tuned to resonance with the QDs, $\Delta_2=0.0$, eventually attaining a broadened peak. This peak shifts slights towards right due to phonon induced effects for T=5K case. At $\Delta_2=10.0g_1$, a peak also appears in the mean photon number of mode 2, $\langle n\rangle_2$ reaching a value equal to that of $\langle n\rangle_1$. Further increase in $\Delta_2$ away from $\Delta_1=10.0g_1$, the mean photon number of the mode 1 decreases sharply and attains a constant value and while, that of mode 2 decreases gradually. In this off-resonant scenario, compared to No EPI case (solid) there is large population transfer and increased mean photon number for T=5K case (dashed). 

In addition to the enhanced cavity mode population at $\Delta_2=10.0g_1$, Fig. \ref{fig:chap3/Fig2}(c) shows the establishment of correlation between both the cavity modes, $\langle a_1^\dagger a_2\rangle$ displaying a sharp peak. This marked enhancement in both mean photon numbers, $\langle a_i^\dagger a_i\rangle$ and inter-mode correlation, $\langle a_1^\dagger a_1\rangle$ can be understood by examining the phonon induced processes in the simplified master equation (SME), Eq. \ref{eqn:SMEtwoModeHyper} (Appendix A). The dominant contributions arise from the processes involving phonon induced cavity mode feeding, $\Gamma_{ii}^{\pm}$ ($i=1,2$) and phonon-assisted photon transfer between the modes, $\Gamma_{kl}^{\pm}$ ($k,l=1,2$ \& $k\neq l$).

The results for intra- and inter-mode zero-time delay second order photon correlation functions are given in Fig. \ref{fig:chap3/Fig2}(d). We can see that for $\Delta_2=\Delta_1=10.0$, there is sharp dip in the intra-mode correlation, $g_{ii}^{(2)}(0)$, here $i=1,2$ driving cavity field from thermal to lasing behavior. We also notice peak in inter-mode correlation function, $g_{12}^{(2)}(0)$ attaining value equal to $g_{ii}^{(2)}(0)$. This implies that each cavity mode is equally correlated with itself and with the other mode when $\Delta_2=\Delta_1$. This correlation contributes to the enhancement of cavity mode photon number and facilitates co-operative two-mode two-photon emission \cite{Verma2020} and is discussed in the latter part of the section. The results show that the influence of the presence of mode 2 on the photon statistics of mode 1 is maximum when both are equally tuned with respect to QDs.

\subsection{Radiance witness}

We present the results for the radiance witness (RW) defined as, RW$=\frac{\langle n\rangle_2-2\times\langle n\rangle_1}{2\times\langle n\rangle_1}$, where $\langle n\rangle_2$ denotes the mean photon number when two emitters are coupled to the cavity mode, and $\langle n\rangle_1$ corresponds to that for a single emitter \cite{Pleinert2017}. We examine the effect of detuning, cavity decay rate ($\kappa$), temperature ($T$), and incoherent pumping rate ($\eta$) on RW in this system. Finally, we also compare how the increase in RW effects the linewidth of the cavity emission.

\begin{figure}
    \centering
    \includegraphics[scale=0.75]{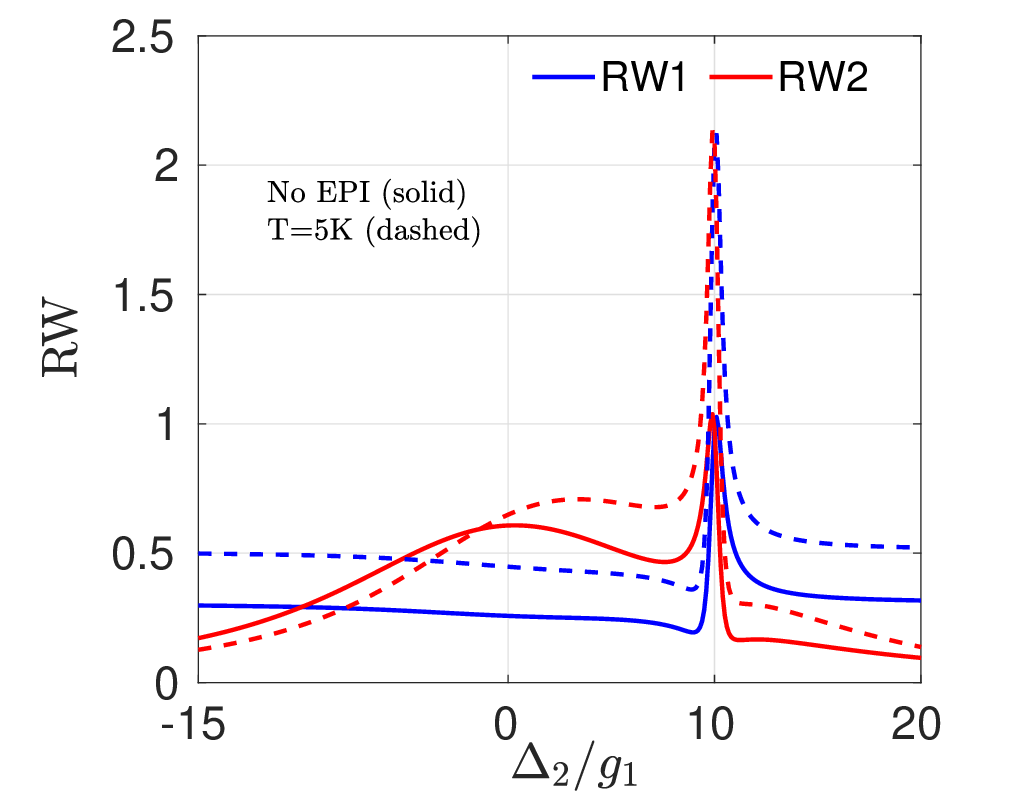}
    \caption{Radiance witness and excess emission rates (EERs) for ``No EPI" (solid) and T=5K (dashed) cases are given.(a) Radiance witness of mode 1, RW1 (blue), mode 2, RW2 (red) for  In off-resonant case, EPI interaction leads to enhanced emission into the cavity modes. (b) Single-photon EER into mode 1 (N1, blue), mode 2 (M1,red) and (c) two-mode two-photon EER (N1M1, black). The parameters are same as in Fig.\ref{fig:chap3/Fig2}.}
    \label{fig:chap3/Fig3}
\end{figure}

\textbf{Effect of detuning}:

In Fig. \ref{fig:chap3/Fig3}, we present the results for the radiance witness and single, two-mode two-photon excess emission rates  varying cavity detuning, $\Delta_2$ and compared the results for No EPI (solid) and temperature, T=5K (dashed) cases. Fig. \ref{fig:chap3/Fig3} presents the result for the radiance witness (RW) defined as, RW$=\frac{\langle n\rangle_2-2\times\langle n\rangle_1}{2\times\langle n\rangle_1}$, where $\langle n\rangle_2$ denotes the mean photon number when two emitters are coupled to the cavity mode, and $\langle n\rangle_1$ corresponds to that for a single emitter \cite{Pleinert2017}. The value of RW characterizes the nature of collective emission: RW$<0$ indicates subradiance, RW$>0$ enhanced emission, RW$=1$ superradiance and RW$>1$ corresponds to Hyperradiance. 

Furthermore, we note that when both the modes are equally detuned, $\Delta_2=\Delta_1=10.0g_1$, the system exhibits Hyperradiant behavior with cavity modes are predominantly populated through single photon emission and cooperative two-mode two-photon emission processes. Fig. \ref{fig:chap3/Fig3} shows the results of the radiance witness (RW) for both the cavity modes using the same parameters as in Fig. \ref{fig:chap3/Fig2}. Similar to the behavior observed in the mean photon number of mode 1, Fig. \ref{fig:chap3/Fig2} (b), the radiance witness of the mode 1 (RW1) also exhibits a peak at the mode 2 detuning, $\Delta_2=10.0g_1$. The radiance witness of mode 2, RW2 shows a gradual increase in its value with a broadened peak around $\Delta_2\approx 0$ and attains sharp peak at $\Delta_2=10.0g_1$. In this off-resonant scheme, phonon-induced effects enhance the radiance witness values, resulting in Hyperradiance with RW$\approx 2.1$ compared to the ``No EPI" case where RW$\approx 1.0$, indicating Superradiance. This enhancement in RW is a result of cooperative two-mode two-photon emission into the cavity modes. 

\begin{figure}
    \centering
    \includegraphics[scale=0.75]{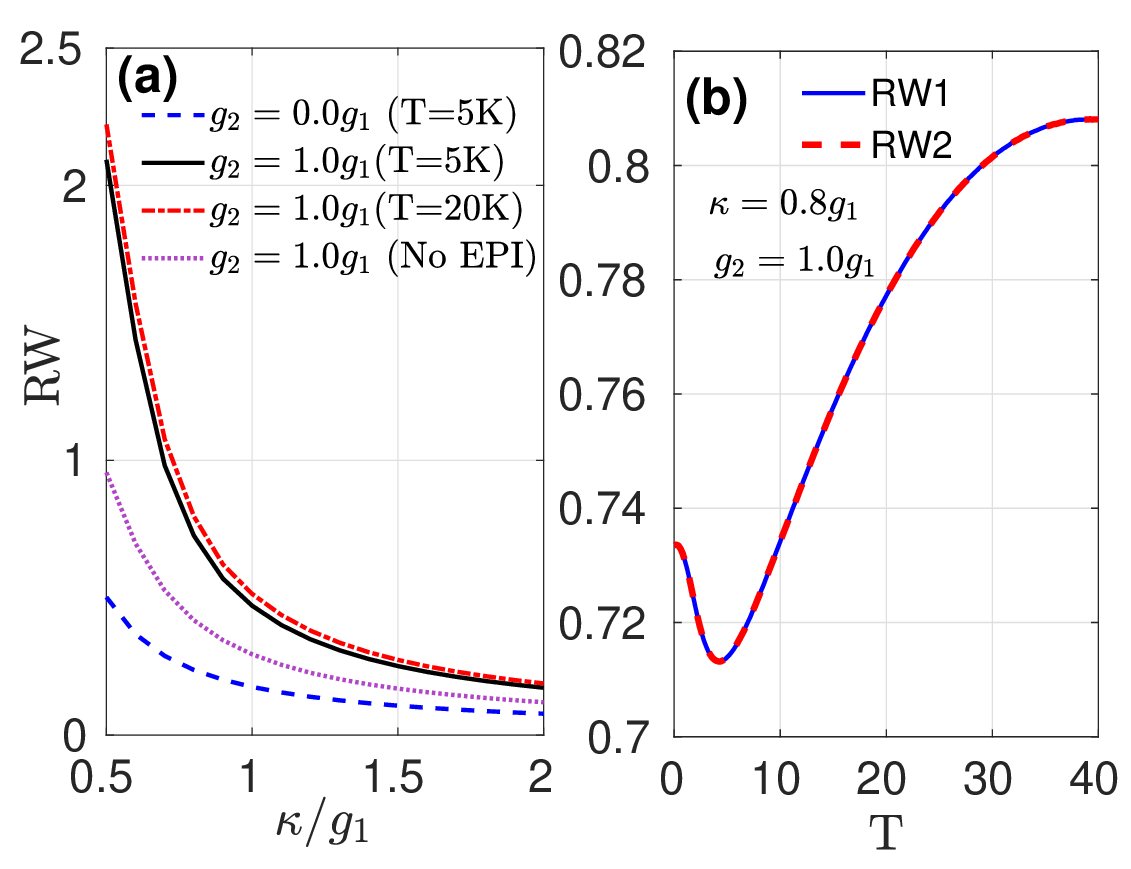}
    \caption{(a) Radiance witness (RW1=RW2=RW) variation with cavity decay rate, $\kappa_1=\kappa_2=\kappa$ is varied. Comparison is made for (i) $g_2=0.0g_1$, single mode T=5K (dashed blue), (ii) $g_2=1.0g_1$, bimodal, T=5K (solid black), (iii) $g_2=1.0g_1$, bimodal T=20K (dash-dotted red) and (iv) $g_2=1.0g_1$, bimodal ``No EPI" (dotted violet) scenarios. In subplot (b), radiance witnesses, RW1 and RW2, and (c) Excess emisssion rates (EERs) are given by varying temperature, T. The cavity mode detunings are fixed at $\Delta_1=\Delta_2=\Delta=10.0g_1$, and cavity decay rates, $\kappa_1=\kappa_2=\kappa=0.8g_1$ considered in subplot (b). The other parameters are same as in Fig. \ref{fig:chap3/Fig2}.}
    \label{fig:chap3/Fig4}
\end{figure}

\textbf{Effect of cavity decay rate, $\kappa$}:

Fig. \ref{fig:chap3/Fig4} (a) shows the effect of cavity decay rate on the RW for (i) single-mode coupling, $g_2=0.0g_1$, for T=5K, bimodal coupling, $g_2=1.0g_1$ for (ii) T=5K, (iii) T=20K and (iv) ``No EPI" scenarios. With the increase in quality of the cavity modes i.e., decrease in $\kappa$, the bimodal cavity system exhibits ``Hyperradiance" (RW$>1$) when compared to the system with QDs coupled to single-mode for decay rates, $\kappa \lessapprox 0.7g_1$ for T=5K. The radiance witness increases slightly for T=20K and is attributed to the increased two-mode two-photon EER discussed in the next subsection. We can see that the RW for ``No EPI" case is far below the T=5K and T=20K cases as the phonon-induced effects dominate.

\textbf{Effect of temperature, $T$}:

In Fig. \ref{fig:chap3/Fig4} (b), we show the results for the variation of RW1 and RW2 with temperature, T  by considering both the cavity modes are equally detuned w.r.t QDs i.e., $\Delta=10.0g_1$ and $\kappa=0.8g_1$. We have considered maximum temperature, T upto 40K and for the considered parameters, the system is within the validity regime of polaron theory \cite{nazir2016}. As the temperature increases, the RW show slight dip initially and rises, which can be attributed to the buildup of correlation between the modes and enhanced contribution of two-mode two-photon emission to the cavity mean photon number.

\begin{figure}
    \centering
    \includegraphics[width=\columnwidth]{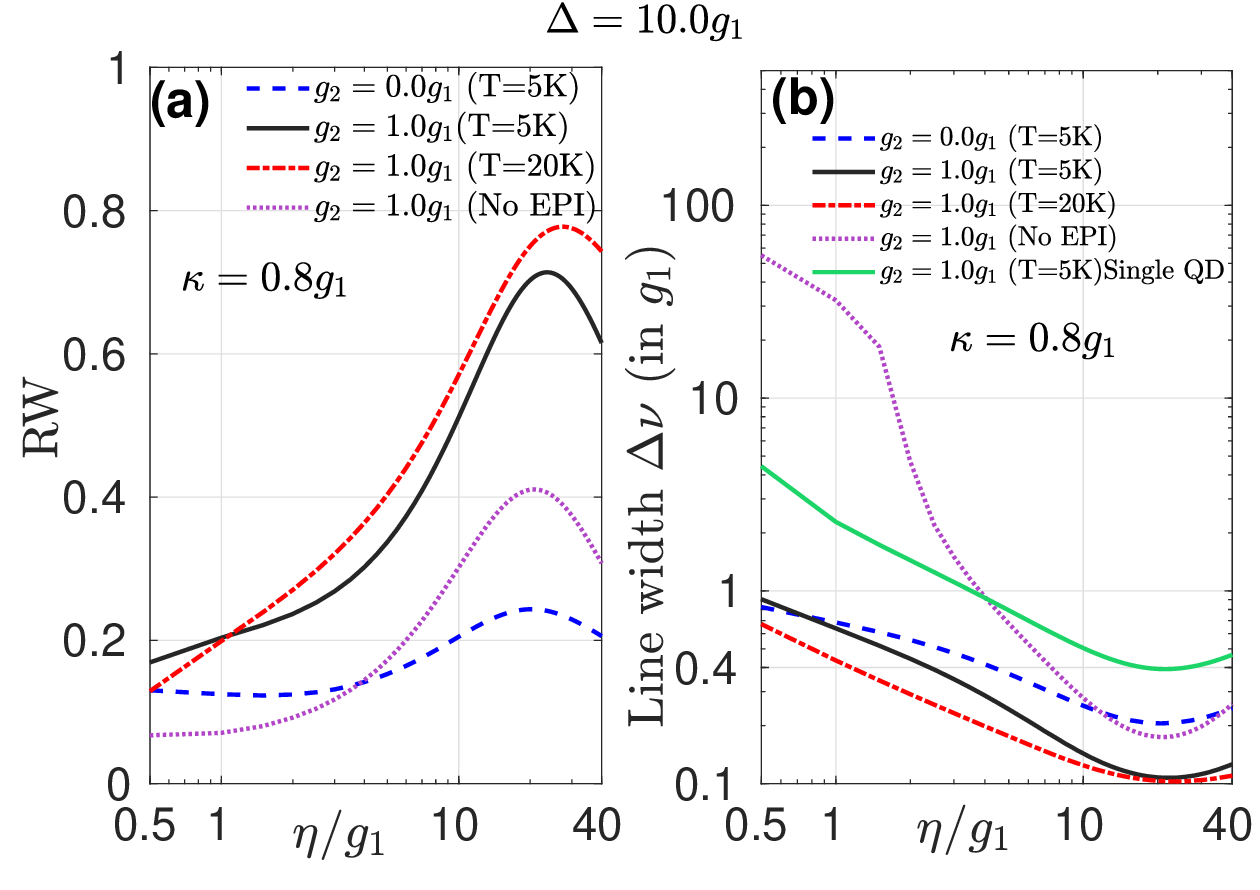}
    \caption{The variation of radiance witness in subplot (a), and linewidth in subplot (b) with increasing incoherent pumping rate, $\eta$. Comparison is made for the scenarios, (i) $g_2=0.0g_1$ (single mode) T=5K, (ii) $g_2=1.0g_1$ (bimodal) T=5K, (iii) $g_2=1.0g_1$ (bimodal), T=20K, (iv) $g_2=1.0g_1$ (bimodal),``No EPI", and (v) $g_2=1.0g_1$ (bimodal) T=5K, single QD . Color scheme is same as in Fig. \ref{fig:chap3/Fig4} (a). The other parameters are same as in Fig. \ref{fig:chap3/Fig4} (b).}
    \label{fig:chap3/Fig5}
\end{figure}

\textbf{Effect of incoherent pumping rate, $\eta$}:

In Fig. \ref{fig:chap3/Fig5}, we show the results of radiance witness and linewidth of the cavity mode emission spectrum with increasing incoherent pumping rate, $\eta$, for equal detuning of the cavity modes $\Delta_1=\Delta_2=\Delta=10.0g_1$ and $\kappa=0.8g_1$. In subplot Fig. \ref{fig:chap3/Fig5} (a) we made comparison of the four cases as we did in Fig. \ref{fig:chap3/Fig4} (a). The results clearly show that the radiance witness is notably higher for the case of QDs coupled to bimodal cavity compared to the coupling with single mode cavity, indicating enhanced cooperative emission. Futher, the peak value of the RW for ``No EPI" case is almost half that of T=20K case. Additionally, in this off-resonant coupling scheme, the dominant phonon induced processes contribute to increased RW with rise in temperature, T, as shown for T=20K (red dash-dotted line). This enhancement in RW leads to the suppression of the linewidth as given in Fig. \ref{fig:chap3/Fig5} (b) as in superradiant lasers \cite{Meiser2009, Kristensen2023}.

\textbf{Radiance witness vs Linewidth}:

The emission spectrum is obtained using the quantum regression theorem applied to the two-time correlation function, $\langle a^\dagger(t) a(0)\rangle$ \cite{SMTan1999} and the linewidth is evaluated by fitting the spectrum with a Lorentzian function. We can see that for T=5K, as the radiance witness increases and attains peak value $\approx 0.7$, the corresponding linewidth decreases. By comparing the minimum linewidth obtained for single QD (green) and two QDs (black) cases, we find $\approx 1/4$ times suppression of linewidth due to the presence of second QD, in agreement with the $\approx 1/N^2$ scaling, a signature of superradiant lasing \cite{Haake1993,Yu2010}. In particular, for T=20K case, the emission linewidth $\Delta\nu$ reduces to approximately 0.1, which is nearly $85\%$ narrower than the bare cavity linewidth, $\kappa = 0.8g_1$. The results clearly demonstrate that the QDs coupled to bimodal cavities outperform both the single-mode coupling case and bimodal coupling``No EPI" case.

\subsection{Laser rate equations}

In the subsection, we analyze the individual contribution of single-photon and two-mode two-photon processes to the cavity mode population by evaluating their emission and absorption rates. We plot the results for the difference between the emission and absorption rates, referred to as the ``Excess Emission Rate (EER)".

We have computed the single and multi-photon absorption and emission rates contributing to the cavity mode mean photon number, as given in Fig. \ref{fig:chap3/Fig3} (b) and (c) for T=5K (dashed), ``No EPI" (solid) cases. We follow the standard procedure of Scully-Lamb quantum theory of lasing to evaluate the emission and absorption rates. Specifically, we employ the simplified master equation (SME) of the system, detailed in Appendix \ref{sec:chap3_Appendix1} to determine the rate equations of the density matrix elements, $\langle i_1,j_2,m,n|\rho_s|i_1,j_2,p,q\rangle$ where $i, j=e,g $ denote QD states and $m, p$ ($n, q$) represent the photon numbers in mode 1 (mode 2). We then trace over the QD states to obtain the rate equation for the cavity fields \cite{Addepalli2024, Hazra2024}.

    \begin{equation}
    \begin{split}
        \dot{P}_{n,m} = &-\alpha_{n,m} P_{n,m}+G^{11}_{n-1,m-1}P_{n-1,m-1} 
        +G^{10}_{n-1,m}P_{n-1,m}+G^{01}_{n,m-1}P_{n,m-1}
        \\&+A^{11}_{n+1,m+1}P_{n+1,m+1}+A^{10}_{n+1,m}P_{n+1,m}+A^{01}_{n,m+1}P_{n,m+1}
        \\&+\kappa_{1}(n+1)P_{n+1,m}- \kappa_{1}nP_{n,m}
        +\kappa_{2}(m+1)P_{n,m+1} - \kappa_{2}mP_{n,m}.
    \end{split}
\end{equation}

The probability of having $n$, $m$ photons in 1st, 2nd cavity modes respectively is given by, $P_{nm}=\Sigma_i \langle i,n,m|\rho_s|i,n,m\rangle$, while  $\alpha_{n,m}=\Sigma_i \alpha_{i,n,m}\langle i,n,m|\rho_s|i,n,m\rangle$. The emission and absorption coefficients are given by,, $G_{n,m}^{ab}P_{n,m}=\Sigma_i G_{i,n,m}^{ab}\langle i,n,m|\rho_s|i,n,m\rangle$ and $A_{n,m}^{ab}P_{n,m}=\Sigma_i A_{i,n,m}^{ab} \langle i,n,m|\rho_s|i,n,m\rangle$ where $i={x,y,g}$. The coefficients $\alpha_{i,n,m}$, $G_{i,n,m}^{ab}$ and $A_{i,n,m}^{ab}$ are obtained numerically. The single-photon emission(absorption) rate for the first and second modes are given by, $\Sigma_{n,m} G_{n,m}^{10} P_{n,m}$($\Sigma_{n,m}A_{n,m}^{10}P_{n,m}$) and $\Sigma_{n,m} G_{n,m}^{01} P_{n,m}$($\Sigma_{n,m}A_{n,m}^{01}P_{n,m}$) respectively and two-mode two-photon emission(absorption) rate is given by $\Sigma_{n,m} G_{n,m}^{11} P_{n,m}$($\Sigma_{n,m}A_{n,m}^{11}P_{n,m}$). Following the definitions used in earlier chapters, we define single-photon and two mode two-photon excess emission rates (EER) as the difference between the corresponding emission and the absorption rates. The sign of EER $>0$ or $<0$ represents net emission or absorption occurring in the cavity mode. 

\textbf{Varying detuning}:

\begin{figure}
    \centering
    \includegraphics[scale=0.75]{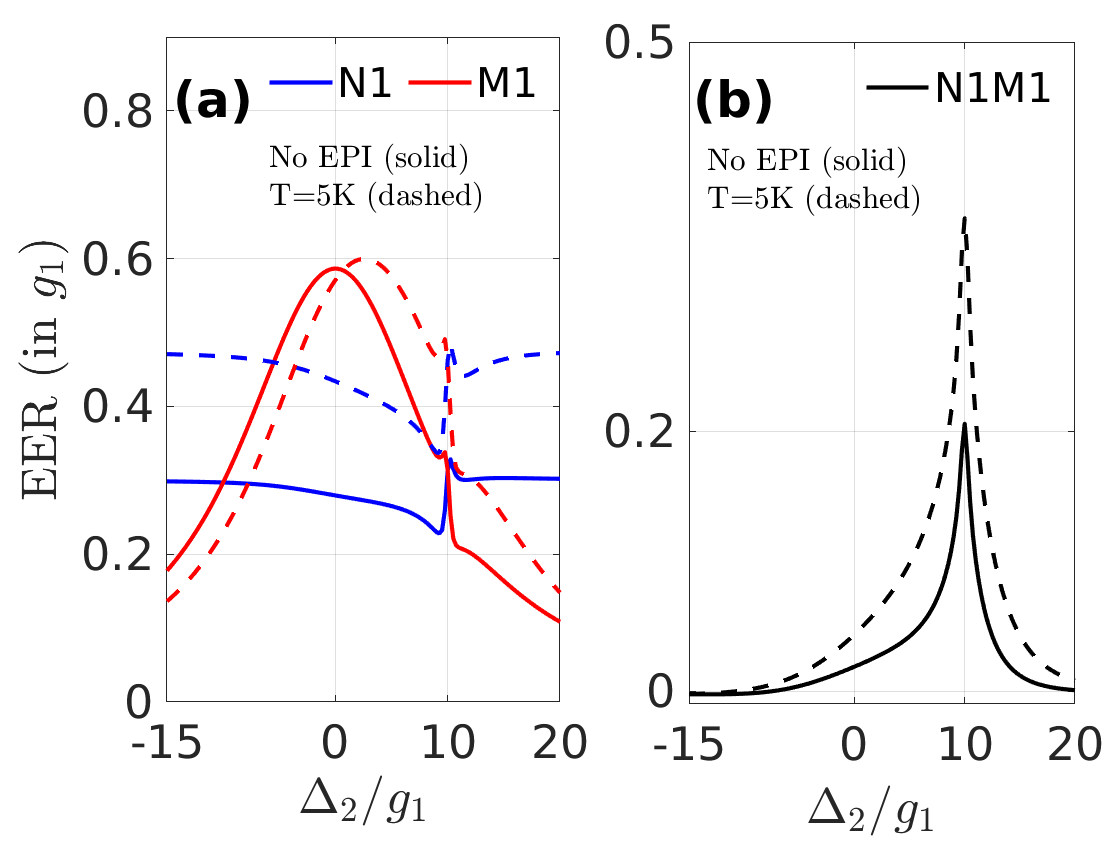}
    \caption{Excess emission rates (EERs) for ``No EPI" (solid) and T=5K (dashed) cases are given.(a) Single-photon EER into mode 1 (N1, blue), mode 2 (M1,red) and (b) two-mode two-photon EER (N1M1, black). The parameters are same as in Fig.\ref{fig:chap3/Fig2}.}
    \label{fig:chap3/Fig6}
\end{figure}

The results for EERs are presented in Fig. \ref{fig:chap3/Fig6} (a) and (b). From Fig.\ref{fig:chap3/Fig3} (a), we observe a broadened peak in single photon excess emission rate into mode 2 (M1 curve) when $\Delta_2\approx0.0g_1$ and for $\Delta_2\neq \Delta_1$, both the modes are predominantly populated by single photon processes (N1, M1). However, as the detuning of the mode 2, $\Delta_2$ varied, and when both modes are equally tuned w.r.t QDs i.e., $\Delta_2=\Delta_1=10.0g_1$, the single photon EERs (N1, M1) cross each other. At $\Delta_2=\Delta_1$, both the cavity modes are not only equally populated by single photon processes (N1, M1) but also show significant contribution from cooperative two-mode two-photon process (N1M1) as shown in Fig. \ref{fig:chap3/Fig6} (b) as we cam see a sharp peak in ``N1M1" curve. This cooperative two-mode two-photon emission process is responsible for the generation of correlation between both the cavity modes, Fig. \ref{fig:chap3/Fig2}(c). We further note that, in comparison to the ``No EPI" case (solid), the two-mode two-photon excess emission rates is twice for T=5K case (dashed) and is attributed to the exciton-phonon interaction in this off-resonantly coupled system.

\textbf{Varying temperature}:

\begin{figure}
    \centering
    \includegraphics[width=0.75\columnwidth]{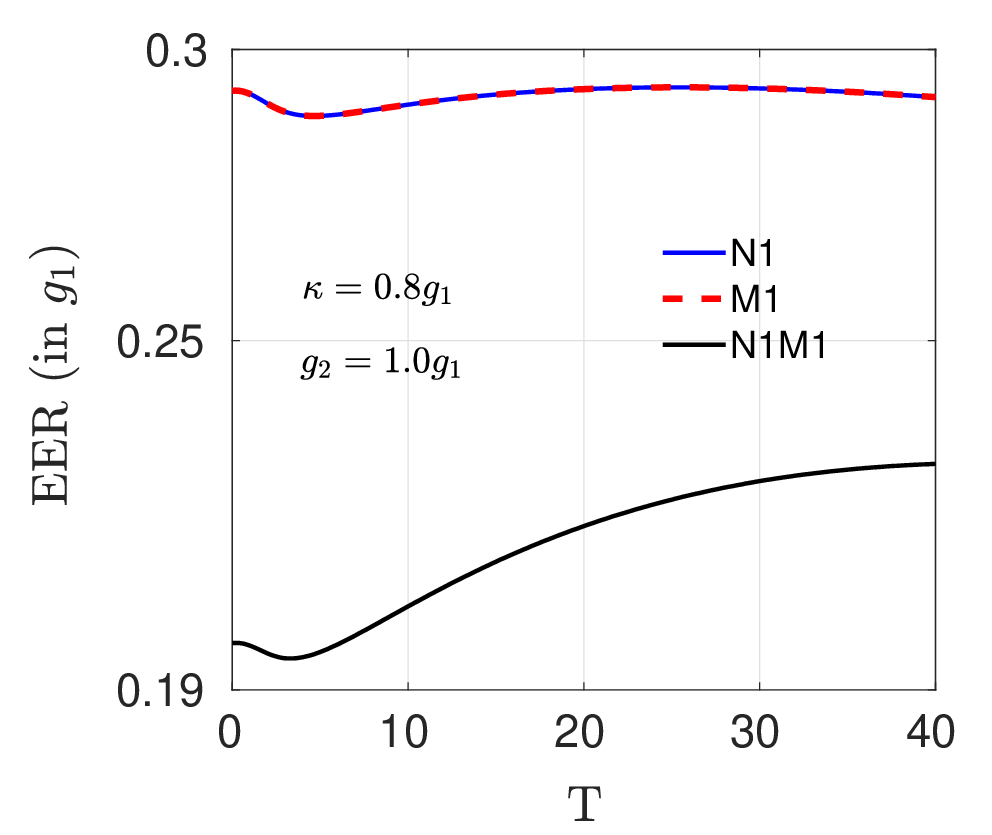}
    \caption{Excess emisssion rates (EERs) are given by varying temperature, T. The cavity mode detunings are fixed at $\Delta_1=\Delta_2=\Delta=10.0g_1$, and cavity decay rates, $\kappa_1=\kappa_2=\kappa=0.8g_1$. The other parameters are same as in Fig. \ref{fig:chap3/Fig2}.}
    \label{fig:chap3/Fig7}
\end{figure}

As the temperature increases, the RW initially dips and rises with further increase in temperature, Fig. \ref{fig:chap3/Fig4} (b). The results for EERs, Fig. \ref{fig:chap3/Fig7} show that as the temperature increases, both single and two-mode two-photon excess emission rates initially show a slight dip. With further increase in T, the single photon EERs show very little variation but the two-mode two-photon EER increases appreciably. Hence, the phonon-assisted two-mode two-photon process strengthens the correlation between the modes, and resulting in the increase of their radiance witnesses, driving the system from superradiance to Hyperradiance.

\section{Resonant coupling}

\subsection{Radiance witness}

\begin{figure}
    \centering
    \includegraphics[width=0.75\columnwidth]{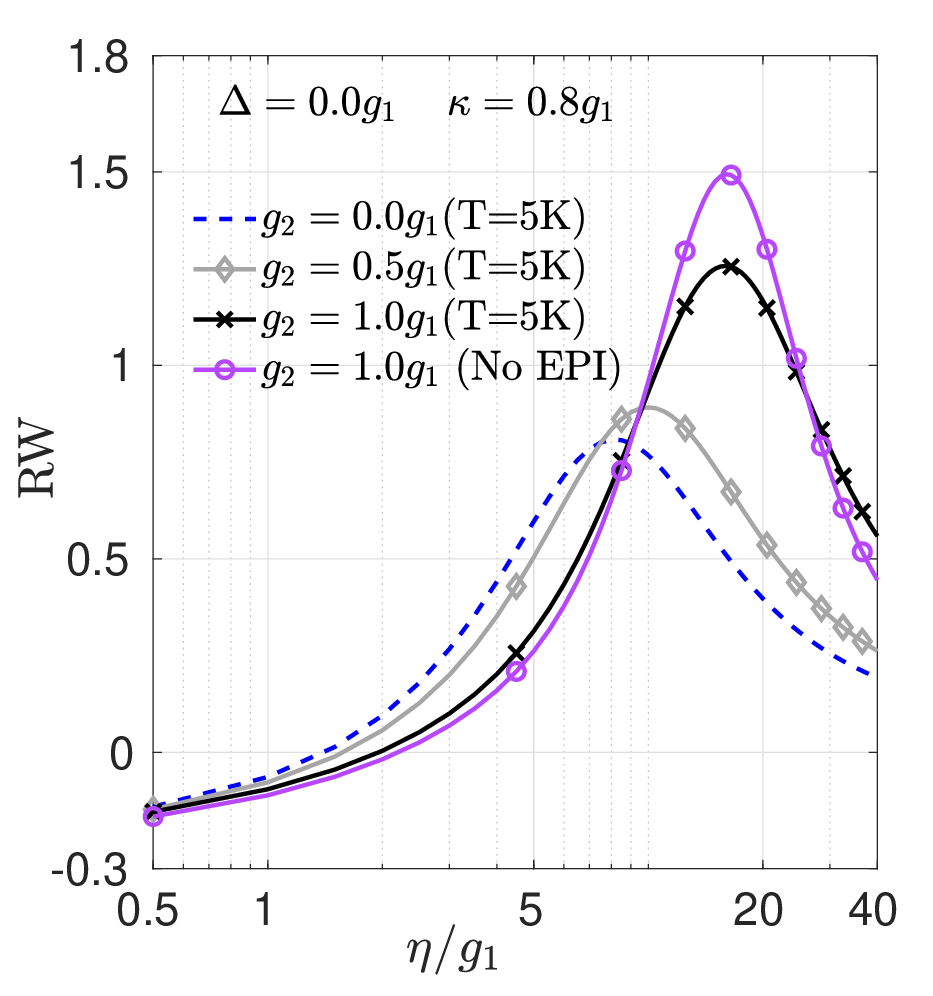}
    \caption{Radiance witness for resonant case, $\Delta_1=\Delta_2=\Delta=0.0g_1$ by varying incoherent pumping rate, $\eta$. Comparison is made between the cases: T=5K  (i) $g_2=0.0g_1$ (Single mode, dashed blue) (ii) $g_2=0.5g_1$ ($\diamond$), (iii) $g_2=0.2g_1$ ($\times$) and (iv) $g_2=1.0g_1$ ``No EPI" ($\circ$). Cavity decay rates, $\kappa=0.8g_1$ and the other parameters are same as in Fig. \ref{fig:chap3/Fig2}.}
    \label{fig:chap3/Fig8}
\end{figure}

Fig. \ref{fig:chap3/Fig8} shows results for the radiance witness of mode 1 (RW1) and mode 2 (RW2), for the case of QDs resonantly coupled to the cavity modes i.e., $\Delta_1=\Delta_2=0.0g_1$. We vary the incoherent pumping rate, $\eta$, considering both the cavity modes have equal quality factors, $\kappa_2=\kappa_1=0.8g_1$. The results are given for different second mode coupling strengths, (i) $g_2=0.0g_1$ (solid brown) corresponds to single mode scenario, (ii) $g_2=0.5g_1$ (`$\diamond$') at T=5K, (iii) $g_2=1.0g_1$ (`$\times$') at T=5K and (iv) $g_2=1.0g_1$, ``No EPI" (`$\circ$') case. We note that the radiance witness for both the modes overlap even when the second mode is coupling strength, $g_2=0.5g_1$. This is attributed to the presence of strong correlation between the modes when they are equally detuned, $\Delta_1=\Delta_2$ and having same cavity decay rates, $\kappa_1=\kappa_2$. Increase in the incoherent pumping rate drives the system from subradiant to Superradiant or Hyperradiant regime. For the considered parameters, the peak value of RW1 increases in the presence of the second mode and exceeds unity (RW1$>1$) indicating ``Hyperradiance" over a range of incoherent pumping strengths, $10g_1 \leq\eta\leq 25.0g_1$ as compared to the single mode case whose peak value is RW$\approx 0.8$ (below Superradiant phase). Furthermore, by comparing the results for (ii) and (iii) cases, we can say that with increase in the mode 2 coupling strength, $g_2$, the peak value of RW increases. For resonant case, the phonon induced decoherence leads to the suppression in mean photon number and the peak value of the radiance witness as shown for the ``No EPI" (`$\circ$') and T=5K (`$\times$') cases unlike the off-resonant case where the phonon induced effects enhance the radiance witness, cf., Fig. \ref{fig:chap3/Fig3} (a), Fig. \ref{fig:chap3/Fig4} (a) and Fig. \ref{fig:chap3/Fig5} (a).

\section{Conclusions}

In conclusion, we have demonstrated that when both cavity modes are equally detuned from the quantum dot (QD) transition frequencies, there is a pronounced enhancement in inter-mode correlations and the cooperative two-mode two-photon emission rate, leading to two-mode hyperradiant lasing. In the case of off-resonant coupling, exciton-phonon interactions (EPI) play a crucial role in enhancing the emission rate, as confirmed by comparisons with scenarios where EPI is neglected. Moreover, we have shown that bimodal coupling, in contrast to the single-mode configuration, significantly enhances the radiance witness and supports the emergence of hyperradiant lasing. This enhancement is also associated with a suppression of the laser linewidths. Finally, we examined the resonant coupling regime where exciton-phonon interactions are negligible. In this regime, cavity-mediated two-mode two-photon emission continues to enable hyperradiant lasing. However, in the presence of exciton-phonon interactions at T = 5\,\text{K}, the peak value of the radiance witness is slightly reduced compared to the "No EPI" case, attributed to phonon-induced decoherence.

%% file: chap4.tex
\chapter{Correlated emission lasing }\label{chap4}
{\small This chapter is based on our work ``\textsc{Correlated emission lasing in a single quantum dot embedded inside a bimodal photonic crystal cavity,}"; \textbf{Lavakumar Addepalli} and P. K. Pathak; \textit{Phys. Rev. B} 111, 125422 (2025).}

\section{Introduction}

In this chapter, we investigate the phenomenon of correlated emission lasing in a coherently driven single quantum dot coupled to a bimodal photonic crystal cavity. In a correlated emission laser (CEL), coherence between the upper levels in a three-level atomic system leads to correlated spontaneous emissions into the cavity modes. This correlation suppresses quantum noise in the laser, driving it towards the vacuum noise limit (VNL). CEL has significant applications in laser gyroscopes \cite{Scully1982, Bergou1991, Mecozzi2023} and gravitational wave detectors \cite{Scully1986, Schnabel2010}, where detecting ultrasmall phase shifts in laser modes is crucial. CEL is also valuable in fields like quantum metrology, sensing and high-resolution spectroscopy \cite{Giovannetti2006,Degen2017}, where noise from spontaneous emission often imposes limitations.

\begin{figure}
    \centering
    \includegraphics[width=0.75\linewidth]{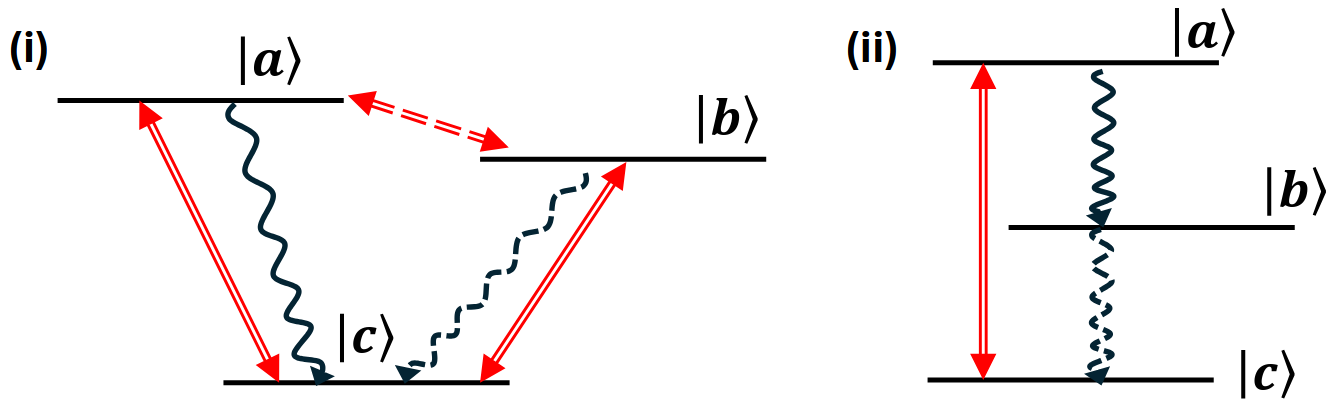}
    \caption{Correlated Emission Laser (CEL) schemes. (a) `V'-type configuration: Coherence between higher energy states, $|a\rangle$ and $|b\rangle$ is generated using external coherent drives coupled to either $|c\rangle\leftrightarrow|a\rangle$, $|c\rangle\leftrightarrow|b\rangle$ or $|a\rangle\leftrightarrow|b\rangle$ transitions. The correlated emission from $|a\rangle$ and $|b\rangle$ states to the common ground state leads to suppression of quantum noise. (b) Cascaded configuration: Coherence between the lowest and highest energy levels ($|c\rangle\leftrightarrow|a\rangle$) is generated using an external coherent drive. The transitions, $|a\rangle\rightarrow|b\rangle$ and $|b\rangle\rightarrow|c\rangle$ coupled to degenerate or nondegenerate radiation modes lead to quantum noise suppression. \cite{ScullyZubairyQOT} }
    \label{fig:CELschemes}
\end{figure}

CEL has been proposed in various atomic-level configurations, such as the ‘V’ type system employed in quantum beat lasers \cite{Bergou1988QBL, Zaheer1988QBL}, Hanle lasers \cite{Bergou1988hanle, Lu1990JOSAB}, and three-level cascade systems \cite{Scully1988cascade, Lu1990cascade}, Fig. \ref{fig:CELschemes}. There have been some interesting proposals for single emitter CEL\cite{Kiffner2007,Dey2023}, and non-degenerate two-photon CEL has been realized in a single superconducting three-level artificial atom \cite{Peng2015} with relative phase diffusion noise observed $10^{-4}$ times the Schawlow-Townes limit. The nonlinear quantum theories for quantum beat lasers and CEL-based Hanle lasers have been developed specifically for atomic systems\cite{Lu1989,Bergou1988QBL}. 

With ongoing important developments in single emitter lasers, there have been some interesting proposals for single emitter CEL\cite{Kiffner2007,Dey2023}. Further, using a single superconducting three-level artificial atom, non-degenerate two-photon CEL has been realized and relative phase diffusion noise has been observed $10^{-4}$ times the Schawlow-Townes limit which is much smaller than the relative phase diffusion noise observed in CEL using ensemble of atoms\cite{Peng2015}. Thus a single emitter CEL is an ideal candidate for suppressing relative phase noise. CEL have been realized in a few remarkable experiments and a large reduction in phase diffusion noise has been observed \cite{Winters1990,Steiner1995,Abich2000}. CEL also facilitates discrete variable (DV) entanglement generating polarization entangled photon pairs \cite{Leppenen2024} and continuous variable (CV) entanglement generation between lasing modes, which provides method for generating entanglement between large number of photons\cite{Ikram2007, Ge2013}.

Realizing the CEL in semiconductor cavity QED systems is particularly exciting and is interesting to investigate the effects of exciton-phonon interactions on relative phase noise in a single QD CEL. Here, we propose CEL using a single quantum dot (QD) embedded in a photonic bimodal cavity. When excited, a QD \textit{x}-polarized or \textit{y}-polarized excitons can be generated. Together, the \textit{x}- and \textit{y}-polarized excitons and the ground state (no electron-hole pair) create a `V' type energy configuration. When these transitions are coupled with two orthogonally polarized cavity modes, the system resembles a Hanle laser \cite{ScullyZubairyQOT}. We analyze the steady-state dynamics of the system and examine fluctuations in the relative and average phase Hermitian operators. We also calculate the phase drift and diffusion coefficients, and finally, the presence of continuous variable (CV) entanglement between the cavity modes is discussed.

\section{Model system}

\begin{figure}
    \centering
    \includegraphics[width=\columnwidth]{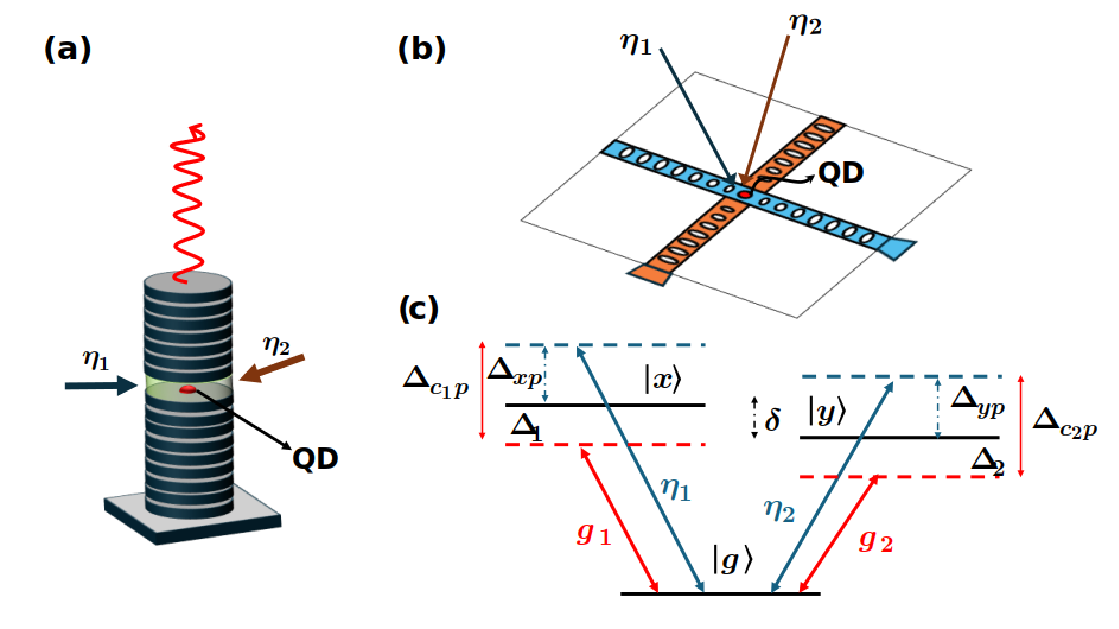}
    \caption{(a) Schematic figure showing a single QD embedded inside an elliptical bimodal micropillar cavity \cite{Leppenen2024,Mehdi2024,Heermeier2022,Reitzenstein2010} and in (b) a single QD placed at the center of a bimodal photonic crystal cavity \cite{Pasharavesh2025,Rivoire2011}. (c) QD level structure with coherent pump ($\eta_1,\eta_2$) and cavity mode ($g_1,g_2$) coupling between $|x\rangle$, $|y\rangle$ and $|g\rangle$ states. \cite{Addepalli2025correlated}.}
    \label{fig:chap4/Fig1}
\end{figure}

We consider a single quantum dot in a three-level configuration with a ground state \( |g\rangle \) and two exciton states, \( |x\rangle \) and \( |y\rangle \), embedded in a bimodal cavity. The transitions \( |g\rangle \leftrightarrow |x\rangle \) and \( |g\rangle \leftrightarrow |y\rangle \) are selectively driven by two orthogonally polarized external coherent fields, and are also coupled to two orthogonally polarized cavity modes. A schematic diagram of the system is shown in Fig. \ref{fig:chap4/Fig1}.

The Hamiltonian of the system is given by,
\begin{equation}
    \begin{split}
         H = & \omega_x \sigma_1^+\sigma_1^-+\omega_y \sigma_2^+\sigma_2^-+ \omega_{c_1} a_1^\dag a_1 + \omega_{c_2} a_2^\dag a_2 
         \\&+g_1(\sigma_1^+a_1+ a_1^\dag\sigma_1^-)+g_2(\sigma_2^+a_2+ a_2^\dag\sigma_2^-)
         \\&+\eta_1(\sigma_1^+e^{-i\omega_{Lx}t}+\sigma_1^-e^{i\omega_{Lx}t})\\&+\eta_2(\sigma_2^+e^{-i\omega_{Ly}t}+\sigma_2^-e^{i\omega_{Ly}t}) + H_{ph},
    \end{split}
\end{equation}

where $a_i$ is the annihilation operator for the i-th cavity mode, $g_1$, $g_2$ are the $|g\rangle\leftrightarrow|x\rangle$, $|g\rangle\leftrightarrow|y\rangle$ transitions coupling strength to the 1st and 2nd cavity modes. QD operators are given by, $\sigma_1^+=|x\rangle\langle g|$, $\sigma_2^+=|y\rangle\langle g|$. The coherent pumping strengths are $\eta_1$, $\eta_2$ for $|g\rangle \leftrightarrow |x\rangle$, $|g\rangle \leftrightarrow |y\rangle$ transitions respectively. The exciton-phonon interaction Hamiltonian is given by, 

\begin{equation}
    H_{ph} = \hbar \Sigma_k\omega_k b_k^\dagger b_k + \hbar\Sigma_{i=x,y}\Sigma_k \lambda_k^i |i\rangle\langle i| (b_k^\dagger + b_k)
\end{equation}

Here, $b_k$ is the annihilation operator for the k-th phonon bath mode. Now, we write the above Hamiltonian in the laser frequencies rotating frame, $H'=e^{iH_0 t} H e^{-iH_0 t}$ where, $H_0=\omega_{Lx}\sigma_1^+\sigma_1^- +\omega_{Ly}\sigma_2^+\sigma_2^- +\omega_{Lx} a_1^\dag a_1 + \omega_{Ly}a_2^\dag a_2$.

\begin{equation}
    \begin{split}
         H' = & \delta_{xp} \sigma_1^+\sigma_1^-+\delta_{yp} \sigma_2^+\sigma_2^- + \Delta_{c_1p} a_1^\dag a_1 + \Delta_{c_2p} a_2^\dag a_2
         \\&+g_1(\sigma_1^+ a_1+ a_1^\dag \sigma_1^-)+g_2(\sigma_2^+ a_2+ a_2^\dag\sigma_2^-) 
         \\&+\eta_1(\sigma_1^+ + \sigma_1^-)+\eta_2(\sigma_2^+ + \sigma_2^-) + H_{ph},
    \end{split}
\end{equation}

where, the detunings of the QD states and the cavity modes w.r.t. pump frequencies are given by $\delta_{xp}=\omega_x-\omega_{Lx}$, $\delta_{yp}=\omega_y-\omega_{Ly}=\omega_x-\delta-\omega_{L_y}$, where $\delta$ is the fine structure splitting and $\Delta_{c_1p}=\omega_{c_1}-\omega_{Lx}$, $\Delta_{c_2p}=\omega_{c_2}-\omega_{Ly}$ respectively.
To include exciton-phonon effects to all orders non-perturbatively, we make polaron transformation\cite{Mahan1990,Xu2016,nazir2016}, $\Tilde{H}= e^S H' e^{-S}$, where $S=\Sigma_{i=1,2}\sigma_i^+\sigma_i^-\Sigma_k \frac{\lambda_k^i}{\omega_k}(b_k^\dagger-b_k)$. $\lambda_k^1$, $\lambda_k^2$ is the $|x\rangle$, $|y\rangle$ exciton and k-th phonon bath mode coupling strength respectively. 

The transformed Hamiltonian, $\Tilde{H}$ has form,

\begin{equation}
    \Tilde{H} = H_s+H_b+H_{sb}
\end{equation}

where, the system Hamiltonian, $H_s$

\begin{equation}
    \begin{split}
        H_s =&\hbar \Delta_{xp}\sigma_1^+\sigma_1^- + \hbar \Delta_{yp}\sigma_2^+\sigma_2^- + \\& \hbar \Delta_{c_1p}a_1^\dagger a_1+\hbar \Delta_{c_2p}a_2^\dagger a_2+\langle B\rangle X_g 
    \end{split}
\end{equation}

the bath Hamiltonian, $H_b$

\begin{equation}
    H_b = \hbar \Sigma_k \omega_k b_k^\dagger b_k
\end{equation}

the system-bath interaction Hamiltonian, $H_{sb}$

\begin{equation}
    H_{sb} = \zeta_g X_g + \zeta_u X_u
\end{equation}

The polaron shifts, $\Sigma_k \frac{(\lambda_k^1)^2}{\omega_k}$, $\Sigma_k \frac{(\lambda_k^2)^2}{\omega_k}$ are absorbed in the $\Delta_{xp}$, $\Delta_{yp}$. The phonon displacement operators are given by, $B_{\pm} = \exp[\pm \Sigma_{i=1,2}\Sigma_{k} \frac{\lambda_k^{i}}{\omega_k}(b_k - b_k^\dagger]$, with $\langle B_{\pm} \rangle = \langle B \rangle$. For simplification, we have considered equal coupling strengths $\lambda_k^{1} = \lambda_k^{2}$. The system operators are given by, $X_g = \hbar(g_1\sigma_1^+a + g_2\sigma_2^+a)+\hbar (\eta_1 \sigma_1^+ + \eta_2 \sigma_2^+)+H.C.$, $X_u = i\hbar(g_1\sigma_1^+ a+g_2\sigma_2^+ a + \eta_1 \sigma_1^+ + \eta_2 \sigma_2^+)+H.C.$ and bath fluctuation operators are, $\zeta_g = \frac{1}{2}(B_++B_- -2\langle B \rangle)$, $\zeta_u = \frac{1}{2i}(B_+ - B_-)$. Using the polaron transformed Hamiltonian, $\Tilde{H}$ and Born-Markov approximation, we derive the master equation for the QD-cavity system \cite{roy2011,nazir2016}.

\section{Master equation}

The density matrix master equation (ME) for the system is given by,

\begin{equation}
    \begin{split}
    \dot{\rho_s} = &-\frac{i}{\hbar}[H_s,\rho_s]-L_{ph}\rho_s-\Sigma_{j=1,2}\frac{\kappa_j}{2}L[a_j]\rho_s\\&-\Sigma_{i=1,2}(\frac{\gamma_i}{2}L[\sigma_i^-]+\frac{\gamma_i'}{2}L[\sigma_i^+\sigma_i^-])\rho_s
    \end{split}
    \label{eqn:ME}
\end{equation}

where $L[\hat{O}]\rho_s = \hat{O^\dagger}\hat{O}\rho_s - 2\hat{O}\rho_s\hat{O^\dagger}+\rho_s\hat{O^\dagger}\hat{O}$ is the Lindblad superoperator. The second term in the master equation $L_{ph}\rho_s$ represents the Liouvillian capturing the effect of system-bath interaction is given by, 

\begin{equation}
    \begin{split}
        L_{ph}\rho_s = &\frac{1}{\hbar^2}\int_{0}^{\infty}d\tau \Sigma_{j=g,u}G_j(\tau)\\&\times[X_j(t),X_j(t,\tau)\rho_s(t)]+H.C.
    \end{split}
\end{equation}

where $X_j(t,\tau)=e^{-iH_s\tau/\hbar}X_j(t)e^{iH_s\tau/\hbar}$, and polaron Green's functions, $G_j(\tau)=\langle\zeta_j(t)\zeta_j(t,\tau)\rangle_{bath}$, $G_g(\tau)=\langle B \rangle^2{\cosh(\phi(\tau)-1)}$, $G_u(\tau)=\langle B \rangle^2\sinh(\phi(\tau))$. The phonon correlation function is given by,
\begin{equation}
    \phi(\tau)=\int_{0}^{\infty}d\omega\frac{J(\omega)}{\omega^2}[\coth(\frac{\hbar\omega}{2k_BT})\cos(\omega\tau)-i\sin(\omega\tau)],
\end{equation}
\par
where $k_B$ and $T$ are the Boltzmann constant and temperature of the phonon bath, respectively. The super-ohmic spectral density function of phonon bath is given by $J(\omega)=\Sigma_k(\lambda_k^{i})^2\delta(\omega-\omega_k)=\alpha_p\omega^3\exp[-\frac{\omega^2}{2\omega_b^2}]$, takes the latter form in continuum limit\cite{Wilson2002}. In our calculations, the phonon bath parameter values considered are, electron-phonon coupling strength, $\alpha_p=2.36$ ps$^2$, and the cut-off frequency, $\omega_b=1$ meV which provide experimentally compatible values of the mean phonon displacement, $\langle B \rangle$=0.9, 0.84 and 0.73 for $T$= 5K, 10K and 20K, respectively\cite{hughes2011mollow}. The polaron transformation is valid for the coupling strengths less than the cut-off frequency i.e., $g_i<\omega_b$ and $\eta_i<\omega_b$ and for the temperature below $75K$\cite{nazir2016}. Further, we also include the Lindblad terms corresponding to cavity damping with decay rates $\kappa_i$, spontaneous exciton decay with rate $\gamma_i$ and pure dephasing with rate $\gamma_i'$. The master equation (\ref{eqn:ME}) is solved numerically using quantum optics toolbox under steady state condition $\dot{\rho_s} =0$, and we obtain the steady-state populations (SSP) and cavity field photon statistics. It is important to note that, in all numerical calculations, we have selected a sufficiently high photon number cutoff to ensure convergence of the results.

\section{Steadystate populations and cavity photon statistics}

\begin{figure}
    \centering
    \includegraphics[width=\columnwidth]{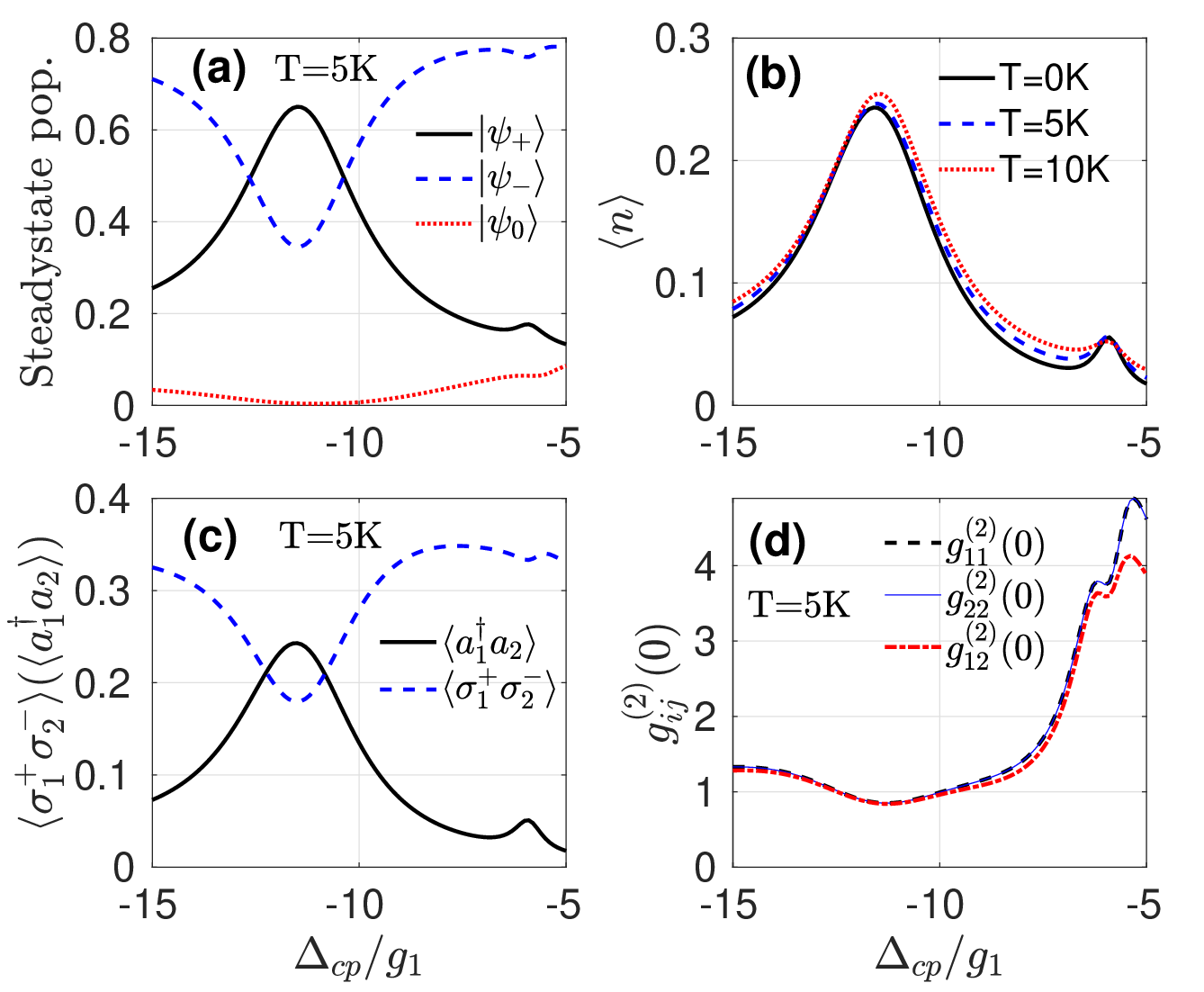}
    \caption{(a)Steady state populations of $|\psi_+\rangle$(solid black), $|\psi_-\rangle$(dashed blue) and $|\psi_0\rangle$(dotted red) for T=5K. (b) Mean cavity photon number $\langle n\rangle$ for T=0K (solid black), T=5K(dashed blue) and T=10K(dotted red). (c) Coherence between $|x\rangle$, $|y\rangle$ excitonic states, $\langle \sigma_1^+\sigma_2^-\rangle$ (dashed blue) and coherence between cavity modes, $\langle a_1^\dagger a_2\rangle$ (solid black). (d) Zero time delay second order intra and inter cavity mode correlations, $g_{11}^{(2)}(0)$ (dashed black), $g_{22}^{(2)}(0)$ (solid blue) and $g_{12}^{(2)}(0)$ (dash-dotted red) respectively. The parameters considered are QD states detuning, $\Delta_{xp}=\Delta_{yp}=\Delta=-10.0g_1$, coherent pumping rate, $\eta_1=\eta_2=\eta=2.0g_1$, cavity decay rates, $\kappa_1=\kappa_2=\kappa=0.5g_1$, spontaneous decay rate for $|x\rangle$, $|y\rangle$ states are $\gamma_1=0.01g_1$, $\gamma_2=0.01g_1$ respectively and the pure dephasing rates are $\gamma_1'=\gamma_2'=0.01g_1$.}
    \label{fig:chap4/Fig2}
\end{figure}

In Fig. \ref{fig:chap4/Fig2} we provide the results for steady state populations and cavity photon statistics. For a Hanle-type laser system, the two-photon CEL condition is given by, $|g_1|=|g_2|$, $\Delta_1=\Delta_2$ \cite{Scully1985}. Hence, we also consider QD states coupled to cavity modes with equal coupling strengths $g_2=g_1=100meV$ and are off-resonant w.r.t. coherent drives, $\Delta_{xp}=\Delta_{yp}=\Delta=-10.0g_1$. We have studied both the strong ($\kappa<g_1$) and weak QD-cavity coupling regimes ($\kappa>g_1$) of the system. The exciton-phonon effects play a significant role in such off-resonantly driven systems populating the excitonic states and thereby transitions can lead to population of coupled cavity modes. We define symmetric and anti-symmetric states of the systems as $|+\rangle=\frac{|x\rangle+|y\rangle}{\sqrt{2}}$, $|-\rangle=\frac{|x\rangle-|y\rangle}{\sqrt{2}}$, assuming equal coherent pumping strengths, $\eta_1=\eta_2=\eta$. Thus, the symmetric state $|+\rangle$ is driven with strength $\sqrt{2}\eta$, leading to pump-dressed states given by $|\psi_+\rangle= \cos\alpha |+\rangle + \sin\alpha |g\rangle$, $|\psi_0\rangle=|-\rangle$, $|\psi_-\rangle=-\sin\alpha|+\rangle + \cos\alpha |g\rangle$. Here $\sin2\alpha=\sqrt{2}\eta/\Omega$, $\cos2\alpha=\Delta/\Omega$ and the generalized Rabi frequency, $\Omega=\sqrt{\Delta^2+8\times \eta^2}$. In Fig. \ref{fig:chap4/Fig2}(a) we plot the steady state populations of these pump dressed states, $|\psi_\pm\rangle$, $|\psi_0\rangle$ for the coherent pumping strengths, $\eta_2=\eta_1=\eta=2.0g_1$. We vary the cavity detunings, $\Delta_{c_1p}=\Delta_{c_2p}=\Delta_{cp}$ equally, such that $\Delta_1=\Delta_2$ satisfying the CEL condition. It is observed that for $\Delta_{cp}=-\Omega=-\sqrt{\Delta^2+8\times\eta^2}$, single-photon resonance condition, the cavity modes are tuned with the transition between the dressed states. These transitions are accompanied with the population of both the cavity modes. The mean cavity photon numbers, $\langle n_1\rangle=\langle a_1^\dagger a_1\rangle$, $\langle n_2\rangle=\langle a_2^\dagger a_2\rangle$ are given in Fig. \ref{fig:chap4/Fig2}(b) for T=5K (blue-dashed). It should be noted that both the cavity modes are equally populated. We notice that the state $|\psi_0\rangle=|-\rangle$ is decoupled from the transitions and is not populated. The presence of a small peak at $\Delta_{cp}=-\Omega/2$ corresponds to two-photon emission process. The results for the single and two-photon emission rates are provided in section \ref{sec:laserRateEq}. Further, with increase in the temperature (T), the curves are broadened and the peak at $\Delta_{cp}=-\Omega/2$ is diminished due to enhanced phonon-induced decoherence suppressing the multi-photon processes. In addition to the polaron shifts, phonon-induced effects introduce to stark shifts ($\delta_i^\pm$, $\delta_{ip}^\pm$ where $i=1,2$) mentioned in the effective Hamiltonian, $H_{eff}$, (Eq. 13) of the simplified master equation, discussed latter in the subsection D. These stark shifts (Eq. A1 $\&$ A2) depend on $\langle B\rangle$, and vary with temperature, resulting in the slight shift of the peaks.

In Fig. \ref{fig:chap4/Fig2}(c) we show the results for the correlation between the modes, $Re(\langle a_1^\dagger a_2\rangle)$ and the coherence between upper QD states, $|x\rangle$ and $|y\rangle$ i.e., $\langle\sigma_1^+\sigma_2^-\rangle=Re(\langle y|\rho|x\rangle)$. We observe that the correlation, $\langle a_1^\dagger a_2\rangle$ rises with decrease in $\langle\sigma_1^+\sigma_2^-\rangle$ at $\Delta_{cp}=-\Omega$ implying transfer of coherence from the upper levels to the cavity modes. This correlation between the modes leads to quenching of relative or average phase fluctuations.
In Fig. \ref{fig:chap4/Fig2} (d), the results for the zero time delay photon-photon correlation functions of the modes, $g_{ij}^{(0)}=\frac{\langle a_i^\dagger a_j^\dagger a_i a_j\rangle}{\langle a_i^\dagger a_i\rangle \langle a_j^\dagger a_j\rangle}$ where $i,j={1,2}$, are given. The value of the inter-mode correlation function, $g_{12}^{(0)}$ is almost equal to that of the intra-mode correlation, $g_{ii}^{(2)}(0)$ where $i={1,2}$ at $\Delta_{cp}=-\Omega$. This shows that both the cavity modes are correlated with each other as much as with themselves.

\begin{figure}
    \centering
    \includegraphics[width=\columnwidth]{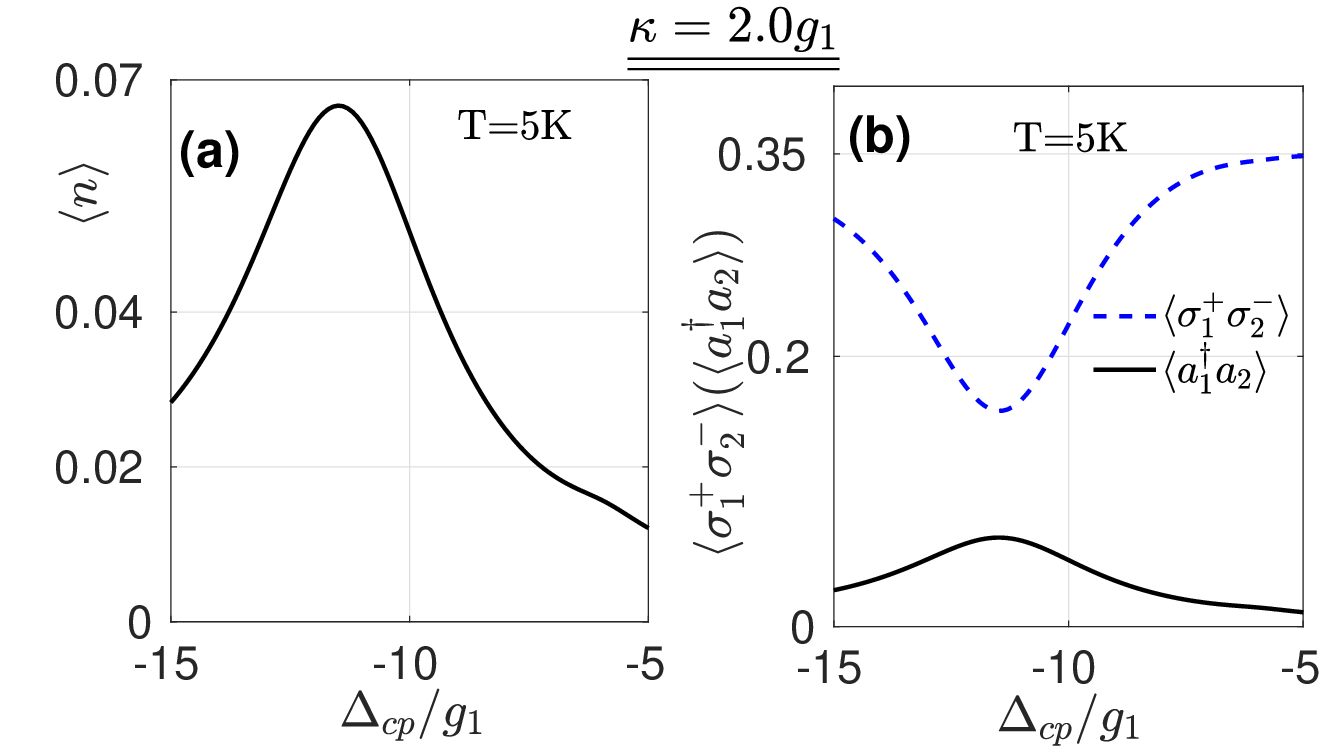}
    \caption{Weak coupling regime, $\kappa=2.0g_1$,  mean photon number of both the cavity modes, $\langle n\rangle$ given in (a) and in  (b) the coherence between $|x\rangle$, $|y\rangle$ states (dashed blue), correlation between cavity modes, $\langle a_1^\dagger a_2\rangle$ (solid black) is provided. In the bad cavity limit also, the coherence transfer to the cavity modes accompanied by mean photon peak at $\Delta_{cp}=-\Omega$ is present. Other parameters are same as in Fig. \ref{fig:chap4/Fig2}.}
    \label{fig:chap4/Fig3}
\end{figure}

Further, we have provided the results for the weak coupling regime in Fig.\ref{fig:chap4/Fig3}. The results show with increase in cavity decay rate, $\kappa$ to $2g_1$, the mean photon number, $\langle n\rangle$ diminishes. Fig.\ref{fig:chap4/Fig3}(b) shows still there is a slight transfer of coherence from emitter states to cavity modes and suggests the possibility of suppression of the noise.

\section{Variances of the Hermitian operators}

\begin{figure}
    \centering
    \includegraphics[width=0.75\columnwidth]{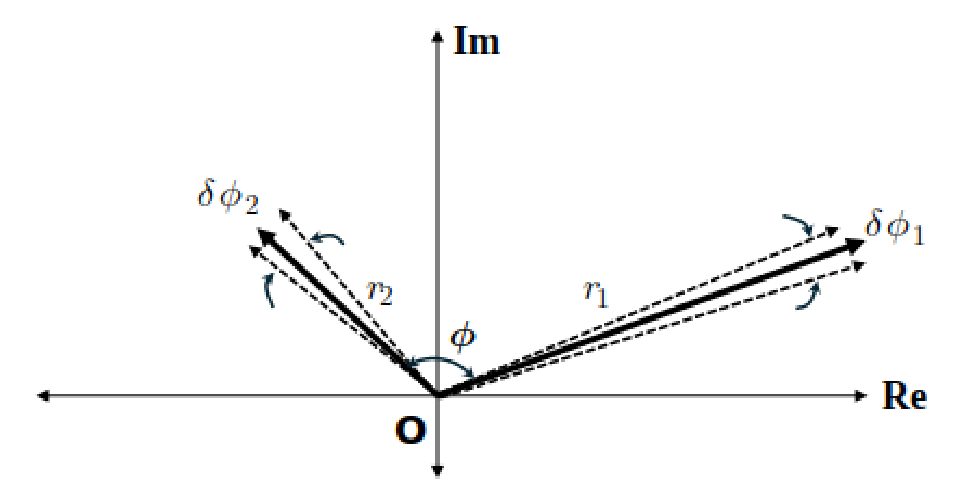}
    \caption{Showing the cavity field phasors, $\langle a_i\rangle=r_ie^{i\phi_i}$ and the fluctuations in their respective phases, $\delta\phi_i$. Correlation between the cavity modes can lead to cancel out these fluctuations and maintain relative phase, $\phi$.}
    \label{fig:chap4/Fig4}
\end{figure}

 The spontaneous emission and other incoherent processes present in the system add fluctuations in the amplitude and phase of the cavity fields. These fluctuations are suppressed by the transfer of coherence to the cavity fields from the QD-states. Intuitively, we can understand this by looking at the diffusion of the tip of the cavity field phasor, $\langle a_i\rangle = r_ie^{i\phi_i}$ in an amplitude-phase plane, Fig.\ref{fig:chap4/Fig4}. The emission into both the cavity modes is correlated so that the relative or average phase does not diffuse. In the present subsection, these fluctuations are calculated using the master equation for the system and in the next subsection using Fokker-Planck equation in Glauber-Sudarshan P representation.
 
 Here, the phase and amplitude fluctuations in the system can be evaluated by defining the Hermitian operators for the relative phase, $\phi=\phi_1-\phi_2$, the average phase, $\Phi=\frac{\phi_1+\phi_2}{2}$, the relative amplitude $r=r_1-r_2$ and the mean amplitude $R=\frac{r_1+r_2}{2}$ as given below \cite{Lu1990JOSAB},

\begin{subequations}
\begin{align}
    B_\phi &= \frac{i}{2}[a_1^\dagger e^{i\phi_1}-a_1 e^{-i\phi_1}]-\frac{i}{2}[a_2^\dagger e^{i\phi_2}-a_2 e^{-i\phi_2}]\label{subeq:Bphi}\\
    B_\Phi &= \frac{i}{4}[a_1^\dagger e^{i\phi_1}-a_1 e^{-i\phi_1}]+\frac{i}{4}[a_2^\dagger e^{i\phi_2}-a_2 e^{-i\phi_2}]\label{subeq:BPhi}\\
    B_r &= \frac{1}{2}[a_1^\dagger e^{i\phi_1} + a_1 e^{-i\phi_1}]-\frac{1}{2}[a_2^\dagger e^{i\phi_2} + a_2 e^{-i\phi_2}] \label{subeq:Br}\\
    B_R &= \frac{1}{4}[a_1^\dagger e^{i\phi_1} + a_1 e^{-i\phi_1}]+\frac{1}{4}[a_2^\dagger e^{i\phi_2} + a_2 e^{-i\phi_2}]
    \label{subeq:BR}
\end{align}
\end{subequations}

\begin{figure*}
    \centering
    \includegraphics[width=\textwidth]{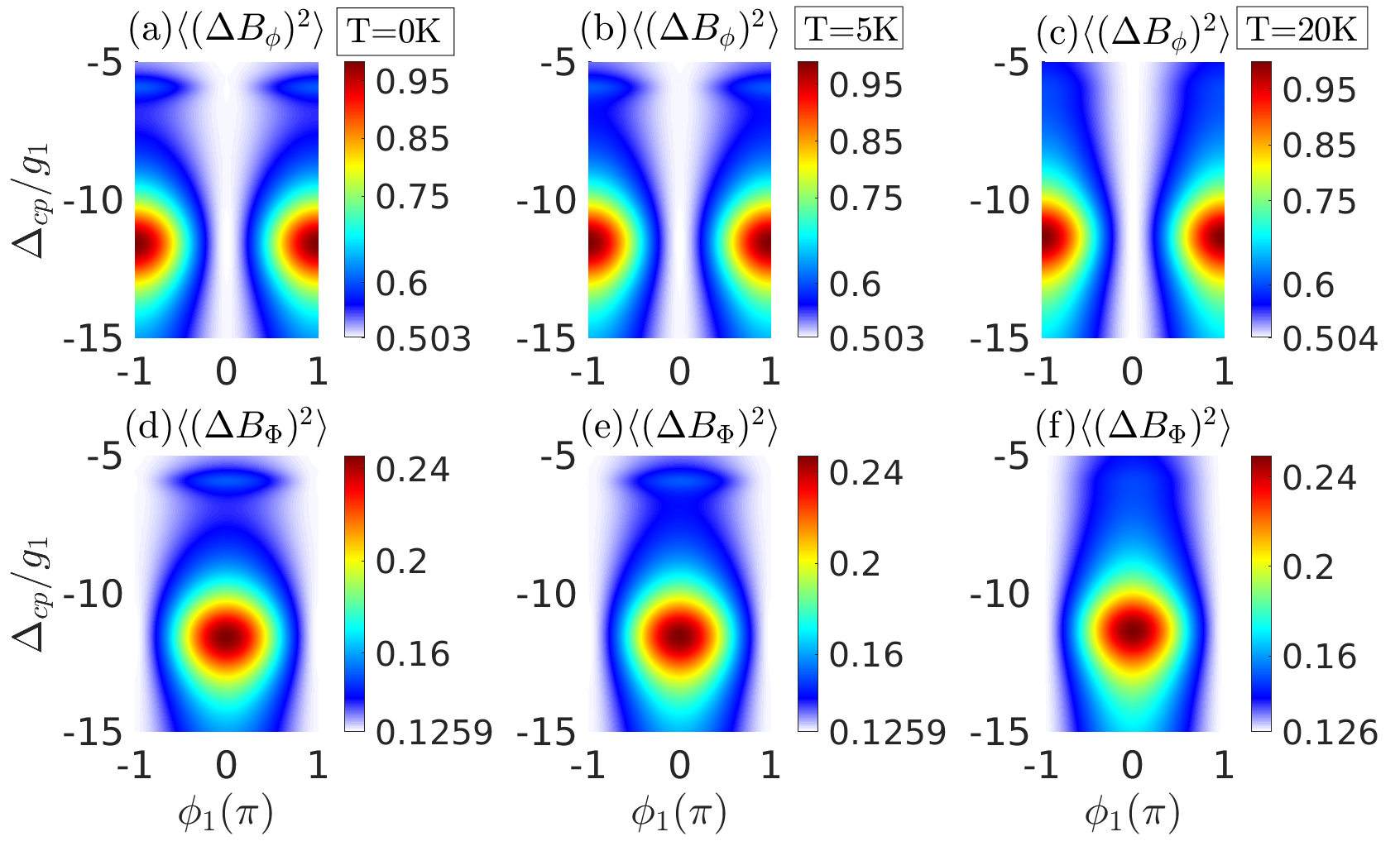}
    \caption{Variances of the Hermitian operators,  $B_\phi$ (a, b, c) and $B_\Phi$ (d, e, f) for T=0K, 5K, 20K are given. The other parameters are same as in Fig. \ref{fig:chap4/Fig2}. The fluctuations attain minimum at $\Delta_{cp}=-\Omega$ at $\phi_1=0$, $\phi_2=0$ for $B_{\phi}$ and at $\phi_1=\pi$, $\phi_2=0$ for $B_\Phi$. With increase in temperatures, there is a slight increase in the minima due to phonon-induced decoherence.}
    \label{fig:chap4/Fig5}
\end{figure*}

Further, in VNL, the average value of the variances take the values, $\langle (\Delta B_{\phi,r})^2\rangle=1/2$, $\langle (\Delta B_{\Phi,R})^2\rangle=1/8$\cite{Lu1990JOSAB}. Here, we investigate the fluctuations in the relative and average phase of the cavity mode fields and the results are given in Fig. \ref{fig:chap4/Fig5}.

In Fig.\ref{fig:chap4/Fig5}, we fix the phase of second cavity mode, $\phi_2=0$ and varied the first cavity mode phase, $\phi_1$ and the cavity detuning, $\Delta_{c_1p}=\Delta_{c_2p}=\Delta_{cp}$. Fig. \ref{fig:chap4/Fig5}(b,e) shows the results for T=5K. We can see that, for $\Delta_{cp}=-\Omega$ at $\phi_1=0$, fluctuations in relative phase Hermitian operator, $B_\phi$, $\langle (\Delta B_\phi)^2\rangle\approx 0.503$ reaches VNL (0.5) implying the presence of correlated emission in the system whereas, fluctuations in the average phase, $B_\Phi$ increases above the VNL(0.125). For $\phi_1=\pm\pi$, the fluctuations in $B_\phi$ increases and in $B_\Phi$ decreases and reaches the value $\approx0.1259$. This reduction in fluctuations is manifested in the vanishing of phase diffusion coefficients shown in the following subsection. The offset in the value of variances from the VNL is attributed to phonon induced noise present in the system. We also find noise quenching at $\Delta_{cp}=-\Omega/2$, where the cavity modes are populated by two-photon emission processes. With increase in temperature, the phonon-induced incoherent scattering increases and lead to broadening and rise in the quantum fluctuations. We can compare the results for T=5K with those of T=0K $\&$ T=20K provided in Fig. \ref{fig:chap4/Fig5}(a,d) $\&$ (c,f) respectively.

\begin{figure}
    \centering
    \includegraphics[width=\columnwidth]{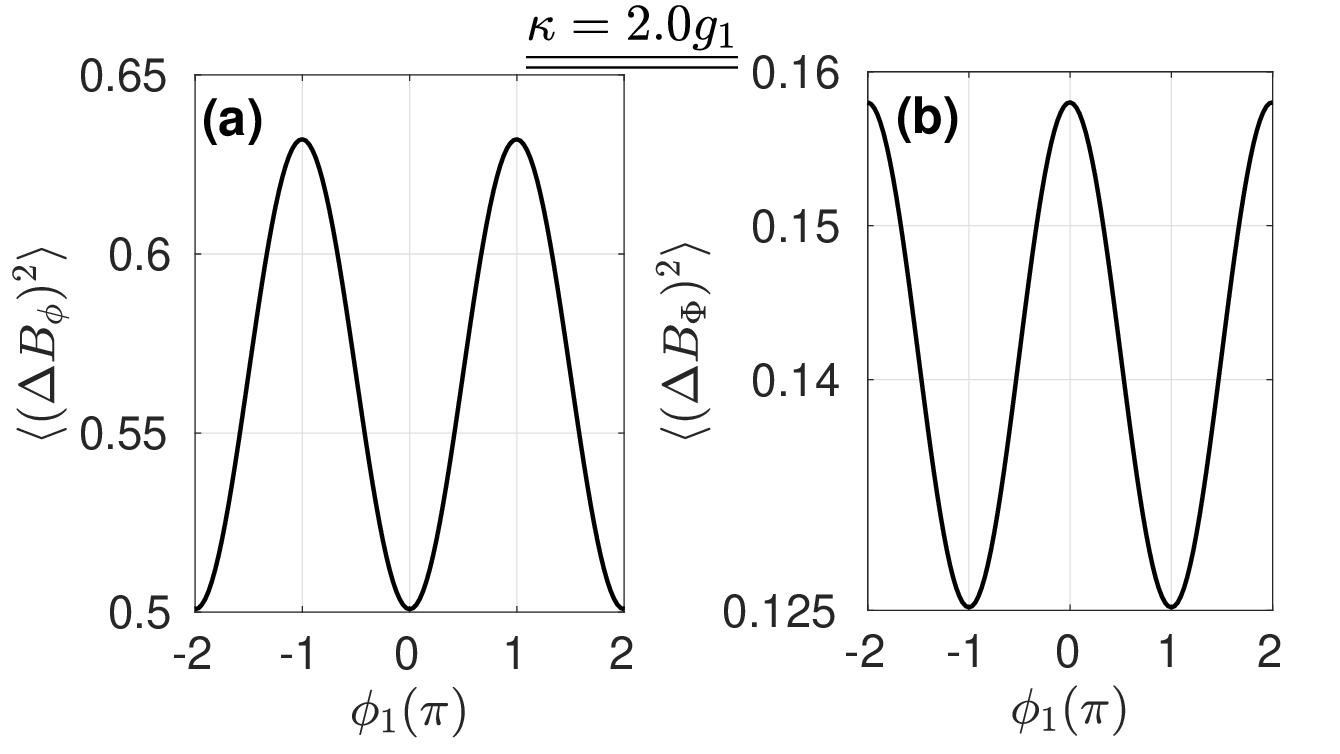}
    \caption{Varying $\phi_1$ in weak coupling regime, $\kappa=2.0g_1$ and $\phi_2=0$. The variances of the Hermitian operators, $B_\phi$ and $B_\Phi$ are provided in (a) and (b) respectively. The coherence transfer from QD-states to the cavity modes at $\Delta_{cp}=-\Omega$ resulted in the suppression of the noise in relative or average phase phase at $\phi_1=0$ or $\phi=\pm\pi$. The other parameters are same as in Fig. \ref{fig:chap4/Fig3}.}
    \label{fig:chap4/Fig6}
\end{figure}

Further, we notice that even in the bad cavity limit, $\kappa=2.0g_1$, variances $B_\phi$ and $B_\Phi$ attains vacuum noise limited values as shown in Fig.\ref{fig:chap4/Fig6}(a) $\&$ (b). Considering the cavity detuning, $\Delta_{cp}=-\Omega$, and $\phi_2=0$, the results show that the variance of $B_\phi$ reaches a value of 0.5 (VNL) for $\phi_1=0$ and that of $B_\Phi$ attains 0.125 (VNL) value for $\phi_1=\pm\pi$ similar to the strong coupling case.

\section{The phase drift and diffusion coefficients}

In this subsection, we demonstrate the quantum-noise quenching in phase fluctuations of the cavity fields by deriving the steady-state drift and diffusion coefficients in Glauber-Sudarshan P representation for both the scenarios: without and with phonons. By casting the master equation into the Fokker-Planck equation in P representation (Eq. \ref{eqn:FPeqn2}), we have obtained the drift coefficients, $D_\phi$, $D_\Phi$ and the diffusion coefficients, $D_{\phi\phi}$ and $D_{\Phi\Phi}$ of the relative and average phase of the cavity fields. These coefficients provide the information describing the evolution of phase of the cavity field. Drift coefficient, $D_\phi$ or $D_\Phi$ determines the evolution of the mean of relative or average phase respectively. Here, $D_\phi=0$ or $D_\Phi=0$ gives the phase locking condition. The spread or variance of the fluctuations are determined by the diffusion coefficients, $D_{\phi\phi}$ and $D_{\Phi\Phi}$. Therefore, $D_{\phi\phi}\leq 0$ or $D_{\Phi\Phi}\leq 0$ suggests the quenching(squeezing) of the relative or average phase diffusion.

\subsection{Without phonons: }Here we derive the Fokker-Planck equation of motion for the system without including exciton-phonon interactions and calculate the average and relative phase diffusion coefficient rates following the standard procedure\cite{ScullyZubairyQOT}. The master equation for the system ($\rho_s$) for the without exciton-phonon interaction case is given by,

\begin{equation}
 \begin{split}
  \dot{\rho_s}=& -\frac{i}{\hbar}[H_s,\rho_s]-\frac{\kappa_1}{2}L[a_1]\rho_s    -\frac{\kappa_2}{2}L[a_2]\rho_s
    \\& -\Sigma_{i=1,2}\Big[\frac{\gamma_i}{2}L[\sigma_i^-]\Big]
    \label{eqn:ME_WP}
 \end{split}
\end{equation}

Using the above equation, (\ref{eqn:ME_WP}), we obtain the master equation for the reduced density operator for cavity fields, $\rho_f$ by tracing over the quantum dot states, $|x\rangle$, $|y\rangle$ and $|g\rangle$.

\begin{equation}
    \begin{split}
    \dot{\rho_f}=& -i[\Delta_{c_1p} a_1^\dagger a_1 + \Delta_{c_2p} a_2^\dagger a_2 , \rho_f]
              -ig_1 a_1\rho_{gx} + ig_1\rho_{xg}a_1^\dagger 
              \\&-ig_2 a_2\rho_{gy}+ ig_2\rho_{yg}a_2^\dagger + ig_1\rho_{gx}a_1- ig_1 a_1^\dagger\rho_{xg} 
              \\&+ ig_2\rho_{gy}a_2- ig_2a_2^\dagger\rho_{yg}  
     \end{split}
     \label{eqn:reducedME_WP}
\end{equation}

\begin{figure}
    \centering
    \includegraphics[width=\columnwidth]{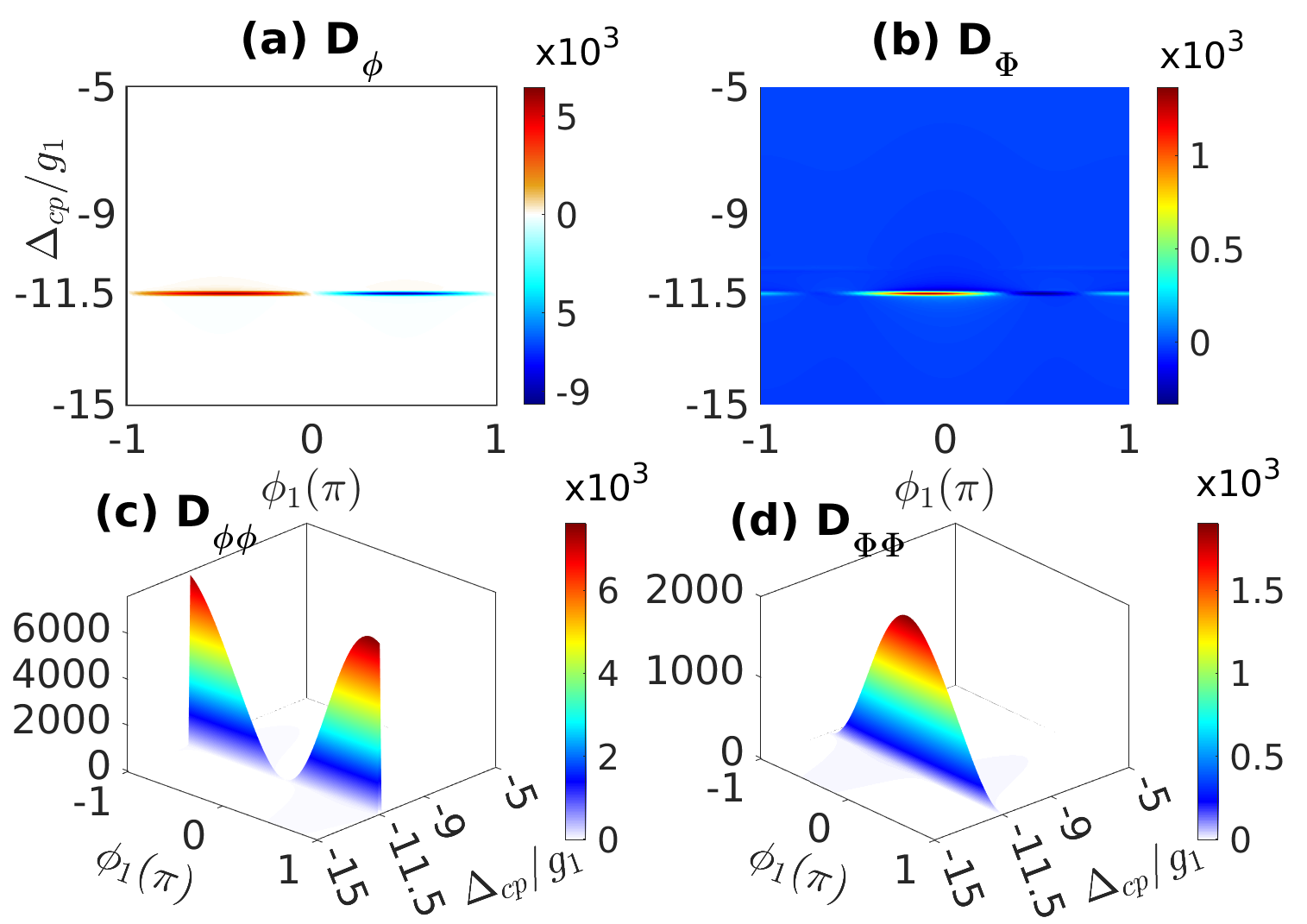}
    \caption{Without phonons: The relative and average phase drift (a,b) and diffusion (c,d) coefficients are provided. The system parameters are same as in Fig. \ref{fig:chap4/Fig2}(b). At $\phi_1=0$ and $\Delta_{cp}=-\Omega$, drift coefficient, $D_\phi=0$ indicating phase locking and the relative phase diffusion coefficient, $D_{\phi\phi}=0$ reaching VNL due to noise reduction. At $\phi_1=\pi$, average phase fluctuations are diminished and is reflected in $D_{\Phi\Phi}$ value going to VNL.}
    \label{fig:chap4/Fig7}
\end{figure}

The density matrix elements, $\rho_{ij}=\langle i|\rho_s|j\rangle$ are calculated and substituted back in the equation as shown in Appendix \ref{sec:chap4_Appendix2} and derived the rate equation for $\rho_f$ upto second order in the coupling strength, $g_i$. The phase diffusion coefficients are given in Fig. \ref{fig:chap4/Fig7} for the case of no exciton-phonon interactions. From Fig. \ref{fig:chap4/Fig7}(a), we can see that for the cavity detunings, $\Delta_{cp}=-\Omega$ and the phase angle, $\phi_1 = 0$, the relative drift coefficient ($D_\phi$) goes to `0'(phase locking condition) and from Fig. \ref{fig:chap4/Fig7}(c) $\&$ (d), the presence of correlated emission into the cavity modes leads to the suppression of the relative phase diffusion coefficient, $D_{\phi\phi}$ for $\phi_1 = 0$ and of the average phase diffusion coefficient, $D_{\Phi\Phi}$ for $\phi_1 = \pm\pi$ attaining VNL values.

\subsection{With phonons:} To obtain the Fokker-Planck equation for the case that includes exciton-phonon interactions, we derive a simplified master equation(SME) for this off-resonantly driven system following the approximations, $|\Delta_{xp}|$,$|\Delta_{yp}|$, $|\Delta_{c_1p}|$, $|\Delta_{c_2p}|$ $>>g_1, g_2, \eta_1, \eta_2$. The SME gives clear view of various phonon induced processes present in this off-resonantly driven system. We write $L_{ph}\rho_s$ terms in the Lindblad form proportional to $g_i^2$, $g_ig_j$, $\eta_i^2$ and $\eta_i\eta_j$. Similar procedure has been used earlier for obtaining SME for QD-cavity QED systems\cite{roy2011}. We also include terms not having Lindblad form such that the results of the SME converges with ME. The comparison between the steadystate results obtained from ME and SME is made in the Appendix\ref{sec:chap4_Appendix3} and the SME mimics the behavior of ME for the parameters considered here. The SME is given below,

\begin{equation}
    \begin{split}
        \dot{\rho_s}=& -\frac{i}{\hbar}[H_{eff},\rho_s]-\frac{\kappa_1}{2}L[a_1]\rho_s    -\frac{\kappa_2}{2}L[a_2]\rho_s-\Sigma_{i=1,2}\Big[\frac{(\gamma_i+\Gamma_{ip}^+)}{2}L[\sigma_i^-] +\frac{\gamma_i'}{2}L[\sigma_i^+\sigma_i^-]\\&+(\frac{\Gamma_i^-}{2}L[\sigma_i^+a_i]+\frac{\Gamma_i^+}{2}L[\sigma_i^-a_i^\dagger]) 
        + \frac{\Gamma_{ip}^-}{2}L[\sigma_i^+] \Big]\rho_s-\Big[\frac{\Gamma_{12}}{2}L[a_2^\dagger\sigma_2^- ,\sigma_1^+a_1]\rho_s
        \\&+\frac{\Gamma^p_{12}}{2}L[\sigma_1^+,\sigma_2^-]\rho_s + ( \Lambda_1^+\sigma_1^+a_1\rho_s\sigma_1^+a_1 + \Lambda_{12}^{++}\sigma_1^+a_1\rho_s \sigma_2^+a_2 
        + H.C.) \\& +\Lambda_{12}^{+-}\sigma_1^+ a_1\rho_s\sigma_2^-a_2^\dagger + (\Lambda_{1p}^+ \sigma_1^+\rho_s\sigma_1^+ + \Lambda_{12p}^{++}\sigma_1^+\rho_s\sigma_2^+ + H.C.) +\Lambda_{12p}^{+-}\sigma_1^+\rho_s\sigma_2^- 
        \\&+ 1\leftrightarrow 2\Big]
    \label{eqn:SME}
    \end{split}
\end{equation}

Here, $L[\hat{O_1},\hat{O_2}]=\hat{O_2}\hat{O_1}\rho_s-2\hat{O_1}\rho_s\hat{O_2}+\rho_s\hat{O_1}\hat{O_2}$. The effective Hamiltonian, $H_{eff}$ signify the coherent evolution of the system. It includes the terms arising due to exciton-phonon interaction and coherent drive-phonon interaction. The phonon induced stark shifts are given by $\delta_i^{\pm}$, $\delta_{ip}^{\pm}$ and the exciton exchange-cavity photon exchange process is given by $\Omega_{12}$ and the exciton-exchange process by $\Omega_{12}^p$.

\begin{equation}
    \begin{split}
        H_{eff} = & H_S + \hbar\delta_1^+\sigma_1^+\sigma_1^- a_1a_1^\dagger + \hbar \delta_2^+\sigma_2^+\sigma_2^-a_2a_2^\dagger \\& + \hbar\delta_1^-\sigma_1^-\sigma_1^+a_1^\dagger a_1 + \hbar \delta_2^- \sigma_2^-\sigma_2^+ a_2^\dagger a_2 + \hbar \delta_{1p}^+ \sigma_1^+\sigma_1^- \\& + \hbar \delta_{2p}^+ \sigma_2^+\sigma_2^- + \hbar \delta_{1p}^- \sigma_1^-\sigma_1^+ + \hbar\delta_{2p}^- \sigma_2^-\sigma_2^+ \\& - (i\hbar \Omega_{12}\sigma_1^+\sigma_2^- a_1a_2^\dagger + i\hbar \Omega_{12}^p \sigma_1^+\sigma_2^- + H.C.)
    \end{split}
\end{equation}
 
\begin{figure}
    \centering
    \includegraphics[width=\columnwidth]{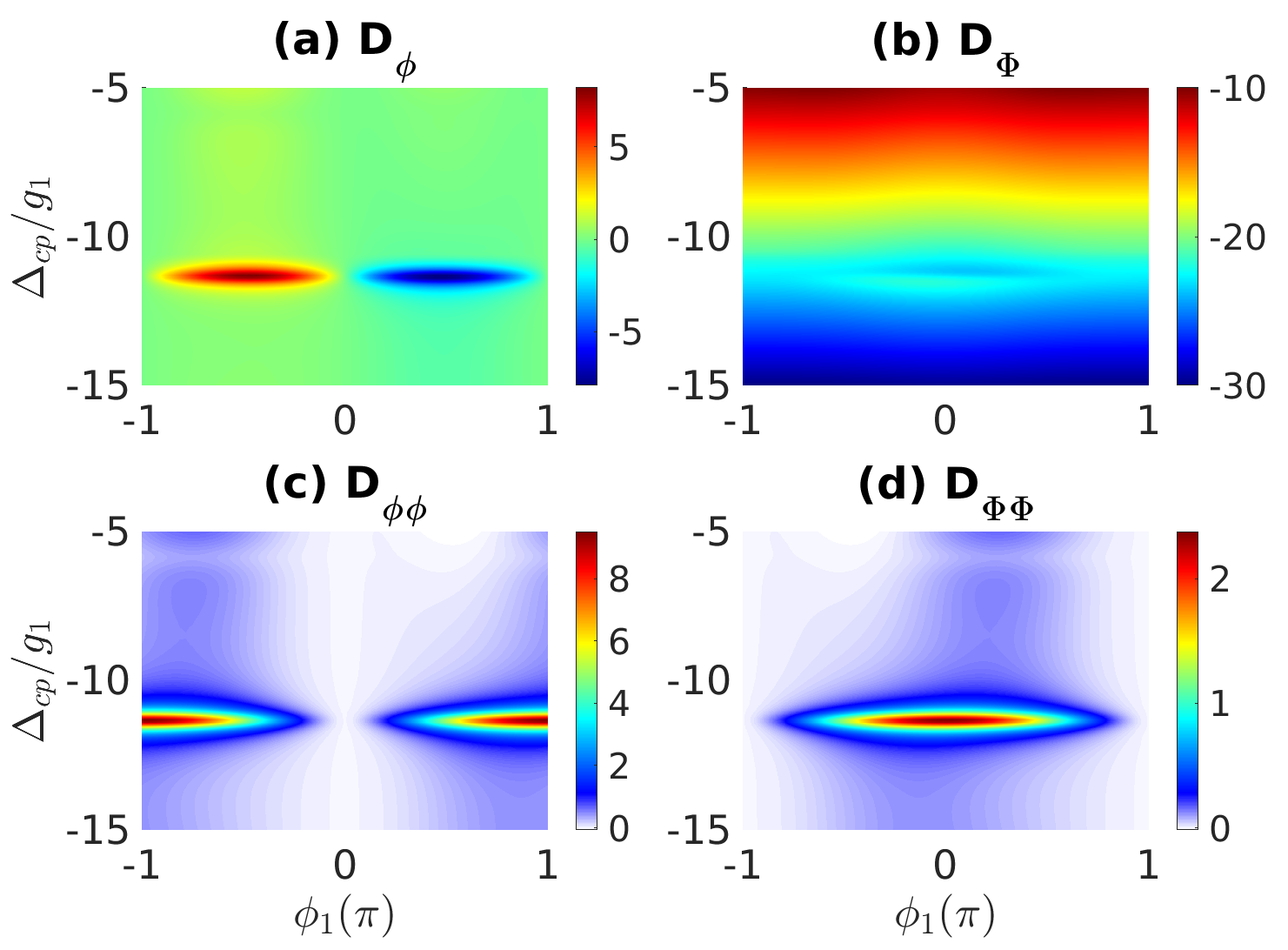}
    \caption{With phonons: The relative and average phase drift (a,b) and diffusion (c,d) coefficients for T=5K. The parameters are same as in Fig. \ref{fig:chap4/Fig2}(b). Exciton-phonon interactions lead to broadening of the peaks compared to the without phonon case. At $\Delta_{cp}=-\Omega$, the phase fluctuations are reduced.}
    \label{fig:chap4/Fig8}
\end{figure}

Additionally, the exciton-phonon interaction induces incoherent processes with scattering rates, $\Gamma_i^+$ ,($\Gamma_{i}^-)$) corresponding to QD excitation(de-excitation) by absorption(emission) of the corresponding cavity mode photon. $\Gamma_{ip}^+$ corresponding to the incoherent pumping of the QD and $\Gamma_{ip}^-$ to the enhanced radiative decay. The phonon induced incoherent exciton exchange processes rates are given by $\Gamma_{ij}$, $\Gamma_{ij}^p$. $\Lambda$'s are the coefficients for the residue terms. The form of these phonon induced scattering rates are given in Appendix \ref{sec:chap4_Appendix1}. Following a similar procedure detailed above for the without phonon case, we derive a Fokker-Planck equation for the system. Thereafter, the phase drift and diffusion coefficients are obtained. The results for T=5K are given in Fig. \ref{fig:chap4/Fig8}.

\begin{figure}
    \centering
    \includegraphics[width=\columnwidth]{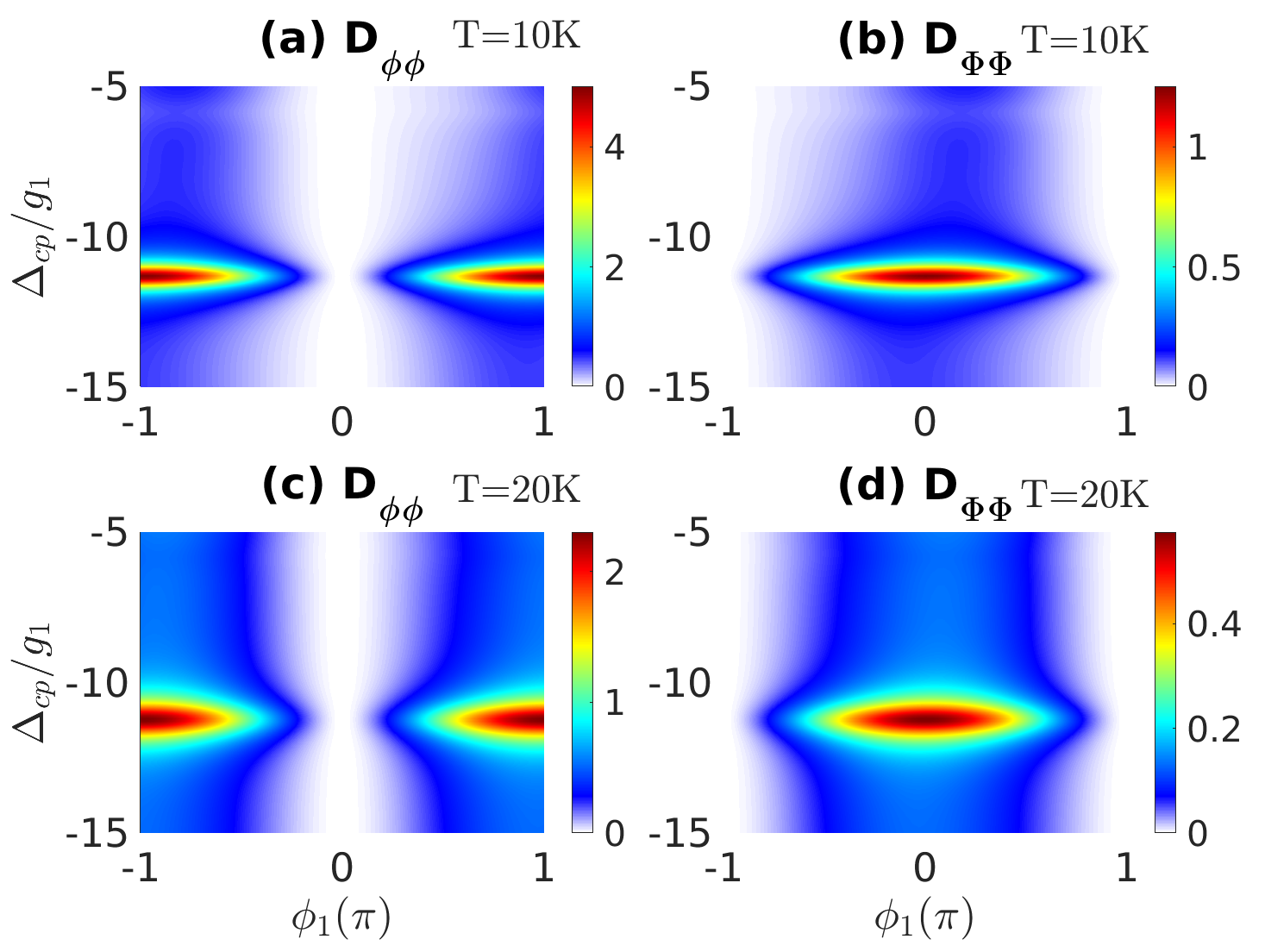}
    \caption{The relative and average phase diffusion coefficients for T=10K (a,b) and T=20K (c,d). The other parameters are same as in Fig. \ref{fig:chap4/Fig2}(b). At higher temperatures, phonon-induced incoherent processes further increase the broadening of the diffusion coefficient peaks.}
    \label{fig:chap4/Fig9}
\end{figure}

We can see that correlated emission into both the cavity modes in this off-resonantly pumped single QD coupled to the bimodal photonic crystal cavity system leads to quenching in the relative or the average phase diffusion coefficients. From Fig. \ref{fig:chap4/Fig8} (a), we can see that for $\phi_1=0,\pm\pi$, at the cavity detunings, $\Delta_{cp}=-\Omega$, the relative phase drift coefficient, $D_\phi=0$ satisfies the phase locking condition. Further, the relative phase diffusion coefficient, $D_{\phi\phi}$ goes to zero at $\phi_1=0$ and the average diffusion coefficient, $D_{\Phi\Phi}=0$ at $\phi_1=\pm\pi$ showing quenching behaviour similar to the without phonon case. The correspondence with the results for variances in the Hermitian operators, $B_\phi$ $\&$ $B_\Phi$ reducing to the VNL as shown earlier (c.f. Fig. \ref{fig:chap4/Fig5}) is evident.

Fig. \ref{fig:chap4/Fig9} shows the diffusion coefficients, $D_{\phi\phi}$, $D_{\Phi\Phi}$ for T=10K in (a,b) and T=20K in (c,d). With increase in the temperature, the phonon-induced decoherence results in broadening of the diffusion coefficients implying that the cavity modes are more noisy. Additionally, we also noticed that these results also hold for the weak QD-cavity coupling case, showing suppression of the diffusion coefficients when the variances of $B_{\phi}$, $B_\Phi$ operators reach VNL. Therefore, we conclude this section by stating that CEL can be observed both in good and bad cavity limits with high and low mean photon numbers respectively. It is appropriate to choose the coupled cavity modes detunings are $\Delta_{cp}=-\Omega$ and the QD states detuning,  $\Delta<0$ w.r.t. external coherent drive.

\section{\label{sec:laserRateEq}Laser rate equations}

In this section, we evaluate the rate of emissions into cavity mode via single and two-photon processes. To obtain the rates, we follow the standard procedure \cite{Scully1967,sargent1974}, using the SME (\ref{eqn:SME}), we trace over the QD states and express the reduced density matrix rate equation for the cavity modes Eq.\ref{eqn:cavityRateEqn}, in terms of ``probability" flowing into and out of the state $|n,m\rangle$ where `$n$' $\&$ `$m$' are the number of photons in 1st $\&$ 2nd cavity modes.

\begin{equation}
    \begin{split}
        \dot{P}_{n,m} = &-\alpha_{n,m} P_{n,m}+G^{11}_{n-1,m-1}P_{n-1,m-1} 
        \\&+G^{10}_{n-1,m}P_{n-1,m}+G^{01}_{n,m-1}P_{n,m-1}
        \\&+G^{20}_{n-2,m}P_{n-2,m}+G^{02}_{n,m-2}P_{n,m-2}
        \\&+A^{11}_{n+1,m+1}P_{n+1,m+1}+A^{10}_{n+1,m}P_{n+1,m}
        \\&+A^{20}_{n+2,m}P_{n+2,m}+A^{01}_{n,m+1}P_{n,m+1}
        \\&+A^{02}_{n,m+2}P_{n,m+2} +\kappa_{1}(n+1)P_{n+1,m}- \kappa_{1}nP_{n,m} 
        \\&+\kappa_{2}(m+1)P_{n,m+1} - \kappa_{2}mP_{n,m}.
    \end{split}
    \label{eqn:cavityRateEqn}
\end{equation}

\begin{figure}
    \centering
    \includegraphics[width=\columnwidth]{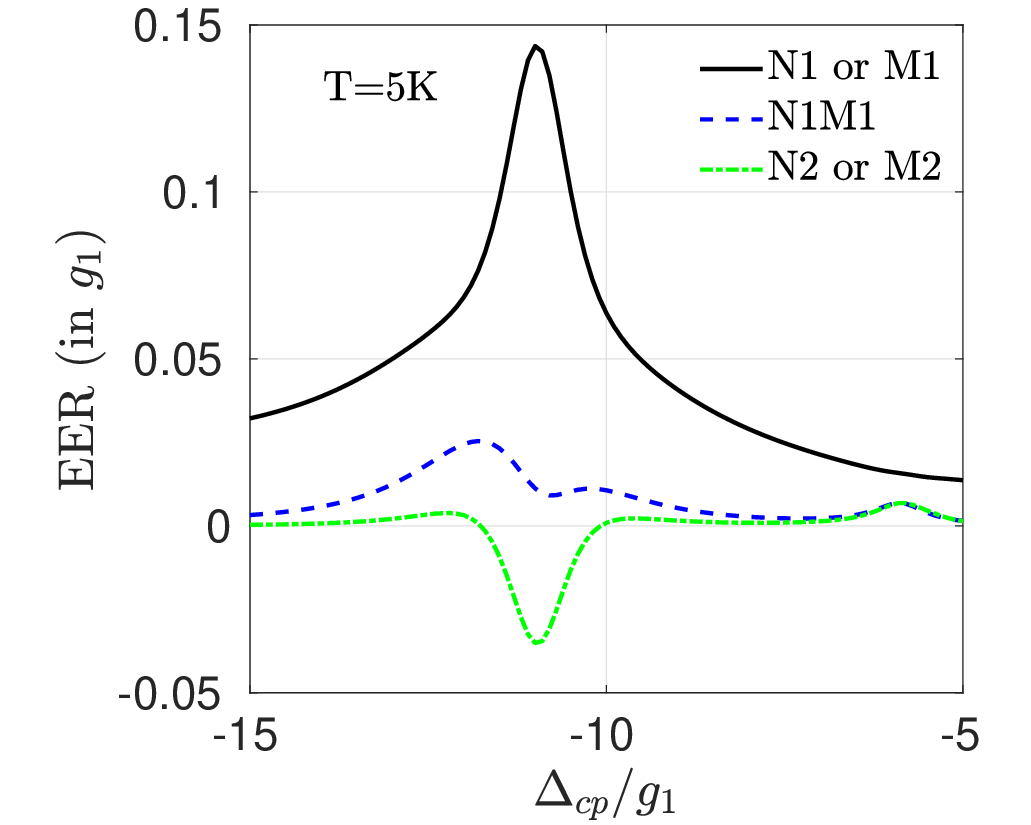}
    \caption{Single photon (first mode N1, second mode M1, solid black), two-mode two-photon (N1M1, dashed blue) and two-photon (first mode N2, second mode M2, dot-dashed green) excess emission rates(EER) for T=5K are given. All other parameters are same as in Fig.\ref{fig:chap4/Fig2}(b). The excess emission rates are equal for both the cavity modes. The peak in N1 or M1 curves at $\Delta_{cp}=-\Omega$ suggest that the single photon emission into 1st cavity mode or the 2nd cavity mode is dominant over the other multi-photon emissions at this resonance condition.}
    \label{fig:chap4/Fig10}
\end{figure}

 The probability of having $n$, $m$ photons in 1st, 2nd cavity modes is given by, $P_{nm}=\Sigma_i \langle i,n,m|\rho_s|i,n,m\rangle$, the coefficients on the right hand side are given by, $\alpha_{n,m}=\Sigma_i \alpha_{i,n,m}\langle i,n,m|\rho_s|i,n,m\rangle$, $G_{n,m}^{ab}P_{n,m}=\Sigma_i G_{i,n,m}^{ab}\langle i,n,m|\rho_s|i,n,m\rangle$ and $A_{n,m}^{ab}P_{n,m}=\Sigma_i A_{i,n,m}^{ab} \langle i,n,m|\rho_s|i,n,m\rangle$ where $i={x,y,g}$. The coefficients $\alpha_{i,n,m}$, $G_{i,n,m}^{ab}$ and $A_{i,n,m}^{ab}$ are obtained numerically. The k-photon emission (absorption) rate for the first and second modes are given by, $\Sigma_{n,m} G_{n,m}^{k0} P_{n,m}$($\Sigma_{n,m}A_{n,m}^{k0}P_{n,m}$) and $\Sigma_{n,m} G_{n,m}^{0k} P_{n,m}$($\Sigma_{n,m}A_{n,m}^{0k}P_{n,m}$) respectively and two-mode two-photon emission(absorption) rate is given by $\Sigma_{n,m} G_{n,m}^{11} P_{n,m}$ ($\Sigma_{n,m}A_{n,m}^{11}P_{n,m}$). The difference between emission and absorption rate is defined as `excess emission rate (EER)' and the sign of EER $>0$ or $<0$ represents net emission or absorption occurring in the cavity mode. In Fig. \ref{fig:chap4/Fig10}, we show results for the single and multi-photon excess emission rates of the system for T=5K. We can see that for $\Delta_{cp}=-\Omega=-11.5g_1$, where the fluctuations in $B_\phi$ attains VNL value and also the relative phase diffusion coefficient is quenched, there is peak in single photon emission (N1 or M1)  of both the modes and dip in other multi-photon excess emission rates. This implies photons are emitted into either of the cavity modes at equal rate and the two-photon emission and the two mode two-photon emission into the cavity modes are suppressed. Therefore, the cavity modes are populated due to single-photon emission predominantly resulting a peak in $\langle n\rangle$ c.f., Fig. \ref{fig:chap4/Fig2}(b). Also for $\Delta_{cp}=-\Omega/2$, at two-photon resonance condition, there are small peaks in both the two-mode two-photon, two-photon EER curves and a slight dip in single-photon EER curves. Further, it is noticed with increase in the temperature, rise in the phonon induced decoherence leads to the suppression in the multi-photon processes and the single photon emission dominates.

 \section{Continuous Variable Entanglement between the cavity modes}

In this section we investigate the quantum correlation between the cavity modes of the system by evaluating the Duan-Giedke-Cirac-Zoller (DGCZ) criterion\cite{DGCZ2000} for CV entanglement. We define two Einstein-Poldosky-Rosen (EPR) like variables, $u$, $v$ given by

\begin{equation}
    u=\hat{x_1}+\hat{x_2}, v=\hat{p_1}-\hat{p_2},
\end{equation}

where $\hat{x_j}$, $\hat{p_j}$ are given below in terms of cavity mode operators,
\begin{align}
    x_{j} &=\frac{1}{\sqrt{2}}(a_{j}^{\dagger}e^{-i\phi_{j}}+a_{j}e^{i\phi_{j}}),\\
    p_{j} &=\frac{i}{\sqrt{2}}(a_{j}^{\dagger}e^{-i\phi_{j}}-a_{j}e^{i\phi_{j}}),j=\{1,2\}.
\end{align}

According to DGCZ criterion, the sum of the variances of $u$, $v$ i.e., $\Delta u^2 + \Delta v^2 = (\langle u^2-\langle u\rangle^2\rangle)^2+ (\langle v^2-\langle v\rangle^2\rangle)^2 < 2$ is the sufficient condition for the presence of CV entanglement between the modes. 

\begin{figure}
    \centering
    \includegraphics[width=\columnwidth]{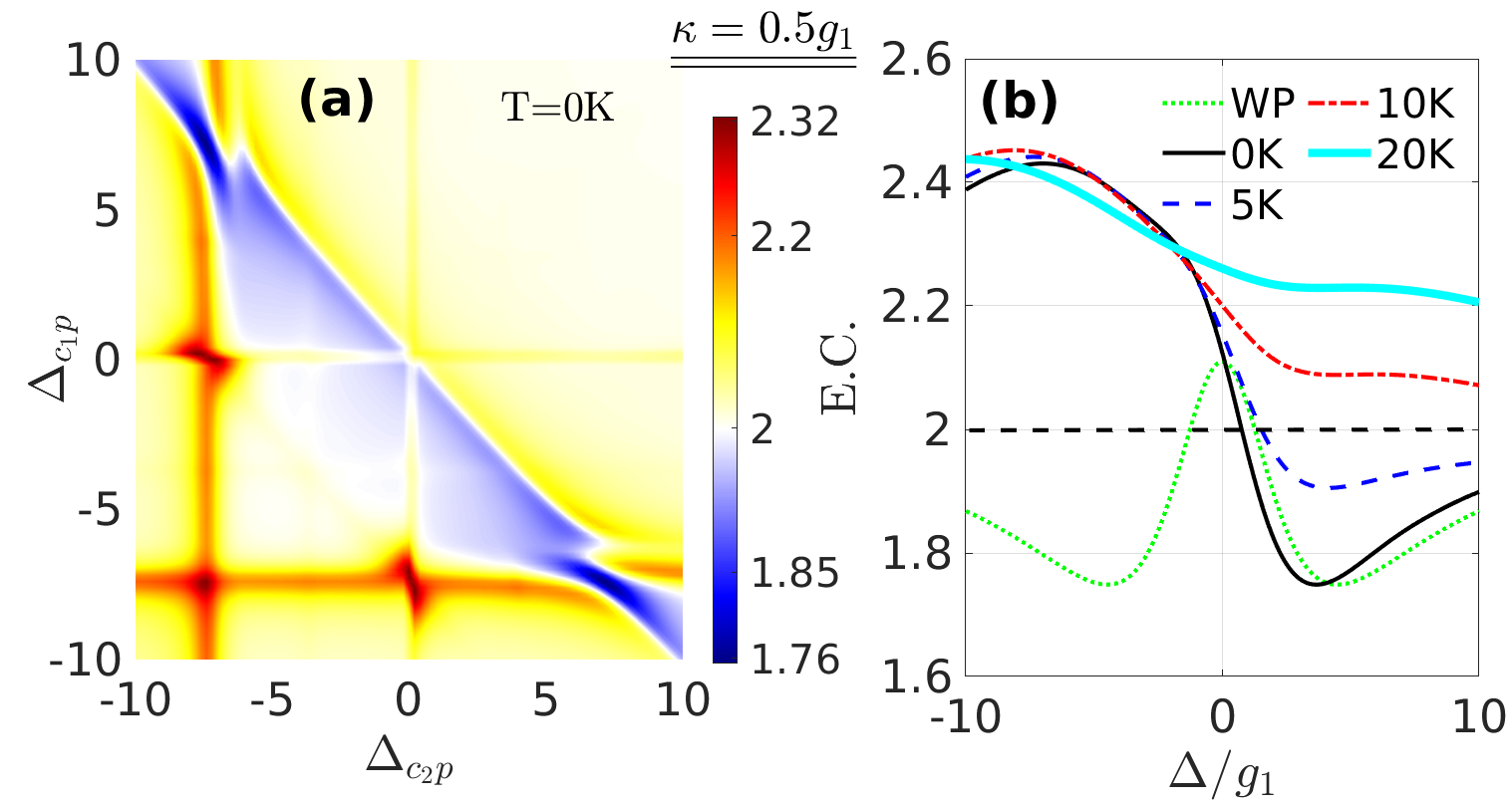}
    \caption{Entanglement criterion (E. C.), $(\Delta u)^2 + (\Delta v)^2$ for $\kappa=0.5g_1$, $\phi_1=\pi$, $\phi_2=0$ and the other parameters are same as in Fig. \ref{fig:chap4/Fig2}(b). The subplot (a) shows the E. C. values by varying $\Delta_{c_1p}$ vs $\Delta_{c_2p}/g_1$  at T=0K. It is clear that for $\Delta_{c_1p}\approx\pm\Omega$ and $\Delta_{c_2p}\approx\mp\Omega$, the sum of the variances of EPR pair like variables attains minimum value. In the subplot (b), varying QD-states detuning ($\Delta_{xp}=\Delta_{yp}=\Delta$) for the cases, without phonons, WP (dotted green), T=0K (solid black), 5K (dashed blue), 10K (dot-dashed red), 20K (thick cyan). The DGCZ criterion is satisfied for the case for $\Delta>0$ or $<0$ in WP scenario and for $\Delta>0$ for T=0K, 5K. This is due to asymmetry in phonon interactions.}
    \label{fig:chap4/Fig11}
\end{figure}

Fig. \ref{fig:chap4/Fig11}(a) presents the E. C. values for T=0K, $\kappa=0.5g_1$ $\&$ $\eta=2.0g_1$ case while varying the cavity detunings. The results indicate that the entanglement between the cavity modes is maintained for the two-photon condition, $\Delta_{c_1p}+\Delta_{c_2p}=0$ and the total variance, $(\Delta u)^2 + (\Delta v)^2$ attains lowest value when the cavity detunings are $\Delta_{c_1p}\approx+\Omega=+7.5g_1$, $\Delta_{c_2p}\approx-\Omega=-7.5g_1$ and vice versa. In this case the QD states are positively detuned w.r.t coherent pumps ($\Delta_{xp}=\Delta_{yp}=\Delta=5.0g_1$), and the cavity field phases are set as, $\phi_1=\pi$, $\phi_2=0$. 
Fig. \ref{fig:chap4/Fig11}(b) illustrates the effect of varying the QD-states detuning, $\Delta$, while fixing the cavity detunings to $\pm\Omega$ for different temperature conditions, T=0K, 5K, 10K, 20K, including without phonons (WP) case. The results demonstrate that when the QD-states are positively detuned w.r.t the coherent pump, $\Delta_{xp}=\Delta_{yp}=\Delta>0$, the entanglement between the cavity modes is preserved. This is attributed to the reduced phonon induced decoherence, which is lower compared to $\Delta<0$ case.

\begin{figure}
    \centering
    \includegraphics[width=\columnwidth]{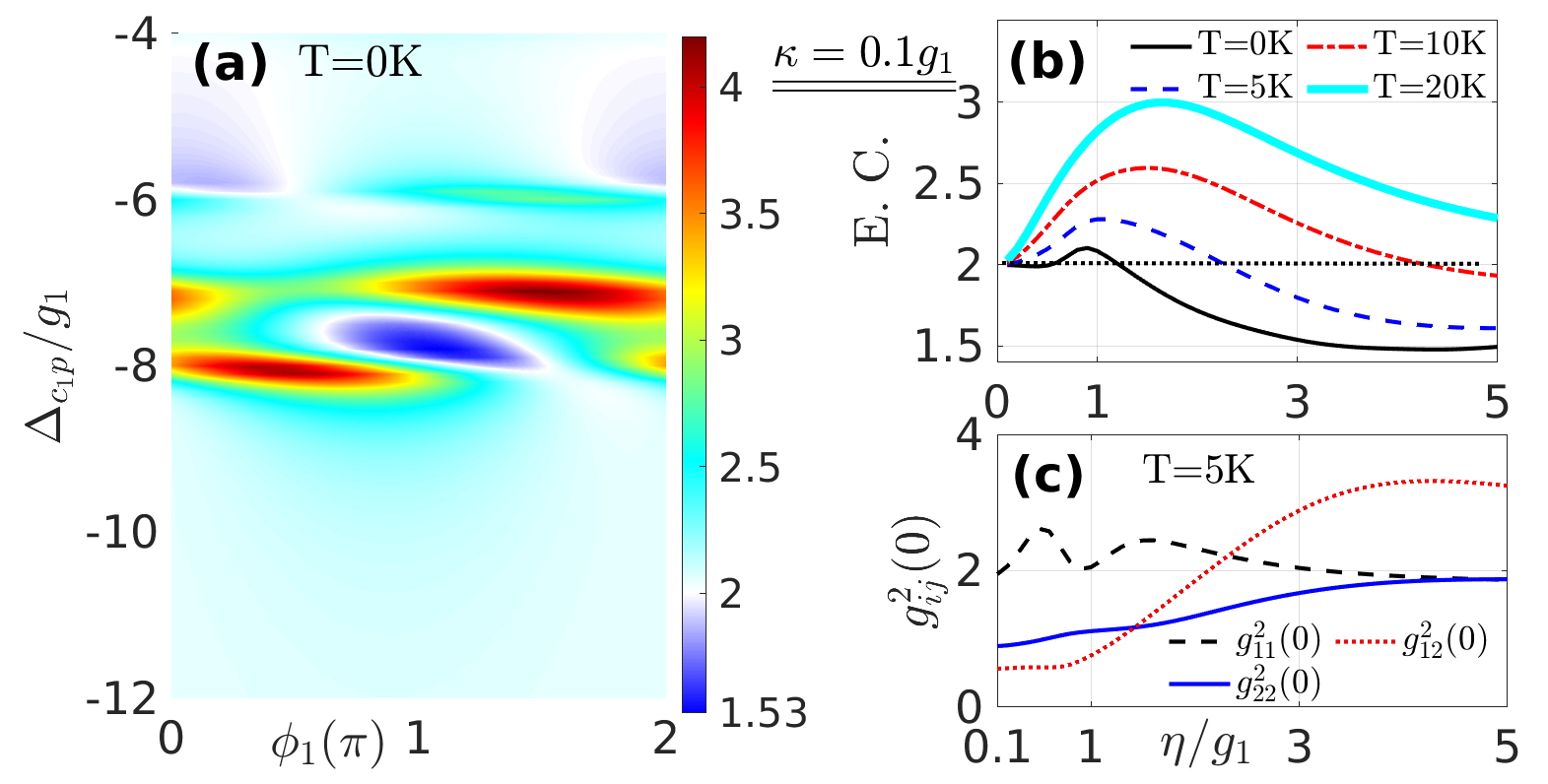}
    \caption{Entanglement criterion (E.C.), $(\Delta u)^2 + (\Delta v)^2$. (a) Varying $\Delta_{c_1p}$ and $\phi_1$ for T=0K and $\phi_2=0$. The E. C. is minimum at $\Delta_{c_1p}\approx -\Omega$ and $\phi_1\approx\pi$. In subplot (b), varying coherent pumping strength $\eta$ for $\Delta_{c_1p}=-\Omega$, $\Delta_{c_2p}=+\Omega$, $\phi_1=\pi$ and $\phi_2=0$. Detunings, $\Delta_{xp}=\Delta_{yp}=\Delta=5.0g_1$, $\kappa=0.1g_1$ and other system parameters are same as in Fig. \ref{fig:chap4/Fig2}(b) and the color coding same as in Fig. \ref{fig:chap4/Fig11} (b). The E. C. initially decreases at a certain pumping rate, reaching a minimum value due to increased inter-mode correlation, $g_{12}^2(0)$ as shown in subplot (c) and increases at higher pumping rates. Also at higher temperatures, E. C. value is amplified.}
    \label{fig:chap4/Fig12}
\end{figure}

\begin{figure}
    \centering
    \includegraphics[width=\columnwidth]{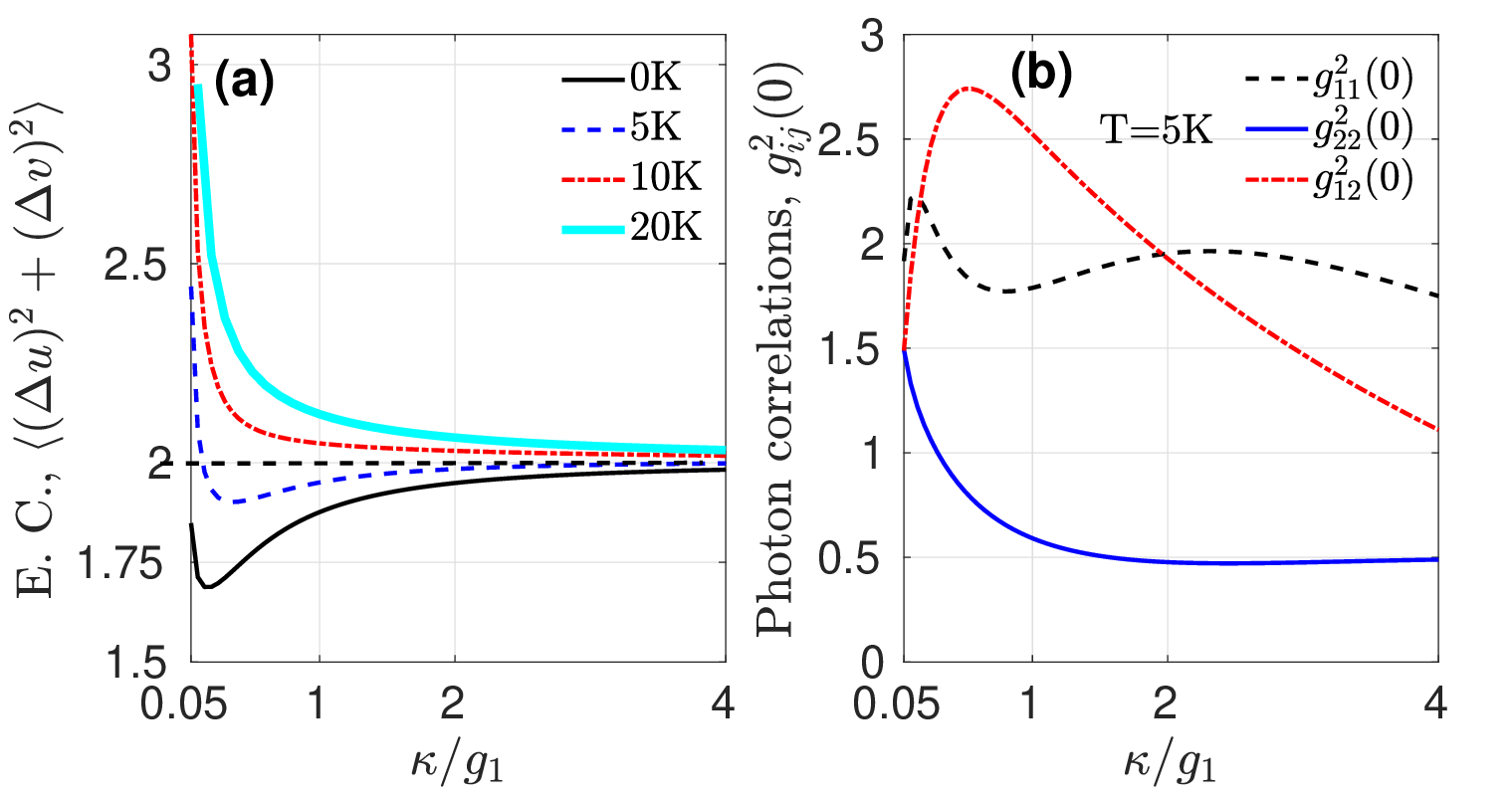}
    \caption{Subplot (a): DGCZ Entanglement criterion dependence on cavity decay rate, $\kappa_1=\kappa_2=\kappa$ for temperatures T=0K, 5K, 10K, 20K. Color scheme is same as in Fig. \ref{fig:chap4/Fig11} (b). Subplot (b): Second order zero time delay photon-photon correlations, $g_{ij}^2(0)$ with increasing $\kappa$. Color coding is same as in Fig.\ref{fig:chap4/Fig2}(d). The parameters considered here are the same as in Fig. \ref{fig:chap4/Fig12}(b) for $\eta=2.0g_1$. The results confirm that cavity fields are strongly quantum correlated in strong coupling regime, $\kappa/g_1<1$ and with increase in $\kappa$, both E. C. and $g_{12}^2(0)$ decrease.}
    \label{fig:chap4/Fig13}
\end{figure}

Fig.\ref{fig:chap4/Fig12} shows the results for QD-state detuning, $\Delta=5.0g_1$, $\kappa=0.1g_1$ and for $\eta=2.0g_1$. In Fig. \ref{fig:chap4/Fig12}(a), we fix the detuning of 2nd cavity mode, $\Delta_{c_2p}=\Omega$ and vary the detuning of the 1st cavity mode, $\Delta_{c_1p}$ and $\phi_1$. We can see that for $\phi_1=\pi$ and $\Delta_{c_1p}\approx-\Omega$, the total variance with its minimum value, $(\Delta u)^2 + (\Delta v)^2=1.525$, the DGCZ criterion for the CV entanglement is satisfied implying quantum correlation between the cavity modes. It should be noted that the EPR variables have the form similar to that of $B_R$, $B_\phi$ c.f. Eq. \ref{subeq:BR}, Eq. \ref{subeq:Bphi} respectively. Therefore we can say, squeezing in either of the operators, $B_R$, $B_\phi$ or both as in cascaded system \cite{Lu1990cascade} leads to the entanglement between the cavity modes. Therefore, the inter-mode two-photon correlation is greater than the intra-mode correlation. In Fig. \ref{fig:chap4/Fig12}(b) we show the change in total variance with increase in temperature and coherent pumping rate by fixing the cavity detunings, $\Delta_{c_1p}=-\Omega$, $\Delta_{c_2p}=\Omega$ and $\phi_1=\pi$. With increase in pumping rate, the cavity modes get populated and the entanglement between them is generated and is also reflected in the dominant inter-mode correlation, $g_{12}^2(0)$ over intra-mode correlation functions, $g_{11}^2(0)$, $g_{22}^2(0)$ as shown in Fig. \ref{fig:chap4/Fig12} (c) for T=5K, implying the non-classical cavity field states\cite{Agarwal_2012}. This is attributed to the domination of the two-mode two-photon excess emission into the cavity mode over other processes. Further increase in pumping rate leads to increased single photon EER  and thereby diminishing entanglement. It is clear that at higher temperature phonon induced incoherent processes dominate and the cavity modes are disentangled. We also noticed that the entanglement between the modes is reduced with increase in cavity decay rates and is lost completely at large decay rates as shown in Fig. \ref{fig:chap4/Fig13} (a). The cavity modes are maximally correlated in the strong coupling regime, $\kappa<g_1$ and the inter-mode correlation decreases with increasing cavity decay rate, Fig. \ref{fig:chap4/Fig13}(b). Therefore, the optimal conditions for observing steady-state CV entanglement between the cavity modes in this coherently pumped QD-photonic crystal cavity system are low temperatures, strong QD-cavity coupling ($\kappa\lesssim g_1)$, and moderate pumping strengths ($\eta<4.0g_1$).

\section{Conclusions}

We have derived a master equation for a coherently pumped single quantum dot (QD) coupled to a bimodal cavity system and investigated its steady-state dynamics including QD-populations, mean cavity photon number and the parameter regime for the coherence transfer from the QD-emitter states to cavity modes. Additionally, we have analyzed the variances of the relative and average phase Hermitian operators. We have derived a Fokker-Planck equation for the system, revealing that the relative and average phase diffusion coefficients vanish due to correlated emission into the cavity modes suppressing the quantum noise in lasing. This feature suggests that such a semiconductor system could serve as a viable platform for on-chip cavity-enhanced light (CEL) and can be useful in detection of weak signals and precision optical measurements. Furthermore, we calculated the single- and two-photon emission rates into the cavity modes using a simplified master equation (SME). In the end we examined the generation of steadystate continuous variable (CV) entanglement between the cavity modes by evaluating the sum of variances of EPR-like variables satisfying the DGCZ criterion and the zero-time delay second order photon-photon correlation functions, $g_{ij}^2(0)$. Therefore, this coherently pumped single QD-bimodal photonic crystal cavity system also holds potential applications in quantum cryptography and communication.

%% file: chap5.tex
\chapter{Nondegenerate two-photon lasing}\label{chap5}
{\small This chapter is based on our work ``\textsc{Nondegenerate two-photon lasing in a single quantum dot,}"; S. K. Hazra, \textbf{Lavakumar Addepalli}, P. K. Pathak, and T. N. Dey \textit{Phys. Rev. B} \textbf{109}, 155428 (2024).}


\section{Introduction}

In the earlier chapter, we discussed a correlated emission laser in the V-type configuration. Degenerate two-photon lasers have been investigated theoretically \cite{Elena2010,Harmanpreet2020} and realized experimentally \cite{Elena2010} in atomic systems with cascade configurations. The theory for a non-degenerate two-photon laser in a cascaded configuration is given by Laughlin and Swain \cite{Laughlin1991twophoton}, P. A. Maia Neto et al., \cite{Neto1991twophoton}. Further, such a laser is shown to act as an entanglement amplifier \cite{Xiong2005}. Recently, a two-mode laser with QD excitons coupled to a bimodal microcavity is realized \cite{Schlottmann2018}, and it is predicted that such lasers to show strong inter-mode correlations and super-thermal photon bunching \cite{Leymann2013photonBunching}. Here, we consider a quantum dot (QD) biexciton driven using $x$-polarized incoherent or coherent pump and the biexcitonic state cascades to ground state via $y$-polarized exciton coupled to two non-degenerate cavity modes. We include the effect of exciton-phonon interactions into the system dynamics by deriving a polaron-transformed master equation.

\section{Model system}
\label{sec:model}

We consider a single quantum dot with ground state, $|g\rangle$, $x$-polarized exciton, $|x\rangle$, $y$-polarized exciton, $|y\rangle$ and biexciton state, $|u\rangle$ forming a four-level system. Further, the transitions $|g\rangle \leftrightarrow |x\rangle$ and $|x\rangle\leftrightarrow|u\rangle$ are driven using external incoherent or coherent drives, and $|g\rangle\leftrightarrow |y\rangle$, $|y\rangle\leftrightarrow |u\rangle$ transitions are coupled to a vertically polarized non-degenerate bimodal cavity as shown in Fig. \ref{fig:chap5/Fig1}. Hamiltonian for the system is given by,

\begin{align}
H &= \hbar\omega_x\sigma_{xx}+\hbar \omega_y\sigma_{yy}- \hbar\omega_{u} \sigma_{uu}+ \hbar\omega_1 a_{1}^{\dagger}a_{1} - \hbar\omega_2a_{2}^{\dagger}a_{2} \nonumber\\
&+\hbar g_{1}(\sigma_{yg} a_{1}+ \sigma_{gy} a_{1}^{\dagger})+\hbar g_{2}(\sigma_{uy} a_{2} + \sigma_{yu} a_{2}^{\dagger})
\end{align}

where $\hbar\omega_x$ ,$\hbar\omega_y$, and $\hbar\omega_u$ are the energies of the QD states $x$-exciton ($|x\rangle$),  $y$-exciton ($|y\rangle$) and bi-exciton ($|u\rangle$). $\omega_i, \, g_i$ where $i=1,2$ are the frequency, coupling strength of $i$-th cavity mode. 

\begin{figure}
   \includegraphics[scale=0.45]{Figschap5/Fig1.eps}
    \caption{\label{fig:epsart}The energy level diagram for QD-cavity model. Left blue arrows indicating incoherent (coherent) pumping from $|g\rangle \rightarrow |u\rangle$ with pumping rates $\eta_{1}~(\Omega_1)$ and $\eta_{2}~(\Omega_2)$. Right green and red arrows indicating cavity coupling with $|y\rangle \rightarrow |g\rangle$ and $|u\rangle \rightarrow |y\rangle$ with coupling constant $g_{1}$ and $g_{2}$ and  frequencies $\omega_{1}$ and $\omega_{2}$. Here $\Delta_{xx}$ and $\delta_{x}$ stand for the biexciton binding energy and fine structure splitting energy between two exciton states. The $\Delta_{1}$ and $\Delta_{2}$ are the effective detunings for the first and second cavity modes \cite{Hazra2024}.}
    \label{fig:chap5/Fig1}
\end{figure}

\section{Incoherently pumped single QD-bimodal cavity system}

We consider both the transitions $|g\rangle\rightarrow|x\rangle$ and $|x\rangle\rightarrow|u\rangle$ are incoherently pumped with equal rates, $\eta_1=\eta_2$. The QD states can be incoherently pumped by the excitation of the charge carriers in the wetting layer \cite{Ota2011TPE} and followed by non-radiative transitions.

The Hamiltonian of the system is given in the rotating frame with $|y\rangle$ exciton frequency, $\omega_y$,

\begin{eqnarray}
H &&= \hbar\delta_{x}\sigma_{xx}- \hbar(\Delta_{xx}-\delta_{x})\sigma_{uu}- \hbar\delta_{1}a_{1}^{\dagger}a_{1} - \hbar(\Delta_{xx}-\delta_{x}+\delta_{2})a_{2}^{\dagger}a_{2} \nonumber\\
 &&+\hbar g_{1}(\sigma_{yg} a_{1}+ \sigma_{gy} a_{1}^{\dagger})+\hbar g_{2}(\sigma_{uy} a_{2} + \sigma_{yu} a_{2}^{\dagger})\label{uhu},
\end{eqnarray}

Here, the fine structure splitting (FSS) between $|x\rangle$ and $|y\rangle$ states, $\hbar\delta_x=\hbar(\omega_x-\omega_y)$, where $\omega_x$ is the $|x\rangle$ frequency. The biexciton binding energy, $\hbar\Delta_{xx}=\hbar(\omega_x+\omega_y-\omega_u)$, where $\omega_u$ is the biexciton frequency. Cavity mode 1 detuning, $\delta_1=\omega_y-\omega_1$ and cavity mode 2 detuning, $\delta_2=\omega_u-\omega_y-\omega_2$. The QD operators are represented by $\sigma_{ij}=| i\rangle \langle j|$. The phonon bath and the exciton-phonon interaction part of the Hamiltonian are given by,

\begin{equation}
H_{ph}=\hbar\sum_k\omega_kb_k^{\dag}b_k+\sum_{i=x,y,u}\lambda_{ik}\sigma_{ii}(b_k+b_k^{\dag}),
\end{equation}

the exciton phonon coupling strengths are given by $\lambda_{ik}$ and $b_k$($b_k^{\dag}$) are the annihilation operator of the $k$-th phonon bath mode having frequency $\omega_k$. The total Hamiltonian is given by,
\begin{equation}
    H'=H+H_{ph}
\end{equation}

Further, we apply the polaron transformation as implemented in the earlier chapters to include the multi-phonon effects. The polaron transformed Hamiltonian is given by,  $\tilde H=e^SH'e^{-S}$ where
\begin{equation}
P=\sum_{i=x,y,u}\frac{\lambda_{ik}}{\omega_k}\sigma_{ii}\left(b_k^{\dag}-b_k\right).\nonumber
\end{equation}
The transformed Hamiltonian $\tilde H$ is given by,

\begin{equation}
    \tilde H = H_s+H_b+H_{sb}
\end{equation}
where 
\begin{eqnarray}
\text{System Hamiltonian}:  H_{s} &&= \hbar\delta_{x}\sigma_{xx}- \hbar(\Delta_{xx}-\delta_{x})\sigma_{uu} - \hbar\Delta_{1}a_{1}^{\dagger}a_{1}\nonumber\\ 
&&- \hbar(\Delta_{xx}-\delta_{x}+\Delta_{2})a_{2}^{\dagger}a_{2}+\langle B\rangle X_{g},\\
\text{Bath Hamiltonian}:
H_b &&=\hbar\sum_k\omega_k b_k^{\dagger}b_k, \label{eqn:Hb}\\
\text{System-bath Hamiltonian}:
H_{sb} &&=\xi_gX_g+\xi_uX_u.\label{eqn:Hsb}
\end{eqnarray}

The phonon-bath induced polaron shifts, $\sum_k\lambda_{ik}^2/\omega_k$ are included by redefining the detuning of the cavity modes to $\Delta_1$ and $\Delta_2$.

The system operators $X_g\,, X_u$ :
\begin{eqnarray}
X_{g}&&=\hbar\left( g_{1}\sigma_{yg} a_{1} + g_{2}\sigma_{uy} a_{2} \right) + H.c.\\
X_{u}&&=i\hbar\left(g_{1}\sigma_{yg} a_{1} + g_{2}\sigma_{uy} a_{2}\right) + H.c. ,
\end{eqnarray}

The phonon bath fluctuation operators:
\begin{eqnarray}
 \zeta_{g} &&= \frac{1}{2}\left( B_{+} + B_{-} -2\langle B\rangle  \right)\\
 \zeta_{u} &&= \frac{1}{2i}\left( B_{+} - B_{-} \right),
 \end{eqnarray}
where $B_{+}$, $B_{-}$ are the phonon displacement operator.

The phonon displacement operators: 
\begin{equation}
B_{\pm} =  \exp\left[{\pm \sum_{k} \frac{\lambda_{k}}{\omega_{k}}\left( b_{k} - b_{k}^{\dagger}\right)}\right].\nonumber 
\end{equation}
The mean value phonon displacement operator, $\langle B_{+}\rangle =\langle B_{-}\rangle = \langle B\rangle$ where
\begin{equation}
\langle B\rangle= \text{exp}\left[-\frac{1}{2}\int_0^{\infty}d\omega\frac{J(\omega)}{\omega^2}
\coth\left(\frac{\hbar\omega}{2K_bT}\right)\right].
\end{equation}
Here, we consider super-ohmic phonon spectral density, $J(\omega)= \alpha_{p}\omega^3\exp[-\omega^2/2\omega_b^2]$ \cite{Wilson2002}, where the $\alpha_p$ and $\omega_b$ are the electron-phonon coupling strength and cutoff frequency, respectively. The parameter values are  $\alpha_p=2.36\, ps^2$ and $\omega_b=1\,meV$ are chosen, which match up the values of $\langle B\rangle=0.90,\, 0.84$, and $0.73$ for $T=5K$, $10K$, and $20K$ in agreement with experimental observations \cite{hughes2011mollow}.

\subsection{Master equation}

The master equation for reduced density matrix  for QD-cavity system $\rho_s$ is obtained after making Born-Markov approximation for $H_{sb}$, chapter 1, Sec. \ref{sec:Phonons} and tracing over the bath states \cite{Dara2010rabiRotations,roy2011,hughes2011mollow},
\begin{eqnarray}
\dot{\rho_s} &&=-\frac{i}{\hbar}[H_s,\rho_s]-{\cal L}_{ph}\rho_s-\frac{\kappa_{1}}{2}{\cal L}[a_{1}]\rho_s\nonumber
-\frac{\kappa_{2}}{2}{\cal L}[a_{2}]\rho_s\nonumber\\
&&-\sum_{i=x,y}\left(\frac{\gamma_1}{2}{\cal L}[\sigma_{gi}]
+\frac{\gamma_2}{2}{\cal L}[\sigma_{iu}]\right)\rho_s
 -\sum_{i=x,y,u}\frac{\gamma_d}{2}{\cal L}[\sigma_{ii}]\rho_s\nonumber\\
&&-\left(\frac{\eta_1}{2}{\cal L}[\sigma_{xg}]
+\frac{\eta_2}{2}{\cal L}[\sigma_{ux}]\right)\rho_s\label{meq},
\end{eqnarray}

the cavity modes decay rates $\kappa_{1}$, $\kappa_{2}$, and  spontaneous emission rate of excitons and biexciton states are $\gamma_1$, $\gamma_2$ respectively. The zero phonon line broadening (ZPL) \cite{Ortner2004ZPL,Rudin2006ZPL} is incorporated by the pure dephasing rates,  $\gamma_d$. The QD states are incoherently pumped with rates $\eta_{1}$ and $\eta_{2}$.  All these incoherent processes are phenomenologically included in the master equation using the Lindblad superoperator ($\cal L$) defined as 
${\cal L}[{\cal O}]\rho_s ={\cal O}^{\dagger}{\cal O} \rho_s - 2 {\cal O} \rho_s {\cal O}^{\dagger} +\rho_s \cal O^{\dagger}\cal O $. 

The second term in the Eq. \ref{meq}, ${\cal L}_{ph}\rho_s$ denotes the phonon-induced processes on the system dynamics. The form of ${\cal L}_{ph}\rho_s$ is given in the Born-Markov approximation,

\begin{equation}
{\cal L}_{ph}\rho_s=\frac{1}{\hbar^2}\int_0^{\infty}d\tau\sum_{j=g,u}G_j(\tau)[X_j(t),X_j(t,\tau)\rho_s(t)]+H.c.
\label{lph}
\end{equation}

where $X_j(t,\tau)=e^{-iH_s\tau/\hbar}X_j(t)e^{iH_s\tau/\hbar}$,  $G_g(\tau)=\langle \zeta_g(t)\zeta_g(t,\tau)\rangle= \langle B\rangle^2\{\cosh[\phi(\tau)]-1\}$ and $G_u(\tau)=\langle \zeta_u(t)\zeta_u(t,\tau)\rangle=\langle B\rangle^2\sinh[\phi(\tau)]$. Here, $\phi(\tau)$ is the phonon bath correlation function given by,

\begin{equation}
\phi(\tau)=\int_0^{\infty}d\omega\frac{J(\omega)}{\omega^2}
\left[\coth\left(\frac{\hbar\omega}{2K_bT}\right)\cos(\omega\tau)-i\sin(\omega\tau)\right],
\end{equation}

where $K_b$ is the Boltzmann constant and $T$ is the temperature of the phonon bath. The master equation (\ref{meq}) is integrated numerically using Quantum Optics Toolbox by S. M. Tan\cite{SMTan1999}. We consider that both the cavity modes are coupled to $|g\leftrightarrow|y\rangle$ and $|y\rangle\leftrightarrow|u\rangle$ transitions with equal strength, $g_2=g_1=g$. The values of biexciton binding energy, $\Delta_{xx}=15.0g_1$ and the FSS, $\delta_x=-1.0g_1$ are considered throughout the chapter.

\subsection{Steadystate populations}

\begin{figure}
   \includegraphics[scale=0.5]{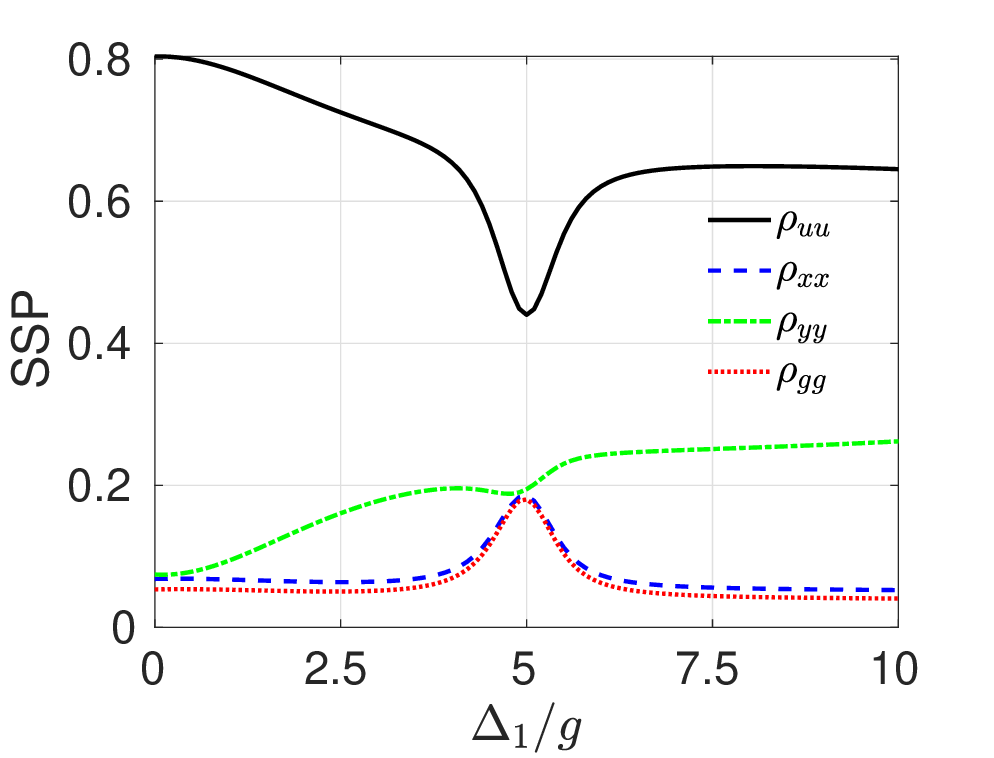}
    \caption{Steady state populations (SSP) of QD states $|u\rangle$(solid black),$|x\rangle$(dashed blue),$|y\rangle$(dash dotted green), $|g\rangle$(dotted red) for phonon bath temperature T=5K. Other system parameters are cavity leakage $\kappa_{1}=\kappa_{2}=0.1g$, QD cavity field couplings $g_{1}=g_{2}=g$, exciton and biexciton spontaneous decay rates $\gamma_{1}=\gamma_{2}=0.01g$, pure dephasing rate $\gamma_{d}=0.01g$, biexciton binding energy $\Delta_{xx}=15.0g$, fine structure splitting between excitons $\delta_{x}=-1.0g$, detuning of second cavity mode $\Delta_{2}= -5.0g$ and $\eta_{1}=\eta_{2}=0.5g$.}
    \label{fig:chap5/Fig2}
\end{figure}

\begin{figure}
   \includegraphics[scale=0.5]{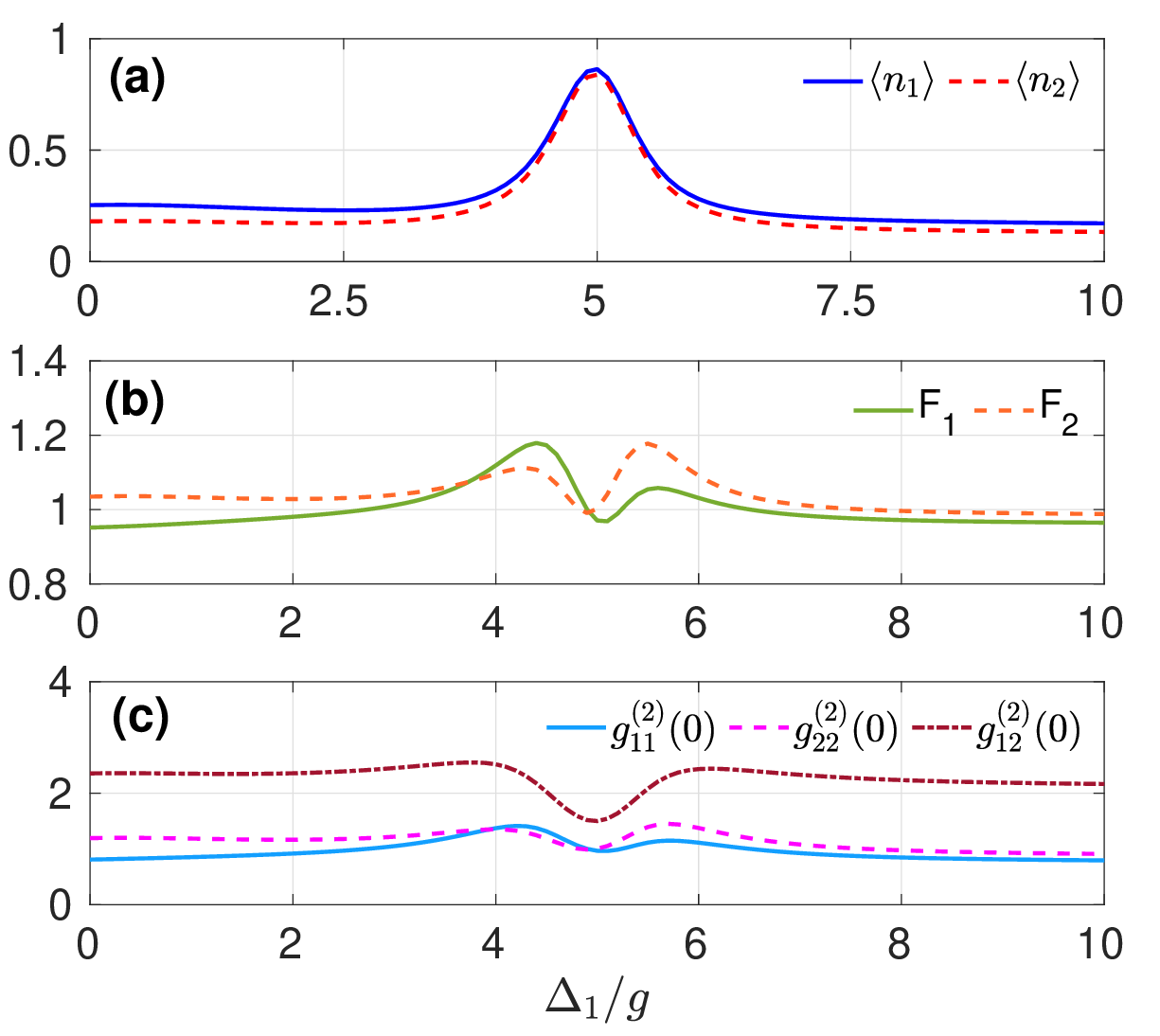}
    \caption{ Cavity photon statistics a) The mean photon number of first and second mode $\langle n_{1}\rangle$(blue solid), $\langle n_{2}\rangle $(red long dash), b) Fano factors $F_{1}$ (solid green), $F_{2}$ (dashed orange), and c) Second order photon correlation function with zero time delay $g_{11}^{(2)}(0)$ (solid cyan), $g_{22}^{(2)}(0)$ (dashed magenta), $g_{12}^{(2)}(0)$ (dash dotted brown) as a function of first cavity mode detuning $\Delta_{1}$ for temperature T=5K. All other parameters are the same as Fig.$\ref{fig:chap5/Fig2}$.
    }
    \label{fig:chap5/Fig3}
\end{figure}

In general, two-photon processes are nonlinear and are weaker in comparison to the single-photon process. For the realization of the dominant two-photon process, the single-photon transitions can be suppressed by considering a far detuned cavity with respect to QD transitions, $\Delta_1,\,\Delta_2\gg g_1$. In Fig. $\ref{fig:chap5/Fig2}$ we provide the results for the steady-state population of the QD states, $|g\rangle$, $|x\rangle$, $|y\rangle$ and $|u\rangle$ by fixing the mode 2 detuning $\Delta_2=5.0g_1$ and varying the mode 1 detuning, $\Delta_1$. We can see that the biexciton state is populated the most compared to other states because the $|u\rangle\rightarrow|y\rangle$ transition is inhibited since mode 2 is off-resonant. As we vary the first mode detuning, $\Delta_1$, at the two-photon resonance condition, $\Delta_1=-\Delta_2$, i.e., $\omega_u=\omega_1+\omega_2$, resulting in transfer of population from the biexciton state to $|g\rangle$ via $|y\rangle$.  When the mode 1 detuning, $\Delta_1$ is increased, the transition $|g\rangle\leftrightarrow|y\rangle$ becomes off-resonant, which leads to the population in $|u\rangle$ decreasing and in $|y\rangle$ increasing monotonically. At the two-photon resonance, $\Delta_1=-\Delta_2=5.0g$, there is a dip in the population of $|u\rangle$ and $|y\rangle$ states with an increase in population of $|x\rangle$ and $|g\rangle$ states. This population transfer is accompanied by an increase in the mean photon number of both the cavity modes, as shown in Fig. \ref{fig:chap5/Fig3} (a).  

We notice that in both the cavity modes the mean photon number increases and attain an identical peak value of $\langle n\rangle\approx 0.9$. This suggests that the photons are emitted in pairs into both modes as the mode 1 is tuned to two-photon resonance. The Fano factor compares the cavity mode state with the coherent state of the same mean value. 

\begin{eqnarray}
\text{Fano Factor}:\\
F_{1} &&= (\langle n_{1}^{2}\rangle-\langle n_{1}\rangle^{2})/\langle n_{1}\rangle \\
F_{2} &&= (\langle n_{2}^{2}\rangle-\langle n_{2}\rangle^{2})/ \langle n_{2}\rangle.
\end{eqnarray}

In Fig. \ref{fig:chap5/Fig3} (b), the results for the Fano factor ($F_i$) are given. We can see a sharp dip in the values of $F_i$ at the two-photon resonance. This suppression of noise is the result of cascaded emission from biexciton to ground state. In Fig. \ref{fig:chap5/Fig3} (b), we present the results of the intensity fluctuations of both the modes by plotting the inter and intra-mode zero-time delay second-order photon-photon correlations. 

\begin{eqnarray}
\text{Intra-mode correlations}:\\
g_{11}^{(2)}(0) &&= \langle a_{1}^{\dagger2}a_{1}^{2}\rangle/\langle a_{1}^{\dagger}a_{1}\rangle^{2} \\
g_{22}^{(2)}(0) &&= \langle a_{2}^{\dagger2}a_{2}^{2}\rangle/\langle a_{2}^{\dagger}a_{2}\rangle^{2} \\
\text{Inter-mode correlation}:\\
g_{12}^{(2)}(0) &&= \langle a_{1}^{\dagger} a_{2}^{\dagger}a_{2}a_1\rangle/\langle a_{1}^{\dagger}a_{1}\rangle \langle a_{2}^{\dagger}a_{2}\rangle.
\end{eqnarray}

From the results, we can see $g_{12}^{(2)}(0)$ is greater than the inter-mode correlations, $g_{ii}^{(2)}(0)$ ($i=1,2$) for the entire range of detuning, $\Delta_1$ stating . Also, at the two-photon resonance where the mean photon number peaks, the correlations satisfy $g_{12}^{(2)}(0) > \sqrt{(g_{11}^{(2)}(0)\times g_{22}^{(2)}(0)}$ condition for non-classicality of the bimodal cavity state \cite{Agarwal_2012} and suggests two-mode two-photon emission into the cavity modes. 

\begin{figure}[h]
   \includegraphics[scale=0.5]{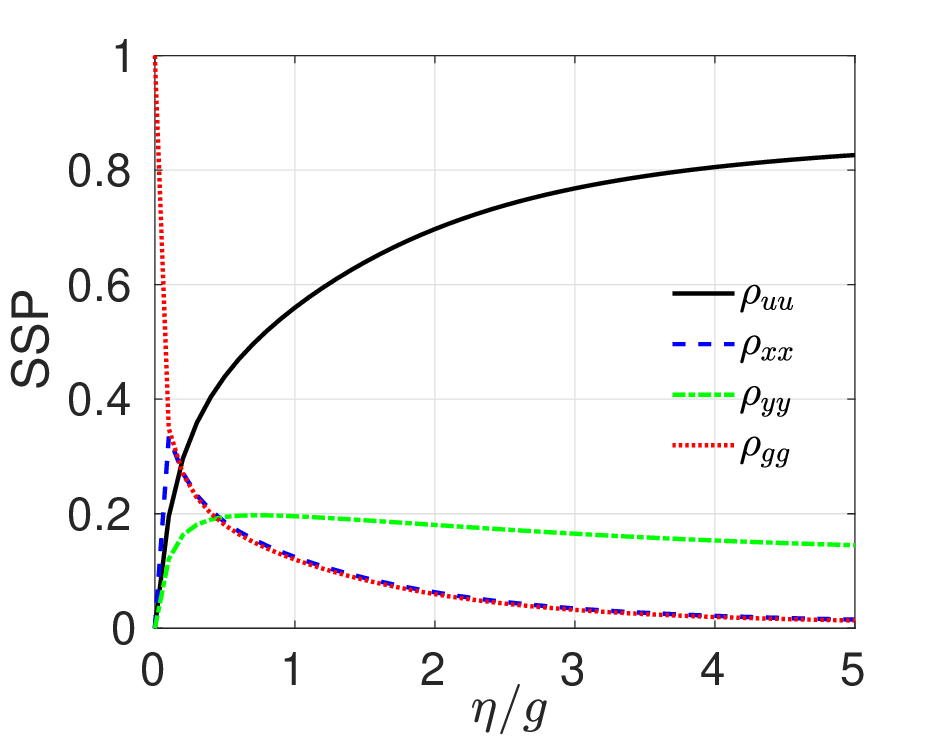}
    \caption{Steady state populations (SSP) in QD states $|u\rangle$(solid black), $|x\rangle$(dashed blue), $|y\rangle$(dash dotted green), $|g\rangle$(dotted red) vs incoherent pumping rate $\eta$ for phonon bath temperature T=5K. All other parameters are same as Fig.$\ref{fig:chap5/Fig2}$ except $\Delta_{1}= 5.0g$ and $\eta_{1}=\eta_{2}= \eta$.}
    \label{fig:chap5/Fig4}
\end{figure}

In Fig. \ref{fig:chap5/Fig4} we present the results for the steady-state population variation at the two-photon resonance, $\Delta_1+\Delta_2=0$, as the incoherent pumping rate, $\eta_1=\eta_2=\eta$ is increased. With an increase in the incoherent pumping rate, the population in $|x\rangle$ and $|g\rangle$ decreases and is transferred to the biexciton $|u\rangle$ state, resulting in the population inversion. The population in the $|y\rangle$ remains constant once the inversion between $|g\rangle$ and $|y\rangle$ is achieved. 

\begin{figure}
   \includegraphics[scale=0.5]{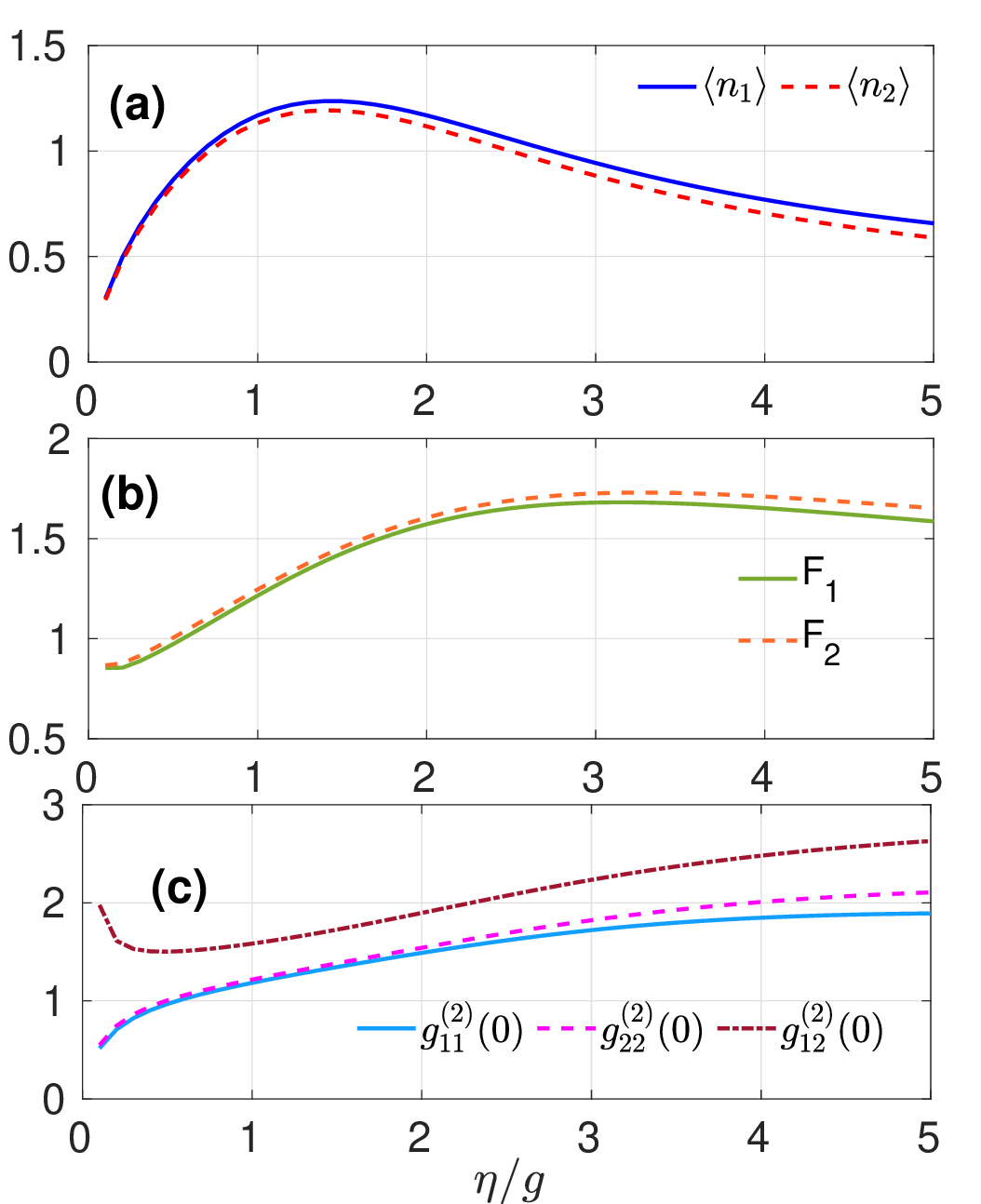}
    \caption{Cavity photon statistics: a) Mean photon number for first and second mode $\langle n_{1}\rangle$(blue solid), $\langle n_{2}\rangle $(red long dash), b)Fano factors $F_{1}$(solid green), $F_{2}$(dashed orange), c) Second order correlations function with zero time delay $g_{11}^{(2)}(0)$(solid cyan), $g_{22}^{(2)}(0)$(dashed magenta), $g_{12}^{(2)}(0)$(dash dotted brown) as a function of $\eta$ for temperature T=5K. All other parameters are same as Fig.($\ref{fig:chap5/Fig2}$) except $\Delta_{1}= 5.0g$ and $\eta_{1}=\eta_{2}= \eta$.
    }
    \label{fig:chap5/Fig5}
\end{figure}

In Fig. \ref{fig:chap5/Fig5}, the cavity photon statistics are presented. With an increase in the incoherent pumping rate, $\eta$, the mean photon number increases in both the cavity modes almost equally and attains a peak value of $\sim 1.3$ around $\eta=1.5g$, Fig. \ref{fig:chap5/Fig5} (a). Further increase in the incoherent pumping rate leads to self-quenching effect \cite{mu1992}, resulting in the decline of the mean photon number. At high pump rates, the mean photon number in mode 2 is greater than that of mode 1 due to asymmetry in the phonon-induced cavity mode feeding rate. It is greater when cavity mode is negatively detuned, $\Delta_2<0$, compared to the positive detuning, $\Delta_1>0$ case. Fano Factor ($F_i$) results are given in Fig. \ref{fig:chap5/Fig5} (b) shows that at low incoherent pumping rate, $F_i<1$, cavity modes show non-classical behavior. As $\eta$ is increased, the cavity field shows coherent and eventually super-Poissonian behavior. The inter and intra-mode correlations in Fig. \ref{fig:chap5/Fig5} (c) show that the cavity modes state is highly non classical $g_{12}^{(2)}(0)>\sqrt{g_{11}^{(2)}(0)g_{22}^{(2)}(0)}$ at low pumping rate, $\eta<0.5g_1$. This is due to the dominant two-photon emission into the cavity modes as shown in the next section. As $\eta$ is increased, the cavity fields become coherent for moderate pumping, $0.5g<\eta<1.0g$, while for $\eta>1.0g$, thermal in nature, showing bunching statistics, $g_{ii}^{(2)}(0)\approx 2$ and enhanced intensity fluctuations.

\subsection{Laser rate equations}

To obtain the laser rate equations for the system using quantum theory of lasers developed by Scully and Lamb \cite{Scully1967,sargent1974}, we derive a simplified master equation (SME) \cite{Verma2020} following the approach used in the earlier chapters. Here, we assume $\vert\Delta_1\vert\gg g_1$ and $\vert\Delta_2\vert\gg g_2$ and simplify the term ${\cal L}_{ph}\rho_s$ in the master equation, Eq. \ref{meq}. SME is given in Appendix \ref{sec:chap5_Appendix1}.

The laser rate equation is given by,
\begin{align}
\dot{P}_{n,m} &=-\alpha_{n,m} P_{n,m}+G^{11}_{n-1,m-1}P_{n-1,m-1}  \nonumber\\
&+G^{10}_{n-1,m}P_{n-1,m}+G^{01}_{n,m-1}P_{n,m-1} \nonumber\\
&+A^{11}_{n+1,m+1}P_{n+1,m+1}+A^{10}_{n+1,m}P_{n+1,m}\nonumber\\
&+A^{01}_{n,m+1}P_{n,m+1}+ \kappa_{1}(n+1)P_{n+1,m}\nonumber\\& - \kappa_{1}nP_{n,m} + \kappa_{2}(m+1)P_{n,m+1} - \kappa_{2}mP_{n,m}, 
\label{inco_rate}
\end{align}
where $P_{n,m}=\sum_i\langle i,n,m|\rho_s|i,n,m\rangle$,  $\alpha_{n,m}P_{n,m}=\sum_i\alpha_{i,n,m}\langle i,n,m|\rho_s|i,n,m\rangle$, $G^{\alpha\beta}_{n,m}P_{n,m}=\sum_iG^{\alpha\beta}_{i,n,m}\langle i,m,n|\rho_s|i,m,n\rangle$, $A^{\alpha\beta}_{m,n}P_{n,m}=\sum_iA^{\alpha\beta}_{i,n,m}\langle i,m,n|\rho_s|i,m,n\rangle$ with $i=g,x,y,u$ and $\alpha,\beta=0,1$. 

Apart from single and two-photon processes, there can be other multi-photon processes contributing to cavity mode population. However, their contributions are insignificant in comparison at two-photon resonance. In Eq.(\ref{inco_rate}), the single-photon emission rates into the first mode, second mode and the two-mode two-photon emission rate are given by  $\sum_{n,m} G_{n,m}^{10}P_{n,m}$, $\sum_{n,m} G_{n,m}^{(01)}P_{n,m}$ and $\sum_{n,m} G_{n,m}^{11}P_{n,m}$, respectively. Similarly, absorption rates corresponding to single-photon processes in first mode, second mode and two-mode two-photon process are given by  $\sum_{n,m} A_{n,m}^{10}P_{n,m}$, $\sum_{n,m} A_{n,m}^{01}P_{n,m}$ and $\sum_{n,m} A_{n,m}^{11}P_{n,m}$ respectively. The rates are obtained in steady-state condition. Next, we define the net single, two-mode two-photon excess emission rates (EER) are obtained by subtracting the absorption rate from the emission rate.

\begin{figure}
   \includegraphics[scale=0.5]{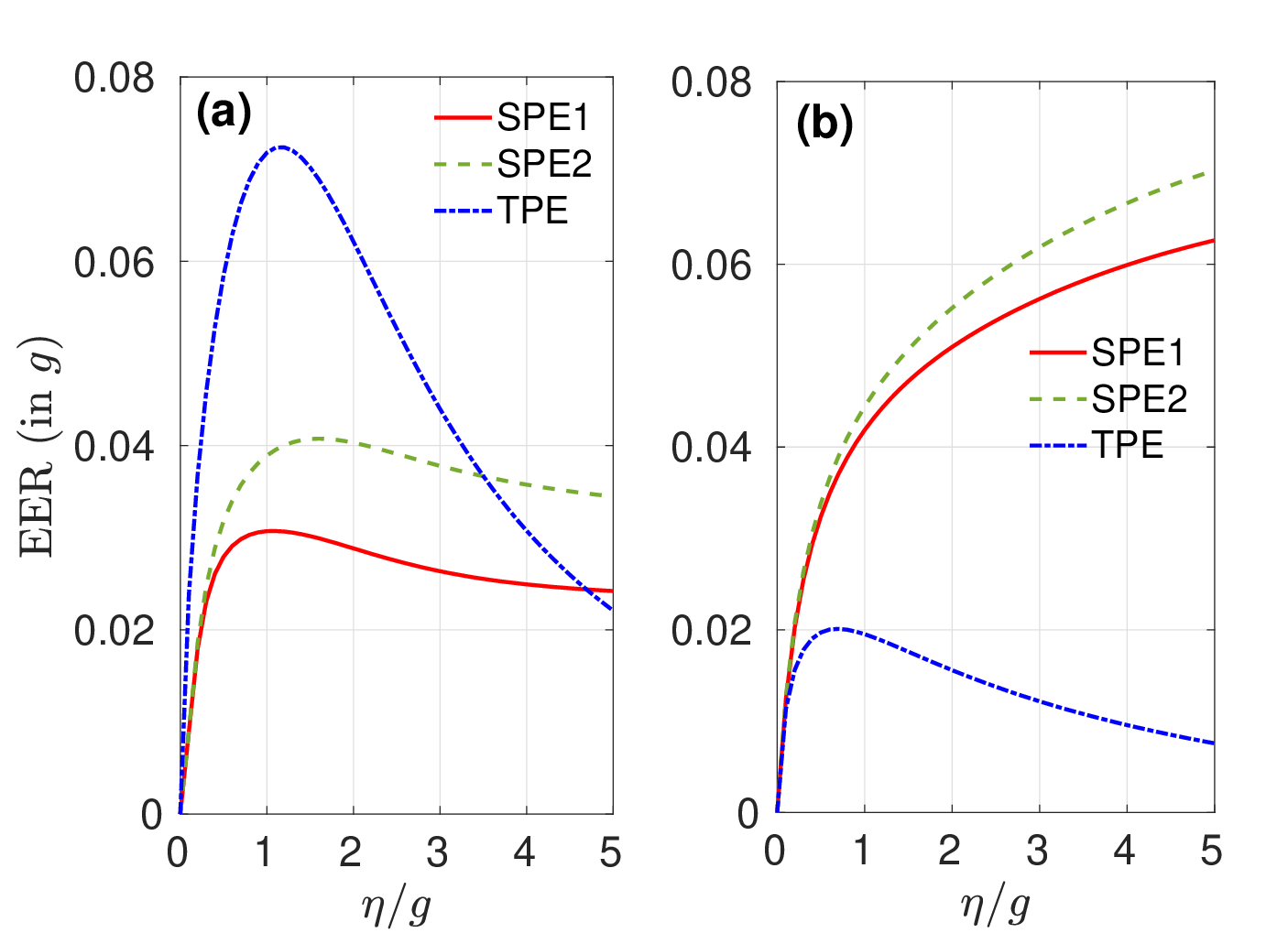}
    \caption{ Excess emission rates (EER). First mode single-photon excess emission rate(solid red), second mode single-photon excess emission rate(dashed green) and two-mode two-photon excess emission rate(dash dotted blue) for the detuning $\Delta_{1}= 5.0g_{1}$, varying incoherent pumping rate,$\eta_{1}=\eta_{2}=\eta$ for a)T=5K and b) T=20K with all other parameters same as Fig.\ref{fig:chap5/Fig2}.
   }
    \label{fig:chap5/Fig6}
\end{figure}

We plot net excess emission rates, Fig.\ref{fig:chap5/Fig6}, for single and two-mode two-photon processes at different temperatures. In Fig.\ref{fig:chap5/Fig6}(a), the two-mode two-photon net excess emission rate is dominant over the single-photon emission into modes 1 and 2. The two-mode two-photon rate attains maxima for low pumping rate ($\eta\approx g$) for T=5K. 
However, with an increase in the incoherent pumping rate,  the two-mode two-photon emission rate decreases due to increased decoherence, and the single-photon emission corresponding to mode 1 and mode 2 starts increasing. Due to asymmetry in phonon-induced cavity mode feeding, the single photon emission rate for mode 2 is greater than that of mode 1. Hence, at a low incoherent pumping rate, the system exhibits two-mode two-photon laser behavior. In Fig.$\ref{fig:chap5/Fig6}$(b), we present the result for $T=20K$ case. With an increase in the system temperature, the phonon-induced decoherence results in the suppression of multi-photon processes. The two-mode two-photon gain decreases, and the cavity modes are populated dominantly by the single-photon processes. Therefore, incoherent pumping rates, $\eta\leq 1.0g$, and low temperatures are preferable for the realization of two-mode two-photon lasing.

\section{Coherently pumped single QD-bimodal cavity system}

In this section, we study the system's steady-state dynamics when the excitonic state, $|x\rangle$ is resonantly excited using external $x$-polarized coherent drive coupled to the $|g\rangle \rightarrow |x\rangle$ transition. This leads to off-resonant coupling to the biexciton, $|u\rangle$, due to large binding energy, $\Delta_{xx}=15.0g$. Therefore, the biexciton state is mostly populated via phonon-induced excitation \cite{Findeis2000biexciton}. Earlier works used this method to generate a QD biexciton state with high fidelity \cite{VMAxt2013fidelity}.

\subsection{Master equation}

The Hamiltonian for the system is given in the rotating frame of coherent pump frequency $\omega_{p}$.

\begin{eqnarray}
H &&= \hbar\Delta_{p}\sigma_{xx}+\hbar(2\Delta_{p}-\delta_{x}-\Delta_{xx})\sigma_{uu}+\hbar(\Delta_{p}-\delta_{x})\sigma_{yy} \nonumber\\ &&+\hbar(\Delta_{p}-\delta_{x}-\Delta_{1})a_{1}^{\dagger}a_{1}\nonumber +\hbar(\Delta_{p}-\Delta_{xx}-\Delta_{2})a_{2}^{\dagger}a_{2} \nonumber\\
&&+\hbar \Omega_{1}(\sigma_{xg} + \sigma_{gx})+\hbar \Omega_{2}(\sigma_{ux}+ \sigma_{xu}) + \hbar g_{1}(\sigma_{yg} a_{1}\nonumber\\
&&+ \sigma_{gy} a_{1}^{\dagger})+\hbar g_{2}(\sigma_{uy} a_{2} + \sigma_{yu} a_{2}^{\dagger}) + H_{ph},
\label{hhcoh}
\end{eqnarray}

where $\Delta_{p}= \omega_{x} - \omega_{p}$ is the detuning between $|g\rangle\leftrightarrow|x\rangle$ transition and coherent drive. The symbols are the same as in the previous section.

The polaron transformed Hamiltonian $\tilde H = H_s+ H_b + H_{sb}$ :

\begin{eqnarray}
\text{System Hamiltonian}:\\
H_{s} &&= \hbar\Delta_{p}\sigma_{xx}+\hbar(2\Delta_{p}-\delta_{x}-\Delta_{xx})\sigma_{uu}+\hbar(\Delta_{p}-\delta_{x})\sigma_{yy}    \nonumber\\ &&+\hbar(\Delta_{p}-\delta_{x}-\Delta_{1})a_{1}^{\dagger}a_{1} +\hbar(\Delta_{p}-\Delta_{xx}-\Delta_{2})a_{2}^{\dagger}a_{2}+\langle B\rangle X_{g},\label{hcoh}\\
X_{g}&&=\hbar\left(\Omega_{1}\sigma_{xg}+\Omega_{2}\sigma_{ux}+g_{1}\sigma_{yg} a_{1} + g_{2}\sigma_{uy} a_{2}\right) + H.c\label{xgcoh}\\
X_{u}&&=i\hbar\left(\Omega_{1}\sigma_{xg}+\Omega_{2}\sigma_{ux}+g_{1}\sigma_{yg} a_{1} + g_{2}\sigma_{uy} a_{2}\right)+H.c.\label{xucoh}
\end{eqnarray}

The bath Hamiltonian, $H_b$ is the same as Eq. \ref{eqn:Hb} and the system-bath interaction Hamiltonian is the same as Eq. \ref{eqn:Hsb}. We use the master equation (\ref{meq}) with $H_s$ given by Eq. \ref{hcoh} and neglecting the incoherent pumping terms in this coherently pumping case.

\subsection{Steadystate populations and cavity photon statistics}

We solve the master equation numerically to obtain the steady-state populations and cavity photon statistics.

\begin{figure}[h]
   \includegraphics[scale=0.5]{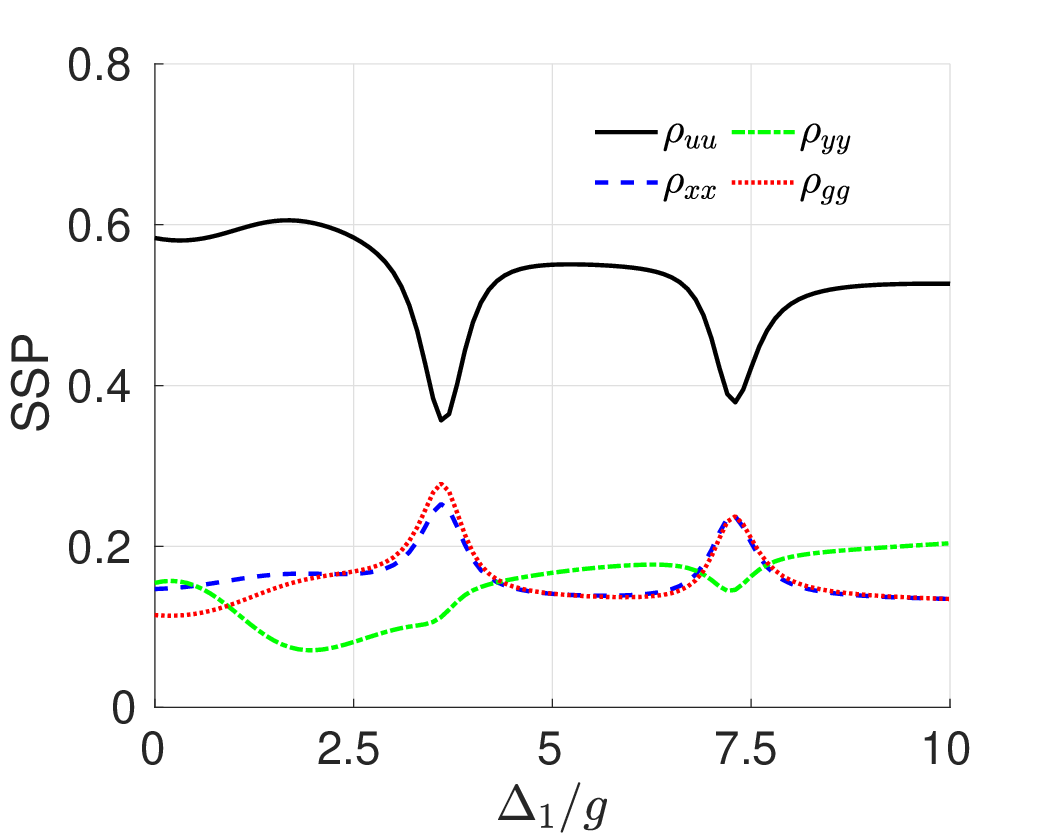}
    \caption{Steady state populations (SSP) of QD states $|u\rangle$(solid black),$|x\rangle$(dashed blue),$|y\rangle$(dash dotted green), $|g\rangle$(dotted red) for phonon bath temperature T=5K are presented by varying first mode detuning, $\Delta_1$. Other parameters are same as Fig. \ref{fig:chap5/Fig2} except detuning of $|x\rangle$ state from coherent pump frequency, $\Delta_{p} =0$ and coherent pumping rate, $\Omega_{1} = \Omega_{2} = 2g$.}
    \label{fig:chap5/Fig7}
\end{figure}

\begin{figure}
   \includegraphics[scale=0.5]{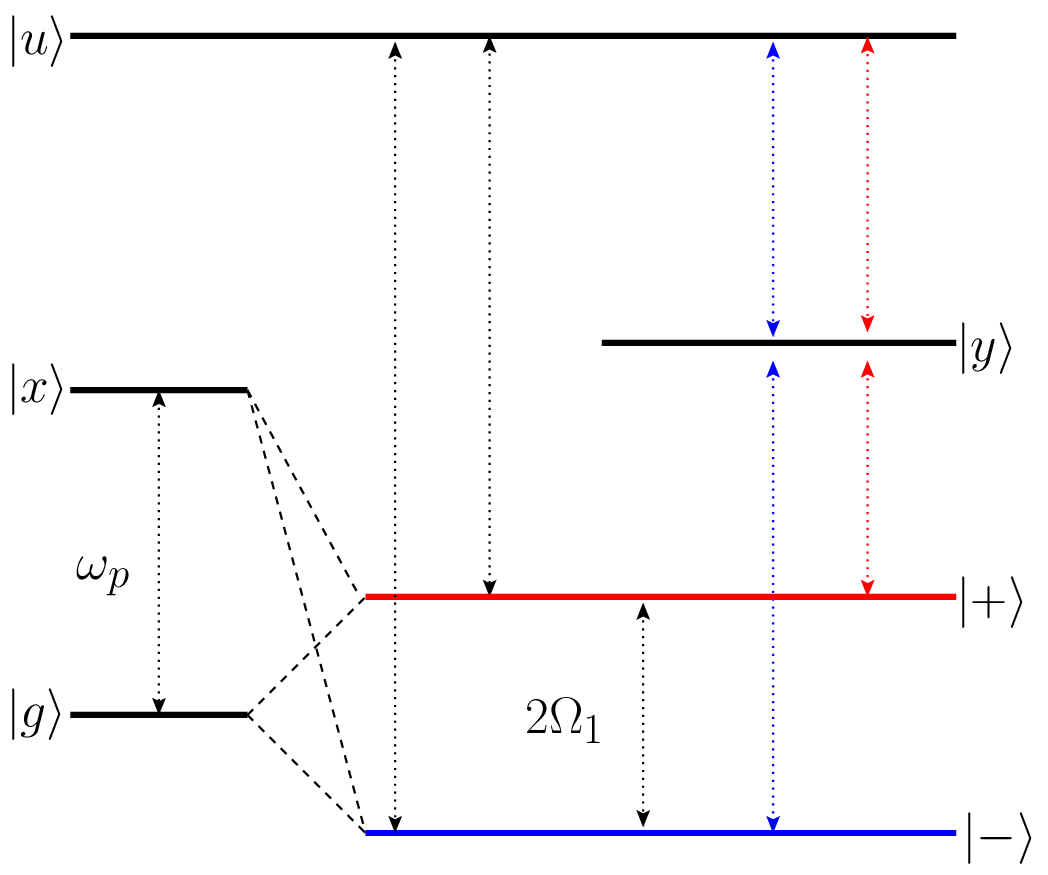}
    \caption{The strong optical field coupled with the $\vert g\rangle\leftrightarrow\vert x \rangle$ transition produces dressed states $\vert \pm\rangle$ with energy separation 2$\Omega_{1}$. Then these dressed states make a phonon-assisted pumping to the biexciton state, denoted by black dotted lines. Now we can get two different values of $\Delta_{1}$ satisfying the two-photon resonance condition, denoted by red and blue dotted lines \cite{Hazra2024}.}
    \label{fig:chap5/Fig8}
\end{figure}

In Fig. \ref{fig:chap5/Fig7}, we plot the steady-state population of the QD states by varying the mode 1 detuning, $\Delta_1$, for $\Omega_1=\Omega_2=2g$. The population in the biexciton state $|u\rangle$ is higher than in the other QD states. At resonance, $\Delta_p=0$, the pump dressed QD states, $\vert \pm\rangle = \frac{1}{\sqrt{2}}\left(\vert x\rangle\pm \vert g\rangle\right)$ are separated by $2\Omega_1$. Fig. \ref{fig:chap5/Fig8} shows the dressed state picture of the system and the allowed transitions in the presence of strong coherent pumping.

The population of the biexciton state $\vert u\rangle$ has a sharp dip for two detunings of mode 1, $\Delta_1=-\Delta_2\pm\Omega_1$, which correspond to the population transfer from the bare biexciton state $\vert u\rangle$ to dressed states  $\vert +\rangle$ and $\vert -\rangle$. Since $|\pm\rangle$ states are comprised of both the $|x\rangle$ and $|g\rangle$ states, they are almost equally populated. The population of the $y$-polarized exciton, $|y\rangle$ remains minimum at two-photon resonance conditions, $\Delta_1=-\Delta_2\pm\Omega_1$. 

\begin{figure}
   \includegraphics[scale=0.5]{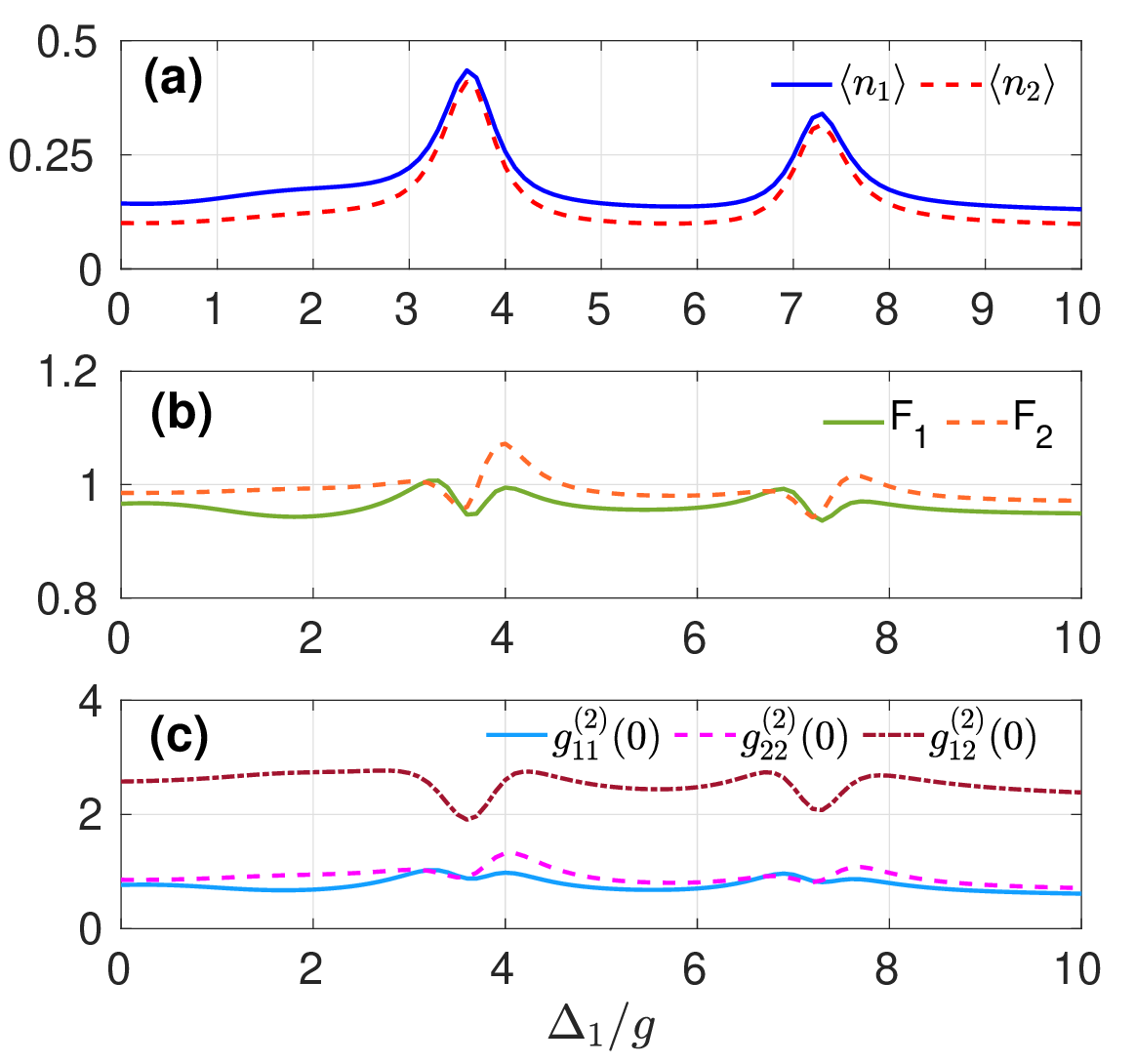}
    \caption{Cavity photon statistics: a) mean photon number $\langle n_{1}\rangle$(blue solid), $\langle n_{2}\rangle $(red long dash), b) Fano factors $F_{1}$(solid green), $F_{2}$(dashed orange), and c) Second order correlation $g_{11}^{(2)}(0)$(solid cyan), $g_{22}^{(2)}(0)$(dashed magenta), $g_{12}^{(2)}(0)$(dash dotted brown) with first mode detuning, $\Delta_{1}$ for temperature T=5K. All other parameters are same as Fig. \ref{fig:chap5/Fig7}.}
    \label{fig:chap5/Fig9}
\end{figure}

The transfer of population from $|u\rangle$ to $|\pm\rangle$ states results in two-mode two-photon emission into the cavity modes. The mean photon number curves of both the cavity modes have a peak when mode 1 detuning is $\Delta_1=-\Delta_2\pm\Omega_1$. The peaks are slightly shifted towards the right and are due to the Stark effect \cite{Banerjee2005Stark}. In Fig. \ref{fig:chap5/Fig9} (b), the Fano factor $F_i$ results show that the cavity fields are coherent in nature with $F_i\approx 1$ and also have dips in the variances at the two-mode two-photon resonance and also, the second order zero-time delay correlation functions, $g_{ij}^{(2)}(0)$ attain minimum values, Fig. \ref{fig:chap5/Fig9} (c). The inter-mode correlation $g_{12}^{(2)}(0)$ is greater than the intra-mode correlations and satisfying the non-classicality condition, $g_{12}^{(2)}(0)>\sqrt{g_{11}^{(2)}(0)g_{22}^{(2)}(0)}$.

\subsection{Emission and absorption rates}

In order to get the laser rate equations, we follow the approach similar to the previous section to obtain the simplified master equation given in Appendix \ref{sec:chap5_Appendix2}.

\begin{figure}
   \includegraphics[scale=0.5]{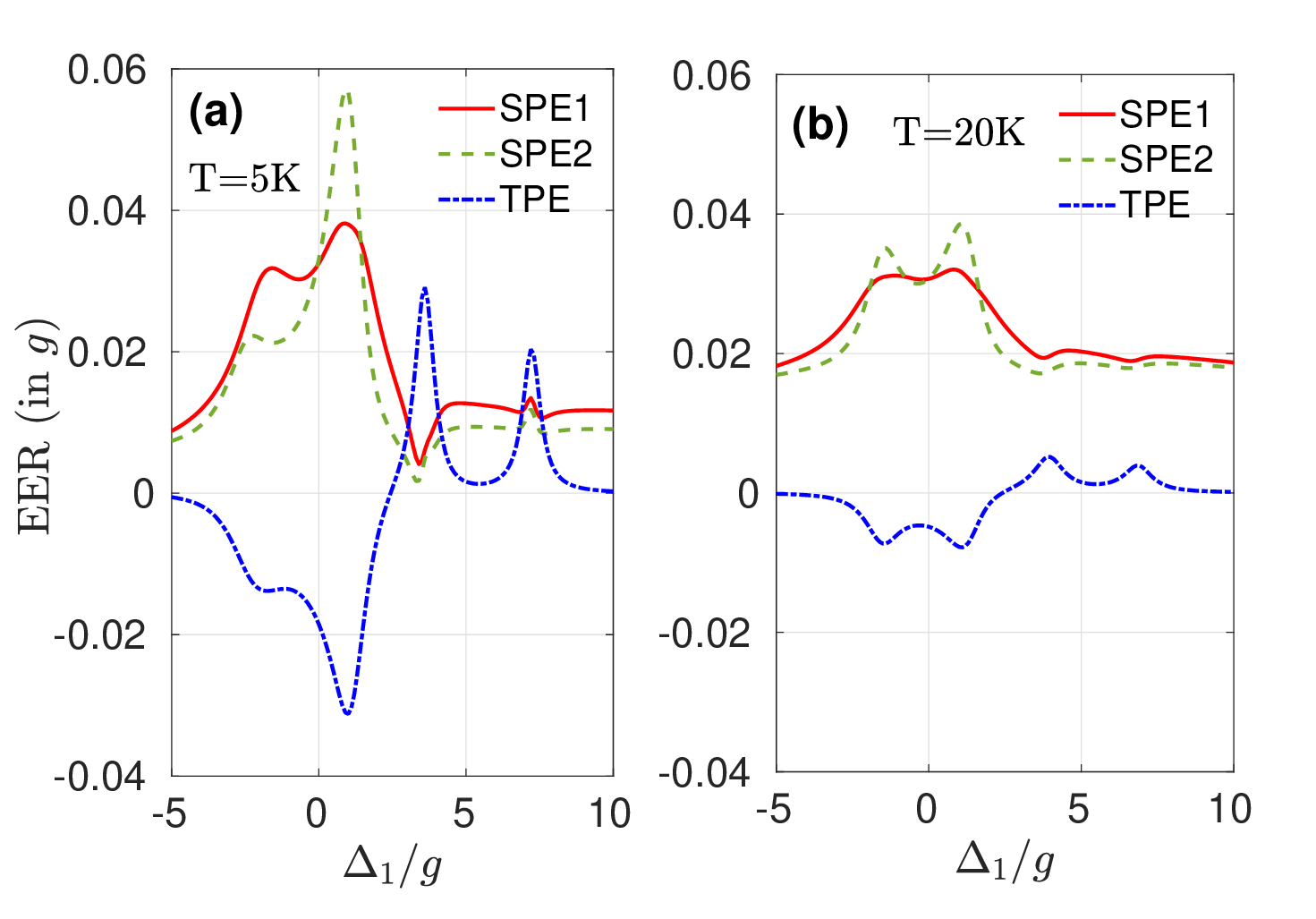}
    \caption{ Excess emission rates (EER). First mode single-photon excess emission rate(solid red), second mode single-photon excess emission rate(dashed green) and two-mode two-photon excess emission rate(dash dotted blue) for $\Omega_{1}=\Omega_{2}= 2.0g_{1}$, for a)T=5K and b) T=20K with all other parameters same as Fig.$\ref{fig:chap5/Fig7}$.}
    \label{fig:chap5/Fig10}
\end{figure}

After obtaining the rate equation for $\dot P_n$ Eq. \ref{inco_rate} for the coherent pumping case,  we plot the net excess emission rates of single and two-mode two-photon processes by varying the mode 1 detuning, $\Delta_1$, in Fig. \ref{fig:chap5/Fig10} for $T=5K, \, 20K$. 

We can see from Fig. \ref{fig:chap5/Fig10} (a) that the net excess emission rate (EER) of single photons into the cavity modes is dominant for $\Delta_1=\pm\Omega_1$, where mode 1 is resonant with $|g\rangle\leftrightarrow|y\rangle$ transition. This results in the decrease of $|y\rangle$ population and the population inversion, $\langle u|\rho_s|u\rangle-\langle y|\rho_s|y\rangle$ increases, leading to the single-photon emission into mode 2. At these detunings, $\Delta_1=\pm\Omega_1$, two-mode two-photon excess emission shows a dip in its value. Further, when mode 1 is tuned to two-photon resonance, $\Delta_1\approx-\Delta_2\pm\Omega_1$, the cavity mode is dominantly populated by the two-mode two-photon process and is clear from the peaks in net excess emission rate curves. The slight shift in the peaks is attributed to the stark effect. Finally, with an increase in temperature, the two-photon process is suppressed over the single-photon process as shown in Fig. \ref{fig:chap5/Fig10} (b) due to the increase in phonon-induced decoherence in the system. We conclude this section by noting that the two-mode two-photon laser is possible when the system is maintained at low temperatures.

\section{Generation of steady state two-mode entangled state}

In this section, we investigate the feasibility of the generation of steady-state continuous variable (CV) entanglement between the cavity modes. Using an external coherent drive, we consider the two-photon resonant transition $\vert g\rangle\rightarrow\vert x\rangle\rightarrow\vert u\rangle$. Therefore, the detuning between $|x\rangle$ exciton and pump, $\Delta_p=\left(\Delta_{xx}+\delta_{x}\right)/2$ is considered such that two-photon resonance, $\omega_u=2\omega_p$ condition is satisfied.

\begin{figure}[h]
   \includegraphics[scale=0.5]{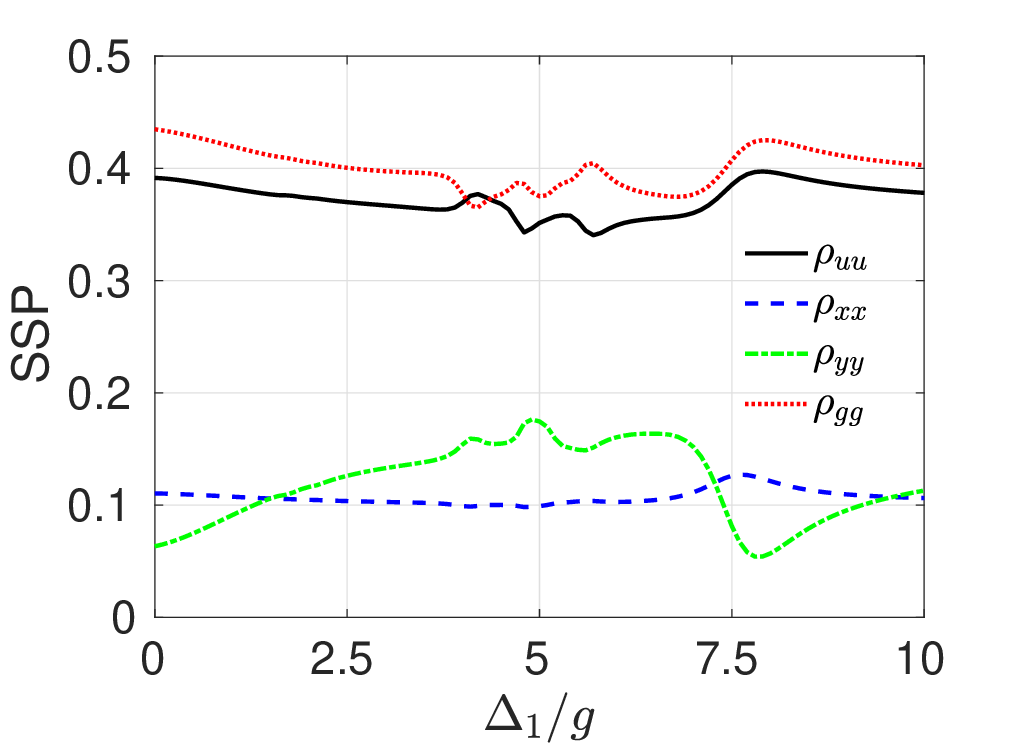}
    \caption{ Steady state populations (SSP) of QD states $|u\rangle$(solid black),$|x\rangle$(dashed blue),$|y\rangle$(dash dotted green), $|g\rangle$(dotted red) for phonon bath temperature T=5K as a function of $\Delta_{1}$ for same parameters as Fig. \ref{fig:chap5/Fig7} except $\Delta_{p}=7.0g$.}
    \label{fig:chap5/Fig11}
\end{figure}

Fig. \ref{fig:chap5/Fig11} shows the steady-state population of the QD states by varying the mode 1 detuning for $\Omega_1=\Omega_2=2g$. Both $|u\rangle$ and $|g\rangle$ states are largely populated compared to those of excitonic states, $|x\rangle, \, |y\rangle$. We also note that the ground state population is slightly higher than the biexciton population due to spontaneous decay. Further, when the mode 1 detuning, $\Delta_1$ is varied and tuned to two-mode two-photon resonance, $\Delta_1=-\Delta_2$, there is a transfer of population from the $|u\rangle\rightarrow|g\rangle$. This results in non-degenerate correlated emission of photons into the cavity modes. This correlated emission can suppress the phase or amplitude fluctuations.

\begin{figure}
   \includegraphics[scale=0.5]{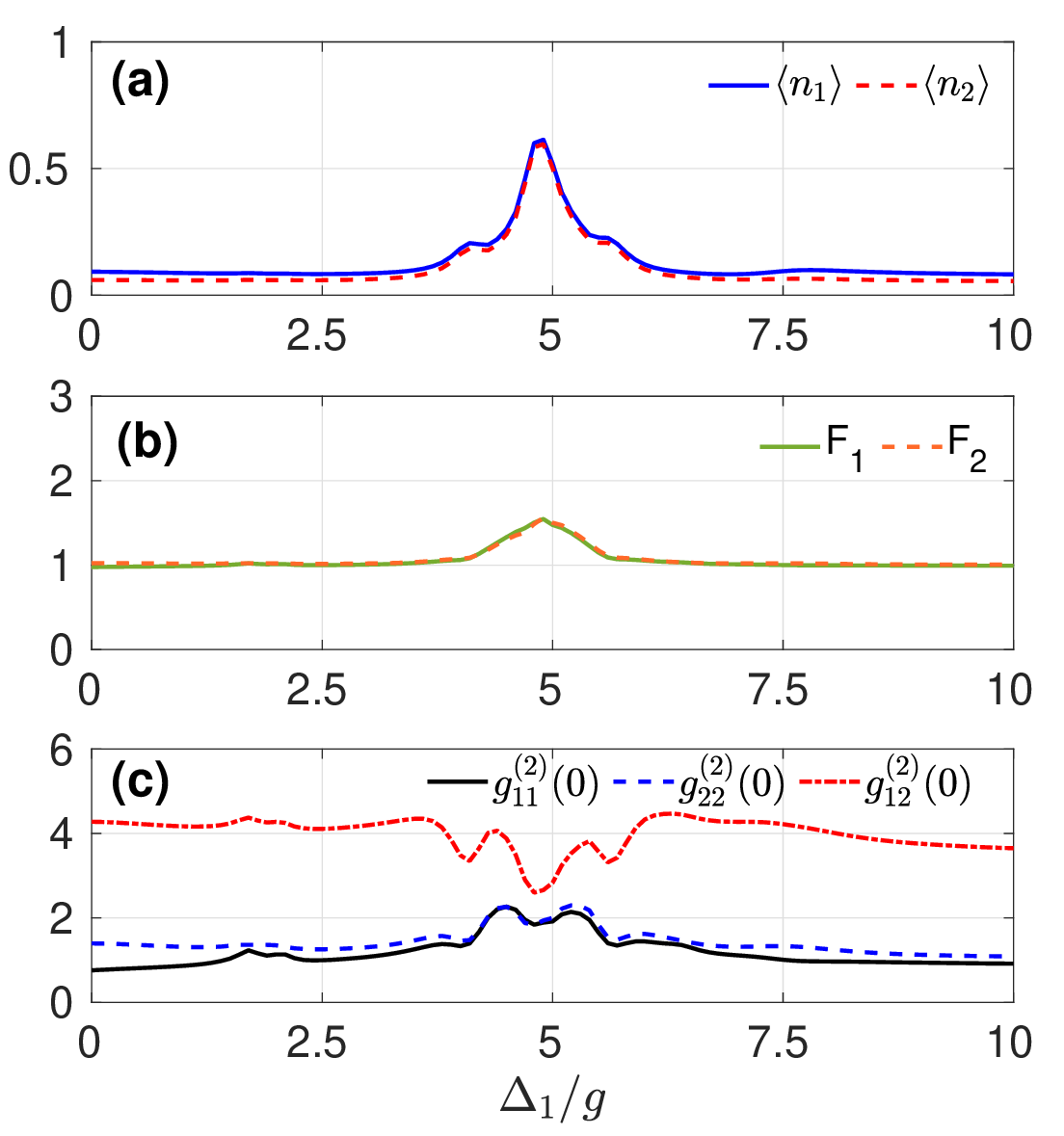}
    \caption{Cavity photon statistics: a) mean photon number $\langle n_{1}\rangle$(blue solid), $\langle n_{2}\rangle $(red long dash), b) Fano factors $F_{1}$(solid green), $F_{2}$(dashed orange), and c) Second order correlation $g_{11}^{(2)}(0)$(solid cyan), $g_{22}^{(2)}(0)$(dashed magenta), $g_{12}^{(2)}(0)$(dash dotted brown) as a function of $\Delta_{1}$ for same parameters as Fig. \ref{fig:chap5/Fig9} except $\Delta_{p}=7.0g$.}
    \label{fig:chap5/Fig12}
\end{figure}

Fig. \ref{fig:chap5/Fig12} shows the results for the cavity photon statistics. The mean photon number attains a maximum value when $\Delta_1=-\Delta_2$ for both the cavity modes, Fig. \ref{fig:chap5/Fig12} (a). Fano factor, $F_i$ values represent that the cavity modes are super-Poissonian in nature at the two-photon resonance. The second order inter and intra-mode correlation function results show that the intensity fluctuations are suppressed at the two-mode two-photon resonance condition, $\Delta_1=-\Delta_2$ and the cavity field is non-classical in nature with $g_{12}^{(2)}(0)>\sqrt{g_{11}^{(2)}(0)g_{22}^{(2)}(0)}$.

Next, examine the possibility of steady-state CV entanglement between the cavity modes by verifying the DGCZ criterion explained in Chapter 1, Sec.\ref{sec:Entang}. 

We define Einstein-Poldosky-Rosen (EPR) like variables for the cavity modes,
 
\begin{equation}
 u =x_{1}+x_{2},v=p_{1}-p_{2},\label{EPRvar}
\end{equation}
where $x_{k},p_{k}$ correspond to the k-th cavity mode, and in terms of cavity modes creation and annihilation operators,

\begin{align}
 x_{j} &=\frac{1}{\sqrt{2}}(a_{j}^{\dagger}e^{-i\phi_{j}}+a_{j}e^{i\phi_{j}}),\\
 p_{j} &=\frac{i}{\sqrt{2}}(a_{j}^{\dagger}e^{-i\phi_{j}}-a_{j}e^{i\phi_{j}}),j=\{1,2\}.
\end{align}

DGCZ criterion states that if,

\begin{equation}
\langle \Delta u^{2} + \Delta v^{2}\rangle = \langle (u-\langle u\rangle)^{2}+(v-\langle v\rangle)^{2} \rangle \geq 2.
\label{eqn:ECcriterion}
\end{equation}
is violated, then there is a presence of CV entanglement between the cavity modes.

\begin{figure}
   \includegraphics[scale=0.5]{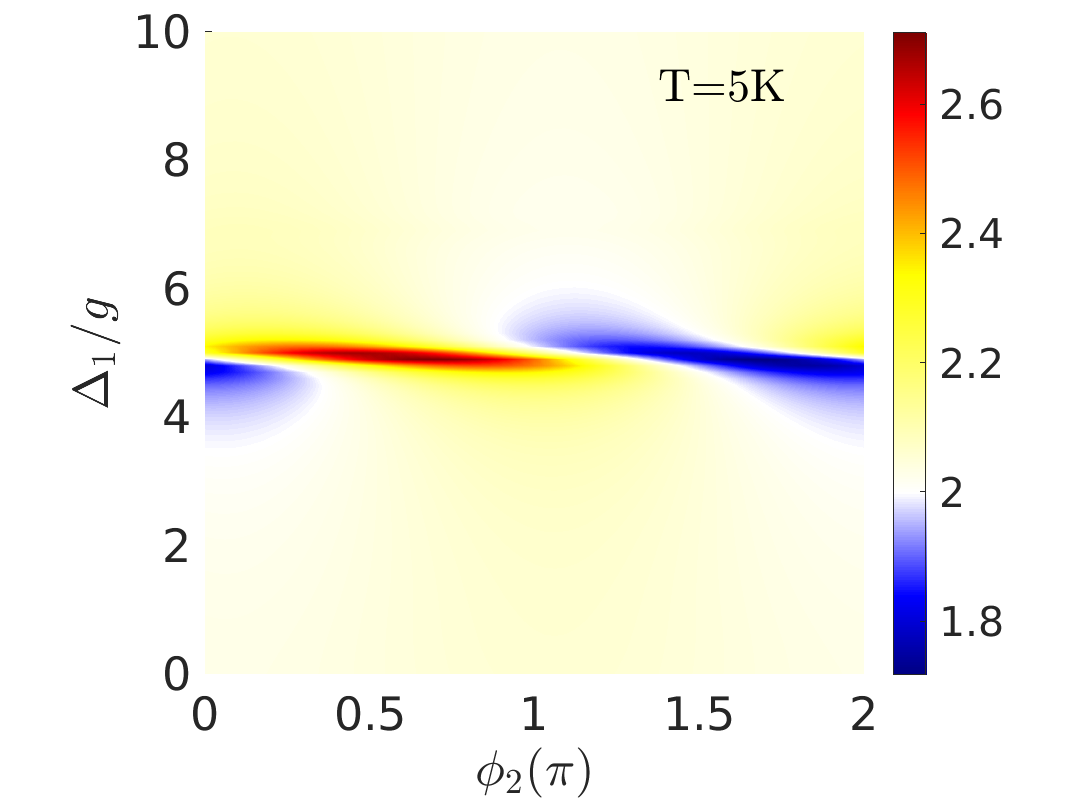}
    \caption{Entanglement criterion, the sum of the variance of EPR-like variable pair, $E.C.=\langle (\Delta u)^2+(\Delta v)^2\rangle$ is plotted by varying the first mode detuning $\Delta_{1}$ and $\phi_2$, and fixing $\phi_1=0$. The coherent pumping rates are $\Omega_1=\Omega_2=0.5g$, and all other parameters are the same as in Fig. \ref{fig:chap5/Fig11}.
   }
    \label{fig:chap5/Fig13}
\end{figure}

Figure \ref{fig:chap5/Fig13} presents the result of the entanglement criterion (E. C.) as a function of first cavity mode detuning, $\Delta_1$, and phase, $\phi_2$, for $\phi_1=0.0$ and coherent pumping rates, $\Omega_1=\Omega_2=0.5g$. It is clear that for $\Delta_1=5$, at two-photon resonance, the transition from biexciton to ground state leads to the generation of CV entanglement between the cavity modes. Over a wide range of $\phi_2$ values, the system is found to violate Eq. \ref{eqn:ECcriterion}, with the strongest violation occurring for $\phi_1+\phi_2=-1.0$.

\begin{figure}
   \includegraphics[scale=0.5]{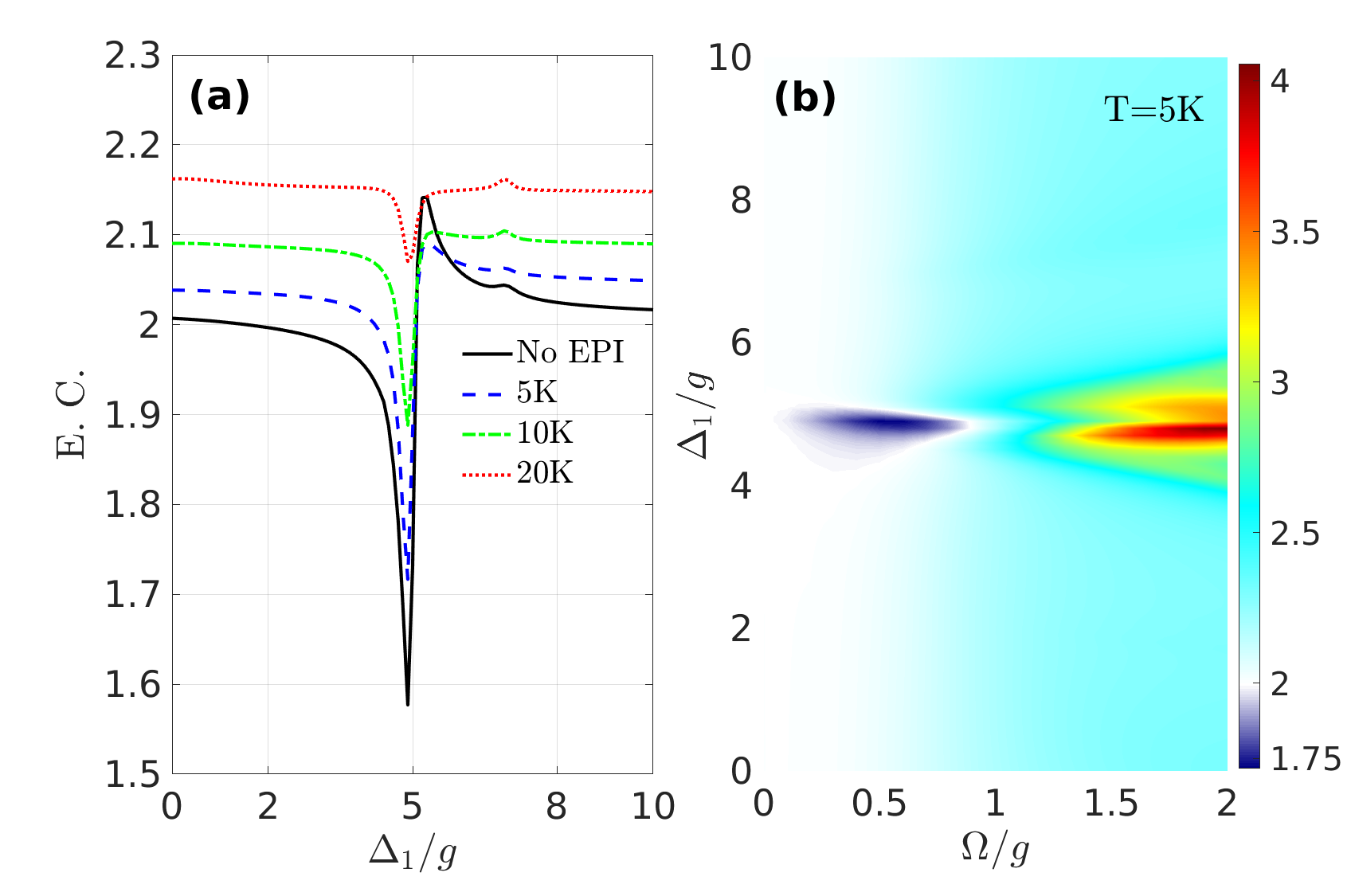}
    \caption{Entanglement criterion, the sum of the variance of EPR like variable pair, $E.C.=\langle (\Delta u)^2+(\Delta v)^2\rangle$ is plotted by varying the first mode detuning $\Delta_{1}$ for different bath temperatures, T.(a) with coherent pumping Rabi frequencies, $\Omega_{1}=\Omega_{2}=0.5g$ and (b) varying the coherent pumping rate $\Omega_{1}=\Omega_{2}=\Omega$ and the first mode detuning $\Delta_1$ for T=5K.  All other parameters are the same as Fig. \ref{fig:chap5/Fig11} with $\phi_{1} =-0.5,\,  \phi_{2} = -0.5$ 
   }
    \label{fig:chap5/Fig14}
\end{figure}

Fig. \ref{fig:chap5/Fig14} (a) shows the result of the entanglement criterion (E.C.), defined as the sum of the variances of EPR-like variables for different temperatures. At the two-photon resonance, $\Delta_1=-\Delta_2$, for coherent pumping rate, $\Omega_1=\Omega_2=\Omega=0.5g_1$, $\phi_1=-0.5$ and $\phi_2=-0.5$, there is a sharp dip in the mean value of $\Delta u^{2} + \Delta v^{2}$ and attains value $<2$ violating Eq. \ref{eqn:ECcriterion}. As the temperature increases, the entanglement is destroyed due to increased decoherence. Figure \ref{fig:chap5/Fig14} (b) further demonstrates that CV entanglement is preserved at low pumping rates, $\Omega\leq g$, and with an increase in pumping rate, the entanglement between the cavity modes is destroyed. 

\section{Conclusions}

In conclusion, we have shown the possibility of a non-degenerate two-mode two-photon laser in the semiconductor QD biexciton coupled to a bimodal cavity. The exciton-phonon effects are included using the polaron transformation technique and the reduced density matrix master equation using the Born-Markov approximation. Using Scully-Lamb theory of quantum laser, we have obtained the laser rate equations without mean field approximation and found the single and two-mode two-photon emission and absorption rates. For the case of incoherent pumping of the quantum dots, the results show that the two-mode two-photon emission is dominant over single photon processes at low pumping and low temperatures $T\sim 5K$. For the case of coherent pumping, the two-mode two-photon excess emission rate is large with a negligible single-photon emission rate, implying a non-degenerate two-photon laser. Further, when the quantum dot biexciton is coherently pumped at two-photon resonance, the system is able to generate continuous variable entanglement between the cavity modes. This non-degenerate two-photon laser can see applications in quantum computation, quantum information, quantum communication, and quantum sensing.

%% file: Conclusion.tex
\chapter{\label{conclusions} Conclusions and Perspectives}

In this thesis, we have investigated multi-photon lasing in various semiconductor QD-cavity QED systems. Such on-chip integrable and scalable quantum lasers can play a major role in this era of quantum technology with applications in quantum sensing, quantum computation, quantum communication, quantum metrology, etc. 

In Chapter 2, we proposed cooperative two-photon lasing in the system with two identical quantum dots (QDs) driven incoherently or coherently and coupled to a single-mode photonic crystal cavity. We have shown that at low pumping rates, the two-photon lasing is accompanied by hyperradiance (subradiance) in the incoherent (coherent) pumping case. We have incorporated exciton-phonon coupling using the polaron-transformed master equation and derived the master equation using the Born-Markov approximation. Further, we derive a simplified master equation (SME) to understand various phonon-induced coherent and incoherent processes. Using SME and Scully-Lamb quantum laser theory, we obtained the laser rate equations without mean-field approximation and evaluated the excess emission rates of the single and multi-photon processes contributing towards cavity mode population. Since Calic et al. \cite{Calic2017} deterministically coupled two quantum dots to a single-mode cavity, and with the inclusion of an external pump, the current results are experimentally realizable with the current state of technology. 

In Chapter 3, we investigated the phenomenon of two-mode Hyperradiant lasing in the system with incoherently pumped two QDs coupled strongly and off-resonantly to a bimodal photonic crystal cavity. We showed that in off-resonant coupling, exciton-phonon interactions (EPI) enhance the inter-mode correlations, leading to an enhancement in the cavity modes mean photon number. Using the radiance witness parameter and laser rate equations, we have shown that the system transits to the Hyperradiant regime due to enhanced cooperative two-mode two-photon emission. Further, we have compared the results for the cases with and without EPI, single cavity mode, weakly coupled second cavity mode by varying the incoherent pumping, cavity decay rate, and temperature. The results showed that the enhanced cavity mode emission resulted in the suppression of the linewidth of the emission spectrum as in the superradiant lasers. By embedding two QDs inside coupled micropillar cavities—similar to the experimental setup of Dousse et al.,\cite{Dousse2010ultrabright} and employing external incoherent pumping, it is possible to realize two-mode hyperradiant lasing with the parameters considered in this work.

In Chapter 4, we explored the phenomenon of correlated emission lasing (CEL) in a coherently driven semiconductor single quantum dot coupled to a bimodal photonic crystal cavity. We showed that the transfer of coherence from the QD states to the coupled cavity modes leads to the suppression of the associated quantum noise. Using the polaron transformed master equation, we analyzed the variances of the Hermitian operators corresponding to the relative and average phase of the cavity modes. Further, we extended our analysis by deriving the Fokker-Planck equation in Glauber-Sudarshan $P$ representation and evaluated the associated drift ($D_\phi,\, D_\Phi$) and diffusion coefficients ($D_{\phi\phi},\, D_{\Phi\Phi}$). To realize CEL, the mode-locking condition ($D_\phi=0$ or $D_\Phi=0$) must be satisfied, together with diffusion coefficients fulfilling $D_{\phi\phi}\leq 0$ ($D_{\Phi\Phi}\leq 0$), which indicates quenching or squeezing of the relative (average) phase noise. Such a correlated emission laser in a semiconductor QD–cavity QED system can be experimentally implemented by coupling a single quantum dot to an elliptical micropillar cavity \cite{Reitzenstein2010, Mehdi2024} or an H1 photonic crystal cavity \cite{Painter1999, Takagi2012}, while coherently driving the quantum dot excitons ($|x\rangle, |y\rangle$). Additionally, we investigated the feasibility of generating continuous variable (CV) entanglement between the cavity modes and provided the details of an optimal parameter regime where one can look for.

In Chapter 5, we analyzed the non-degenerate two-photon lasing in a system with a single QD biexciton coupled to a non-degenerate bimodal cavity and driven incoherently or coherently. We showed that when the cavity modes are tuned to a two-photon resonant condition, the population transfer from the biexciton to the ground state via $y$-exciton is accompanied by peaks in the mean cavity modes photon number. Using laser rate equations without mean-field approximations, we showed that the cavity modes are indeed dominantly populated by a two-mode two-photon emission process and the system behaves like a non-degenerate two-photon laser. Further, by considering QD biexciton driven at two-photon resonance with a coherent pump, we explored the creation of CV entanglement between the cavity modes and showed that the entanglement is preserved at low pumping rates $\Omega\leq g$ and temperatures ($T\leq 10K$). Embedding a QD inside a bimodal micropillar cavity with non-degenerate modes \cite{Reitzenstein2010} provides a feasible route to realizing a non-degenerate two-photon laser and also CV entanglement generation.

\section{Future outlook}

In the present study, we employed the polaron transformation approach, which provides a more accurate description than the weak coupling theory but is valid only when the cavity coupling strengths or coherent pumping strengths remain below the phonon bath cutoff frequency. To extend this regime, one may adopt the variational polaron transformation theory, where the QD–phonon coupling strength is treated as a free parameter. This method broadens the range of accessible cavity and pump strengths while yielding more accurate predictions of the system dynamics. Moreover, the inclusion of non-Markovian dynamics could provide deeper insights into the various phenomena explored here.

Future investigations could also explore the modification of cooperative effects (discussed in Chapter 1) when non-identical quantum dots are coupled to cavity modes, and determine the optimum conditions for realizing two-photon lasing. Similarly, following the discussion in Chapter 4, one could examine the quenching or squeezing of phase fluctuations in QD cascaded configurations, where biexciton decay is coupled to single or bimodal cavities.

These non-classical light sources can enable a wide range of applications in quantum computing, secure communications, high-precision spectroscopy, non-linear optics, and biophotonics. With the advent of advanced fabrication and tuning techniques, semiconductor QD–cavity QED systems can now be integrated into photonic circuits, enabling scalable on-chip quantum technologies \cite{Osada2019,Zhu2025}. In particular, cooperative two-photon lasing accompanied by hyperradiance, which leads to suppressed cavity field fluctuations and narrowed emission linewidth, has potential applications in atomic clocks \cite{Meiser2009} and quantum metrology \cite{Giovannetti2011}. On-chip correlated emission lasers—generating degenerate or non-degenerate correlated photon pairs with phase noise reduced to the vacuum noise limit—offer exciting opportunities for gravitational wave detection \cite{Scully1986, Schnabel2010}, quantum sensing \cite{Degen2017}, quantum key distribution \cite{Margarida2020}, and quantum networks.

Furthermore, continuous-variable (CV) entanglement holds key advantages over discrete-variable entanglement, especially in supporting deterministic quantum operations. Generating CV entanglement between non-degenerate modes allows frequency conversion and information transfer across different spectral regions \cite{Villar2005}, and quantum teleportation \cite{Furusawa1998}

Finally, the biomedical applications of the non-classical two-photon lasers studied in this work are noteworthy. These include nonlinear spectroscopy, which reveals vibrational molecular states hidden from linear absorption \cite{Janina2006}, and two-photon microscopy \cite{Winfried1990, Helmchen2005}, which provides reduced scattering, deep-tissue imaging, and minimal photodamage. In particular, non-degenerate two-photon lasers can simultaneously excite multiple fluorophores, improving spatial resolution and penetration depth compared to degenerate two-photon lasers \cite{Sadegh2019}.
 

%% file: Publications.tex
\chapter*{List of publications and presentations}

\section*{Journal publications}	
\begin{enumerate}[label=(\roman*), noitemsep]

   \item Cooperative two-photon lasing in two quantum dots embedded inside a photonic microcavity
    \item[] \textbf{Lavakumar Addepalli} and P. K. Pathak; \textit{Phys. Rev. B} \textbf{110}, 085408 (2024).

    \item Correlated emission lasing in a single quantum dot embedded inside a bimodal photonic crystal cavity
    \item[] \textbf{Lavakumar Addepalli} and P. K. Pathak; \textit{Phys. Rev. B} 111, 125422 (2025).
    
   \item Nondegenerate two-photon lasing in a single quantum dot
    \item[] S. K. Hazra, \textbf{Lavakumar Addepalli}, P. K. Pathak, and T. N. Dey \textit{Phys. Rev. B} \textbf{109}, 155428 (2024).

    \item Hyperradiant lasing in the incoherently driven two quantum
    dots coupled to a single mode cavity
    \item[] \textbf{Lavakumar Addepalli} and P. K. Pathak; \textit{The European Physical Journal Special Topics}, 1-4, Jul. 2025.

    \item Two-mode hyperradiant lasing in a system of two quantum dots embedded in a bimodal photonic crystal cavity
    \item[] \textbf{Lavakumar Addepalli}, P. K. Pathak, arXiv:2506.21202 [quant-ph] (Under review).
   
\end{enumerate}

\section*{Presentations in conferences and workshops}
\begin{enumerate}[label=(\roman*)] 

    \item Oral presentation: \textit{``Lasing in Incoherently pumped Quantum dot- Cavity system"} at Student Conference on Photonics \textbf{(SCOP)}, 2023, PRL Ahmedabad, India.
    \vspace{0.5cm}
    \item Oral presentation: \textit{``Correlated emission lasing in a single quantum dot embedded inside a bimodal photonic crystal cavity"} at International Topical meeting on Classical optics and Quantum optics \textbf{(INTOCQ)}, 2024, IIST Thiruvanthapuram, India.
    \vspace{0.5cm}
    \item Poster presentation: \textit{``Exciton-Phonon Effects in the Coherently Driven Two Quantum Dots-Photonic Microcavity System Showing Cooperative Two-photon Lasing"} at Frontiers of Quantum and Mesoscopic Thermodynamics \textbf{(FQMT)}, 21-27 July, 2024, Prague, Czech Republic.
    \vspace{0.5cm}
    \item Poster presentation: \textit{``Two photon lasing: Coherently and incoherently pumped two quantum dots-photonic crystal cavity system} at Conference on Optics Photonics and Quantum optics \textbf{(COPaQ)}, 2022, IIT Roorkee, India, Nov 10-13, 2022.
    \vspace{0.5cm}
    \item Attended online \textit{International Conference on Quantum Information and Quantum Technology"} \textbf{(QIQT)} 2023, hosted by the Indian Institute of Science Education and Research Kolkata.
	 	
\end{enumerate}

%% file: app_chap2.tex
\chapter{}\label{app:chap2_Appendix}

\section{Comparison of some results obtained from ME, SME $\&$ EQ.  \ref{eqn:meanphotonNumber}}\label{sec:chap2_Appendix1}

\begin{figure}
    \centering
    \includegraphics[width=\columnwidth]{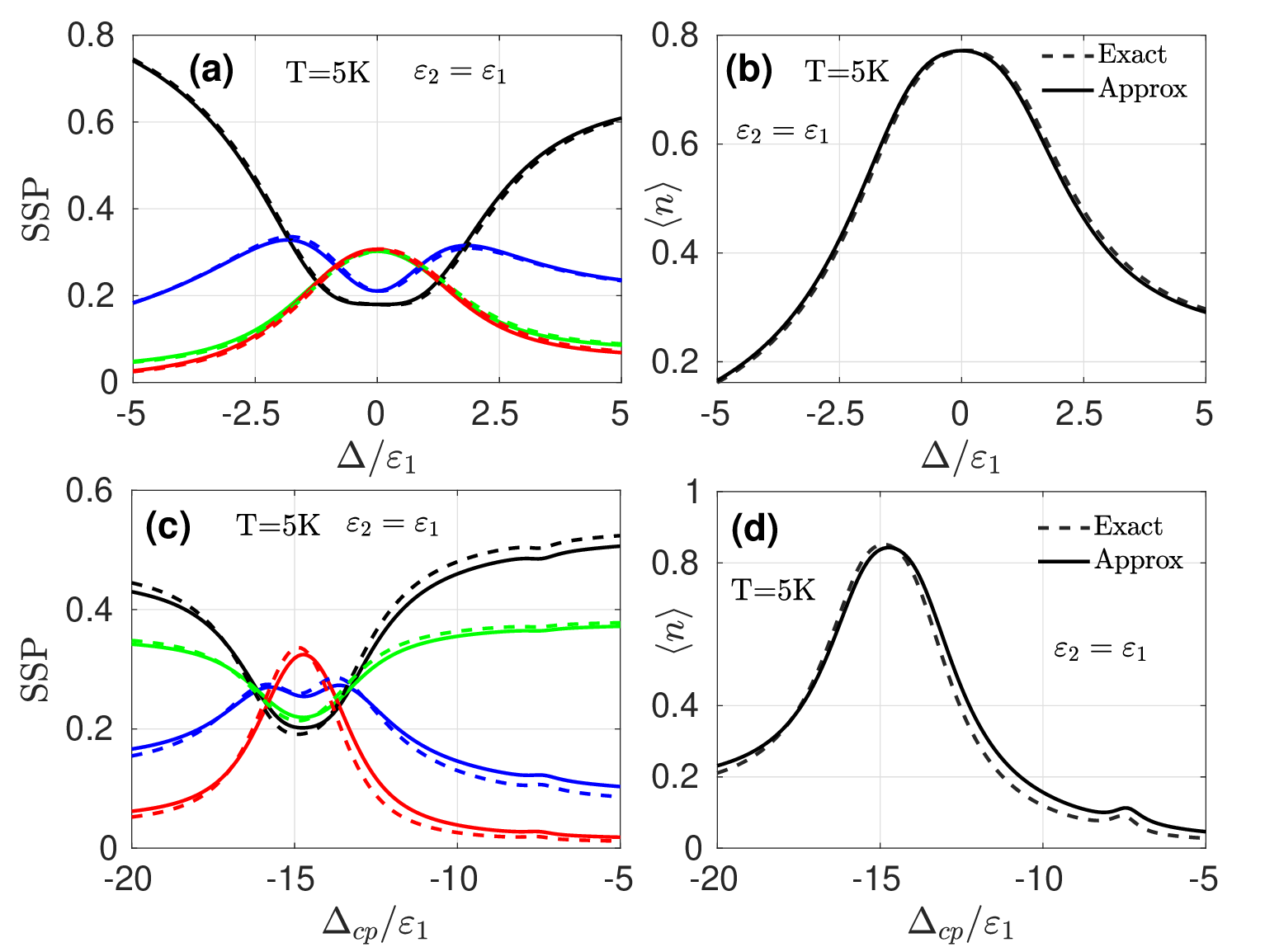}
    \caption{Comparison of steady-state populations of collective QD states calculated using exact master equation (dashed) and simplified master equation (solid). In (a)$\&$(b) parameters are same as in Fig. \ref{fig:chap2/Fig2}(c) and in (c)$\&$(d) parameters are same as in Fig. \ref{fig:chap2/Fig6}(a). Colour scheme is same as Fig. \ref{fig:chap2/Fig2}.}
    \label{fig:chap2/Fig12}
\end{figure}

\begin{figure}
    \centering
    \includegraphics[width=\columnwidth]{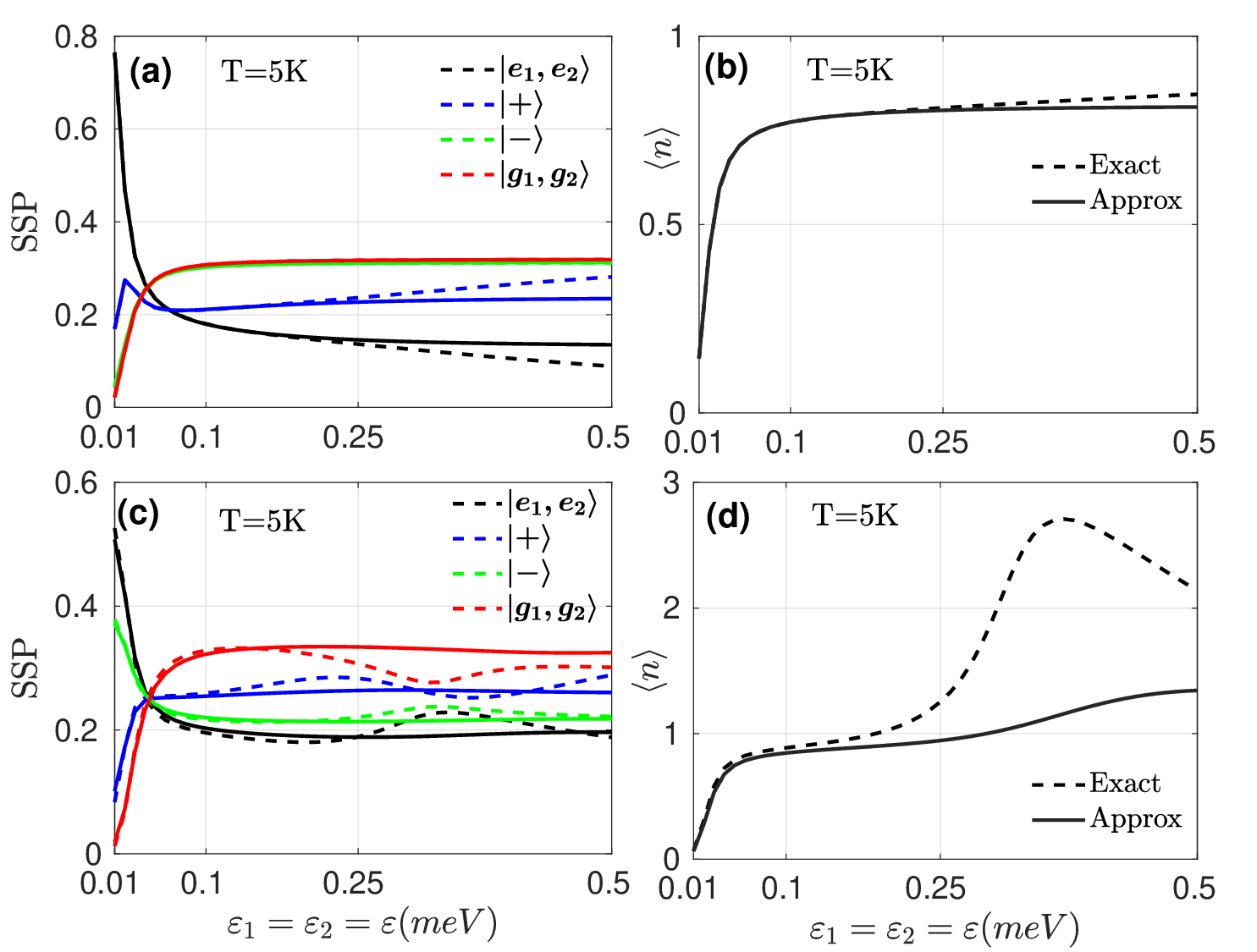}
    \caption{Comparison of steady-state populations (SSP) of collective QD states and mean cavity photon number $\langle n\rangle$ calculated using exact master equation (dashed) and simplified master equation (solid) for incoherent case in (a) $\&$ (b), coherent case in (c) $\&$ (d) by varying cavity coupling strength ($\varepsilon_1=\varepsilon_2=\varepsilon$). The parameters are for (a) $\&$ (b), $\Delta=0.0meV$ and others are same as in Fig. \ref{fig:chap2/Fig2} (c). For (c) $\&$ (d), $\Delta_{cp}=-1.48 meV$ and others are same as in Fig. \ref{fig:chap2/Fig6}(a). Coloring scheme same as in Fig. \ref{fig:chap2/Fig2}.}
    \label{fig:chap2/Fig13}
\end{figure}

\begin{figure}
    \centering
    \includegraphics[width=\columnwidth]{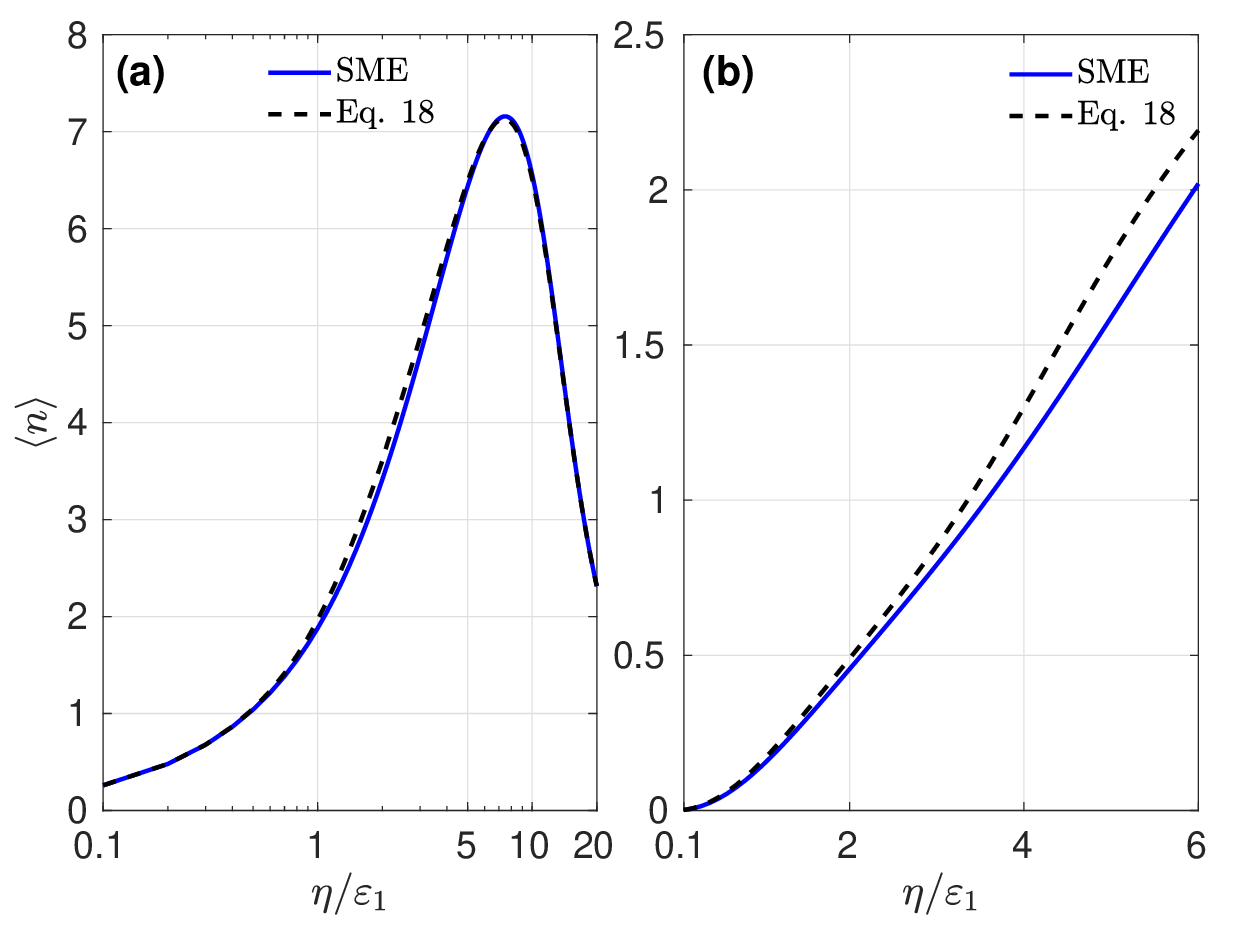}
    \caption{Comparison of mean photon number obtained from simplified master equation (solid blue) and the mean photon number obtained from laser rate equation \ref{eqn:meanphotonNumber} (dashed black) for a) incoherent pump and b) coherent pump, with variation in pump strength. The other parameters are same as in Fig. \ref{fig:chap2/Fig5}(a) for a) and Fig. \ref{fig:chap2/Fig9}(a).}
    \label{fig:chap2/Fig14}
\end{figure}

In Fig. \ref{fig:chap2/Fig12} $\&$ Fig. \ref{fig:chap2/Fig13}, we compare the results of steady-state populations and mean photon number obtained from the exact master equation (\ref{eqn:incohME}) and the simplified master equations, (\ref{eqn:incohSME}),(\ref{eqn:cohSME}) for both incoherent and coherent pumping cases using same parameter considered in Fig. \ref{fig:chap2/Fig2} and Fig.\ref{fig:chap2/Fig5} respectively. By comparing the results of steady-state populations in collective QD states, $\ket{e_1,e_2}, \ket{+}, \ket{-}, \ket{g_1,g_2}$, and mean photon number in cavity mode it confirms the approximations made to obtain simplified master equations are very good to work in the range of parameters considered. For QD-cavity coupling strengths, $\varepsilon\leq0.2meV$, the SME is in good agreement with ME for both incoherent and coherent pumping cases and the approximations made are valid.
 
We have also compared the mean photon number calculated from the simplified master equation for the incoherent pumping and the coherent pumping with the mean photon number equation (\ref{eqn:meanphotonNumber}) derived from the laser rate equation in Fig. \ref{fig:chap2/Fig14}. We can see the results are converging for the incoherent pump case exactly and are in good agreement for the coherent pump case. Here, we have considered terms up to $k=4$ in Eq. \ref{eqn:meanphotonNumber}. The results converge precisely if higher contributions are considered for coherent pump rates.

%% file: app_chap3.tex
\chapter{}\label{app:chap3_Appendix}

\section{Simplified master equation}\label{sec:chap3_Appendix1} 

The simplified master equation for the system is given below after making the approximations, $\Delta_1,\Delta_2>>g_i$, and is used to calculate the density matrix element rate equations, thereby calculating single and multi-photon emission and absorption rates.

\begin{equation}
    \begin{split}
        \dot{\rho_s}=& -\frac{i}{\hbar}[H_{eff},\rho_s]-\sum_{i=1}^2\Big(\frac{\kappa_i}{2}L[a_i] -\frac{\gamma_i}{2}L[\sigma_i^-]-\frac{\eta_i}{2}L[\sigma_i^+]-\frac{\gamma_i'}{2}L[\sigma_i^+\sigma_i^-]\Big)\rho_s\\&-\sum_{i,j,k,l=1,i\neq j}^2 \frac{\Gamma_{kl}^{--}}{2}(a_l^\dagger \sigma_j^- a_k^\dagger\sigma_i^-\rho_s-2a_k^\dagger\sigma_i^-\rho_s a_l^\dagger\sigma_j^-+\rho_s a_l^\dagger \sigma_j^- a_k^\dagger \sigma_i^-)\\&+\frac{\Gamma_{kl}^{++}}{2}(\sigma_j^+ a_l \sigma_i^+a_k\rho_s - 2\sigma_i^+a_k\rho_s\sigma_j^+a_l+\rho_s\sigma_j^+a_l\sigma_i^+a_k) \\& -\sum_{i,j,k,l=1}^2 \frac{\Gamma_{kl}^-}{2}(a_l^\dagger\sigma_j^-\sigma_i^+a_k\rho_s - 2\sigma_i^+a_k\rho_s a_l^\dagger\sigma_j^- + \rho_s a_l^\dagger \sigma_j^-\sigma_i^+ a_k)\\&+\frac{\Gamma_{kl}^+}{2}(\sigma_j^+ a_l a_k^\dagger\sigma_i^-\rho_s - 2 a_k^\dagger\sigma_i^-\rho_s\sigma_j^+ a_l + \rho_s\sigma_j^+ a_l a_k^\dagger \sigma_i^-)
    \label{eqn:SMEtwoModeHyper}
    \end{split}
\end{equation}

Where the effective Hamiltonian is,

\begin{equation}
    H_{eff}=H_s-i\hbar\sum_{i,j,k,l=1}^2 \Omega_{kl}^- a_l^\dagger \sigma_j^-\sigma_i^+a_k + \Omega_{kl}^+\sigma_j^+a_l a_k^\dagger\sigma_i^- + \sum_{i,j,k,l=1, i\neq j}^2 i\hbar \Omega_{kl}^{--}a_l^\dagger\sigma_j^-a_k^\dagger\sigma_i^- + H.C.
\end{equation}

The phonon-induced scattering rates are given by,

\begin{equation}
    \Omega_{kl}^{\pm}=\frac{g_k g_l}{2}\int_0^\infty d\tau (G_+e^{\pm i\Delta_k\tau}-G_-^* e^{\mp i\Delta_l\tau})
\end{equation}

\begin{equation}
    \Omega_{kl}^{--}=\frac{g_k g_l}{2}\int_0^\infty d\tau(G_- e^{i\Delta_k\tau}-G_-^* e^{i\Delta_l\tau})
\end{equation}

\begin{equation}
    \Gamma_{kl}^{\pm}=g_k g_l\int_0^\infty d\tau(G_+ e^{\pm i\Delta_k\tau} + G_+^* e^{\mp i\Delta_l\tau})
\end{equation}

\begin{equation}
    \Gamma_{kl}^{--/++}= g_k g_l\int_0^\infty d\tau (G_- e^{\pm i\Delta_k\tau}+G_-^* e^{\pm i\Delta_l\tau})
\end{equation}

%% file: app_chap4.tex
\chapter{}\label{app:chap4_Appendix}

\section{PHONON INDUCED SCATTERING RATES}\label{sec:chap4_Appendix1}

The phonon induced scattering rates are given below,

\begin{equation}
    \delta_{i}^{\pm} = g_i^2 Im\Big[\int_0^\infty d\tau G_+e^{\pm i\Delta_i\tau}\Big]
\end{equation}

\begin{equation}
    \delta_{ip}^{\pm} = \eta_i^2 Im\Big[\int_0^\infty d\tau G_+e^{\pm i\Delta_{c_ip}\tau}\Big]
\end{equation}

\begin{equation}
    \Omega_{ij} = \frac{g_ig_j}{2}\int_0^\infty d\tau(G_+e^{ i\Delta_j\tau}-G_+^*e^{-i\Delta_i\tau})
\end{equation}

\begin{equation}
    \Omega_{ij}^p = \frac{\eta_i\eta_j}{2}\int_0^\infty d\tau(G_+e^{ i\Delta_{jp}\tau}-G_+^*e^{-i\Delta_{ip}\tau})
\end{equation}

\begin{equation}
    \Gamma_i^{\pm}=g_i^2\int_0^\infty d\tau (G_+ e^{\pm i\Delta_i\tau}+G_+^* e^{\mp i\Delta_i\tau)}
\end{equation}

\begin{equation}
    \Gamma_{ip}^{\pm}=\eta_i^2\int_0^\infty d\tau (G_+ e^{\pm i\Delta_{c_ip}\tau}+G_+^* e^{\mp i\Delta_{c_ip}\tau})
\end{equation}

\begin{equation}
    \Gamma_{ij} = g_ig_j\int_0^\infty d\tau(G_+e^{ i\Delta_i\tau}+G_+^*e^{-i\Delta_j\tau})
\end{equation}

\begin{equation}
    \Gamma_{ij}^p = \eta_i\eta_j\int_0^\infty d\tau(G_+e^{ i\Delta_{ip}\tau}+G_+^*e^{-i\Delta_{jp}\tau})
\end{equation}

\begin{equation}
    \Lambda_i^\pm = g_i^2\int_0^\infty d\tau (G_- + G_-^*)e^{\mp i\Delta_i\tau})
\end{equation}

\begin{equation}
    \Lambda_{ij}^{\pm\pm} = g_ig_j\int_0^\infty d\tau (G_-e^{\mp i\Delta_i\tau} + G_-^*e^{\mp i\Delta_j\tau})
\end{equation}

\begin{equation}
    \Lambda_{ij}^{+-} = g_ig_j\int_0^\infty d\tau (G_- e^{-i\Delta_i\tau}+G_-^* e^{i\Delta_j\tau})
\end{equation}

\begin{equation}
    \Lambda_{ip}^{\pm} = \eta_i^2\int_0^\infty d\tau (G_- + G_-^*)e^{\mp i\Delta_{ip}\tau}
\end{equation}

\begin{equation}
    \Lambda_{ijp}^{\pm\pm} = \eta_i\eta_j\int_0^\infty (G_- e^{\mp i\Delta_{ip}\tau} + G_-^* e^{\mp i\Delta_{jp}\tau})
\end{equation}

\begin{equation}
    \Lambda_{ijp}^{+-} = \eta_i\eta_j\int_0^\infty (G_-e^{-i\Delta_{ip}\tau} + G_-^* e^{i\Delta_{jp}\tau})
\end{equation}

\section{DERIVATION OF DRIFT AND DIFFUSION COEFFICIENTS}\label{sec:chap4_Appendix2}

For the without phonons case, we obtain the density matrix elements, $\rho_{ij} = \langle i|\rho_s|j\rangle$ upto zeroth  in coupling strength, $g_i$ by solving the matrix rate equation given by, $\dot{R^{(0)}}=-MR^{(0)}+X$,  Here, $R=\begin{bmatrix}\rho_{xg} & \rho_{yg} & \rho_{gx} & \rho_{gy} & \rho_{xx} & \rho_{yy} & \rho_{xy} & \rho_{yx}\end{bmatrix}^T$ , \\
$X=\begin{bmatrix}i\eta_1 & i\eta_1 & -i\eta_2 & -i\eta_2 & 0 & 0 & 0 & 0\end{bmatrix}^T$

$M=\scriptsize\begin{bmatrix}
-i\Delta_{xp}-\frac{\gamma_1}{2} & 0 & 0 & 0 & 2i\eta_1 & i\eta_1 & i\eta_2 & 0\\
0 & -i\Delta_{yp}-\frac{\gamma_2}{2} & 0 & 0 & i\eta_2 & 2i\eta_2 & 0 & i\eta_1\\
0 & 0 & i\Delta_{xp}-\frac{\gamma_1}{2} & 0 & -2i\eta_1 & -i\eta_1 & 0 & -i\eta_2\\
0 & 0 & 0 & i\Delta_{yp}-\frac{\gamma_1}{2} & -i\eta_2 & -2i\eta_2 & -i\eta_1 & 0\\
i\eta_1 & 0 & -i\eta_1 & 0 & -\gamma_1 & 0 & 0 & 0\\
0 & i\eta_2 & 0 & -i\eta_2 & 0 & -\gamma_2 & 0 & 0\\
i\eta_2 & 0 & 0 & -i\eta_1 & 0 & 0 & -i(\Delta_{xp}-\Delta_{yp})-\frac{\gamma_1+\gamma_2}{2} & 0\\
0 & i\eta_1 & -i\eta_2 & 0 & 0 & 0 & 0 & i(\Delta_{xp}-\Delta_{yp})-\frac{\gamma_1+\gamma_2}{2}
\end{bmatrix}$

and 1st order in $g_i$ are obtained by integrating the rate equation, $\dot{R}=-MR-ig_1A_1e^{i\Delta_{c_1p}t}-ig_2A_2e^{i\Delta_{c_2p}t}-ig_1A_3e^{-i\Delta_{c_1p}t}-ig_2A_4e^{-i\Delta_{c_2p}t}$.

\begin{align}
    A1 = \begin{bmatrix} ig_1\rho_{xx}^{(0)}a_1-ig_1a_1\rho_{gg}^{(0)} \\ ig_1\rho_{yx}^{(0)}a_1 \\ 0 \\ 0 \\ -ig_1a_1\rho_{gx}^{(0)} \\ 0 \\ -ig_1a_1\rho_{gy}^{(0)}\\ 0 \end{bmatrix}; 
    A2 = \begin{bmatrix} ig_2\rho_{xy}^{(0)}a_2 \\ ig_2\rho_{yy}^{(0)}a_2 - ig_2a_2\rho_{gg}^{(0)} \\ 0 \\ 0 \\ 0 \\ -ig_2a_2\rho_{gy}^{(0)} \\ 0 \\ -ig_2a_2\rho_{gx}^{(0)} \end{bmatrix} ;
\end{align}

\begin{align}
    A3 = \begin{bmatrix} 0 \\ 0 \\ -ig_1a_1^\dagger \rho_{xx}^{(0)} + ig_1\rho_{gg}^{(0)}a_1^\dagger \\ -ig_1a_1^\dagger \rho_{xy}^{(0)} \\ ig_1\rho_{xg}^{(0)}a_1^\dagger \\ 0 \\ 0 \\ ig_1\rho_{yg}^{(0)}a_1^\dagger\end{bmatrix};
    A4 = \begin{bmatrix} 0 \\ 0 \\ -ig_2a_2^\dagger\rho_{yx} \\ -ig_2a_2^\dagger\rho_{yy}^{(0)} + ig_2\rho_{gg}^{(0)}a_2^\dagger \\ 0 \\ ig_2\rho_{yg}^{(0)}a_2^\dagger \\ ig_2\rho_{xg}^{(0)}a_2^\dagger \\ 0 \end{bmatrix}
\end{align}

\begin{equation}
    \begin{split}
        R^{(1)}(t) = & (M-i\Delta_{c_1p})^{-1}A1 + (M-i\Delta_{c_2p})^{-1}A2 +(M+i\Delta_{c_1p})^{-1}A3 + (M+i\Delta_{c_2p})^{-1}A4
    \end{split}
\end{equation}

After substituting the matrix elements, $\rho_{ij}^{(1)}$ into the Eq. (7), the reduced density matrix for the cavity field takes the following form,

\begin{equation}
    \begin{split}
         \dot{\rho_f} = & -i\Delta_{c_1p}a_1^\dagger a_1\rho_f + i\Delta_{c_1p}\rho_f a_1^\dagger a_1 -i\Delta_{c_2p}a_2^\dagger a_2\rho_f +i\Delta_{c_2p}\rho_f a_2^\dagger a_2 \\&+ \Big[\alpha_{11}\rho_f a_1a_1^\dagger - \alpha_{11}a_1^\dagger \rho_f a_1 + \alpha_{12}\rho_f a_2a_1^\dagger - \alpha_{12}a_1^\dagger \rho_f a_2 \\&+ \alpha_{13}\rho_f a_1^\dagger a_1^\dagger - \alpha_{13}a_1^\dagger \rho_f a_1^\dagger + \alpha_{14}\rho_f a_2^\dagger a_1^\dagger - \alpha_{14}a_1^\dagger \rho_f a_2^\dagger \\& + \alpha_{15}a_1 \rho_f a_1^\dagger - \alpha_{15} a_1^\dagger a_1\rho_f + \alpha_{16}a_2 \rho_f a_1^\dagger - \alpha_{16} a_1^\dagger a_2\rho_f \\& + \alpha_{17}a_1^\dagger \rho_f a_1^\dagger - \alpha_{17} a_1^\dagger a_1^\dagger\rho_f + \alpha_{18}a_2^\dagger \rho_f a_1^\dagger - \alpha_{18} a_1^\dagger a_2^\dagger\rho_f
         \\&+ \nu_{21}\rho_f a_1a_2^\dagger - \nu_{21}a_2^\dagger \rho_f a_1 + \nu_{22}\rho_f a_2a_2^\dagger - \nu_{22}a_2^\dagger \rho_f a_2 \\&+ \nu_{23}\rho_f a_1^\dagger a_2^\dagger - \nu_{23}a_2^\dagger \rho_f a_1^\dagger + \nu_{24}\rho_f a_2^\dagger a_2^\dagger - \nu_{24}a_2^\dagger \rho_f a_2^\dagger \\& + \nu_{25}a_1 \rho_f a_2^\dagger - \nu_{25} a_2^\dagger a_1\rho_f + \nu_{26}a_2 \rho_f a_2^\dagger - \nu_{26} a_2^\dagger a_2\rho_f \\& + \nu_{27}a_1^\dagger \rho_f a_2^\dagger - \nu_{27} a_2^\dagger a_1^\dagger\rho_f + \nu_{28}a_2^\dagger \rho_f a_2^\dagger - \nu_{28} a_2^\dagger a_2^\dagger\rho_f + H.C.\Big]
    \end{split}
\end{equation}

the coefficients, $\alpha_{ij},\nu_{ij}$ are obtained numerically after the substitution of $\rho_{ij}^{(1)}$ . The Fokker-Planck equation in the P-representation is given below after making the substitutions,  $a_i\rho_f \rightarrow q_iP$; $a_i^\dagger\rho_f \rightarrow (q_i^*-\frac{\partial}{\partial q_i})P$; $a_i^\dagger a_i\rho_f \rightarrow (q_i^*-\frac{\partial}{\partial q_i})q_i P$; $a_ia_i^\dagger\rho_f\rightarrow q_i(q_i^*-\frac{\partial}{\partial q_i})P$; $a_i^\dagger a_j^\dagger \rho_f\rightarrow (q_i^*-\frac{\partial}{\partial q_i})(q_j^*-\frac{\partial}{\partial q_j})P$; $a_j^\dagger\rho_f a_i\rightarrow (q_j^*-\frac{\partial}{\partial q_j})(q_i-\frac{\partial}{\partial q_i^*})P$.

\begin{equation}
    \begin{split}
        \frac{\partial P}{\partial t} = & \Big[ (\Delta_{c_1p}+\alpha_{11}+\alpha_{15})\frac{\partial}{\partial q_1}q_1 + (\Delta_{c_2p}+\nu_{22}+\nu_{26})\frac{\partial}{\partial q_2}q_2 +  H.C.\Big]P
        \\& + \Big[ (\alpha_{13}+\alpha_{17})\frac{\partial}{\partial q_1}q_1^* + (\nu_{24}+\nu_{28})\frac{\partial}{\partial q_2}q_2^* + H.C. \Big] P
        \\& + \Big[ (\alpha_{12}+\alpha_{16})\frac{\partial}{\partial q_1}q_2 + (\nu_{21}+\nu_{25})\frac{\partial}{\partial q_2}q_1 + H.C. \Big] P
        \\& + \Big[ (\alpha_{14}+\alpha_{18})\frac{\partial}{\partial q_1}q_2^* + (\nu_{23}+\nu_{27})\frac{\partial}{\partial q_2}q_1^* + H.C. \Big] P
        \\& - \Big[ \alpha_{17}\frac{\partial^2}{\partial q_1^2} + \nu_{28}\frac{\partial^2}{\partial q_2^2} + \alpha_{18}\frac{\partial^2}{\partial q_1\partial q_2} + \nu_{27}\frac{\partial^2}{\partial q_2\partial q_1} + H.C. \Big]P
        \\& - \Big[ \alpha_{11}\frac{\partial^2}{\partial q_1\partial q_1^*} + \nu_{22}\frac{\partial^2}{\partial q_2\partial q_2^*} + \alpha_{12}\frac{\partial^2}{\partial q_1\partial q_2^*} + \nu_{21}\frac{\partial^2}{\partial q_2\partial q_1^*} + H.C. \Big]P
    \end{split}
    \label{eqn:FokkerPlanckEq}
\end{equation}

Further, defining $q_i$ in polar coordinates, $q_i = r_i e^{i\phi_i}$, the average and relative phase are given by $\phi=\phi_1-\phi_2$ and $\Phi=\frac{\phi_1+\phi_2}{2}$ respectively. We also have,

\begin{equation}
    \frac{\partial}{\partial q_l} = \frac{e^{-i\phi_l}}{2}(\frac{\partial}{\partial r_l} + \frac{1}{ir_l}\frac{\partial}{\partial \phi_l})
\end{equation}

\begin{eqnarray}
    \frac{\partial}{\partial \phi_1} = \frac{1}{2}\frac{\partial}{\partial \Phi}+\frac{\partial}{\partial \phi}\\
    \frac{\partial}{\partial \phi_2} = \frac{1}{2}\frac{\partial}{\partial \Phi}-\frac{\partial}{\partial \phi}
\end{eqnarray}

Considering negligible variations in the mean photon number, we get

\begin{equation}
    \frac{\partial}{\partial q_l} = \frac{e^{-i\phi_l}}{2}(\frac{1}{ir_l}\frac{\partial}{\partial \phi_l})
\end{equation}

After rewriting the Eq. \ref{eqn:FokkerPlanckEq} in terms of average and relative phases ($\Phi$ and $\phi$), 

\begin{equation}
    \frac{\partial P}{\partial t} = \frac{\partial}{\partial \phi} D_\phi P + \frac{\partial}{\partial \Phi} D_\Phi P + \frac{\partial^2}{\partial \theta^2}D_{\phi\phi} P + \frac{\partial^2}{\partial \Phi^2}D_{\Phi\Phi}P + \frac{\partial^2}{\partial \phi \partial \Phi} D_{\phi\Phi}P
    \label{eqn:FPeqn2}
\end{equation}

The drift and the diffusion coefficients are given by, 

\begin{equation}
    \begin{split}
        D_\phi = & \Delta_{c_1p}-\Delta_{c_2p}\\&-\frac{i}{2}\Big[ (\alpha_{11}-\nu_{22}) + (\frac{\alpha_{12}r_2}{r_1}e^{-i\phi}-\frac{\nu_{21}r_1}{r_2}e^{i\phi})-(\frac{\alpha_{12}}{r_1r_2}e^{-i\phi}-\frac{\nu_{21}}{r_1r_2}e^{i\phi})\\&+(\alpha_{13}e^{-i(2\Phi+\phi)}-\nu_{24}e^{-i(2\Phi-\phi)})
        +(\alpha_{14}\frac{r_2}{r_1}e^{-i2\Phi}-\nu_{23}\frac{r_1}{r_2}e^{-i2\Phi})+(\alpha_{15}-\nu_{26})\\&+(\alpha_{16}\frac{r_2}{r_1}e^{-i\phi}-\nu_{25}\frac{r_1}{r_2}e^{i\phi})+(\alpha_{17}e^{-i(2\Phi+\phi)}-\nu_{28}e^{-i(2\Phi-\phi)})\\&-(\frac{3\alpha_{17}}{2r_1^2}e^{-i(2\Phi+\phi)}-\frac{3\nu_{28}}{2r_2^2}e^{-i(2\Phi-\phi)})+(\alpha_{18}\frac{r_2}{r_1}e^{-i2\Phi}-\nu_{27}\frac{r_1}{r_2}e^{-i2\Phi}) \Big] + H.C.
    \end{split}
\end{equation}

\begin{equation}
    \begin{split}
        D_\Phi = & \frac{\Delta_{c_1p}}{2}+\frac{\Delta_{c_2p}}{2}
        \\& -\frac{i}{4}\Big[ (\alpha_{11}+\nu_{22})+(\alpha_{12}\frac{r_2}{r_1}e^{-i\phi} + \nu_{21}\frac{r_1}{r_2}e^{i\phi}) + (\alpha_{13}e^{-i(2\Phi+\phi)}+\nu_{24}e^{-i(2\Phi-\phi)}) \\&+ (\alpha_{14}\frac{r_2}{r_1}e^{-i2\Phi}+\nu_{23}\frac{r_1}{r_2}e^{-i2\Phi})
         + (\alpha_{15}+\nu_{26})+(\alpha_{16}\frac{r_2}{r_1}e^{-i\phi}+\nu_{25}\frac{r_1}{r_2}e^{i\phi})\\&+(\alpha_{17}e^{-i(2\Phi+\phi)}+\nu_{28}e^{-i(2\Phi-\phi)})-(\frac{3\alpha_{17}}{2r_1^2}e^{-i(2\Phi+\phi)}+\frac{3\nu_{28}}{2r_2^2}e^{-i(2\Phi-\phi)})
        \\& + (\alpha_{18}\frac{r_2}{r_1}e^{-i2\Phi}+\nu_{27}\frac{r_1}{r_2}e^{-i2\Phi})-(\frac{\alpha_{18}}{r_1r_2}e^{-i2\Phi}+\frac{\nu_{27}}{r_1r_2}e^{-i2\Phi}) \Big] + H.C.
    \end{split}
\end{equation}

\begin{equation}
\begin{split}
    D_{\phi\phi} &= \frac{1}{4}\Big[ -(\frac{\alpha_{11}}{r_1^2}+\frac{\nu_{22}}{r_2^2})+(\frac{\alpha_{12}}{r_1r_2}e^{-i\phi}+\frac{\nu_{21}}{r_1r_2}e^{i\phi})+(\frac{\alpha_{17}}{r_1^2}e^{-i(2\Phi+\phi)}+\frac{\nu_{28}}{r_2^2}e^{-i(2\Phi-\phi)})\\&-(\frac{\alpha_{18}}{r_1r_2}e^{-i2\Phi}+\frac{\nu_{27}}{r_1r_2}e^{-i2\Phi})\Big]+H.C.
\end{split}
\end{equation}

\begin{equation}
\begin{split}
    D_{\Phi\Phi} &= \frac{1}{16}\Big[ -(\frac{\alpha_{11}}{r_1^2}+\frac{\nu_{22}}{r_2^2})-(\frac{\alpha_{12}}{r_1r_2}e^{-i\phi}+\frac{\nu_{21}}{r_1r_2}e^{i\phi})+(\frac{\alpha_{17}}{r_1^2}e^{-i(2\Phi+\phi)}+\frac{\nu_{28}}{r_2^2}e^{-i(2\Phi-\phi)})\\&+(\frac{\alpha_{18}}{r_1r_2}e^{-i2\Phi}+\frac{\nu_{27}}{r_1r_2}e^{-i2\Phi})\Big]+H.C.
\end{split}
\end{equation}

\section{COMPARISON BETWEEN ME AND SME}\label{sec:chap4_Appendix3}

\begin{figure}
    \centering
    \includegraphics[width=\columnwidth]{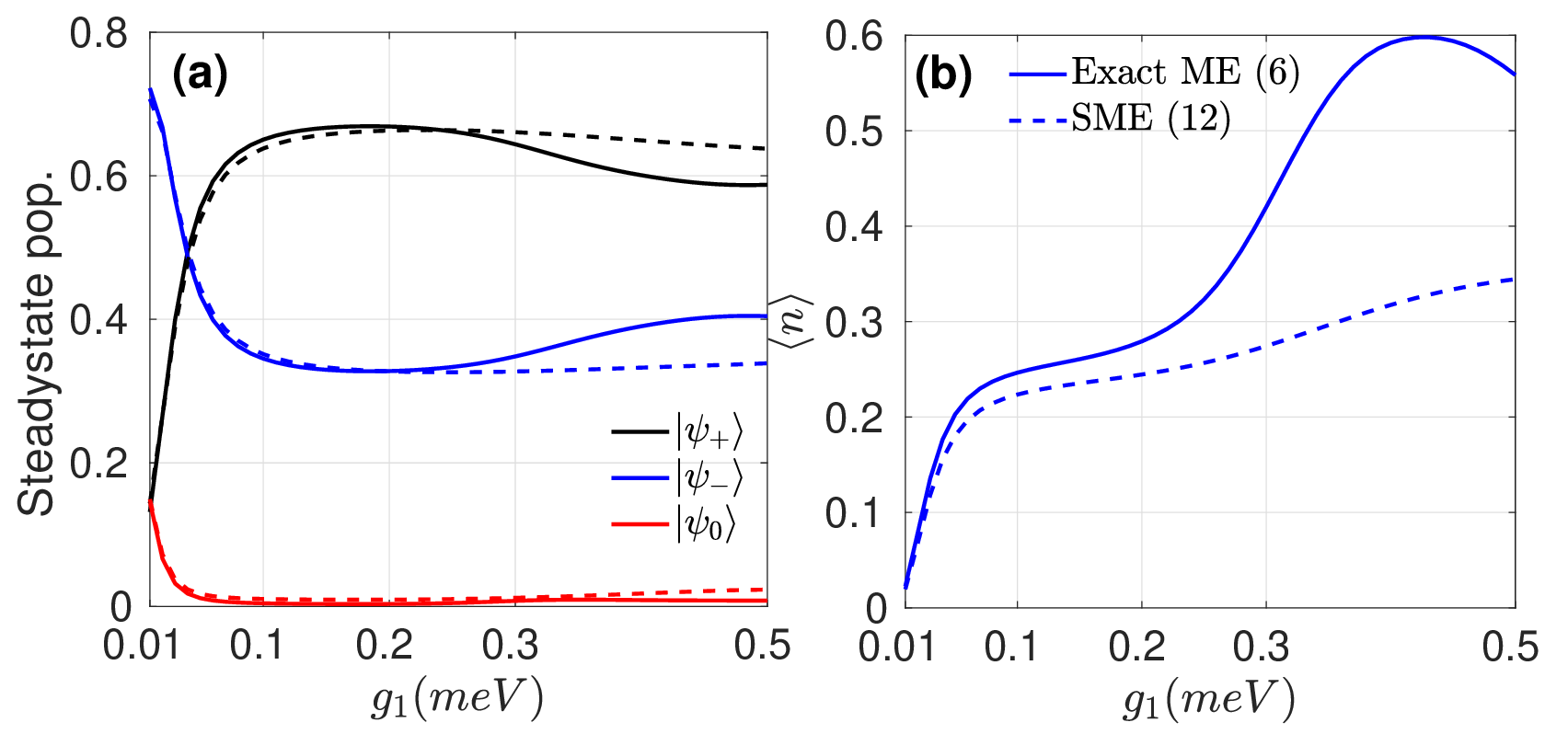}
    \caption{Subplot (a) shows the comparison of steadystate populations evaluated using ME, Eq. \ref{eqn:ME} and SME, Eq. \ref{eqn:SME} and the mean cavity photon number, $\langle n\rangle$ is given in subplot (b). Both ME and SME show similar behavior for the coupling strength, $g_1\leq 0.3meV$. The other parameters are same as in Fig. \ref{fig:chap4/Fig2}.}
    \label{fig:chap4/Fig14}
\end{figure}

\begin{figure}
    \centering
    \includegraphics[width=\columnwidth]{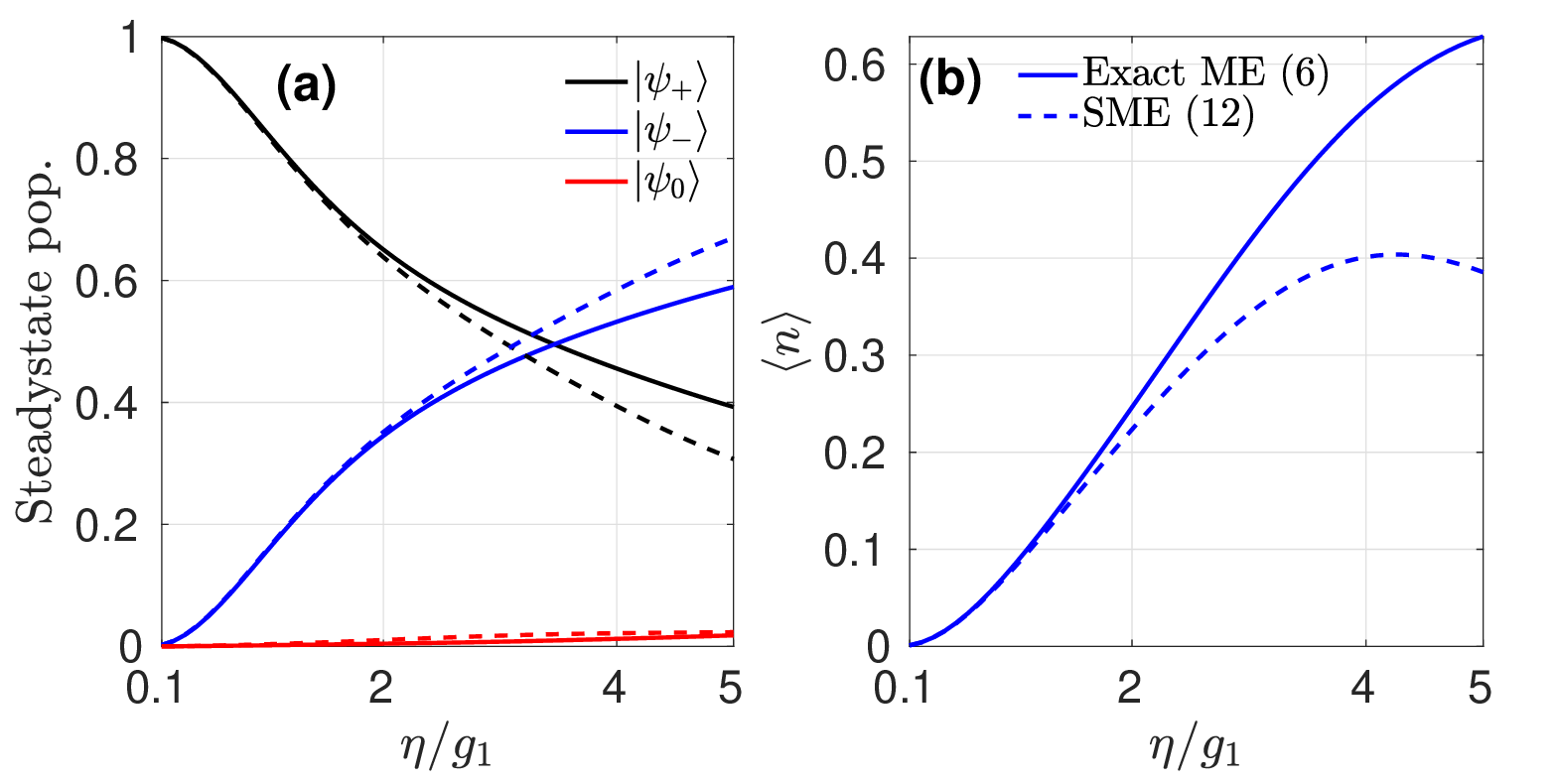}
    \caption{Comparison of steadystate populations in (a) and mean cavity photon number in (b) is shown. For coherent pumping strengths, $\eta\leq 3g_1$, SME mimics ME behavior. The other parameters are same as in Fig. \ref{fig:chap4/Fig2}.}
    \label{fig:chap4/Fig15}
\end{figure}

We compare of the results of steadystate populations and mean cavity photon number from the master equation, Eq. \ref{eqn:ME} and SME, Eq. \ref{eqn:SME}. Since all the parameters are normalized w.r.t cavity coupling strength, $g_1$, we study by varying $g_1$ in the Fig. \ref{fig:chap4/Fig14}. We can see that both SME and ME predict similar behavior for $g_1\leq300\mu eV$. Further, we also made the comparison with increasing coherent pumping strength, $\eta$ in Fig. \ref{fig:chap4/Fig15}. The results show that it is safe to work with SME for the pumping strengths, $\eta\leq 3.0g_1$. 

%% file: app_chap5.tex
\chapter{}\label{app:chap5_Appendix}

\section{Simplified master equation for incoherent pumping case}\label{sec:chap5_Appendix1}

The simplified master equation is given by,
\begin{align}
\dot{\rho_s} &=-\frac{i}{\hbar}[H_{eff},\rho_s]-\sum_{i=x,y}\left(\frac{\gamma_1}{2}{\cal L}[\sigma_{gi}]
+\frac{\gamma_2}{2}{\cal L}[\sigma_{iu}]\right)\rho_s\nonumber-\frac{\kappa_{1}}{2}{\cal L}[a_{1}]\rho_s -\frac{\kappa_{2}}{2}{\cal L}[a_{2}]\rho_s \\&-\sum_{i=x,y,u}\frac{\gamma_d}{2}{\cal L}[\sigma_{ii}]\rho_s\nonumber-\left(\frac{\eta_1}{2}{\cal L}[\sigma_{xg}] +\frac{\eta_2}{2}{\cal L}[\sigma_{ux}]\right)\rho_s -\frac{1}{2}( \Gamma_{2}^{+}{\cal L}[\sigma_{yu}a_{2}^{\dagger}]\nonumber+\Gamma_{2}^{-}{\cal L}[\sigma_{uy}a_{2}]\\& +\Gamma_{1}^{+}{\cal L}[\sigma_{gy}a_{1}^{\dagger}] + \Gamma_{1}^{-}{\cal L}[\sigma_{yg}a_{1}])-\frac{\Gamma_{ug}}{2}(\sigma_{ug}a_{1}a_{2}\rho_s -2\sigma_{yg}a_{1}\rho_s a_{2}\sigma_{uy} + \rho_s\sigma_{ug}a_{1}a_{2})\nonumber\\
&-\frac{\Gamma_{gu}}{2}(\sigma_{gu}a_{1}^{\dagger}a_{2}^{\dagger}\rho_s -2\sigma_{yu}a_{2}^{\dagger}\rho_s a_{1}^{\dagger}\sigma_{gy} + \rho_s\sigma_{gu}a_{1}^{\dagger}a_{2}^{\dagger})
\label{inco_simple_ms}
\end{align}
where the effective Hamiltonian is given by,
\begin{align}
H_{eff} &= H_{s} + \hbar (\delta_{2}^{+}\sigma_{uu}a_{2}a_{2}^{\dagger} +\delta_{2}^{-}\sigma_{yy}a_{2}^{\dagger}a_{2} +\delta_{1}^{+}\sigma_{yy}a_{1}a_{1}^{\dagger}\nonumber\\ 
&+\delta_{1}^{-}\sigma_{gg}a_{1}^{\dagger}a_{1}) +\hbar\Omega_{12}(\sigma_{ug}a_{1}a_{2} + \sigma_{gu}a_{1}^{\dagger}a_{2}^{\dagger}),
\end{align}

The stark shits, $\delta_i^\pm$, phonon induced cavity mode feeding, $\Gamma_i^+$, phonon induced incoherent pumping, $\Gamma_i^-$, phonon mediated two-mode two-photon coherent process, $\Omega_{12}$ and incoherent processes, $\Gamma_{ug},\, \Gamma_{gu}$ rates are given below,

\begin{align}
\delta_{i}^{\pm} &= g_{i}^{2}\langle B\rangle^{2}Im\left[ \int_{0}^{\infty} d\tau (e^{\phi(\tau)}-1)e^{\pm i\Delta_{i}\tau}\right];\nonumber\\
\Gamma_{i}^{\pm} &= 2g_{i}^{2}\langle B\rangle^{2}Re\left[ \int_{0}^{\infty} d\tau (e^{\phi(\tau)}-1)e^{\pm i\Delta_{i}\tau}\right];\nonumber\\
\Omega_{12} &= -\frac{i}{2}g_{1}g_{2}\langle B\rangle^{2}\left[\beta_{1} - \beta_{2}^{*}\right]\nonumber; 
\quad
\Gamma_{ug} = g_{1}g_{2}\langle B\rangle^{2}\left[\beta_{1} + \beta_{2}^{*}\right]\nonumber; \quad
\Gamma_{gu} = g_{1}g_{2}\langle B\rangle^{2}\left[\beta_{1}^{*} + \beta_{2}\right]\nonumber\\
\beta_{1} &= \int_{0}^{\infty} d\tau (e^{-\phi(\tau)}-1)e^{- i\Delta_{1}\tau}\nonumber; \qquad
\beta_{2} = \int_{0}^{\infty} d\tau (e^{-\phi(\tau)}-1)e^{i\Delta_{2}\tau}\nonumber.
\end{align}

\section{Simplified master equation for coherent pumping case}\label{sec:chap5_Appendix2}

\begin{align}
\dot{\rho_s} &=-\frac{i}{\hbar}[H_{eff},\rho_s]-\sum_{i=x,y}\left(\frac{\gamma_1}{2}{\cal L}[\sigma_{gi}]
+\frac{\gamma_2}{2}{\cal L}[\sigma_{iu}]\right)\rho_s\nonumber\\
&-\frac{\kappa_{1}}{2}{\cal L}[a_{1}]\rho_s -\frac{\kappa_{2}}{2}{\cal L}[a_{2}]\rho_s -\sum_{i=x,y,u}\frac{\gamma_d}{2}{\cal L}[\sigma_{ii}]\rho_s\nonumber\\
&-\left(\frac{\eta_1}{2}{\cal L}[\sigma_{xg}] +\frac{\eta_2}{2}{\cal L}[\sigma_{ux}]\right)\rho_s -\frac{1}{2}( \Gamma_{2}^{+}{\cal L}[\sigma_{yu}a_{2}^{\dagger}]\nonumber\\
&+\Gamma_{2}^{-}{\cal L}[\sigma_{uy}a_{2}] +\Gamma_{1}^{+}{\cal L}[\sigma_{gy}a_{1}^{\dagger}] + \Gamma_{1}^{-}{\cal L}[\sigma_{yg}a_{1}])\\
&-\frac{\Gamma_{ug}}{2}(\sigma_{ug}a_{1}a_{2}\rho_s -2\sigma_{yg}a_{1}\rho_s a_{2}\sigma_{uy} + \rho_s\sigma_{ug}a_{1}a_{2})\nonumber\\
&-\frac{\Gamma_{gu}}{2}(\sigma_{gu}a_{1}^{\dagger}a_{2}^{\dagger}\rho_s -2\sigma_{yu}a_{2}^{\dagger}\rho_s a_{1}^{\dagger}\sigma_{gy} + \rho_s\sigma_{gu}a_{1}^{\dagger}a_{2}^{\dagger})\nonumber\\
&-\frac{1}{2}( \Gamma_{2}^{p+}{\cal L}[\sigma_{xu}]+\Gamma_{2}^{p-}{\cal L}[\sigma_{ux}] +\Gamma_{1}^{p+}{\cal L}[\sigma_{gx}] + \Gamma_{1}^{p-}{\cal L}[\sigma_{xg}])\nonumber\\
&-\frac{\Gamma_{ug}^{p}}{2}(\sigma_{ug}\rho_s -2\sigma_{xg}\rho_s\sigma_{ux} + \rho_s\sigma_{ug})\nonumber\\
&-\frac{\Gamma_{gu}^{p}}{2}(\sigma_{gu}\rho_s -2\sigma_{xu}\rho_s \sigma_{gx} + \rho_s\sigma_{gu})\nonumber
\label{coh_simple_ms}
\end{align}
where the effective Hamiltonian with coherent pumping
\begin{align}
H_{eff} &= H_{s} + \hbar (\delta_{2}^{p+}\sigma_{uu} + (\delta_{2}^{p-} + \delta_{1}^{p+})\sigma_{xx} + \delta_{1}^{p-}\sigma_{gg}\nonumber\\
&+ \delta_{2}^{+}\sigma_{uu}a_{2}a_{2}^{\dagger} +\delta_{2}^{-}\sigma_{yy}a_{2}^{\dagger}a_{2} +\delta_{1}^{+}\sigma_{yy}a_{1}a_{1}^{\dagger}\nonumber\\ 
&+\delta_{1}^{-}\sigma_{gg}a_{1}^{\dagger}a_{1}) +\hbar\Omega_{12}(\sigma_{ug}a_{1}a_{2} + \sigma_{gu}a_{1}^{\dagger}a_{2}^{\dagger})\nonumber\\
&+\hbar\Omega^{p}(\sigma_{ug} + \sigma_{gu}),
\end{align}
and new symbols are defined as
\begin{align}
\delta_{2}^{p\pm} &= \Omega_{2}^{2}\langle B\rangle^{2}Im\left[ \int_{0}^{\infty} d\tau (e^{\phi(\tau)}-1)e^{\pm i\Delta_{p}^{'}\tau}\right]\nonumber\\
\delta_{1}^{p\pm} &= \Omega_{1}^{2}\langle B\rangle^{2}Im\left[ \int_{0}^{\infty} d\tau (e^{\phi(\tau)}-1)e^{\pm i\Delta_{p}\tau}\right]\nonumber\\
\Gamma_{2}^{p\pm} &= 2\Omega_{2}^{2}\langle B\rangle^{2}Re\left[ \int_{0}^{\infty} d\tau (e^{\phi(\tau)}-1)e^{\pm i\Delta_{p}^{'}\tau}\right]\nonumber\\
\Gamma_{1}^{p\pm} &= 2\Omega_{1}^{2}\langle B\rangle^{2}Re\left[ \int_{0}^{\infty} d\tau (e^{\phi(\tau)}-1)e^{\pm i\Delta_{p}\tau}\right]\nonumber\\
\Omega^{p} &= -\frac{i}{2}\omega_{1}\omega_{2}\langle B\rangle^{2}\left[\alpha_{1} - \alpha_{2}^{*}\right]\nonumber\\
\Gamma_{ug}^{p} &= \Omega_{1}\Omega_{2}\langle B\rangle^{2}\left[\alpha_{1} + \alpha_{2}^{*}\right]\\
\Gamma_{gu}^{p} &= \Omega_{1}\Omega_{2}\langle B\rangle^{2}\left[\alpha_{1}^{*} + \alpha_{2}\right]\nonumber\\
\alpha_{1} &= \int_{0}^{\infty} d\tau (e^{-\phi(\tau)}-1)e^{- i\Delta_{p}\tau}\nonumber\\
\alpha_{2} &= \int_{0}^{\infty} d\tau (e^{-\phi(\tau)}-1)e^{i\Delta_{p}^{'}\tau}, \Delta_{p}^{'} = \omega_{u} - \omega_{x} - \omega_{p}.\nonumber
\end{align}